\pdfoutput=1

\documentclass[11pt,twoside,a4paper,cmspaper,final,collab]{cms-tdr}
\def\svnVersion{230905}\def\cmsCernNo{2013-037}\def\cmsCernDate{\today}\def\cmsMessage{Published in the Journal of Instrumentation as \href{http://dx.doi.org/10.1088/1748-0221/8/11/P11002}{\doi{10.1088/1748-0221/8/11/P11002.}}}

\begin{document}\cmsNoteHeader{MUO-11-001}

\hyphenation{had-ron-i-za-tion}
\hyphenation{cal-or-i-me-ter}
\hyphenation{de-vices}
\RCS$Revision: 193090 $
\RCS$HeadURL: svn+ssh://alverson@svn.cern.ch/reps/tdr2/papers/MUO-11-001/trunk/MUO-11-001.tex $
\RCS$Id: MUO-11-001.tex 193090 2013-06-28 17:21:58Z alverson $
\hyphenation{tech-nol-o-gies}
\cmsNoteHeader{MUO-11-001}
\title{The performance of the CMS muon detector in proton--proton collisions at $\sqrt{s} = 7$\TeV at the LHC}

\date{\today}

\abstract{
The performance of all subsystems of the CMS muon detector has been studied by using a sample of proton--proton collision data at $\sqrt{s} = 7$\TeV collected at the LHC in 2010 that corresponds to an integrated luminosity of approximately 40\pbinv. The measured distributions of the major operational parameters of the drift tube (DT), cathode strip chamber (CSC), and resistive plate chamber (RPC) systems met the design specifications. The spatial resolution per chamber was 80--120\micron in the DTs, 40--150\micron in the CSCs, and 0.8--1.2\unit{cm} in the RPCs. The time resolution achievable was 3\unit{ns} or better per chamber for all 3 systems. The efficiency for reconstructing hits and track segments originating from muons traversing the muon chambers was in the range 95--98\%. The CSC and DT systems provided muon track segments for the CMS trigger with over 96\% efficiency, and identified the correct triggering bunch crossing in over 99.5\% of such events. The measured performance is well reproduced by Monte Carlo simulation of the muon system down to the level of individual channel response. The results confirm the high efficiency of the muon system, the robustness of the design against hardware failures, and its effectiveness in the discrimination of backgrounds.}

\hypersetup{%
pdfauthor={CMS Collaboration},%
pdftitle={The performance of the CMS muon detector in proton-proton collisions at sqrt(s) = 7 TeV at the LHC},%
pdfsubject={CMS muon detector},%
pdfkeywords={CMS, detector performance, muon system, DT, CSC, RPC, DQM, software, calibration, resolution, efficiency, trigger, timing, synchronization, simulation, alignment, backgrounds}}

\maketitle 

\section{Introduction}

Muon detection is a powerful tool for recognizing signatures of interesting physics processes over the high background rates at the Large Hadron Collider (LHC)~\cite{Evans:2008zzb}.
The muon detector system in the Compact Muon Solenoid (CMS) experiment~\cite{CMSdet} has 3 primary functions: muon triggering, identification, and momentum measurement. Good muon momentum resolution
is provided by the high spatial resolution of the detector and the high magnetic field of the superconducting solenoidal magnet and its flux-return yoke.

The CMS muon system (Figs.~\ref{profile}, \ref{DT_Barrel}, and~\ref{ME1}) is designed to measure the momentum and charge of muons over a large kinematic range in LHC collisions.
Table~\ref{tab:reqs} lists selected design requirements as specified in the Muon Technical Design Report (TDR)~\cite{MUON-TDR} in 1997.
These requirements were determined by assuming a 4-T magnetic field. Later it was decided to operate
the magnet at 3.8\unit{T}, leading to a $\approx$5\% degradation in the achievable momentum resolution.

\begin{table}[!Hhtb]
   \begin{center}
         \topcaption{Requirements for the CMS muon system as originally specified (adapted from Ref.~\cite{MUON-TDR}). ``BX identification" is explained in the text.} \label{tab:reqs}
         \begin{tabular} {|l|l|l|}
         \hline
 {Muon}           &  BX  & Trigger on single- and multi-muon events with            \\
 {trigger}        & identification    & well defined thresholds from a few to 100\GeVc       \\ \hline
 {Momentum}     &  Muon system        & 8--15\% at 10\GeVc, 20--40\% at 1\TeVc  \\ \cline{2-3}
 {resolution }     &  Tracker \& muon system  & 1--1.5\% at 10\GeVc, 6--17\% at 1\TeVc                \\
                        &                    & Track matching between tracker and muon   \\
                        &                    & system at 1\TeVc better than 1\unit{mm} in the bending     \\
                        &                    & plane, better than 10\unit{mm} in the non-bending plane \\ \hline
{Charge}      &                    & Less than 0.1\% at muon  \pt = 100\GeVc \\
{misassignment}   &          &                  \\\hline
          \end{tabular}
   \end{center}
\end{table}

The CMS detector uses 3 types of gas-ionization particle detectors for muon identification.
To be compatible with the geometry of the central solenoidal magnet, it is natural to have a cylindrical barrel region and planar endcaps.
Because of the large area to be covered the muon detectors must be relatively inexpensive, and because the CMS detector is inaccessible when the LHC is running, it is important that the muon systems be
robust.
This technology also allows fast access to hit information from the entire muon detector for use in a hardware muon trigger that is relatively  immune to the high particle densities originating from proton--proton collisions at the center of the detector.

Previous publications~\cite{Chatrchyan:2009ih, Chatrchyan:2009hg, Chatrchyan:2009ig, :2009gz, Chatrchyan:2009si, CMSNOTE:2008003, CSCPerfCRAFT, CRAFT08, dt_fine_synch, CRAFT08_track_alignment_paper, CRAFT08_hw_alignment_paper} have described the performance of the muon system during the long cosmic-ray muon runs of 2008 and 2009.
Here we describe the performance of the CMS muon detectors using a data sample
accumulated at $\sqrt{s} = 7$\TeV during the 2010 LHC proton--proton physics run corresponding to an integrated luminosity of approximately 40\pbinv.
The same data have already been used to study the performance of CMS muon reconstruction~\cite{POG-paper}, whereas the results presented in this paper focus on the performance at the individual detector level.
Following a general description of the muon system and its operation, Section~\ref{section-calibration} describes the calibration of the individual muon subsystems.
This is followed by sections on CSC and DT local triggering, and the measurement of the position and time resolution of all of the muon subsystems.
Local reconstruction efficiency is discussed in Section~\ref{Efficiency}, and the radiation background and alignment issues are explained in Sections~\ref{background} and \ref{sec:Alignment}, respectively.
The important role of data quality monitoring (DQM) is considered in Section~\ref{sec:DQM}.
A summary of the failure rates of muon system electronic components is presented in Appendix~\ref{electronics}.
Finally, the offline simulation of the muon detector is described in Appendix~\ref{simulation}.

\section{Overview of the muon system}

\begin{figure}[htbp]
{\centering
\includegraphics[width=16cm]{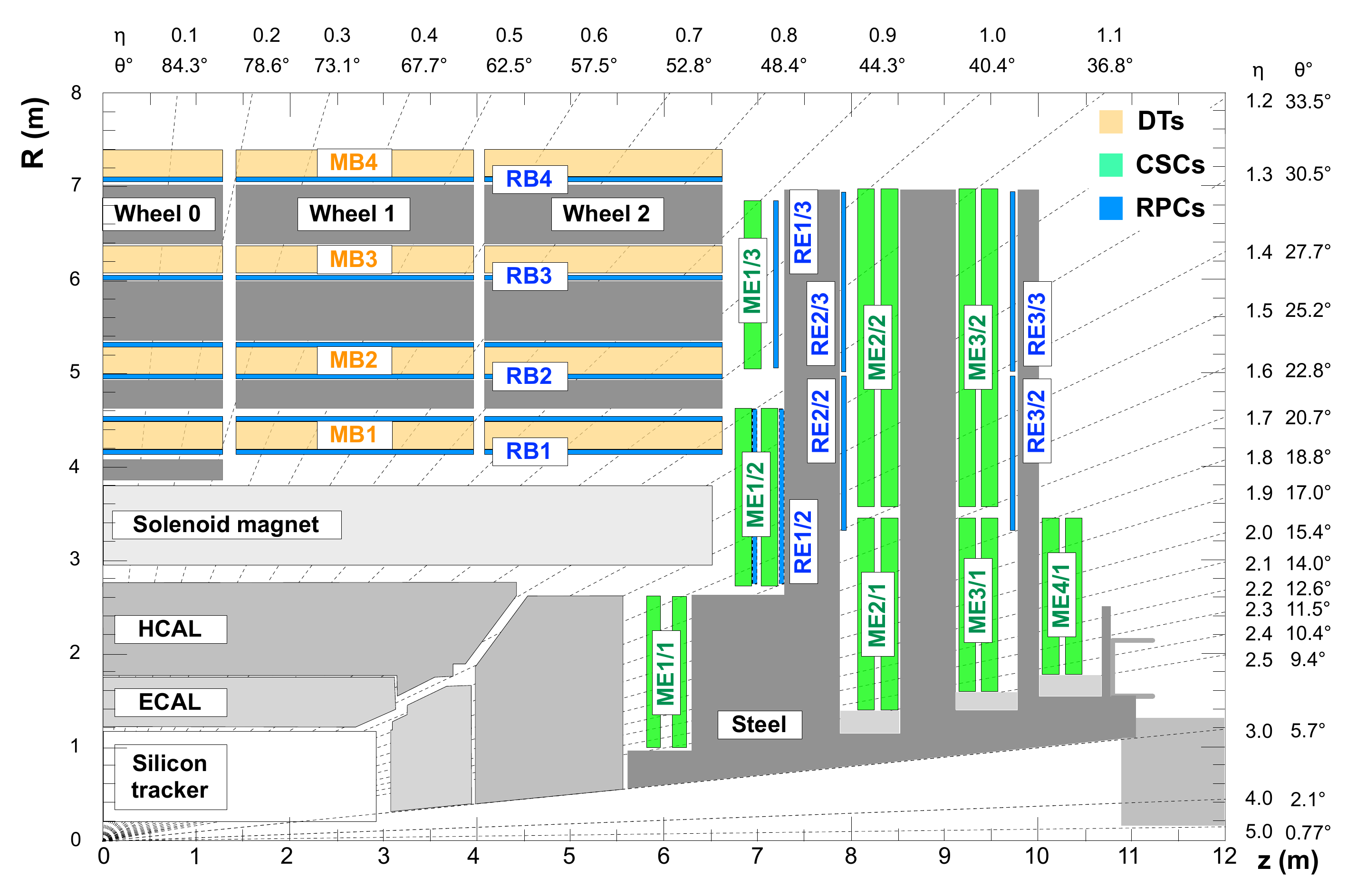}
\caption{\label{profile}
An $R$--$z$ cross section of a quadrant of the CMS detector with the axis parallel to the beam ($z$) running horizontally and radius ($R$) increasing upward. The interaction point is at the lower left corner. Shown are the locations of the various muon stations and the steel disks (dark grey areas). The 4 drift tube (DT, in light orange) stations are labeled MB (``muon barrel") and the cathode strip chambers (CSC, in green) are labeled ME (``muon endcap"). Resistive plate chambers (RPC, in blue) are in both the barrel and the endcaps of CMS, where they are labeled RB and RE, respectively.}
}
\end{figure}

\begin{figure}[th]
{\centering
\includegraphics[width=15.5cm]{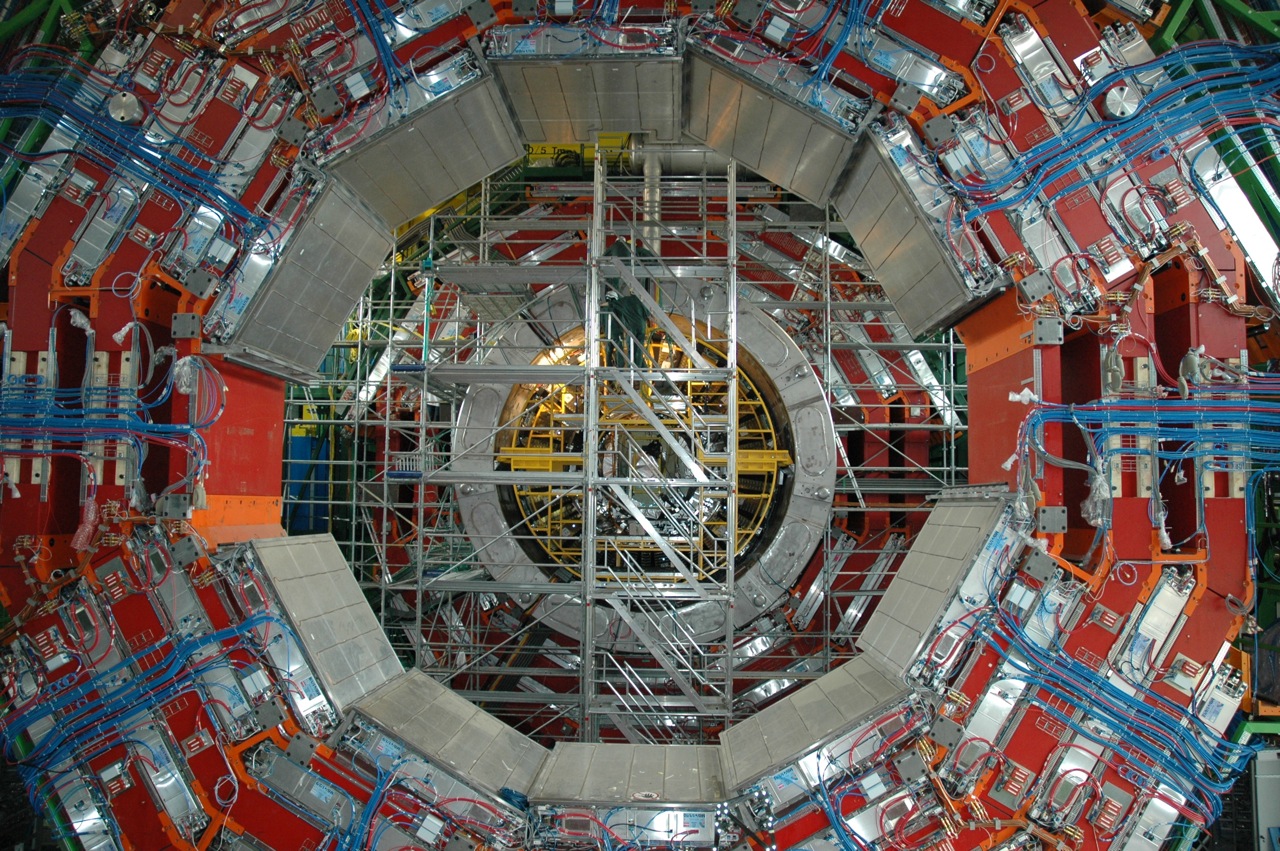}
\caption{\label{DT_Barrel}
Photograph of a barrel wheel during the construction of CMS in June 2006.
The 4 stations of DT chambers are separated by layers of the yoke steel (painted red). Several chambers had not yet been installed.}
}
\end{figure}

\begin{figure}[th]
{\centering
\includegraphics[width=15.5cm]{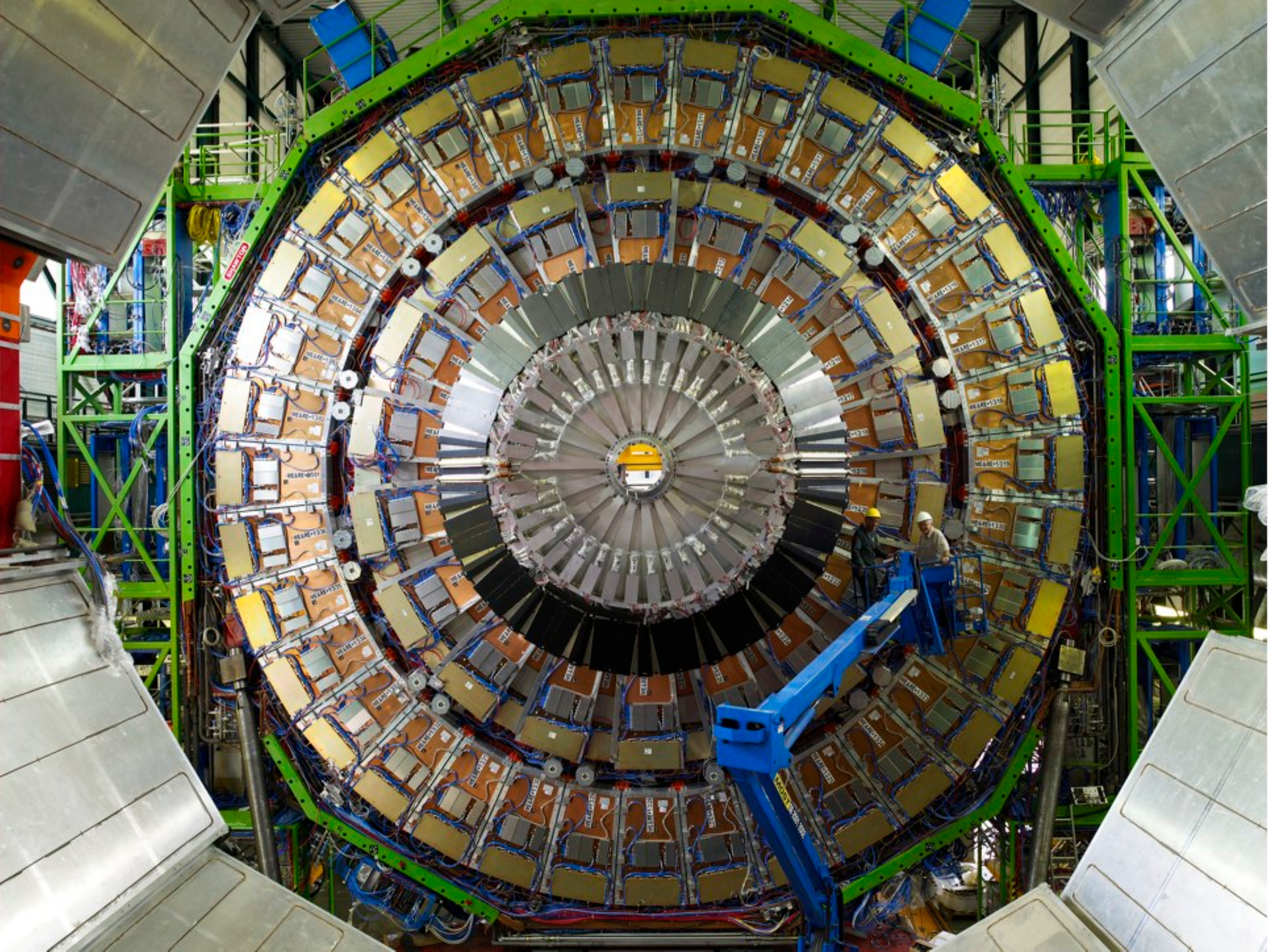}
\caption{\label{ME1}
Photograph of the ME1 muon station of the ``plus'' ($z >0$) endcap during the construction of CMS.
Visible in concentric rings are the CSC types ME1/2 and ME1/3. The ME1/1 chambers are hidden behind the endcap calorimeters closest to the center. The endcap RPCs are in the layer behind the CSCs. }
}
\end{figure}

The basic detector process utilized in the CMS muon systems is gas ionization.
For all the different technologies---drift tubes, cathode strip proportional planes, and resistive plates---the basic physical modules are called ``chambers".
The chambers are independently-operating units, which are assembled into the overall muon detector system of CMS.
The chambers form part of a spectrometer in which the analyzing magnet is the central solenoid together with the flux return yoke of CMS.
To match the cylindrical geometry of the solenoid, the barrel region is instrumented with drift tube chambers, and the 2 endcap regions with cathode strip chambers.
Resistive plate chambers are interspersed in both the barrel and endcap regions.
The muon chambers must detect the traversing track at several points along the track path to utilize the magnet to measure the deflection of muons as they pass through its field.
In the barrel region, this requires chambers to be positioned at several different values of the radial distance $R$  from the beam line, and in the endcap region at several different values of distance along the beam direction $z$.
A ``station" is an assembly
of chambers around a fixed value of $R$ (in the barrel) or $z$ (in the endcap).
There are 4 stations in the barrel and in each endcap (Fig.~\ref{profile}), labeled MB1--MB4 and ME1--ME4, respectively.
Along $z$, the drift tubes and resistive plate chambers in the barrel are divided into 5 ``wheels'', with Wheel 0 centered at $z = 0$ and wheels W+1 and W+2 in the $+z$ direction and W-1 and W-2 in the $-z$ direction.
Similarly in the $R$ direction in the endcaps, there are ``rings'' of endcap resistive plate chambers and cathode strip chambers. The latter are labeled ME1/n--ME4/n, where integer n increases with the radial distance from the beam line.

\begin{figure}[th]
{\centering
\includegraphics[width=15.3cm]{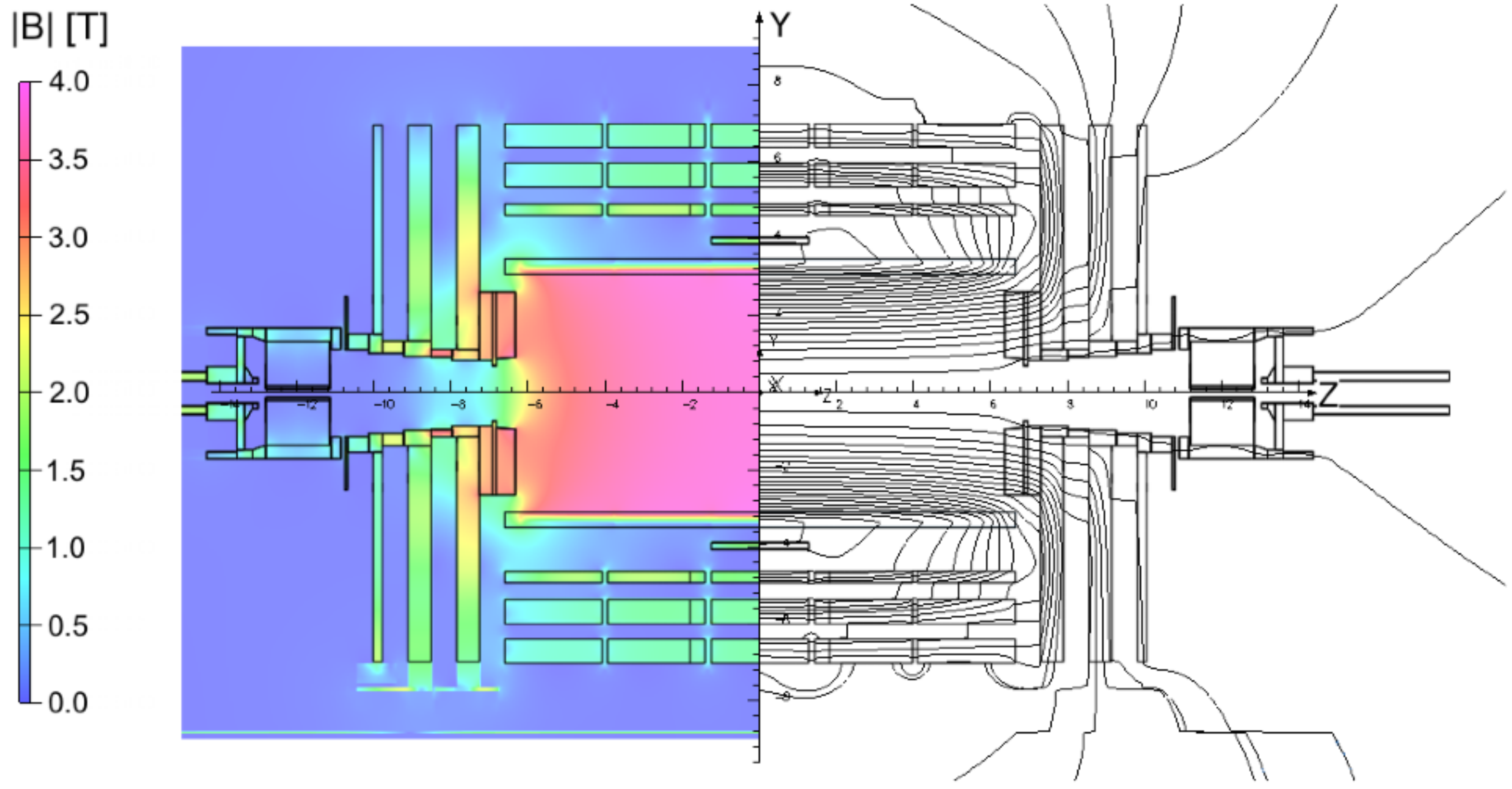}
\caption{\label{MagField}
Map of the $|B|$ field (left) and field lines (right) predicted for a longitudinal section of the CMS detector by a magnetic field model at a central magnetic flux density of 3.8\unit{T}. Each field line represents a magnetic flux increment of 6\unit{Wb}.}
}
\end{figure}

\subsection{Drift tube and cathode strip chamber systems}

In the barrel region, the muon rate is low, the neutron background is relatively small (except in the outermost station MB4), and the magnetic field is mostly uniform with strength below 0.4\unit{T} in between the yoke segments (Fig.~\ref{MagField}).
Here, drift chambers with standard rectangular cells and sophisticated electrical field shaping are employed.
The barrel drift tube (DT) chambers cover the pseudorapidity region $| \Pgh |  < 1.2$, where $\eta=-\ln[\tan(\theta/2)]$ and $\theta$ is the polar angle with respect to the counterclockwise beam direction.
They are organized into 12 $\phi$-segments per wheel, where $\phi$ is the azimuthal angle, forming 4 stations at different radii interspersed between plates of the  magnet flux return yoke.
Each station consists of 8 layers of tubes measuring the position in the bending plane and 4 layers in the longitudinal plane (except MB4).

\begin{figure}[htbp]
 \begin{center}
  \begin{tabular}{c}
   \resizebox{6.5cm}{!}{\includegraphics{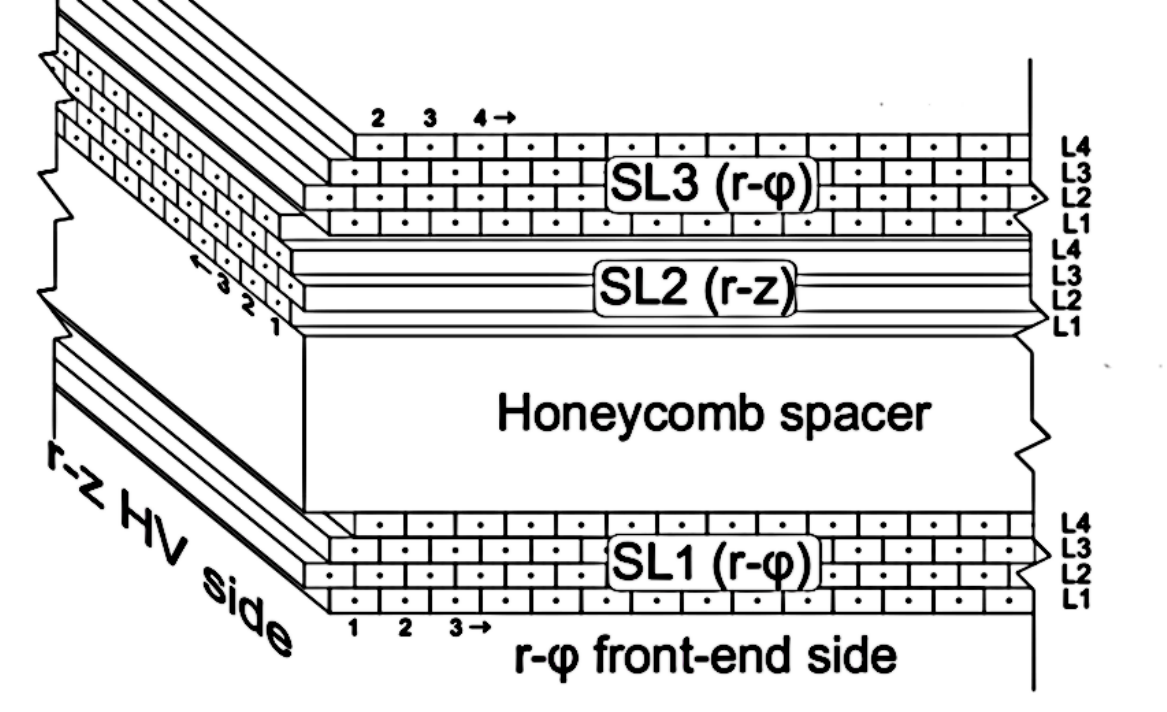}}
   \resizebox{8cm}{!}{\includegraphics{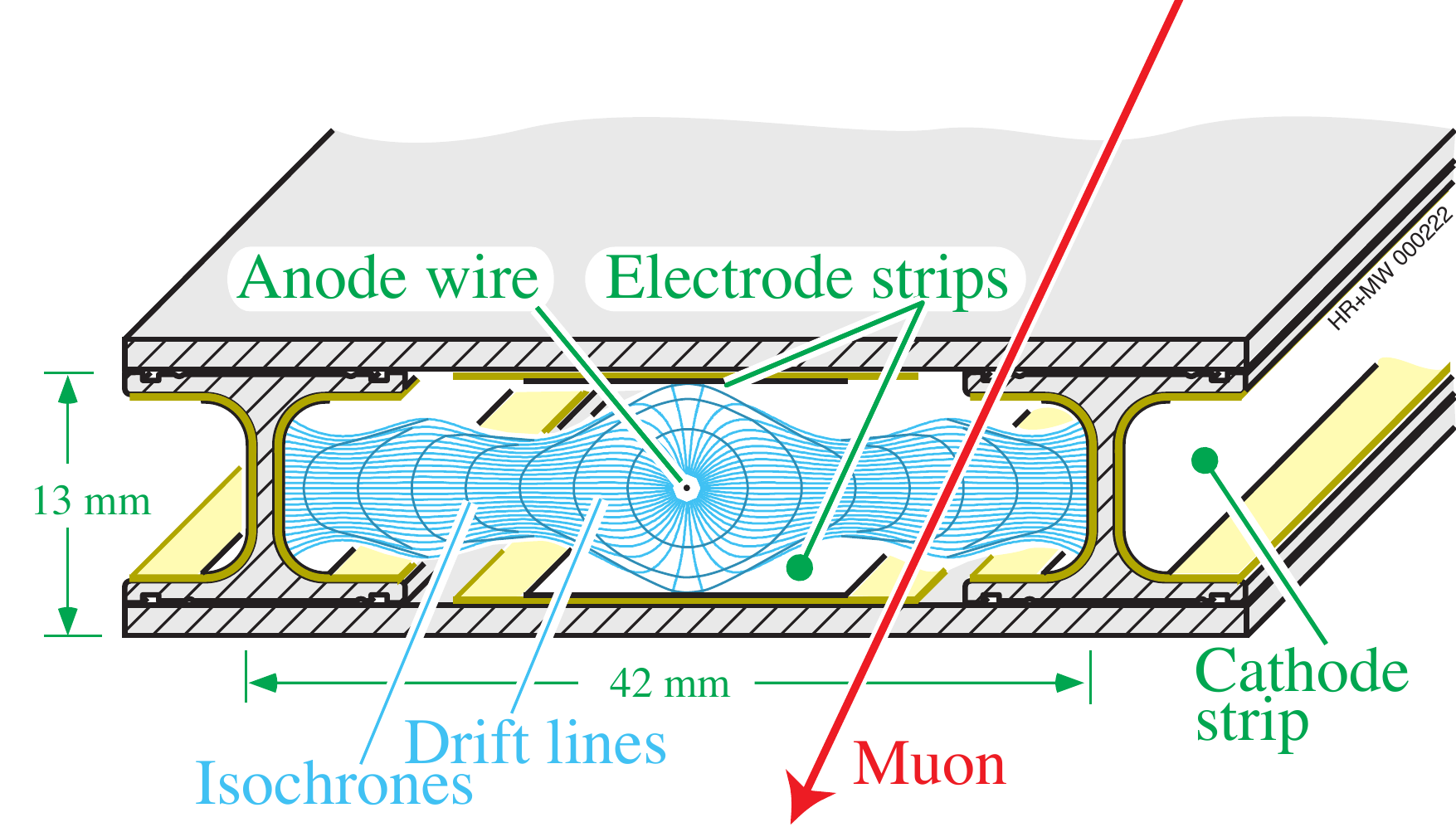}}
 \end{tabular}
 \caption{Left: Schematic view of a DT chamber. Right: Section of a
   drift tube cell showing the drift lines and isochrones.
  }
  \label{fig:dt-chamber-cell}
 \end{center}
\end{figure}

The basic element of the DT system is the drift cell (Fig.~\ref{fig:dt-chamber-cell}, right).
The cell has a transverse size of $42\times 13\unit{mm}^2$ with a 50-$\mu$m-diameter gold-plated stainless-steel
anode wire at the center. The wire operates at a voltage of $+$3600\unit{V}.
The gas mixture (85\%/15\% of Ar/CO$_2$) provides good quenching properties and a saturated
drift velocity of about 55\micron/ns.
The maximum drift time is almost 400\unit{ns}.
The cell design makes use of 4 electrodes (including 2 cathode strips) to shape the effective drift field: 2 on the side walls of the tube, and 2 above and below the wires on the ground planes between the layers.
They operate at $-$1800 and $+$1800\unit{V}, respectively.
Four staggered layers of parallel cells form a superlayer (SL). A chamber consists of 2 SLs
that measure the $r$-$\phi$ coordinates with wires parallel to the beam line, and an orthogonal
SL that measures the $r$-$z$ coordinate, except for MB4, which has only an $r$-$\phi$ SL (Fig.~\ref{fig:dt-chamber-cell}, left).
Here $r$ is the nominal distance from the beam collision point.
The chambers are limited in size in the longitudinal dimension by the segmentation of the barrel yoke, and are about 2.5\,m long.
On the transverse side, their length varies with the station, ranging from 1.9\,m for MB1 to 4.1\,m for MB4.

\begin{figure}
{\centering
\includegraphics{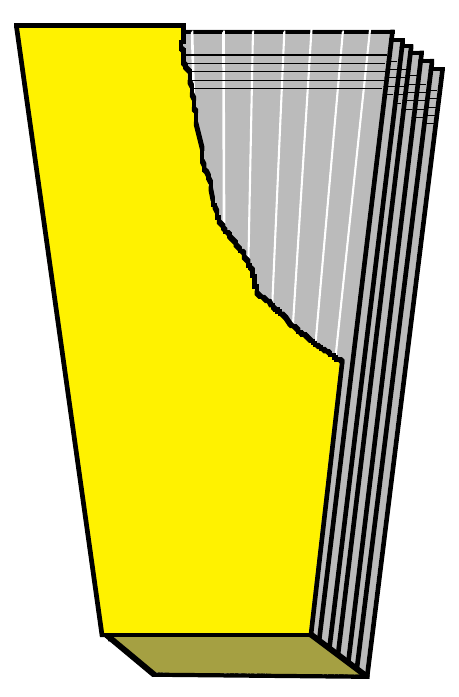}
\includegraphics{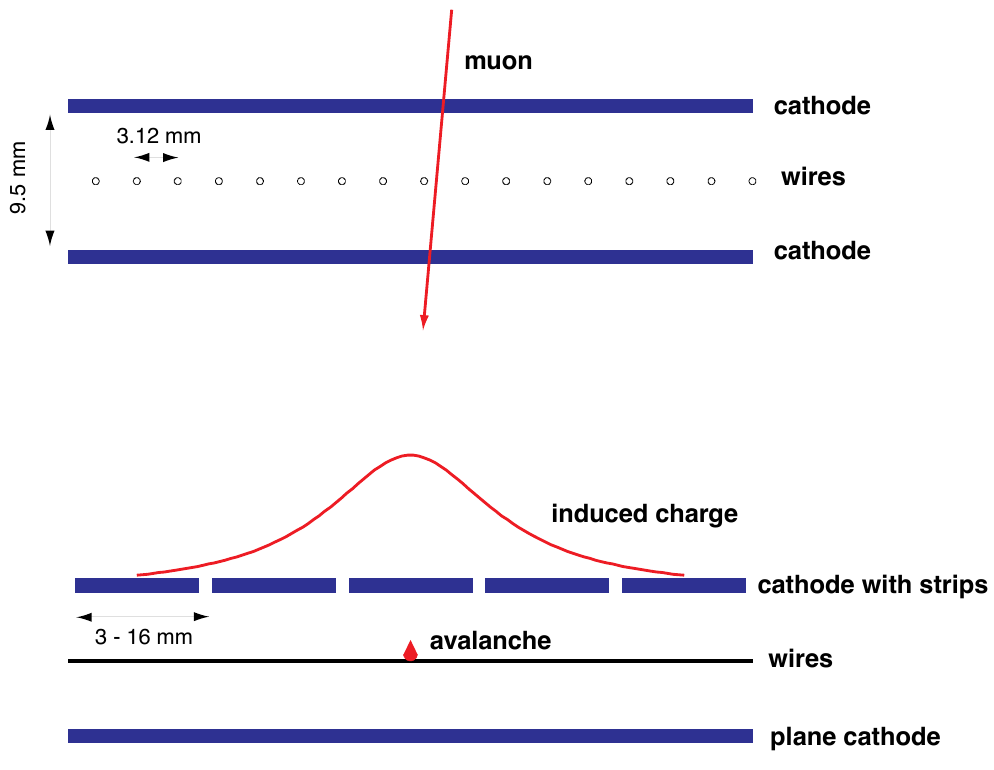}
\caption{\label{csc}
Left: Cut-away diagram of a CSC showing the 6 layers and the orientations of the wires and strips (not all shown).
Right: Cross-sectional views of the gas gap in a CSC showing the anode wires and cathode planes, and a schematic illustration of the gas ionization avalanche and induced charge distribution on the cathode strips.}
}
\end{figure}

In the endcap regions of CMS the muon rates and background levels are higher, and the magnetic field is strong and non-uniform (Fig.~\ref{MagField}).
Here, cathode strip chambers (CSC) are installed since they have fast response time (resulting from a short drift path), they can be finely segmented, and they can tolerate the non-uniformity of the magnetic field.
The CSCs cover the  $| \Pgh |$ region from 0.9 to 2.4.
Each endcap has 4 stations of chambers mounted on the faces of the endcap steel disks perpendicular to the beam.
 A CSC consists of 6 layers, each of which measures the muon position in 2 coordinates. The cathode
 strips run radially outward and provide a precision measurement in the $r$-$\phi$ bending plane (Fig.~\ref{csc}, left).
The wires provide a coarse measurement in the radial direction.

The CSCs operate as standard multi-wire proportional counters (MWPC), but with a cathode strip readout that precisely measures the position at which a muon or other charged particle crosses the gas volume (Fig.~\ref{csc}, right)~\cite{Charpak}.
CSCs of various physical dimensions are used in the system, ranging in length from about 1.7 to 3.4\,m in the radial dimension.
In the inner rings of stations 2, 3, and 4, each CSC subtends a $\phi$ angle of about 20$^{\circ}$; all other CSCs subtend an angle of about 10$^{\circ}$.
Each layer of a CSC contains 80 cathode strips, each of which subtends a constant $\phi$ angle between 2.2 and 4.7\unit{mrad} and projects to the beamline.
The anode wires have a diameter of 50\micron  and are spaced by 3.16 or 3.12\unit{mm} in all chambers except ME1/1 where they have 30\micron diameter  and are  2.5\unit{mm} apart.
They are ganged in groups of 5 to 16 wires, with widths from 16 to 51\unit{mm}, which limits the position resolution in the wire coordinate direction.
All chambers use a gas mixture of 50\% CO$_{2}$, 40\% Ar, and 10\% CF$_{4}$.
The ME1/1 chambers are operated at an anode voltage of 2.9\unit{kV} and all others at 3.6\unit{kV}.
Alternate layers of all CSCs except those in ME1/1 are shifted by half a strip width, and neighboring CSCs within all rings except ME1/3 overlap each other by 5 strip widths to avoid gaps between chambers.

The ME1/1 CSCs in the innermost ring of station 1 have a structure different from those of the other rings.
The chambers have narrower strips, which are divided
into 2 regions at $| \Pgh |$ = 2.1 so that the region closest to the beam line can trigger and be read out independently of the outer region.
The innermost region is labelled ``ME1/1a'' and the outer ``ME1/1b''. The 48 strips in each ME1/1a region of an ME1/1 chamber are ganged in groups of 3, in steps of 16, to give 16 readout channels, to satisfy space and cost constraints for the on-chamber electronics.
This ganging leads to ambiguities in reconstruction and triggering, and will be removed in a future upgrade of the detector.

The B field in the CSC chamber volumes does not exceed 0.5\unit{T} except in ME1. In ME1/1 the field is almost purely axial;
in ME1/2  there is both an axial component of about 1\unit{T} decreasing to 0.5\unit{T} with increasing distance from the magnet axis and a radial component decreasing from about 1\unit{T} close to the magnet axis to zero far from it.
In ME1/1 the anode wires are tilted by 29$^{\circ}$ to compensate for the Lorentz drift of electrons from the gas ionization process that otherwise causes a smearing of the induced charge distribution on the cathode strips and hence a deterioration in position resolution.
In ME1/2 the radial component of the field induces a smearing equivalent to that from muons incident at non-zero $\phi$, but there is no simple way to compensate for this in chamber construction.
However, the degraded resolution is still within the specified requirements.

The DT and CSC muon detector elements together cover the full CMS pseudorapidity interval $| \Pgh | < 2.4$ with no acceptance gaps, ensuring good muon identification over a range corresponding to $10^{\circ}<\theta < 170^{\circ}$.
Offline reconstruction efficiency for the muons is typically 96--99\% except in the gaps between the
5 wheels of the yoke (at $| \Pgh | = 0.25 $ and 0.8) and the transition region between the barrel outer wheels and the endcap disks~\cite{POG-paper}.
The amount of absorbing material before the first muon station reduces the contribution of punch-through particles to about 5\% of all muons reaching the first station and to about 0.2\% of all muons reaching further muon stations.
Crucial properties of the DT and CSC systems are that they can each identify the collision bunch crossing that generated the muon and trigger on the \pt of muons with good efficiency, and that they have the ability to reject background by means of timing discrimination.

The LHC is a bunched machine, in which the accelerated protons are distributed in bunches separated by one (or more) time steps of 25\unit{ns}.
This is therefore also the minimum separation between bunch crossings, in which proton--proton collisions occur.
Thus, a convenient time quantity for both the accelerator and the detectors is the bunch crossing (BX) ``unit'' of 25\unit{ns}, and, because the fundamental readout frequency is 40\unit{MHz}, clock times are often quoted in BX units.
The ability of the muon chambers to provide a fast, well-defined signal is crucial for triggering on muon tracks.
To ensure unambiguous identification (ID) of the correct bunch crossing and the time coincidence of track segments among the many muon  stations, the local signals must have a time dispersion of a few nanoseconds,  much less than the minimum 25\unit{ns} separation of bunch crossings.
A design in which intrinsically slow tracking chambers nevertheless
provide good timing and spatial performance at the trigger level is an important feature of the CMS muon system.

\subsection{Resistive plate chamber system}
\label{Intro:RPC}

In addition to these tracking detectors, CMS includes a complementary, dedicated triggering detector system with excellent time resolution to reinforce the measurement of the correct beam crossing time at the highest LHC luminosities.  The resistive plate chambers (RPC) are located in both the barrel and endcap regions, and they can provide a fast, independent trigger with a looser \pt threshold over a large portion of the pseudorapidity range ($| \Pgh | < 1.6$).  The RPCs are double-gap chambers, operated in avalanche mode to ensure reliable operation at high rates.

\begin{figure}
{\centering
\includegraphics[width=13.7cm]{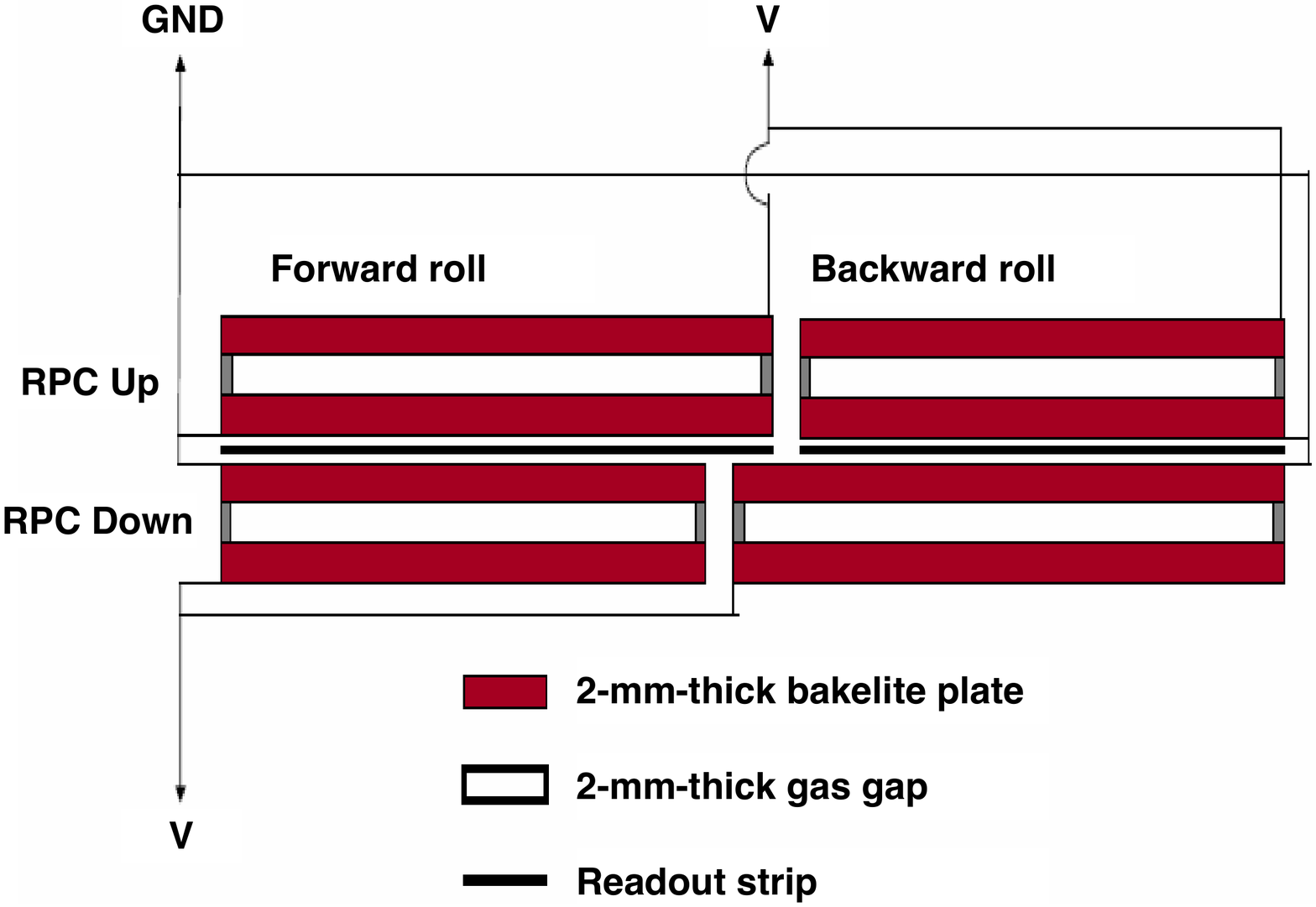}
\caption{\label{rollv2}
Schematic view of a generic barrel RPC with 2 ``roll'' partitions.}
}
\end{figure}

Figure~\ref{rollv2} shows the layout of a double-gap RPC.
Each gap consists of two 2-mm-thick resistive Bakelite plates separated by a 2-mm-thick gas gap.
The outer surface of the bakelite plates is coated with a thin conductive graphite layer, and a voltage of about 9.6\unit{kV} is applied.
The RPCs are operated with a 3-component, non-flammable gas mixture that consists of 95.2\% Freon (C$_2$H$_2$F$_4$, known as R134a), 4.5\% isobutane (i-C$_4$H$_{10}$), and 0.3\% sulphur hexafluoride (SF$_6$).
After mixing, water vapor is added to obtain a mixture with a relative humidity of 40\%--50\%.
Readout strips are aligned in $\eta$ in between the 2 gas gaps.
A charged particle crossing an RPC will ionize the gas in both gas volumes and the avalanches generated by the high electric field will induce an image charge, which is picked up by the readout strips.
This signal is discriminated and shaped by the front-end electronics.

The RPCs are organized in stations following a sequence similar to the DTs and CSCs.
In the RPC barrel (RB) there are 4 stations, namely RB1, RB2, RB3, and RB4, while in the RPC endcap (RE) the 3 stations are RE1, RE2, and RE3.
The innermost barrel stations RB1 and RB2 are instrumented with 2 layers of RPCs facing the innermost (RB1in and RB2in) and outermost (RB1out and RB2out) sides of the DT chambers.
Every chamber is then divided from the readout point of view into 2 or 3 $\Pgh$ partitions called ``rolls'' (Fig.~\ref{rollv2}).
In the endcaps, each station is divided into 3 rings (identified as rings~1, 2, and 3) at increasing radial distance from the beam line.
Ring 1 was not instrumented in 2010; the RPC system therefore covered only the region up to $| \Pgh | = 1.6$.
Each endcap ring is composed of 36 chambers covering the full azimuthal range.
From the readout point of view, each endcap chamber is divided into 3 $\Pgh$ partitions (rolls) identified by the letters A, B, and C.
Thus the endcap RPCs are identified in the following way: REn/r/x where n is the station ($\pm$1,
$\pm$2, $\pm$3), r is the ring (2 or 3), and x is the roll (A, B, or C).

\subsection{Muon triggering, tracking, and reconstruction}

The triggering scheme of the CMS muon system relies on 2 independent and complementary triggering technologies: one based on the precise tracking detectors in the barrel and endcaps, and the other based on the RPCs.
The tracking detectors provide excellent position and time resolution, while the RPC system provides excellent timing with somewhat poorer spatial resolution.

For values of \pt up to 200\GeVc, the momentum resolution is dominated by the large multiple scattering in the steel, combined in the endcaps with the effect of the complicated magnetic field that is associated with the bending of the field lines returning through the barrel yoke.
The detectors designed to meet the required measurement specifications and to operate in this environment are robust, multilayered chambers from which the fine spatial resolution required for good momentum resolution at high muon momenta can be obtained with a modest resolution per layer.

The large number of layers in each tracking chamber is exploited by a trigger hardware processor that constructs track segments within the chambers with a precision sufficient to set sharp transverse momentum (\pt) thresholds at the level-1 (L1) trigger level up to 100\GeVc, and to tag the parent bunch crossing with very good time resolution.
This component of the L1 trigger is called the ``local'' trigger since it operates purely with information local to a chamber.
The local trigger information is combined to form a ``regional'' trigger, one for each of the muon subsystems.
These regional triggers are finally combined to form the ``global'' muon trigger that serves as one of the primary L1 triggers.
A suitably steep transverse momentum threshold is obtained by requiring a local spatial resolution of the segments on the order of a couple of millimeters.
This resolution is necessary to guarantee a high trigger efficiency and defines a lower limit on the accuracy that must be reached by the alignment of the chamber positions.

The geometry of CMS has a deep influence on the performance of the muon system.
The change in the direction of the magnetic field in the return yoke causes the curvature of the muon trajectory to reverse.
Therefore, the first muon detector stations (ME1, MB1) in both the barrel and endcap regions are critical, since they provide the largest sagitta and, hence, the most important contribution to the measurement of the momentum of high momenta (more than a
few hundred\GeVc) muons, for which multiple scattering effects become less significant.

Reconstruction proceeds by first identifying hits in the detection layers of a muon chamber due to the passage of a muon (or other charged particle), and in the DT and CSC systems by then building straight-line track segments from these hits.
This is referred to as ``local'' reconstruction.
The reconstruction of muon tracks from these hits and segments is called ``global'' reconstruction.
Muon tracks can be reconstructed by using hits in the muon detectors alone; the resulting muon candidates are called ``standalone muons".
Alternatively, the reconstruction can combine hits in the muon detectors with those in the central tracker; the resulting candidates are called ``global muons".
The muon system can also be used purely to tag extrapolated tracks from the central tracker; such tracks are called ``tracker muons".
For muons with momenta below $\approx$300\GeVc,  tracker muons have better resolution than global muons.
As the \pt value increases, the additional hits in the muon system gradually improve the overall resolution.
Global muons exploit the full bending of the CMS solenoid and return yoke to achieve the ultimate performance in the\,\TeVc region.

 \begin{table}[!Hhtb]
   \begin{center}
         \topcaption{ Properties and parameters of the muon systems during the 2010 data-taking period.
         The design values of the position and time resolutions are from the CMS Muon TDR (1997) (Ref.~\cite{MUON-TDR}). As discussed in later sections, in general, these specifications were met and in some cases exceeded.}
         \label{tab:props}
         \begin{tabular} {|l|c|c|c|}
         \hline
  { Muon subsystem} &{Drift Tubes}  & {Cathode Strip} & {Resistive Plate }    \\
                           & {(DT)}         &   {Chambers (CSC)}      & {Chambers (RPC)}      \\ \hline \hline
Function         & Tracking, \pt trigger,  & Tracking, \pt trigger,  & \pt trigger,         \\
                         & BX ID  &   BX ID                           &   BX ID      \\   \hline
  $|\eta|$ range  &   0.0--1.2       &  0.9--2.4                   &  0.0--1.6                        \\ \hline
  No. of stations  &       4           &             4                &  Barrel 4;                \\
                        &                    &  (no ME4/2 ring)        & Endcap 3               \\  \hline
  No. of layers &   $r$-$\phi$: 8, $z$: 4  &             6                &           2 in RB1 and RB2;        \\
                        &                                          &                               &  1 elsewhere               \\  \hline
  No. of chambers &       250         &         468                  &  Barrel 480;         \\
                            &                       &                                   & Endcap 432       \\   \hline
  No. of channels &     172\,000      &  Strips 220\,000;   & Barrel 68\,000;  \\
                           &                         &  Wire groups 183\,000 & Endcap 41\,000 \\  \hline
  Design position  & per wire 250\micron;  &  per chamber      $r$-$\phi$ (6 pts)            &  Strip size (on  \\
  resolution ($\sigma$) for    & $r$-$\phi$ (6/8 pts) 100\micron; &  ME1/1, ME1/2 \; 75\micron                    &  the order of a  \\
  perpendicular tracks            &  $z$ (3/4 pts) 150\micron              & other CSCs 150\micron;                          &  centimeter)   \\
                                    &                                                          & $r$ (6 pts) 1.9--6.0\unit{mm} &   \\  \hline
  Design time resolution &   5\unit{ns}                                               &  6\unit{ns}                                                           &         3\unit{ns}    \\ \hline
           \end{tabular}
     \end{center}
 \end{table}

Table~\ref{tab:props} lists the functions and parameters of the muon systems as installed in CMS during the 2010 running period. The design specifications for spatial and time resolution are also listed~\cite{MUON-TDR}.
One of the goals of the present publication is to compare the requirements as outlined in the Muon TDR with the performance achieved in 2010.

\section{Calibration}
\label{section-calibration}

This section describes the procedures used in the extraction of calibration parameters that affect the performance of the CMS muon system, such as the drift velocity for the DTs and the electronic channel gains for the CSCs.

Calibration constants and monitoring parameters are stored as {\it conditions data}. They are produced both online, directly from the front-end
electronics, and by dedicated calibration algorithms run offline. The CMS conditions database~\cite{:2009gz} uses 3
databases (DB) for storing non-event data: the online master database system (OMDS) is in the online network at the detector
site and stores the data needed for the configuration and appropriate detector settings together with the conditions data produced
directly from the front-end electronics; the offline reconstruction condition DB online subset (ORCON), also located at
the detector site, stores all conditions data that are needed for the high-level trigger (HLT) as well as for detector performance studies; and the
offline reconstruction condition DB offline subset (ORCOFF), located at the CERN computing center (tier 0),
contains a copy of the information in ORCON. It is the database used for all offline processing and physics analyses.

To provide a rapid response to changing detector operating conditions, special offline calibration workflows use dedicated calibration data streams, distinct from the main collisions data output stream, as input to the offline calibration procedures.
These can then provide updated calibration information rapidly enough to be used in the full offline reconstruction of the collision data~\cite{:2009gz}.
They are available with very low latency and are analyzed at the CERN analysis facility for a prompt
determination of new constants.

The specific calibration procedures for the DTs and CSCs are described in the next sections, followed by a brief description of the procedure followed for setting the operating voltages and electronics thresholds of the RPCs.

\subsection{DT system calibration}
\label{section-dt-calibration}

\label{section-dt-calibration-introduction}

Charged particles crossing a drift cell in the DTs ionize the gas within the cell.
The drift time of the ionization electrons is obtained by using a high-performance
time-to-digital converter (TDC)~\cite{Christiansen:1067476}, after subtraction of a time pedestal.
The time pedestal contains contributions from the latency of the trigger and from the propagation time of the signal within the detector and the data acquisition chain.
The hit position, \ie, the distance of the muon track with respect to the anode wire, is reconstructed as

\begin{equation}
x_{\text{hit}} = t_\text{drift}\cdot v_{\text{drift}} \equiv \left(t_\text{TDC} - t_{\text{ped}}\right)\cdot v_{\text{drift}} ,
\label{eq:hit-reconstruction}
\end{equation}

where $t_\text{TDC}$ is the measured time, $t_{\text{ped}}$ the time pedestal, and $v_{\text{drift}}$ the effective drift velocity, which is assumed to be approximately constant in the cell volume.

The operating conditions of the chambers are monitored continuously~\cite{MUON-TDR}.
The high voltage supplies have a built-in monitor for each channel.
The temperature of the gas, nominally at room temperature,
is measured on each preamplifier board inside the chamber.
The gas pressure is regulated and measured at the gas distribution rack on each wheel, and is monitored by 4 additional sensors, 2 at the inlets and 2 at the outlets of each chamber.
The flow sharing from a single gas distribution rack to 50 chambers
is monitored at the inlet and outlet lines of each individual chamber. A possible leakage in the gas line can be detected via the flow and pressure measurements.

Small gas chambers called drift velocity chambers~\cite{diplom-altenhoefer,diplom-frangenheim}
are located in the accessible gas room adjacent to the cavern outside of the CMS magnetic field.
They are used to measure the drift velocity in a gas volume with a very homogeneous electric field.
Each of these chambers is able to selectively measure the gas being distributed to, and returned from,
each individual chamber of the wheel, thus providing rapid feedback on any changes due to the gas mixture or contamination of the gas.

\subsubsection{Time pedestal offline calibration}
\label{section-dt-calibration-pedestal}

The drift time $t_\text{drift}$ is obtained from the TDC measurement after the subtraction of a time pedestal. In an ideal
cell, the time distribution from the TDC, $t_\text{TDC}$, would have a box shape
starting at close to 0\unit{ns}
for muon tracks passing near the anode and extending up to 380\unit{ns} for those passing close to the cathode.
In practice, different time delays related to the trigger latency and the length of the cables to the readout electronics
contribute to the time measured by the TDC as follows:

\begin{equation}
t_\text{TDC} = t_\text{drift} + t_0^{\text{wire}} + \underbrace{t_\text{L1} + t_\text{TOF} + t_\text{prop}}_{t_\text{trig}} ,
\label{eq:tdc-time}
\end{equation}

where the different contributions to the time pedestal are classified as
\begin{itemize}
  \item $t_0^{\text{wire}}$, the channel-by-channel signal propagation time to the readout electronics,
      relative to the average value in a chamber; it is referred to as inter-channel synchronization
      since it is used to equalize the response of all the channels within a chamber;
  \item $t_\text{L1}$, the latency of the Level-1 trigger;
  \item $t_\text{TOF}$, the time-of-flight (TOF) of the muon produced in a collision event, from the
    interaction point to the cell;
  \item $t_\text{prop}$, the propagation time of the signal along the
    anode wire.
\end{itemize}

The inter-channel synchronization $t_0^{\text{wire}}$ is determined by test pulse calibration runs.
It is a fixed offset, since it depends only on the cable or fiber lengths.
A test pulse is simultaneously injected in 4 channels of a front-end board,
each from a different layer in an SL, simulating a muon crossing the detector.
The same test pulse signal is also distributed to other 4-channel groups, 16 channels apart,
such that the entire DT system is scanned in as many cycles.

The remaining contribution to the time pedestal is extracted from the data for each SL.
It is computed as the turn-on point of the TDC time distribution~\cite{Abbiendi:2009zzb,Chatrchyan:2009ih},
after correction for the inter-channel synchronization. Channels identified as noisy are not considered.
This contribution is called $t_\text{trig}$, since it is dominated by the Level-1 trigger latency
and includes the contributions from the average time-of-flight, roughly corresponding to the time taken by muons to reach the
center of the SL, as well as the average signal propagation time along the anode wire,
taken from the center of the wire to the front-end board.

The correction for the propagation time along the wire and the muon time-of-flight to the cell is performed at the
reconstruction level, after the 3D position of the track segment is determined~\cite{CMSNOTE:2009008}.
The segment is built in a multi-step procedure, as described in more detail in Section~\ref{Efficiency}.
First, the reconstruction is performed in the $r$-$\phi$ and $r$-$z$ projections independently.
Once the 2 projections are paired, the segment position inside the chamber can be
estimated and the drift time is further corrected for the propagation time along the anode wire and for
the time-of-flight from the center of the SL. The 3D segment is then updated.

\begin{figure}[htbp]
  {\centering
   \includegraphics[width=0.75\textwidth]{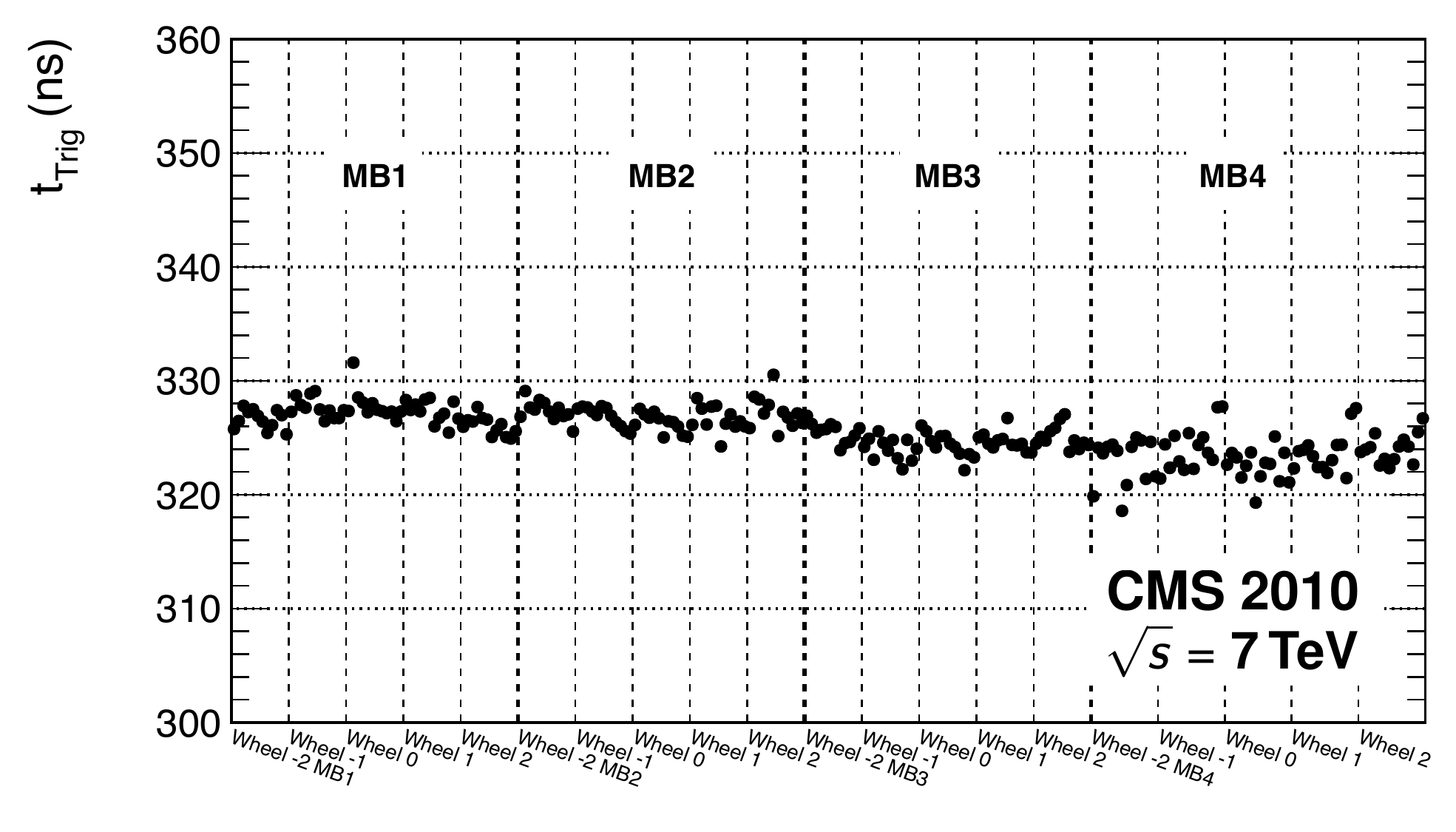}
   \includegraphics[width=0.75\textwidth]{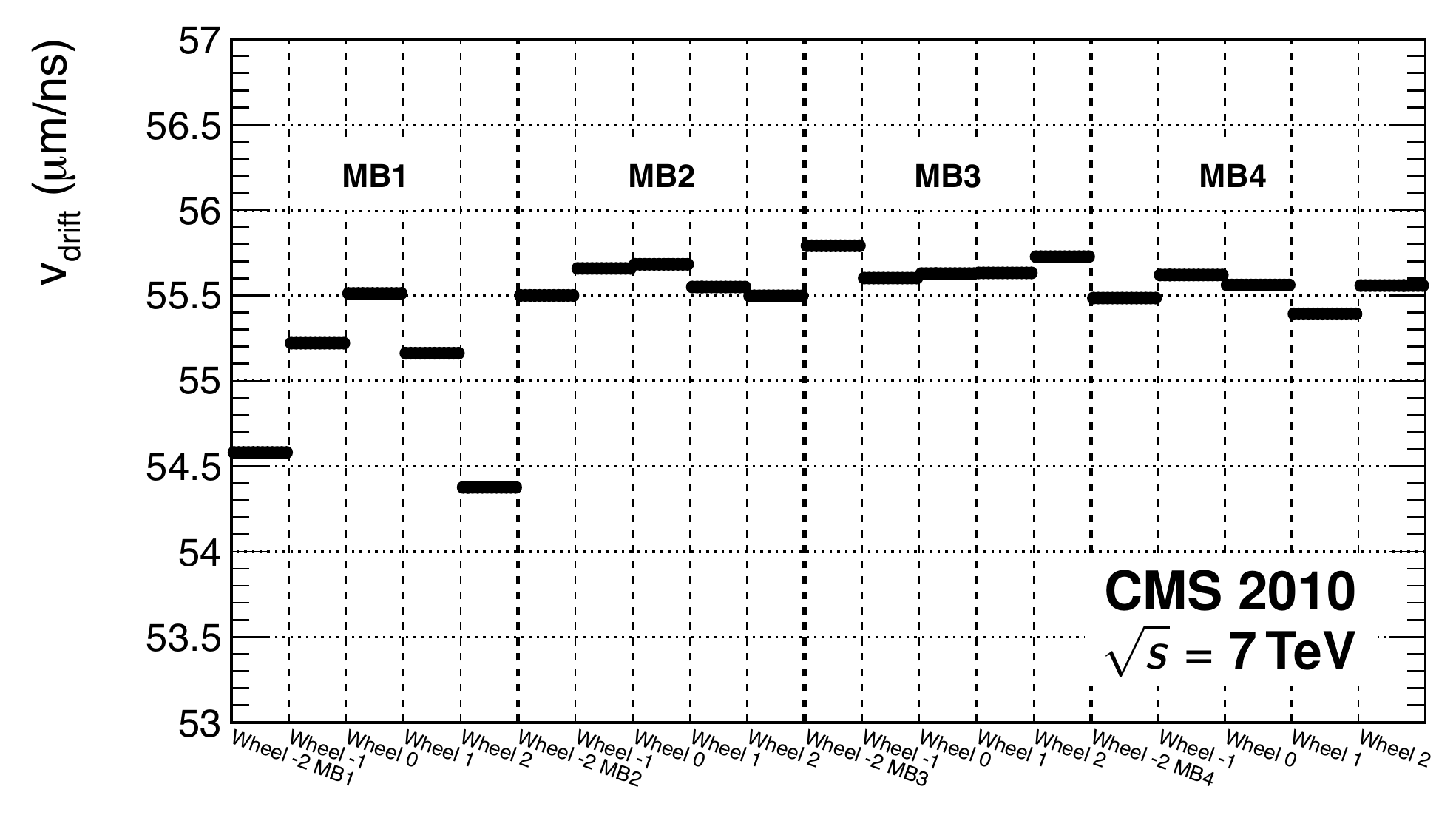}
  \caption{
      Top: Time pedestals $t_\text{trig}$ computed for all chambers in the DT system. Only values corresponding to the first $r$-$\phi$ superlayer (SL1) in each chamber are shown. Bottom: Drift velocities of all the chambers in the DT system. Chambers belonging to the same wheel were merged by station in the computation of the drift velocity.
   }
   \label{fig:ttrig-vdrift-all}
   }
\end{figure}

To define the turn-on point of the TDC time distributions more precisely, a correction to the $t_\text{trig}$ pedestal is calculated by using the hit position residuals.
The residuals are computed as the distance between the hit position and the intersection of the 3D segment
with the layer plane, reconstructed as described above.
The offset in the mean of the residual distribution for each SL (divided by the drift velocity) is used as an estimate of the correction that is added to the time pedestal (subtracted from the drift time reconstruction). This procedure is repeated iteratively.
The $t_\text{trig}$ values derived from a representative subset of the 2010 collision data are shown in Fig.~\ref{fig:ttrig-vdrift-all} (top) for the first $r$-$\phi$ superlayer (SL1) in each chamber of the DT system.
Similar values are obtained for all SLs.

\subsubsection{Drift velocity calibration}
\label{section-dt-calibration-vdrift}

The drift velocity depends on the gas mixture, purity, and the electrostatic configuration of the cell. The effective drift velocity, used in the hit reconstruction (see Eq.~(\ref{eq:hit-reconstruction})), is further affected by the presence of the residual magnetic field and the track incidence angle~\cite{Chatrchyan:2009ih, Abbiendi:2009zzb}. The drift velocity is computed as an average for each SL in the system.

Two methods have been used  to determine the drift velocity: the first is based on the ``mean-time'' technique,
and the second is based on the local muon reconstruction in which a track is fitted to measurements from one chamber at a time.

The mean-time method~\cite{Abbiendi:2009zzb} exploits the staggering of the chamber layers (see also Section~\ref{sec:DTsync}).
Because of the staggering, ionization electrons drift in opposite directions in even and odd layers.
Therefore the maximum drift time $t_\text{max}$ in a semi-cell can be calculated,
by using relations derived from detailed Garfield simulation of an individual drift cell~\cite{GARFIELD},
 from the drift times of hits from the track crossing nearby cells in consecutive layers.
In general, these relations depend on the track inclination and on the pattern of cells crossed by the track.
The appropriate mean-time relation is chosen for each track by using the 3D position and direction of the track segment in the SL.
A linear approximation is used to determine the drift velocity, $v_\text{drift}^\text{eff} = {L_\text{semi-cell}}/{<t_\text{max}>}$, where $L_\text{semi-cell} = 20.3$\unit{mm} is approximately half the width of a drift cell.

The drift velocity obtained with the mean-time method depends directly on the measured drift time and hence on the
time pedestal. Conversely, the time pedestal (see Section~\ref{section-dt-calibration-pedestal}),
corrected by using the mean of the hit residual distributions, is itself dependent on the drift velocity value used in the hit
position computation (Eq.~(\ref{eq:hit-reconstruction})). The 2 parameters cannot be fully disentangled.

An alternative method for computing the drift velocity relies on the full reconstruction of the
trajectory within the muon system~\cite{Chatrchyan:2009ih}. A track is reconstructed by assuming the nominal
drift velocity $v_\text{drift}$
 and, in a second step, is refitted with
the drift velocity and the time of passage of the muon through the chamber as free parameters.
The method is
applied to the $r$-$\phi$ view of the track segment in a chamber, where up to 8 hits
can be assigned to the track. This method cannot, however, be applied to the $r$-$z$ SLs where only 4 points are available because an insufficient number of degrees of freedom is available in the fit to disentangle the drift velocity and synchronization contributions.
Figure~\ref{fig:ttrig-vdrift-all} (bottom) shows the drift velocity values obtained for each
chamber in the DT system (corresponding to $r$-$\phi$ SLs) from a subset of the 2010 collision data.
Since the drift velocity is not expected to vary substantially among different sectors, the distributions corresponding to all
chambers in each wheel were merged by station, thus yielding a constant value for every station and wheel intersect.
A notable reduction ($\approx$2\%) in the value of the drift velocity is observed in the innermost chambers of the outer wheels because of the Lorentz angle induced by the stronger magnetic field.

\subsection{CSC system calibration}
\label{section-csc-calibration}

\label{section-csc-calibration-introduction}

A set of calibrations and related tests of the CSCs and the front-end electronics is performed periodically.
The tests are intended both to monitor the stability of the system and to determine parameters required for configuration of the electronics modules.
Counting rates, chamber noise levels, and channel connectivity are monitored.
The configuration constants include anode front-end board (AFEB) discriminator
thresholds and delays, cathode front-end board (CFEB) trigger primitive thresholds, and
numerous
timing constants required for the peripheral crate electronics.
These values are then uploaded to the electronics modules.
Calibration of electronics channels is required to normalize the measured signals for use in reconstruction offline and in the HLT.
The calibration constants, required for optimal hit reconstruction and for simulation of the CSC detectors, specify strip-to-strip crosstalk, strip channel noise, strip pedestals, and strip channel electronic gains.

\subsubsection{CSC CFEB operation}

The cathode strips are connected to 16-channel amplifier-shaper ASIC (application-specific integrated circuit) chips. The outputs from these ASIC chip channels are sampled every 50\unit{ns}.
The sampled voltage levels are stored in switched capacitor array (SCA)~\cite{SCA} ASICs during the Level-1 trigger latency of 120 BXs.
There are 96 channels per CFEB distributed across six 16-channel SCAs, with each channel containing 96 capacitors (equivalent to 192 BXs).
These samples are digitized and read out when a local charged track (LCT) trigger associated with the CFEB is correlated in time with a Level-1 trigger accept (L1A) signal.

\subsubsection{CSC CFEB calibration}

For calibration purposes, L1As are generated by the local trigger control and are received synchronously by the CSC electronics.
Two internal capacitors are incorporated on the amplifier-shaper ASIC chip for each cathode amplifier channel, and one precision external capacitor is mounted on the CFEB board to service each amplifier-shaper.
Each capacitor can be used to generate a test pulse. These pulses are activated in parallel so that pulses for calibration runs can be completed during beam injection. For more details, see Ref.~\cite{OSU_NIM}.

The linearity, offset, and saturation of the amplifier are determined by incrementally varying the test pulse amplitude applied to every channel and measuring the output.
This also yields the electronic gain for each strip channel.
The raw pulse height measurements from each strip channel are normalized by these gains before use in the reconstruction of muon hit positions.

To determine crosstalk between neighboring strips, a fixed-sized pulse of very short duration (an approximate delta-function pulse) is injected into each amplifier channel.
The output on neighboring channels is then measured.
To obtain the crosstalk fraction (the ratio of the charge induced on a neighboring strip to the charge deposited on a strip) appropriate for an operating CSC, these amplifier responses to a delta-function charge deposition must be convoluted with the expected ion drift time distribution
and the arrival time distribution of electrons (approximately uniform in time).
Then, for a given time bin, the ratio of one side pulse to the central pulse plus both side pulses yields the crosstalk fraction.
This is found to be linear in time for an interval of about 160\unit{ns} around the peak time of the central charge distribution.
This linearity is confirmed by test beam measurements. The typical magnitude of the crosstalk fractions ranges from about 5\% to 10\%.
Thus the crosstalk fractions for a given strip are determined by 4 crosstalk constants associated with a strip: a slope and intercept to describe the straight line representing the crosstalk coupling to each of its 2 neighbors, as a function of the SCA time bin.
This information is stored as a 3$\times$3 matrix for each strip and SCA time bin, with elements that relate the measured charge on the strip and its 2 neighbors to the input charge on the strip.
These matrices are used directly to model crosstalk in simulation, and the inverse matrices can be used to unfold crosstalk from the real data.
This is performed even though the reconstructed position of a muon is largely insensitive to crosstalk because the reconstruction algorithm is robust against symmetric channel-to-channel coupling.

To determine the pedestals on each chamber strip, the amplifier output is sampled continuously at 20\unit{MHz} with no input signal.
Then the charge measured in each SCA time bin is taken as the pedestal. The pedestal noise (pedestal RMS) is correlated between the SCA time bins. Therefore, for each strip, a symmetric ``noise matrix'' is defined to describe the covariance between the pedestals in each time bin, with elements
$C_{ij} = \left<Q_i\cdot Q_j\right> - \left<Q_i\right>\left<Q_j\right>$,
where $Q_i$ is the charge in the time bin $i$, and the averages are over a large number of calibration events. This matrix is used to introduce appropriate pedestal correlations in the simulation.
These ``static'' pedestals are not used in the reconstruction.
Instead, the reconstruction uses a ``dynamic'' pedestal, which is based on the average of the first 2 SCA time bins before the peak of the signal arrives.  This compensates for any baseline shift arising from high-rate operation.
The static pedestals, however, are used in simulation to provide an appropriate baseline to the simulated signals.

\begin{figure}[htb]
  \begin{center}
   {\includegraphics[width=7.5cm]{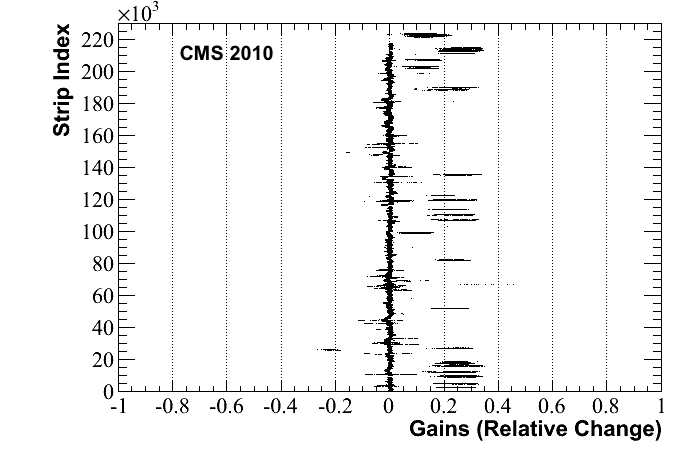}}
   {\includegraphics[width=7.5cm]{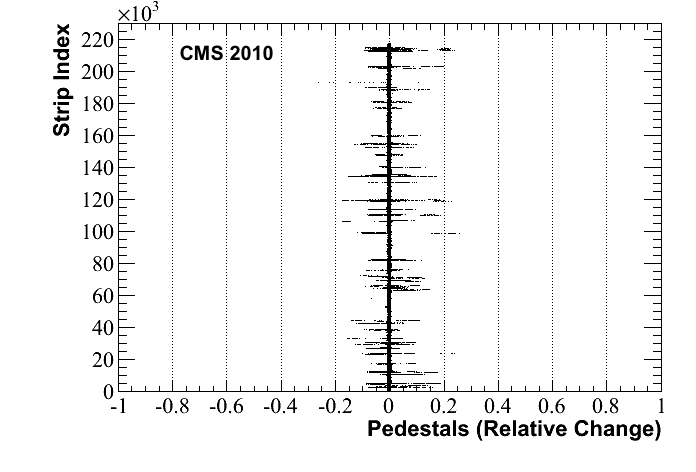}}
   \caption{
      The change in CSC calibration constants calculated from calibration runs taken before and after CMS 2010 $\sqrt{s} = 7$\TeV operation. Left: Relative difference in gains. Right: Relative difference in pedestals. The vertical axis ``strip index'' just indexes each strip in the system counting from 1.
   }
   \label{fig:csc-calibplots}
  \end{center}
\end{figure}

Figure~\ref{fig:csc-calibplots} displays the relative differences for gains and pedestals strip-by-strip between constants calculated before and after the $\sqrt{s} = 7$\TeV operation in 2010. There are about 220\,000 strip channels, and the correlation between values associated with the same front-end boards is visible. Gross changes occurred when electronic boards were replaced during the year.

\subsection{Validation and monitoring of calibration constants}
\label{section-summary-validation}

A detailed validation is performed to assess the quality and monitor the stability of the calibration constants.
For each new calibration set, a comparison to previous constants is carried out and the impact on the reconstruction performance is analyzed.
The CMS data quality monitoring (DQM; see Section~\ref{sec:DQM}) framework~\cite{Tuura:2010zza} is used throughout the validation procedure.

The local hit reconstruction in the DT system is directly dependent on variations in the time pedestal, as well as the drift velocity calibration.
Small changes in the system are accounted for with new calibration constants, thus optimizing the chamber performance.
The calibration procedures discussed in Section~\ref{section-dt-calibration-pedestal} guarantee that the hit position residuals are well centered around 0. The measured resolutions
(Fig.~\ref{fig:DTResolution}) are within 300\micron for $r$-$\phi$ SLs and are slightly worse for the outermost station (MB4) because there is no theta information.
They are larger for $r$-$z$ SLs, when moving towards the external wheels (see Section~\ref{sec:Resolution}).

The impact of the calibration on the local muon reconstruction is studied by analyzing the hit position residuals.
Resolution measurements are discussed in detail in Section~\ref{sec:Resolution}.
The effect of changing calibration constants calculated before and after extended run periods on the hit position distributions in each CSC station has been shown to be very small.
Testing and validation of new constants is performed regularly and possible changes due to the effect of calibration are closely monitored.

\subsection{Setting RPC system operating voltages and thresholds}

Once the gas composition is fixed, the main calibration parameters for the RPCs are the time synchronization (see Section~\ref{sec:RPCsync}), the operating high-voltage points, and the electronics thresholds.
The optimal operating voltages have been determined by means of dedicated runs only in 2011, while work on the electronics thresholds continued.
During 2010,
just 2 different values were chosen: 9550 and 9350\unit{V} for the endcap and barrel chambers, respectively.
These values were derived from measurements made during the construction and commissioning phases, which were averaged over the 2 sets of chambers and extrapolated to the different conditions typical of CMS at the LHC (including the rate, which could not be accounted for previously).

\section{Level-1 trigger}
\label{LocalTrigger}
The purpose of the trigger system is to find event candidates
fulfilling a predefined set of criteria and to assign them to an
appropriate bunch crossing number.
At the LHC many interesting physics signatures contain muons, and the muon
subdetectors are therefore included in the level-1 (L1) trigger system.

The local triggers provide the segments for the L1 trigger from each barrel and endcap muon chamber.
In the barrel, this task is performed by the DT local trigger (DTLT), and in the endcaps by the CSC local trigger (CSCLT).
The RPC trigger is not based on local trigger devices, as muon candidates are constructed from the spatial and temporal coincidence of hits in the RPCs.
The DT and CSC local trigger segments from each muon station are collected by the trigger track finders (TF), which combine them to form a muon track and assign a transverse momentum value.
At least 2 segments in 2 different stations are needed by the TFs to construct a muon candidate.

The DTLT system is described in
detail in Refs.~\cite{Chatrchyan:2009ig, trigTDR, NIMA534_441, NIM2007}.
Only the main functions and characteristics are
summarized here.
The trigger segments are found separately in the transverse plane $x$-$y$
(called the $\phi$ view) and in the plane that contains the $z$
direction (called the $\theta$ view). The maximum drift time in the DT
system is almost 400\unit{ns},
which is much longer than the minimum interval of 25\unit{ns} interval that separates 2
consecutive colliding bunches.
Therefore, the DTLT system must associate each trigger segment to
the bunch crossing at which the muon candidate was produced.
For each BX the system provides up to 2 trigger segments per chamber in the $\phi$ view and 1 in the $\theta$ view.
In the $\phi$ view, each trigger segment is associated with
\begin{itemize}
\item the BX at which the corresponding muon candidate was produced;
\item the position and direction of the trigger segment;
\item a quality word describing how many aligned DT hits were found;
\item a bit that flags the segment for that BX as the
first or second candidate, ordered according to their assigned quality .
\end{itemize}
A set of such quantities is called a ``DT trigger primitive''.

Trigger primitives are provided separately for each station.
During the 2010 data taking, the DTLT electronics was configured in the following way.
A trigger segment was accepted if the
minimum number of aligned hits in a trigger segment was 4 if the
segment contained hits only from a single SL,
or at least 3 in each SL if hits from both $\phi$ SLs were used.
A DT muon trigger candidate is then identified by the DT track finder (DTTF)
if an acceptable spatial and angular matching  is found between at least 2
trigger primitives in 2 different stations~\cite{trigTDR}.

A single muon is expected to produce one and only one
DTLT segment in each station crossed by the particle.
Nevertheless,
an additional ``false'' trigger candidate (``ghost'') can be occasionally produced.
False candidates can arise from the presence of additional misplaced
hits around the muon track, or from the fact that adjacent electronics
units devoted to reconstruct the trigger
segments
share a common group of DT cells,
such that a trigger candidate can be found twice.
A dedicated ghost-suppression
algorithm is used by the DTLT electronics
to discard such duplicate candidates (see Section~\ref{FalseTrig}).

To measure the DTLT performance during the 2010 LHC run, approximately 10$^6$ events from
2 data samples containing reconstructed muons were studied:
a minimum bias sample and a sample of W/Z
events decaying to muons.
Muon candidates in minimum bias events are characterized by
low transverse momenta, $\pt > 3\GeVc$, whereas W/Z decays
produce muon tracks with $\pt > 20\GeVc$.
A sample of approximately  7\ten{5} simulated minimum bias events containing at
least 1 reconstructed muon track was
also used for comparison.

The CSCLT system is described in detail in Ref.~\cite{trigTDR}.
Signals are recorded by the front-end cathode and anode electronic
boards connected to the chambers.
Muon track segments are found separately in the
nearly orthogonal
cathode and anode planes, where the 6-layer redundancy of the system is
used to measure the
muon-segment BX, its pseudorapidity $\eta$, and the azimuthal angle $\phi$.
Up to 2 cathode
and 2 anode local track projections can be found in each chamber at any BX.
These are then combined into 3D tracks by requiring a timing coincidence in the trigger electronics device.
The CSCs are read out in zero-suppressed mode,
requiring a pretrigger that
depends on specific patterns of cathode and anode local tracks.
Hence, the CSC readout is
highly correlated with the
presence of CSCLT segments. This feature is taken into account when
measuring the local trigger efficiency.

The CSCLT efficiency measurements are performed by using
a minimum bias event sample
and a sample of events containing \JPsi and Z decays to dimuons.
A sample of
2.5 million simulated minimum bias
events containing at least 1 muon at the generator level is also used for comparison, as well as samples of simulated events
containing a \JPsi or a Z decaying to a pair of muons.
Muon candidates from \JPsi decays are characterized by  $\pt < 20\GeVc$, whereas muon tracks from Z decays have $\pt > 20\GeVc$.

When a muon passes through an RPC, it creates a pattern of hits that contains information about the bending of the track and thus about the \pt of the muon.
Trigger processors compare the observed pattern within a segment with predefined patterns corresponding to certain \pt values.
The pattern comparator trigger (PACT) allows for coincidences of 4 hits out of 4 stations (4/4) and 3 hits out of 4 stations (3/4).
The latter are assigned a lower quality with respect to the former.
In the barrel, higher quality triggers are also possible with 5/6 and 6/6 coincidences.
If the observed hits match multiple patterns, the muon candidate with the highest quality and highest \pt is selected.
All candidates are first sorted by quality, then by \pt and the L1 RPC regional trigger delivers the 4 best muons in the barrel and the 4 best muons in the endcap to the global muon trigger (GMT).

\subsection{Timing and synchronization}
\label{TimingAndSync}
The L1A signal, which is broadcast to all subdetectors,
initiates the readout of the event.
Trigger synchronization is of great importance because
as simultaneous hits in multiple chambers are required for an L1 trigger,
out-of-time chambers can reduce the overall trigger efficiency.
Moreover, if the L1 muon trigger is generated early or late relative to the collision time, it forces readout of the entire
detector at the wrong BX.
For these reasons, online synchronization of the muon chambers was a
 priority during the early running period.

Trigger synchronization of each subsystem must be achieved at 3 levels: intrachamber
synchronization, chamber-to-chamber relative synchronization, and subsystem-to-subsystem synchronization.
Although each muon subsystem faced unique challenges due to differences
in chamber design, trigger electronics design, and physical position on the CMS detector, the
general synchronization procedures were similar. The general procedure is discussed in
Section~\ref{sec:gensync}.
The details and results of the separate DT, CSC, and RPC trigger synchronization methods are found in
Sections~\ref{sec:DTsync}, \ref{sec:CSCsync}, and \ref{sec:RPCsync}, respectively.  The overall
L1 GMT synchronization results are discussed in Section~\ref{sec:L1sync}.

For physics analyses, the time assigned to the muon hits once the event
has been collected and fully reconstructed is also important.
This is called the "offline time."
For a muon produced in a proton--proton collision and with the correct BX assignment,
the offline time of any muon chambers hit should be reported
as t=0.
Any deviations from 0 may be caused by backgrounds such as cosmic-ray muons, beam
backgrounds, chamber noise, or out-of-time pileup, or it may be an indication of new physics such as a slow moving,
heavy charged particle. In Section~\ref{sec:OfflineTimeAlignment}, the offline
time alignment procedure and results are shown.

\subsubsection{Common synchronization procedure}
\label{sec:gensync}

Track segments are promptly obtained by the local front-end trigger electronics from hits in the 4 layers of a DT chamber superlayer or the 6 layers of strips and wires in a CSC.
Trigger primitives are delivered to the Level-1 trigger at a fixed delay with respect to the chamber local clock, which is a copy of the master LHC clock.
The time of a hit caused by the passage of a particle through the muon chamber with respect to the locally distributed clock signal depends on the following:

\begin {itemize}
\item the muon time-of-flight from the interaction point to the chamber;
\item individual chamber properties and geometrical position;
\item the latency of the trigger electronics;
\item the length of the cables and fibers connecting the chamber electronics to the peripheral crates.
\end{itemize}

The last 3 items are specific to each chamber and were already studied during cosmic data taking, before proton--proton collisions were recorded at CMS.
The synchronization with respect to the master LHC clock (and hence with the rest of CMS) is achieved by moving the phase of the locally distributed BX signal with respect to the master LHC clock.

The tool used by CMS subdetectors for the synchronization of trigger and data
acquisition chains is the trigger and timing control system device (TTC) \cite{ttc}.
The purpose of the TTC is to distribute the machine clock signal to the various parts of
the detector and broadcast the L1A ``strobe'' trigger signal.

Prior to the start of collisions in the LHC, each muon subsystem used cosmic-ray data or early single-beam data to adjust its TTC delays
for a rough chamber-to-chamber synchronization.
Additional adjustments were introduced based on
calculated time-of-flight paths to each chamber.
Once collision data were available, each subsystem
used high-\pt muon data to refine internal delay settings
so that the on-chamber clocks would be in the correct phase with respect to the
LHC machine clock.  This procedure was iterative.
The L1 trigger is different for each muon subsystem, so the subsystem-specific figures of merit
for synchronization are presented in the next sections.

\subsubsection{DT trigger synchronization}
\label{sec:DTsync}
To time the detector to collect collision data, 2 independent synchronization steps ( ``coarse'' and ``fine'' ) are carried out.
The first refers to the chamber-to-chamber adjustment of the overall DTLT latency in terms of BX spacing units and is performed to provide equalized input to the DT regional trigger.
The second refers to the tuning of the sampling phase of the DTLT to a precision of 1--2\unit{ns}
to optimize the system response to muons arriving at a fixed time after
the beam crossings.

Every DT is equipped with a trigger and timing control receiver (TTCrx)
device that
provides a parameter to adjust the clock phase between
on-board electronics and the CMS master clock  (TTCrx delay parameter). The latter is used to
perform the ``fine'' synchronization of the DTLT and is configurable in steps of 0.1\unit{ns}.

Since it is possible to tune the TTCrx delay parameters only chamber-by-chamber, delays
due to signal propagation between the boards equipping a single DT chamber need to
be taken into account.
To compensate for this effect, the DTLT internal timing was equalized using cables of
appropriate lengths.
The maximum skew of the clock distribution after equalization has been
measured to be around 1\unit{ns}, ensuring that each chamber was intrinsically
synchronous within this level of precision.
Such a level of accuracy compares well with the design performance; hence the online DT software
allows timing adjustments to a precision of 1\unit{ns}.

The tuning of the TTCrx delay parameter affects both readout and DTLT boards.
Therefore every adjustment related to trigger timing optimization needed to be followed by an update of the DT calibration pedestals used for local hit reconstruction.

The procedure used to synchronize the DTLT with collision data is extensively described in Ref.~\cite{dt_fine_synch}.
A precise measurement of the particle arrival time in a muon station with respect to the calibration pedestal (see Section~\ref{section-dt-calibration}) can be performed for each local segment.
The method exploits the staggering of the wires in the SLs.
The ionization electrons in the 2 odd layers drift in a direction opposite to that in the 2 even layers. In the case of a 4-hit track,  2 segments can be reconstructed, one from the even hits and the other from the odd hits.
If the reference time does not have the correct phase with respect to the time of passage of the particle, the 2 segments  do not coincide but are separated by a time that depends on the reference pedestal time.
This time difference can be measured precisely via an optimization procedure that uses this difference as a free parameter in the reconstruction.

During commissioning of the detector with cosmic-ray muons, the DTLT response
was studied to characterize the DTLT performance with respect
to the relative phase between the particle's crossing time and the rising edge of the bunch crossing
TTC signal. This allowed identification of the timing phases within a bunch crossing,
where the system has optimal performance. These were collected for each chamber and used as
a startup reference to time in the detector during bunched beams operation.

Relying on the aforementioned method, the timing
distribution of reconstructed segments matched to global muons coming from
the interaction vertex was computed
as soon as sufficient bunched beam data were available.
The timing distribution was then compared to the set of optimal timing phases previously measured from cosmic data.
This was used to estimate additional corrections that optimize the DTLT performance to maximize the BX identification efficiency of global muons.
The procedure was iterated 4 times to reach the final configuration.

\begin{figure}[htp]
  \begin{center}
    \includegraphics[width=0.45\textwidth]{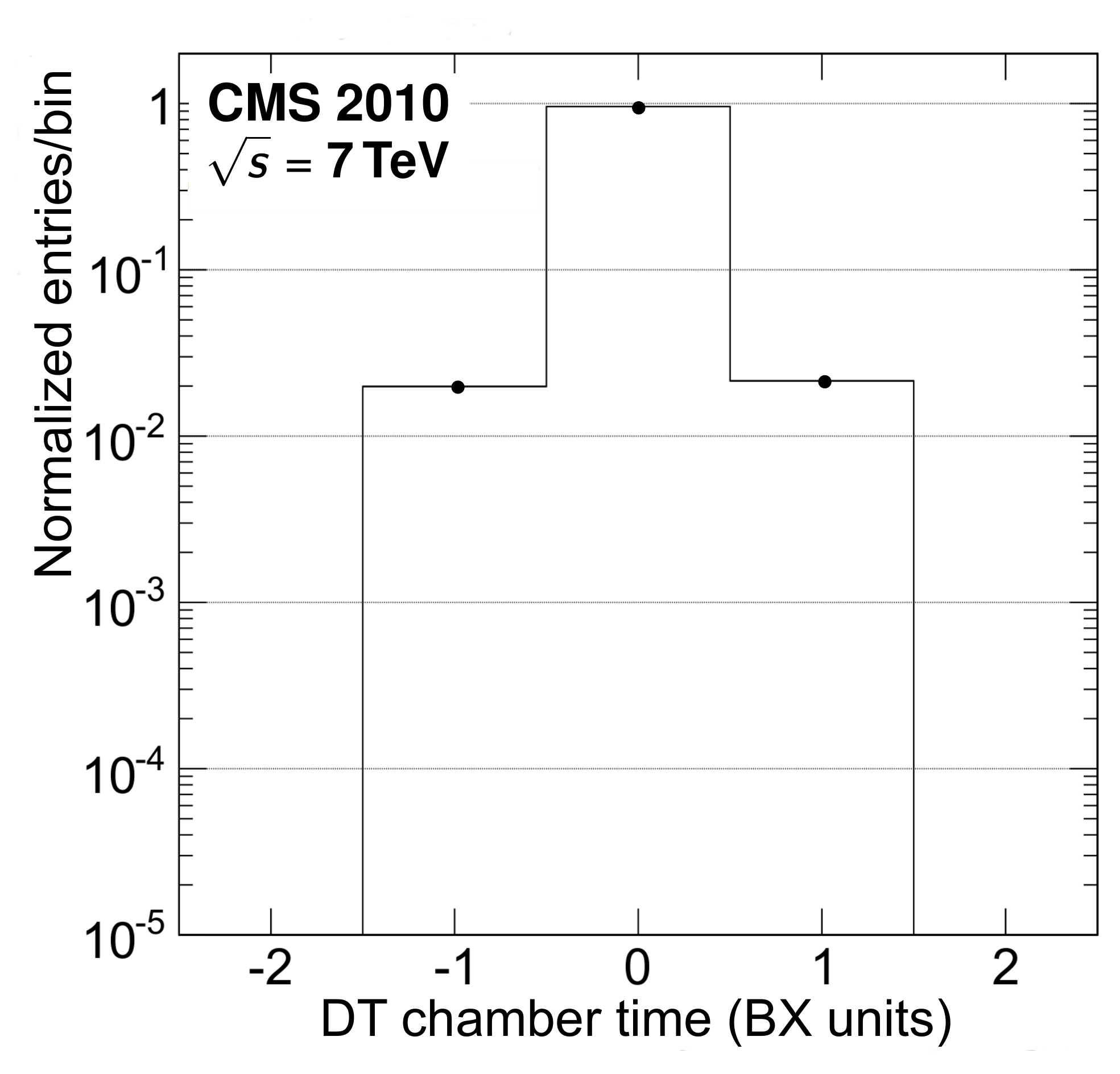}
    \includegraphics[width=0.45\textwidth]{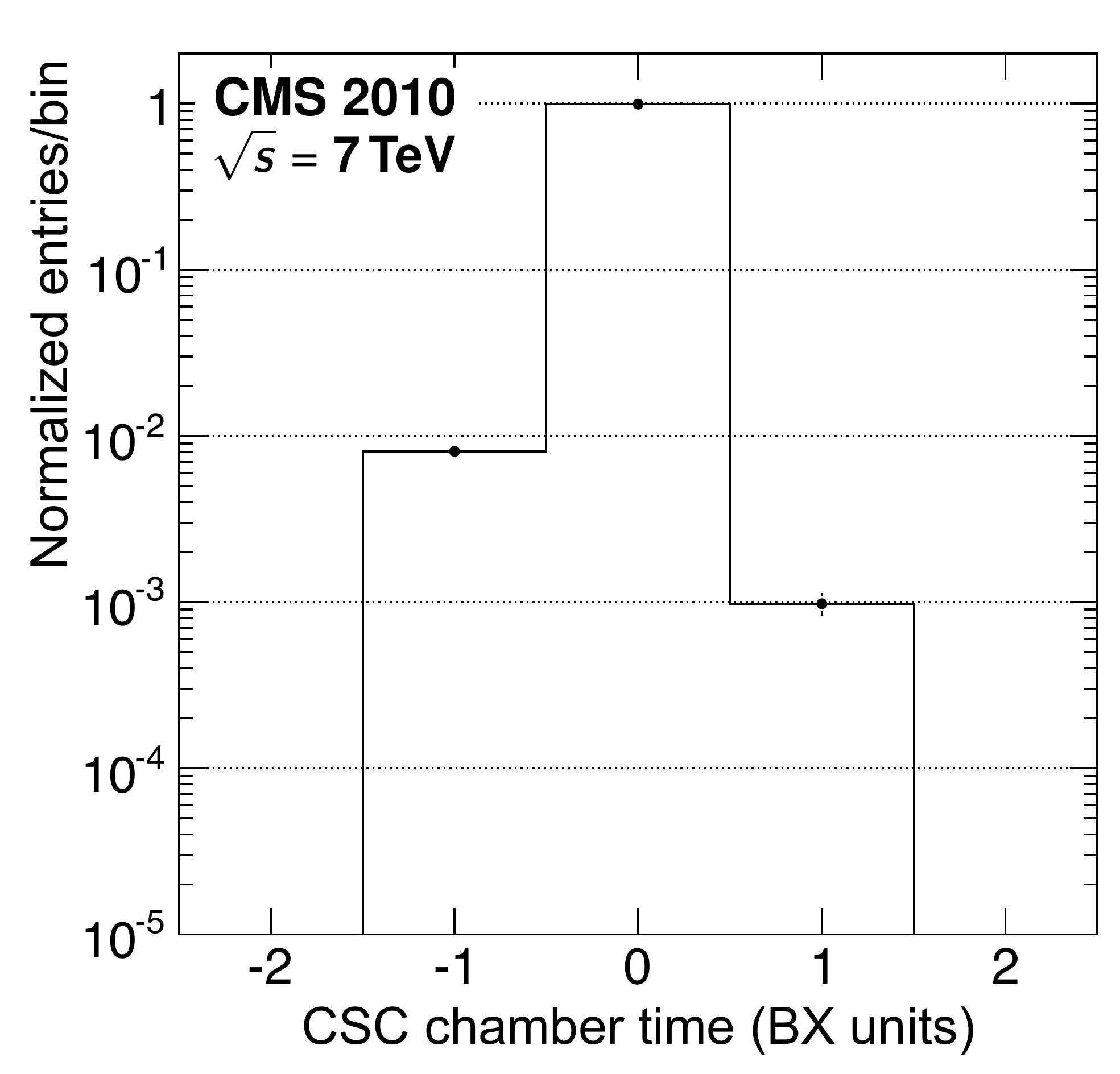} \\
    \includegraphics[width=0.45\textwidth]{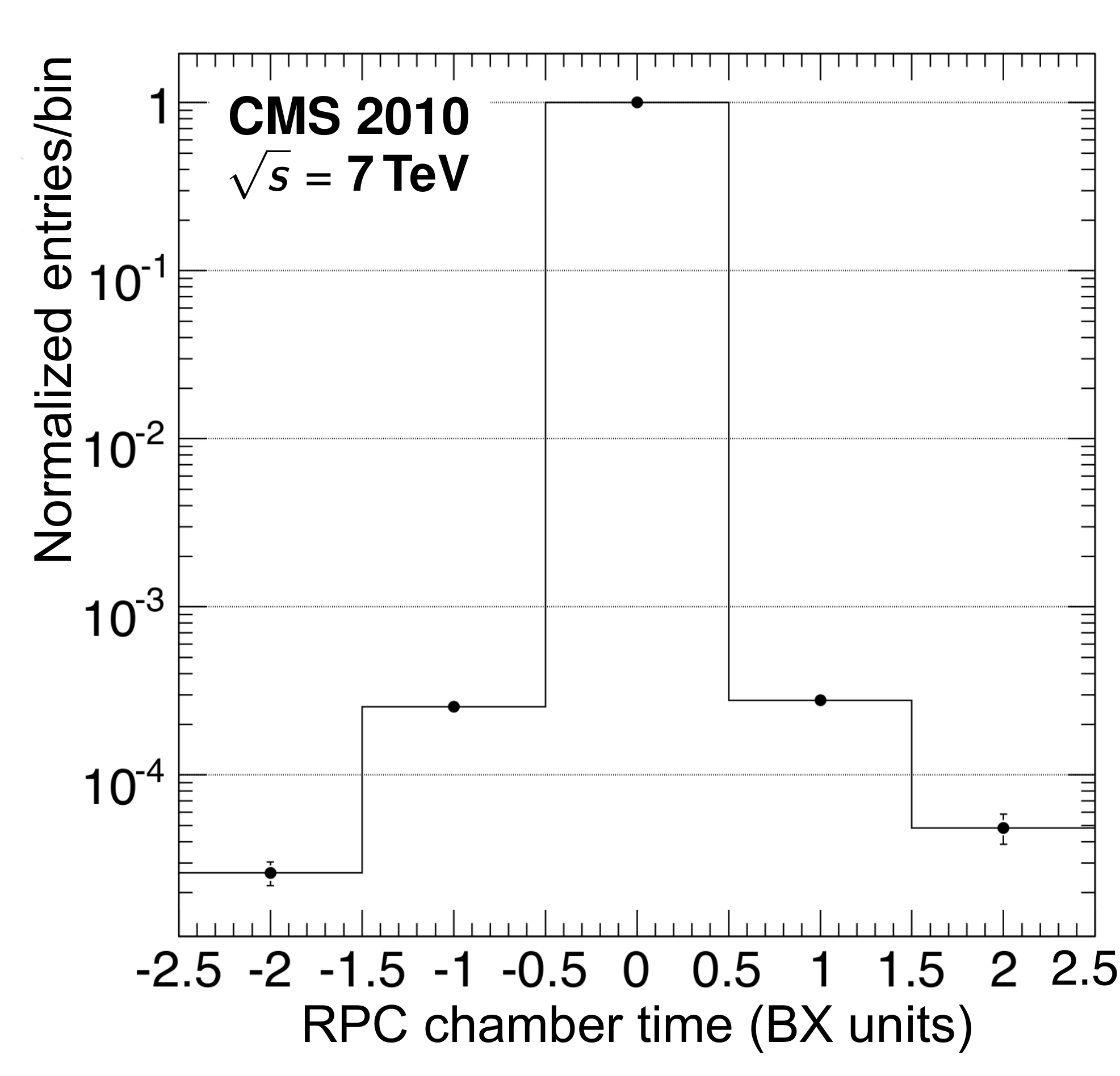}
    \includegraphics[width=0.45\textwidth]{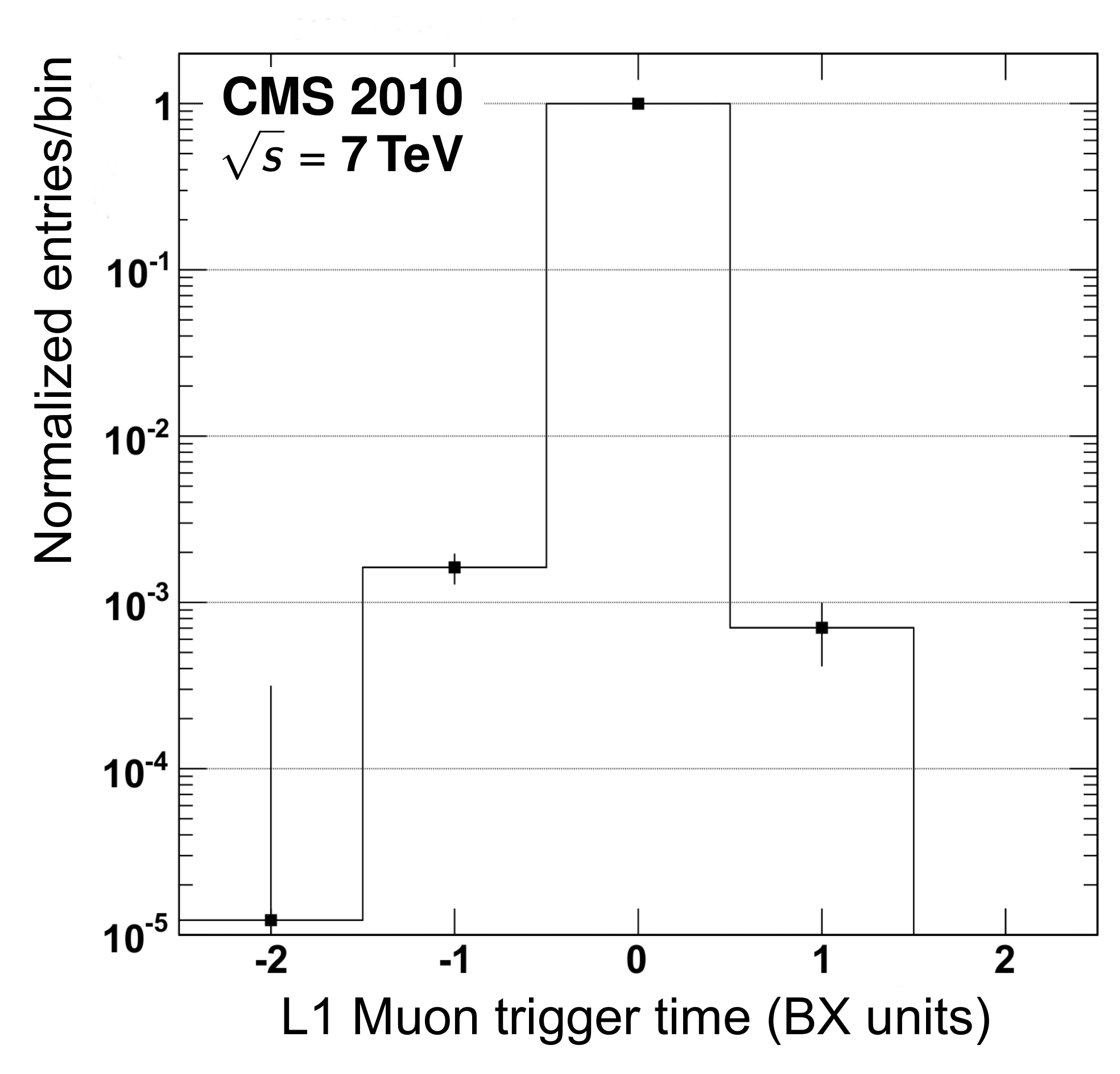}
  \end{center}
  \caption{Time distributions for the chamber-level trigger primitives for (upper left) DT and (upper right) CSC, and for (lower left) RPC hits, relative to the true event BX (1 BX unit = 25\unit{ns}). In each distribution, data from all
chambers were summed together to show the overall subsystem synchronization.  (lower right) Distribution of the combined L1 single-muon trigger.}
  \label{fig:sync}
\end{figure}

Figure~\ref{fig:sync}\,(upper left) shows the time distribution of the highest quality local DT trigger primitives.
 These trigger primitives were constructed with at least 7 out of 8 layers in a chamber and were
found in the DTLT readout window of triggering stations crossed by offline reconstructed muons from LHC collisions.
In this figure, data from all DT chambers were summed together.

Out-of-time primitives symmetrically populate the bins to the right and to the left of the correct
BX. The pre- and post-triggering rates of the highest quality DTLT primitive are both on the order of 2\%.
These are mainly due to the presence of DTLT out-of-time ``ghosts'' that can occasionally
be generated together with in-time trigger primitives. As outlined in Section~\ref{FalseTrig},
this effect has been carefully investigated. Under 2010 timing conditions and LHC luminosities, the efficiency to deliver a trigger primitive at
the correct BX and the low rate of out-of-time triggers are in good agreement
with Muon TDR expectations and simulation studies (see Appendix~\ref{simulation}).

The final set of DTLT timing corrections was applied by the end of August 2010.
Data collected during the
remaining 2010 LHC operation period were analyzed to compute a further set of adjustment parameters,
which were tested at the beginning of the 2011 LHC run.

\subsubsection{CSC trigger synchronization}
\label{sec:CSCsync}

Data from the anode wires and cathode strips of the CSCs are split into 2 paths,  one feeding the readout for data acquisition and the other the trigger.
Within a chamber, the CSC trigger object is called a local charged track (LCT).
An LCT is defined by a pattern of hits on at least 4 layers that is compatible with the straight line segment produced by a  muon from a proton--proton collision~\cite{Hauser:CSCTrigger}.
The LCTs from different chambers are fed into the CSC track finder trigger hardware, which combines them to identify candidate muon tracks.
These candidate tracks are then passed to the main CMS Level-1 trigger system.
As the CMS trigger is a synchronous system, the time associated with these LCTs is crucial for proper operation of the trigger and subsequent synchronization of the front-end readout.
The LCT time is a measure of the BX in which the collision occurred, and the process of identifying this time in BX units is the ``BX assignment''.

The LCT is formed from a coarse time coincidence (within $\pm 3$~BX) of
\begin{itemize}
  \item a cathode LCT (CLCT), formed from the strip hits, and
  \item an anode LCT (ALCT), formed from the wire hits.
\end{itemize}
Since the anode signal timing is more precise than the cathode signal timing, the BX assignment of the LCT is  determined by the ALCT time.
The ALCT signal development is briefly outlined as follows.
When a collision muon passes through a CSC, charge collected on the anode wire is input to a constant fraction discriminator in the anode front-end board.
If the charge is above the detection threshold, a 35-ns pulse is output to the chamber's ALCT board.
Because the pulse is digitized every 25\unit{ns}, the start time of the pulse will determine if the anode hit spans a time period equivalent to 1 or 2~BXs.

In forming an ALCT, the digitized hit pulses are stretched in time to the duration of 6 BXs based only on the leading edge of the pulse.
The ALCT BX assignment is defined as the first BX in which 3 or more layers within the anode pattern contain a hit (to be confirmed by a coincidence of 4 or more layers).
The resulting BX identification efficiency is better than 99\%.

The ``anode hit time'' is defined as either the time of the single BX or the average time of the 2~BXs to which the hit corresponds.
Averaging the chamber anode hit times over several events yields a characteristic ``chamber anode time'', which is sensitive to changes in the clock delays sent to the ALCT electronics boards.
These delays can be adjusted in steps of 2\unit{ns}.

The average chamber anode time in a sample of reconstructed muons was correlated with the fraction of apparently early ($-1$~BX), in-time (BX=0), and late ($+1$~BX) ALCTs.
The ALCT times were then adjusted so that they optimize the fraction of in-time ALCTs by appropriate adjustment of the clock delays for each ALCT board.
After adjustment, the distribution of the difference between the ALCT time and the true event time is shown in Fig.~\ref{fig:sync}\,(upper right).
This distribution is
intentionally
asymmetric:  when 2 ALCTs measure different times, the CSC Track Finder logic chooses the later value, so the optimal performance point is set slightly earlier than the zero of the distribution.

\subsubsection{RPC trigger synchronization}
\label{sec:RPCsync}

The RPCs possess very good intrinsic timing resolution (typically below 2\unit{ns}) \cite{czyrk}, and therefore are very well suited for the task of muon triggering and BX assignment.
Unlike the CSCs and DTs,
the RPC system does not form trigger primitives, but the chamber hits are used directly for muon trigger candidate recognition.

 The signal path can be summarized as follows:
\begin{itemize}
 \item amplification and discrimination at the front-end boards
located on the chambers;
 \item transmission through cables to the link boards situated on the
balconies in the CMS experimental cavern;
 \item zero-suppression and transmission through optical fibers to the
trigger electronics outside of the cavern;
 \item arrival at the electronics room and distribution of signals to
the individual processing elements of the pattern comparator (PAC) trigger system.
 \end {itemize}

Because hit signals are discriminated at the chamber level, there are
different offsets for individual chambers.
Signals coming from a single chamber may be shifted in time in
steps on the order of 0.1\unit{ns} to achieve synchronization.
By using cable length values and muon time-of-flight, it was
possible to produce a first approximation of synchronization constants.
These were further refined during studies of beam halo and beam splash events.
Beam splash events are recorded during intentional beam dumps about 100\unit{m} upstream from the CMS detector that result in large fluxes through the chambers of about 5 synchronous muons$\unit{cm}^{-2}$ that are parallel to the beam line.

After the start of LHC collisions, the recorded data were used
to further improve the synchronization. The experimental
procedure consisted of selecting global muons with tracks that would
cross the RPC system.
Furthermore, only the first hit from any chamber was selected, since a particle crossing an RPC may sometimes provide
afterpulses. The distribution of the hit time, in units of BX, relative to the true collision time was studied for all chambers.
In cases where the distribution was asymmetric or shifted, the synchronization parameters were
adjusted to obtain a symmetric distribution centered at 0. This
procedure was repeated 3 times and the final results are presented in Fig.~\ref{fig:sync}\,(lower left).
The data used here correspond roughly to 1
million muon tracks. Hits outside the central bin
contribute significantly less than 0.1\%.
The fraction of out-of-time hits decreased by roughly 50\% after the third iteration.

\subsubsection{L1 muon trigger synchronization}
\label{sec:L1sync}

L1 muon triggers are formed from trigger primitives (DT, CSC) or hits (RPC) forwarded from multiple chambers and possibly from different muon subsystems.
The BX identified by the overall L1 trigger is assigned by regional triggers~\cite{trigTDR}. It is determined according to a logic that depends on
the muon subsystems involved and combines primitive or hit information, reducing the contribution of
early/late signals from individual chambers.
The HLT filtering biases the use of normally triggered data for studies of out-of-time L1 triggers, since such events may be rejected.

To deal with this effect, a dedicated DAQ stream was developed to collect at high rate
a fraction of the CMS raw data content consisting only of L1 trigger information, before any HLT processing.
By comparing the distributions of L1 muon trigger times to the expected collision times, one can measure the L1 synchronization.

Results obtained this way for L1 muon triggers with no minimum \pt requirement are shown in Fig.~\ref{fig:sync}\,(lower left).
This dedicated data stream did not contain the information required for reconstruction of muons and hence explicit rejection of cosmic and beam halo backgrounds was not possible.
Instead, the L1 trigger rate attributable to cosmic rays was
computed from regions of the LHC orbit that were not populated in a given beam fill pattern.
Likewise, the beam background was estimated from the L1 muon trigger rate
when a single bunch was present in CMS.
After contamination from cosmic-ray muons and beam halo was subtracted, the fraction of pre- and post-triggered events was below
0.2\% and 0.1\%, respectively.  These results exceeded the Physics TDR expectations of 99\% in-time triggering~\cite{TDR}.

\subsection{Measurement of the DT and CSC local trigger efficiency}
\label{Trig_Effic}

To measure the DTLT efficiency, selected events are required to
be triggered by the RPC system,
without any requirement on the presence of the DT trigger, which
could otherwise bias the measurement.
The presence of a reconstructed muon track in the event is required.
To remove contamination from cosmic rays, muon candidates must have an impact parameter in the transverse and longitudinal
planes within
$|d_{xy}| <$ 0.2\unit{cm} and $|d_{z} | <$ 24\unit{cm}, respectively.
Their pseudorapidity must be in the range $ | \eta |<1.2$
to be within the acceptance of the inner stations of the muon barrel.
The number of DT
hits associated to the track $N_\mathrm{DT}$ must be greater than 3.
Such hits can be located anywhere along the track, and not necessarily
in a single muon station.
This requirement does not introduce any bias on the efficiency measurement, as at least
4 aligned hits in 1 SL are necessary to deliver a trigger primitive.
Poorly reconstructed muon tracks are removed by requiring the
normalized $\chi^2$ of the track fit to be less than 10.
The track transverse momentum is required to be $\pt > 7\GeVc$, to allow
the particle to reach the outer station of the muon barrel.

\begin{figure}[hbtp]
  \begin{center}
\includegraphics[width=7.5cm]{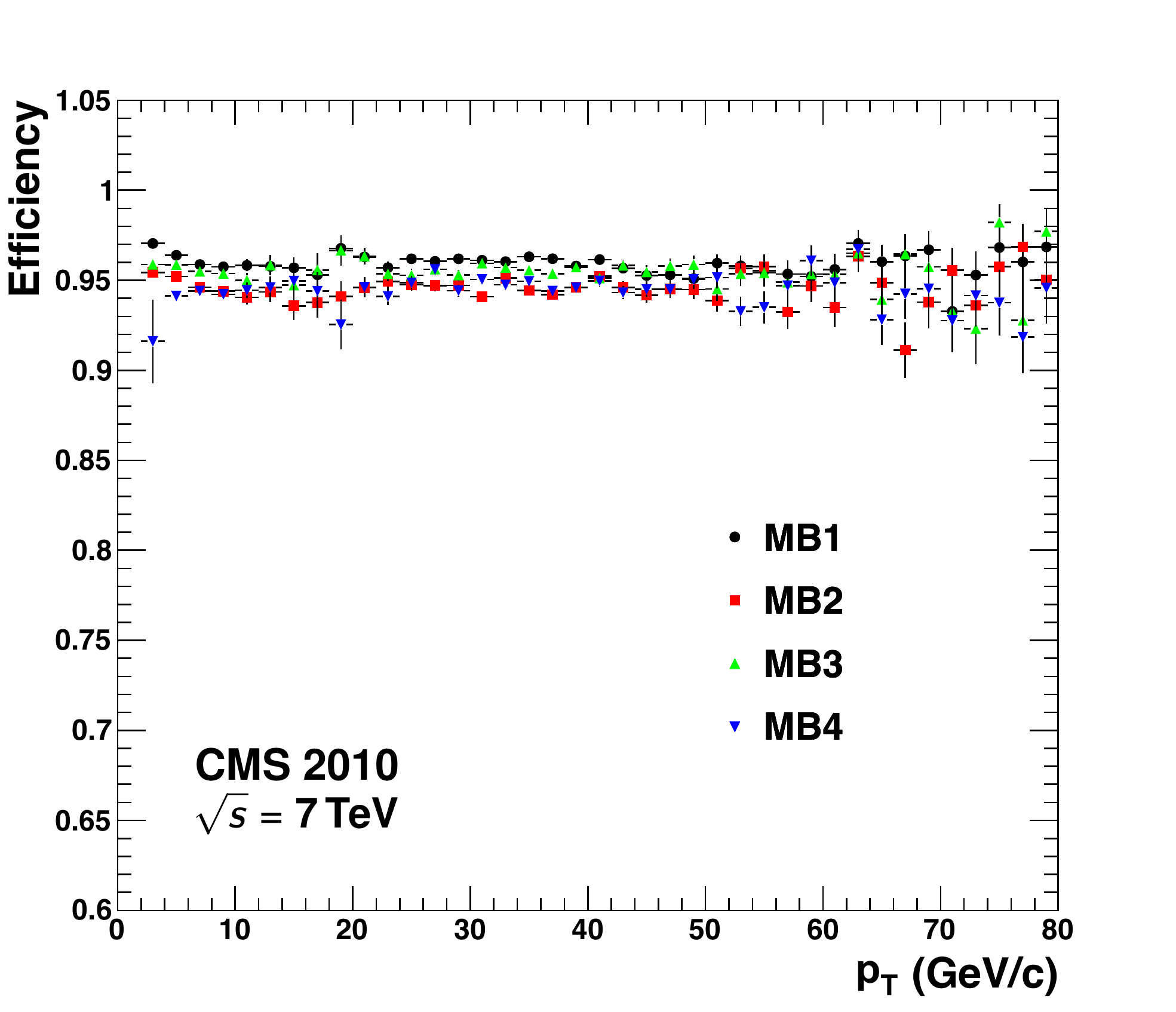}
\includegraphics[width=7.5cm]{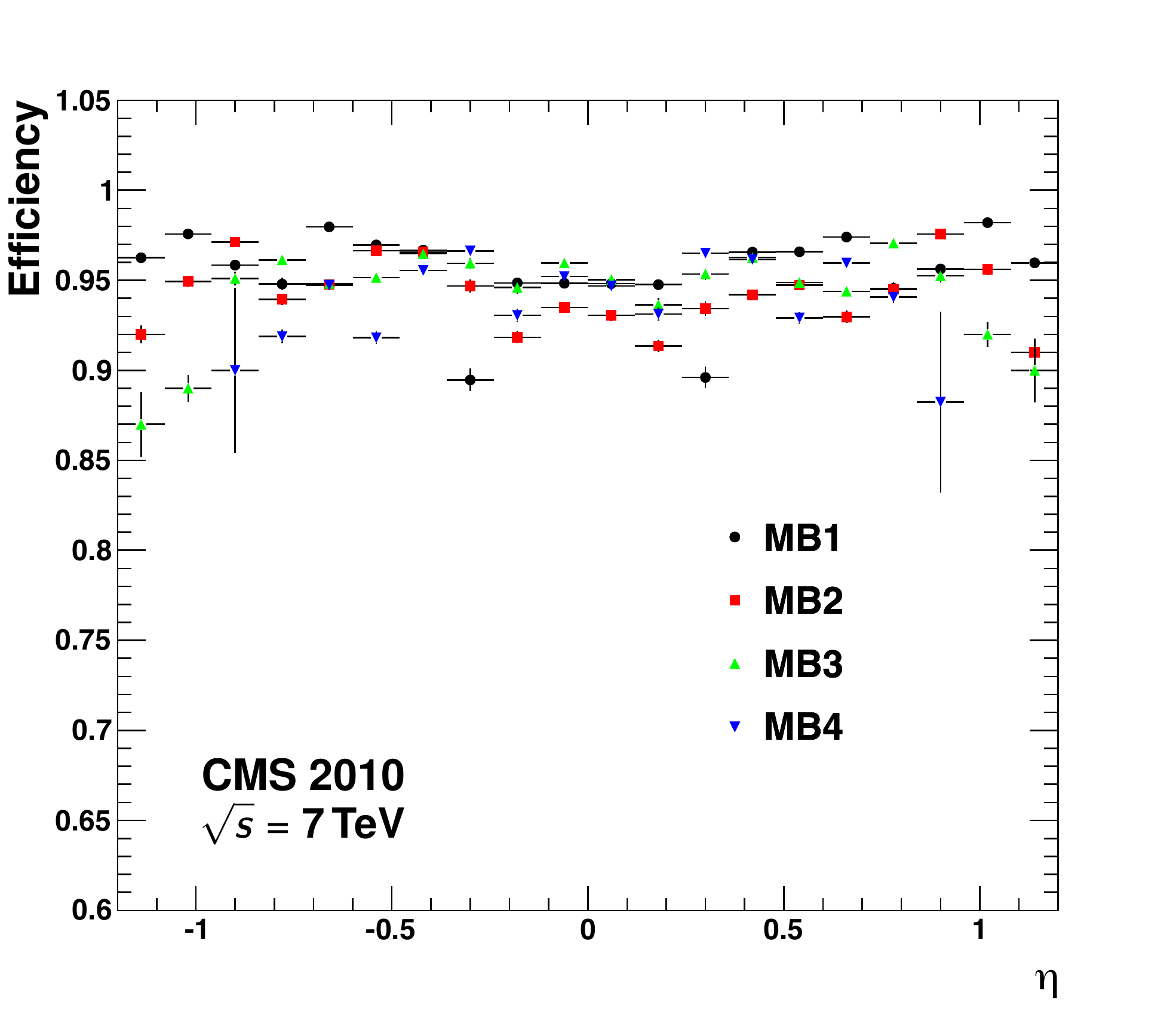}
\includegraphics[width=7.5cm]{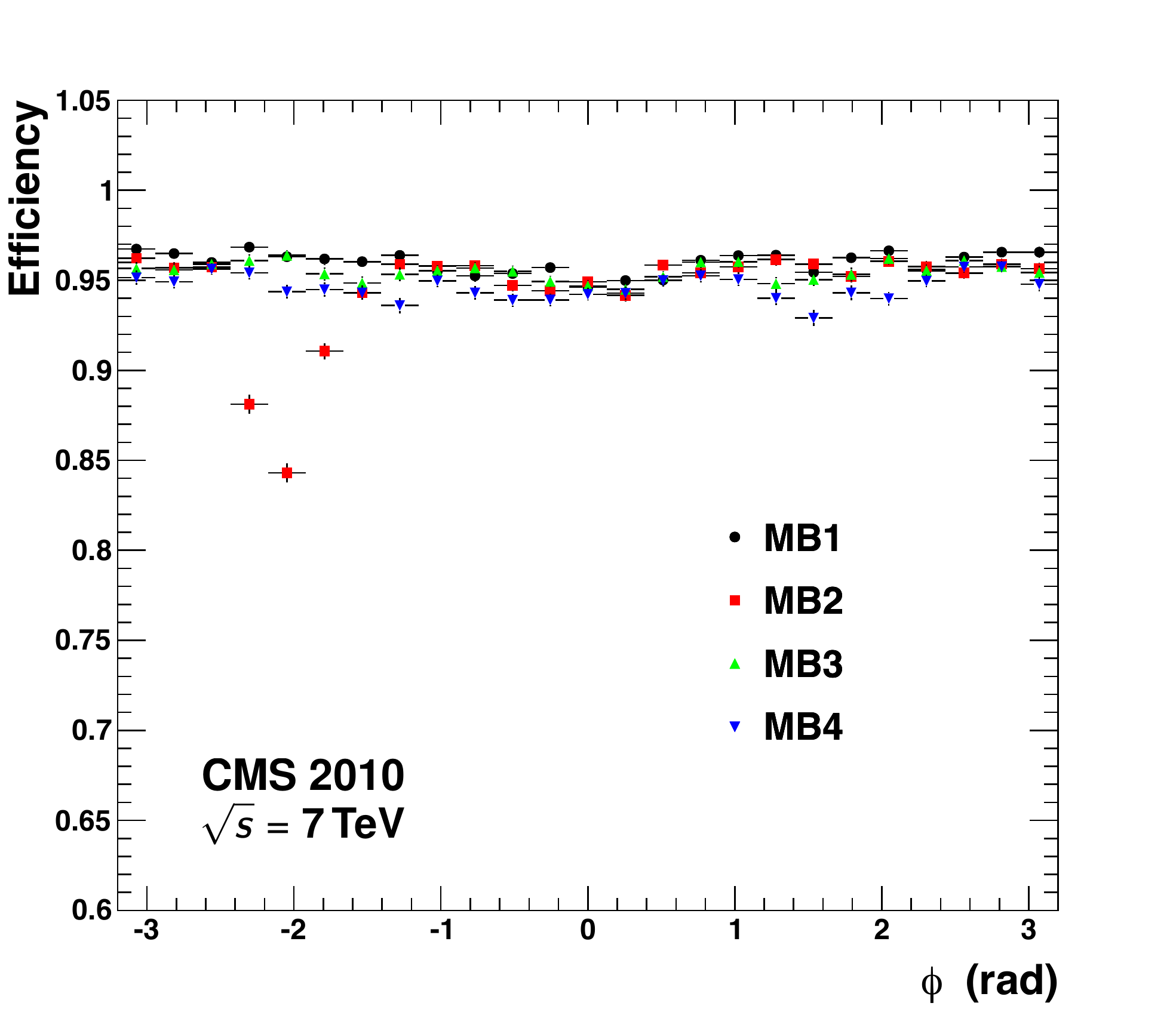}
 \caption{ The measured DTLT efficiency as a function of the muon transverse
momentum \pt, the
pseudorapidity $\eta$, and the azimuthal angle $\phi$.
Results for the 4 stations are superimposed.
}
    \label{fig:DTeffic_vs_pt_eta_phi}
  \end{center}
\end{figure}
To measure the DTLT efficiency in a chamber, the presence of a
track segment associated
with the selected muon track is required.
The segment must be
reconstructed  with at least 4 out of 8 hits in its $\phi$ view.
In addition, it is required that
$| \psi | < 40^\circ$, where $\psi$ is the
local track segment angle with respect to the direction to the interaction point in the CMS bending plane.
This allows the segment to be fully contained in
the angular acceptance of the DTLT units.
If more than a single track segment is found in the chamber,
the event is not used for the efficiency calculation.
The efficiency of the DTLT in a given chamber is defined
as the fraction of
selected track segments
with an associated trigger primitive.
This definition allows the effective trigger efficiency to be measured and eliminates effects related to the geometrical
acceptance, which account for a few percent of the inefficiency.

\begin{table}[hptb]
\begin{center}
\topcaption{Average DTLT efficiencies for the different station types for 2010 collision data and simulation. Results that include the correct BX identification (BXID) are also shown. The uncertainties include both the statistical and systematic components summed in quadrature.}
\begin{tabular}{|c|c|c|c|c|}
\hline
                    & \multicolumn{2}{c|}{DTLT Efficiency  (\%)}  & \multicolumn{2}{c|}{BXID Efficiency (\%)} \\  \cline{2-5}
    Station   &  Data             & Simulation      &  Data             & Simulation  \\ \hline\hline
MB1            & 96.2 $\pm$ 0.1 & 97.9 $\pm$ 0.9   & 94.5 $\pm$ 0.9 & 96.4 $\pm$ 0.9     \\ 
MB2            & 95.7 $\pm$ 0.8 & 98.0 $\pm$ 0.8   & 94.0 $\pm$ 0.8 & 96.7 $\pm$ 0.7    \\ 
MB3            & 95.8 $\pm$ 0.9 & 98.3 $\pm$ 0.8   & 93.8 $\pm$ 0.9 & 96.9 $\pm$ 0.8     \\ 
MB4            & 95.0 $\pm$ 0.1 & 97.1 $\pm$ 0.9  & 93.0 $\pm$ 0.9 & 95.6 $\pm$ 0.9     \\
\hline
\end{tabular}
\label{DT:Trig-Results}
\end{center}
\end{table}

During the 2010 data-taking period, about 3\% of the DT chambers
suffered from hardware failures that affected the DTLT efficiency.
Removing these chambers,
the average DTLT efficiency in a station is 95.7\%, to be compared with 97.8\% obtained with simulated events.
The average DTLT efficiency is shown in Table~\ref{DT:Trig-Results}, for data and simulation, for the 4 barrel stations.
The uncertainties are dominated by systematic effects. Differences in the DTLT efficiency
from station to station are caused by small shifts in the time synchronization
of the local trigger electronics and differences in the average angular incidence
of the muon tracks. The overall systematic uncertainty is estimated from the
observed spread of the measured DTLT efficiencies over the various stations,
after removing stations
with known hardware problems.
The lower efficiency measured in the data compared with the simulation is partially due to small differences in the timing of the muon stations with respect to ideal conditions, and was a subject of further investigation during the 2011 data-taking campaign.
The DTLT efficiency as a function of the muon transverse momentum \pt, the
pseudorapidity $\eta$, and the azimuthal angle $\phi$ is shown in Fig.~\ref{fig:DTeffic_vs_pt_eta_phi}, for the 4 stations.
All the DT chambers are used for this measurement,
and the inefficiency observed in the region   $-2.5 < \phi < -1.5$ radians
is due to a known hardware failure in a single MB2 station.

If the trigger primitive
is also required to correctly assign the BX at which
the muon candidate is produced, the average DTLT efficiency
decreases to 93.8\%. This is more than 1\% better
than the design (L1 Trigger TDR) performance~\cite{trigTDR}.
Results for the DTLT efficiency including the correct
BX assignment are shown in Table~\ref{DT:Trig-Results}.

The important component of the CSCLT efficiency is the efficiency for creating CLCT candidates, since the CLCTs provide the CSC L1 trigger track finder with the critical information about the bending of a muon in the magnetic field. Two methods are used to measure the CLCT efficiency: a ``single-track matching'' method and the standard CMS ``tag-and-probe'' method~\cite{WZcross_section}.
Both methods are based on tracks reconstructed using silicon tracker detector information alone (so-called ``tracker tracks'') to allow efficiency measurements free of any bias from the use of muon detector information in track reconstruction.
Contrary to the DTLT efficiency measurement, the presence of an RPC trigger is not required in the event selection, since the RPCs only cover $|\eta| < 1.6$.
As shown in Section~\ref{TimingAndSync}, the BX identification efficiency of the CSCs could be adjusted to exceed 99\% (even better than the TDR design of 99\%~\cite{trigTDR}), so out-of-time BX assignment is an insignificant contribution to CSCLT inefficiency.

In the single-track matching method, a high-quality tracker track that projects to match a track segment in a CSC station  is selected. The CLCT efficiency is measured in any upstream station through which the track must have passed.
If a track is matched to a segment in station ME2, then station ME1 is examined for the presence of a CLCT. If the track is matched to a segment in ME3, then ME2 is examined.
Downstream of ME3, only ME4/1 exists, so only part of ME3 can be probed. In general the method is effective only for chambers in stations ME1 and ME2.
Only tracker tracks that have $\eta$ within the CSC geometrical coverage, $0.9 < |\eta| < 2.4$, are used.
A high-quality track is identified by requiring that the transverse impact parameter $| d_{xy} |$ be less than 0.2\unit{cm} and the longitudinal impact parameter $| d_z |$ be less than 24\unit{cm}, that there be at least 11 associated hits in the silicon tracker, that the normalized $\chi^2$\ of the track fit be less than 4, that the $\eta$ and $\phi$ uncertainties be less than 0.003, that the momentum be above 15\GeVc, and that the relative \pt uncertainty be $\Delta\pt/\pt < 0.05$.
The track must cross the chamber in which it matches a segment at least 5\unit{cm} away from the chamber's edge to ensure it is well within the geometrical coverage of the system, and it must be the only track to cross that chamber.
The track is required to project to within 10\unit{cm} of a CSC track segment, $D_{\rm trk-seg} < 10$\, cm, where $D_{\rm trk-seg}^2 = (X_{\rm trk-proj}-X_{\rm seg})^2 +(Y_{\rm trk-proj}-Y_{\rm seg})^2$, to confirm that the track indeed reached that station.
In the upstream station in which CLCT efficiency is measured, a track is considered to be associated with a CLCT if it projects to within 40\unit{cm} of a CLCT: $D_{\rm trk-LT} < 40$\unit{cm}, where $D_{\rm trk-LT}^2 = (X_{\rm trk-proj}-X_{\rm LT})^2 +(Y_{\rm trk-proj}-Y_{\rm LT})^2$.

A track that projects into a chamber known to be inoperative because of hardware or electronic board failures is not used in the computation of the efficiency. (In 2010, typically about 8 of the 473 chambers in the system were inoperative at any given time.)

In the tag-and-probe method, dimuons from \JPsi and Z decays are used. They are, however, collected with an inclusive single-muon trigger, so that one of the muons (the probe) is unbiased by the performance of the muon system.
The triggering muon (the tag) is identified as a muon by the standard muon selection criteria, involving selection of a track in the silicon tracker matched to information in the muon detectors.
The probe track is not identified as a muon other than by forming an invariant mass with the tag muon near the resonance mass.
The Z sample was selected from events collected using a single-muon trigger, and the  \JPsi sample from events collected using a dedicated  \JPsi trigger in which only 1 of the 2 muons was required to trigger the muon system, and the other only had to be a track reconstructed in the silicon tracker.
The tag track must fulfill the following silicon tracker track
selection criteria: $| d_{xy} | < 0.2$\unit{cm} and $| d_z | < 24$\unit{cm},
$\pt > 5 \GeVc$, at least 11 hits in the
tracker, and the normalized track $\chi^2 < 4$.
The tag track can be in the barrel or endcap, and
is required to match the HLT object that
triggered the event within an angular distance
$\Delta R < 0.4$, where
${\Delta R}^2=(\eta_\text{trk}-\eta_\text{HLT})^2 +(\phi_\text{trk}-\phi_\text{HLT})^2$,
and $\eta_\text{trk}$,  $\phi_\text{trk}$, $\eta_\text{HLT}$, and $\phi_\text{HLT}$ are
the $\eta$ and $\phi$ (in radians) values of the track and the HLT candidate, respectively.
The tag track is also required to be associated with at least
2 track segments in 2 different muon stations. The probe
is a
tracker track fulfilling the same selection criteria as for the probe in the single-track matching method described earlier. The tag and probe tracks are also required to
share the same primary vertex.

The invariant mass of the tag and probe tracks is fitted to extract the \JPsi and Z signal
event yields for both the denominator
and the numerator of the efficiency ratio, with the requirement that all probes in the numerator match a CLCT candidate.
Several different fitting functions for peak and background are used, and the spread in the results arising from the different choices is assigned as a systematic uncertainty.

The CLCT efficiency is measured for both data and simulated events.
The single-track matching method is applied to minimum-bias data, whereas the tag-and-probe method is applied to the \JPsi and Z samples.
The average efficiencies from both methods are shown in Table~\ref{CSC:Trig} for each station (although single-track matching only provides measurements in
stations ME1 and ME2).
For the tag-and-probe method the data are from the combined \JPsi and Z samples.

Both methods are susceptible to possible systematic effects arising from the choice of selection criteria for the tracks, CLCTs, and reconstructed segments.
The tag-and-probe method involves subtraction of background events not originating from \JPsi and Z decays, and this too can introduce systematic differences.
The robustness of the values for the extracted efficiencies has been examined by varying the selection criteria over a range of reasonable values.
The results are stable to within 1\%--2\% and we therefore estimate the systematic uncertainty on each value to be at this level.
For the single track matching, this dominates the statistical uncertainty.
For the tag-and-probe method the systematic and statistical uncertainties are of similar magnitude.

For the single-track matching method, the simulated and measured efficiencies agree, while for the tag-and-probe method the simulation slightly underestimates the trigger efficiencies.
The efficiencies measured by the tag-and-probe technique tend to decrease with distance from the IP, in both data and simulation.
This is consistent with a few probe muons not actually reaching the probed station because of losses in the intervening magnet yoke steel, since there is no guarantee that a probe muon actually reaches the probed station.
In the single-track matching case, the track always reaches a chamber downstream of the station in which the efficiency is measured.

\begin{table}[hptb]
\begin{center}
\topcaption{Average CLCT efficiencies per station; the statistical uncertainties are shown.}
\begin{tabular}{|l|c|c|c|c|c|}
\hline
& \multicolumn{4}{|c|}{Cathode Trigger Primitive Efficiency (\%)} \\  \cline{2-5}
& \multicolumn{2}{|c|}{Single-Track Matching Method} & \multicolumn{2}{|c|}{Tag-and-Probe Method} \\  \cline{2-5}
Station & Data & Simulation & Data &   Simulation \\ \hline\hline
ME1 & 97.9 $\pm$ 0.1 & 99.0 $\pm$ 0.1 & 98.7 $\pm$ 0.9 & 97.2 $\pm$ 0.1 \\ 
ME2 & 97.0 $\pm$ 0.1 & 96.7 $\pm$ 0.1 & 95.6 $\pm$ 0.9 & 94.2 $\pm$ 0.2  \\ 
ME3 & -- & -- & 96.0 $\pm$ 0.9 & 92.5 $\pm$ 0.2 \\ 
ME4 & -- & -- & 94.5 $\pm$ 1.6 & 89.8 $\pm$ 0.3  \\ \hline
\end{tabular}
\label{CSC:Trig}
\end{center}
\end{table}

\begin{figure}[hptb]
{\centering
\includegraphics[height=6.2cm]{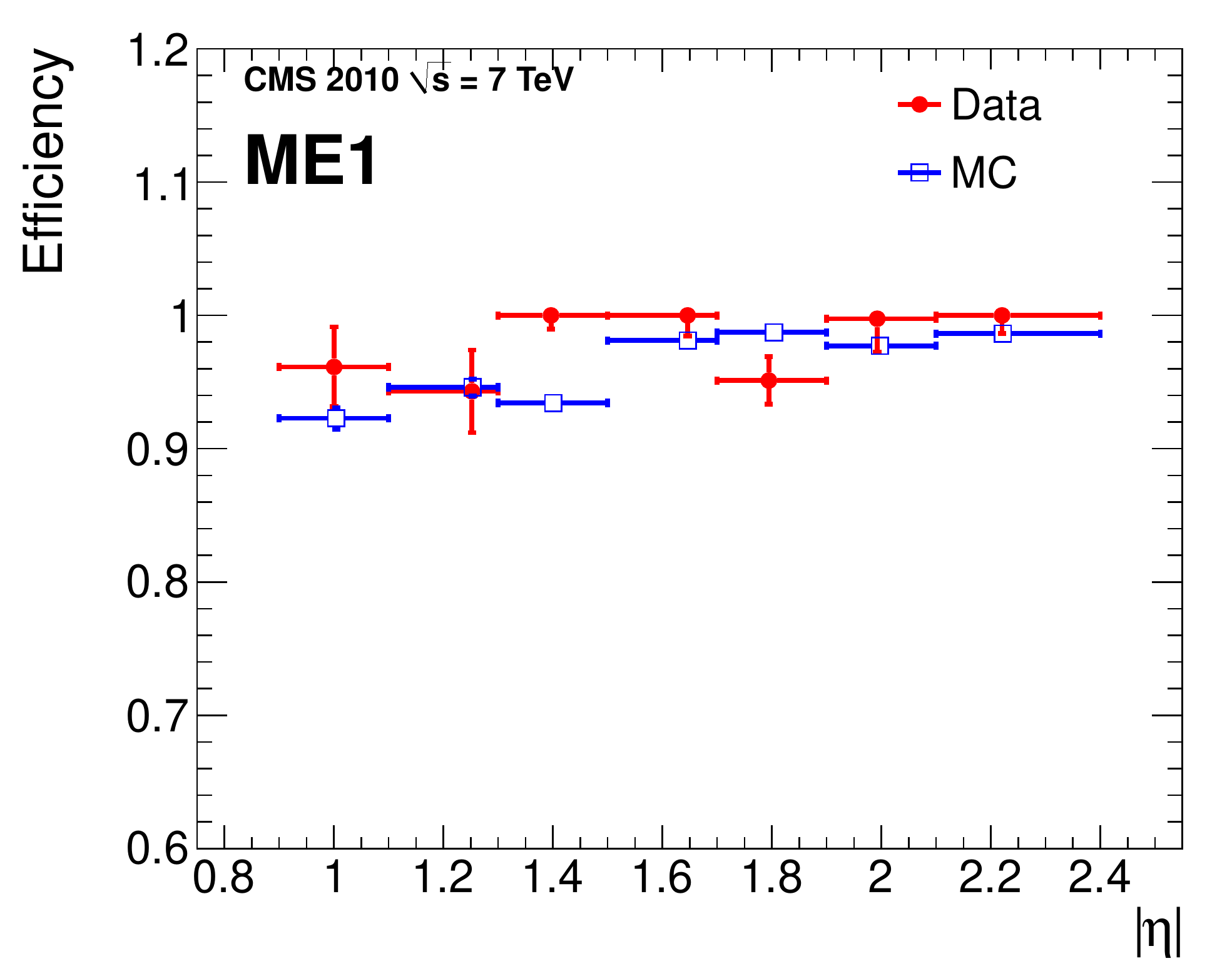}
\includegraphics[height=6.2cm]{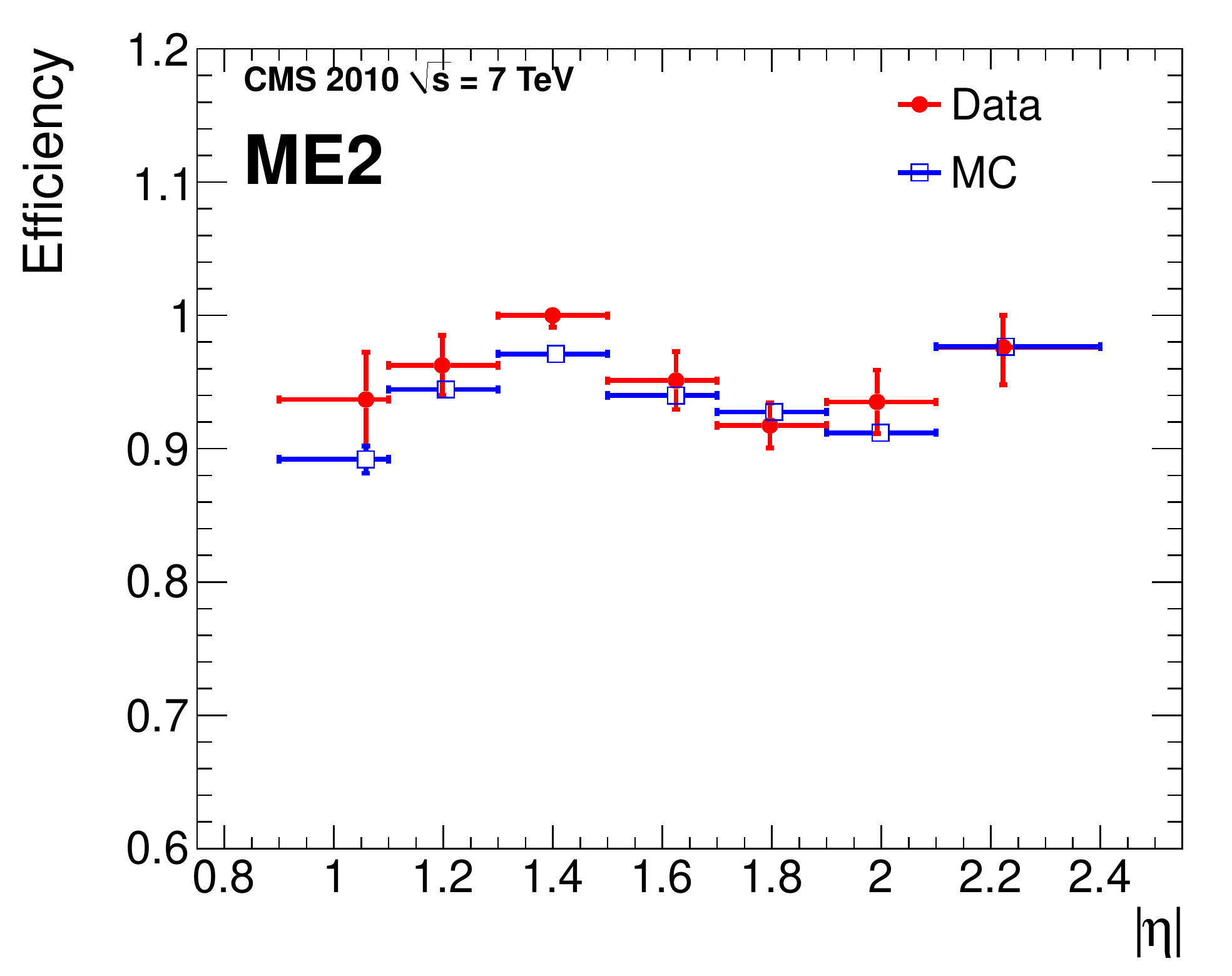}
\includegraphics[height=6.2cm]{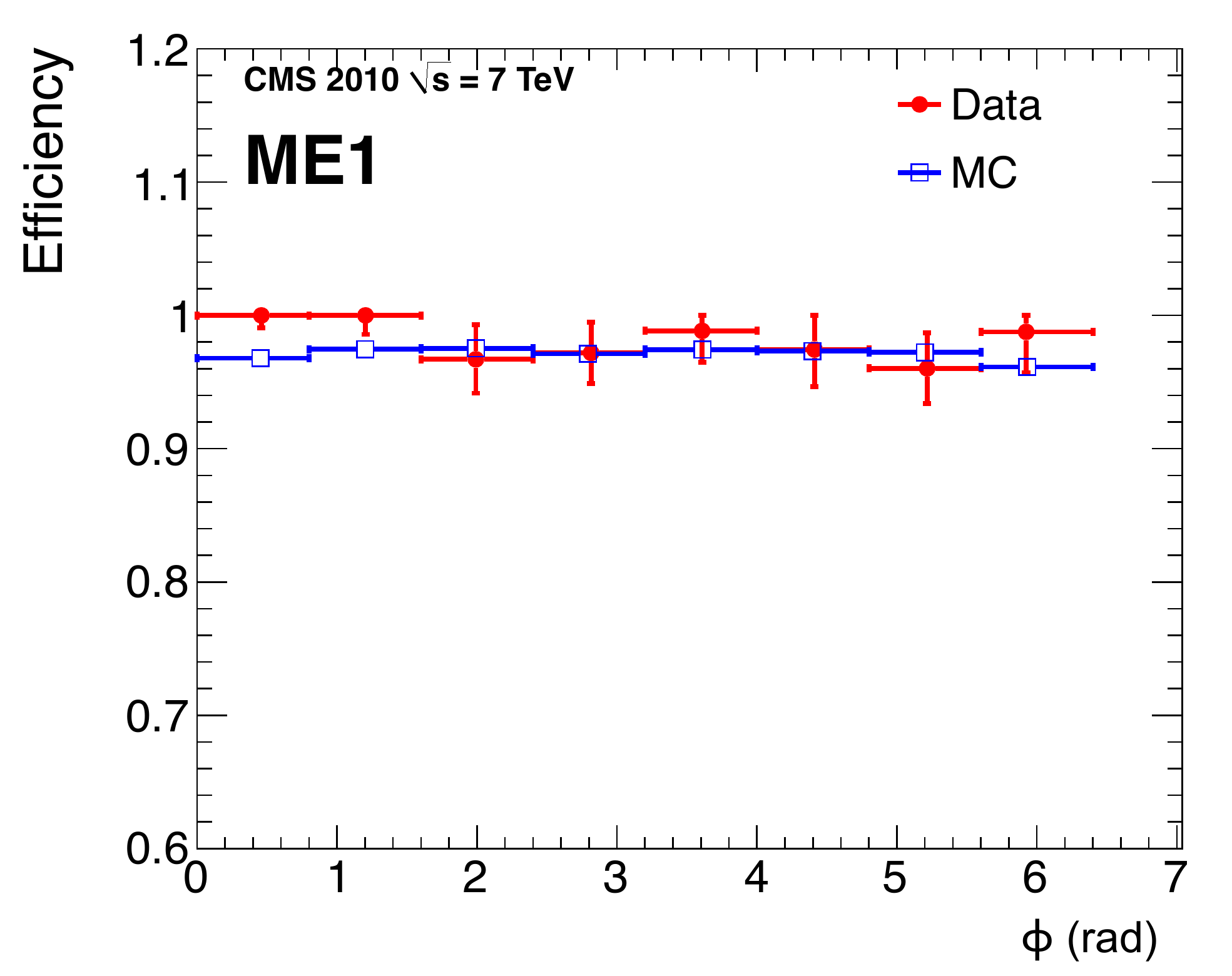}
\includegraphics[height=6.2cm]{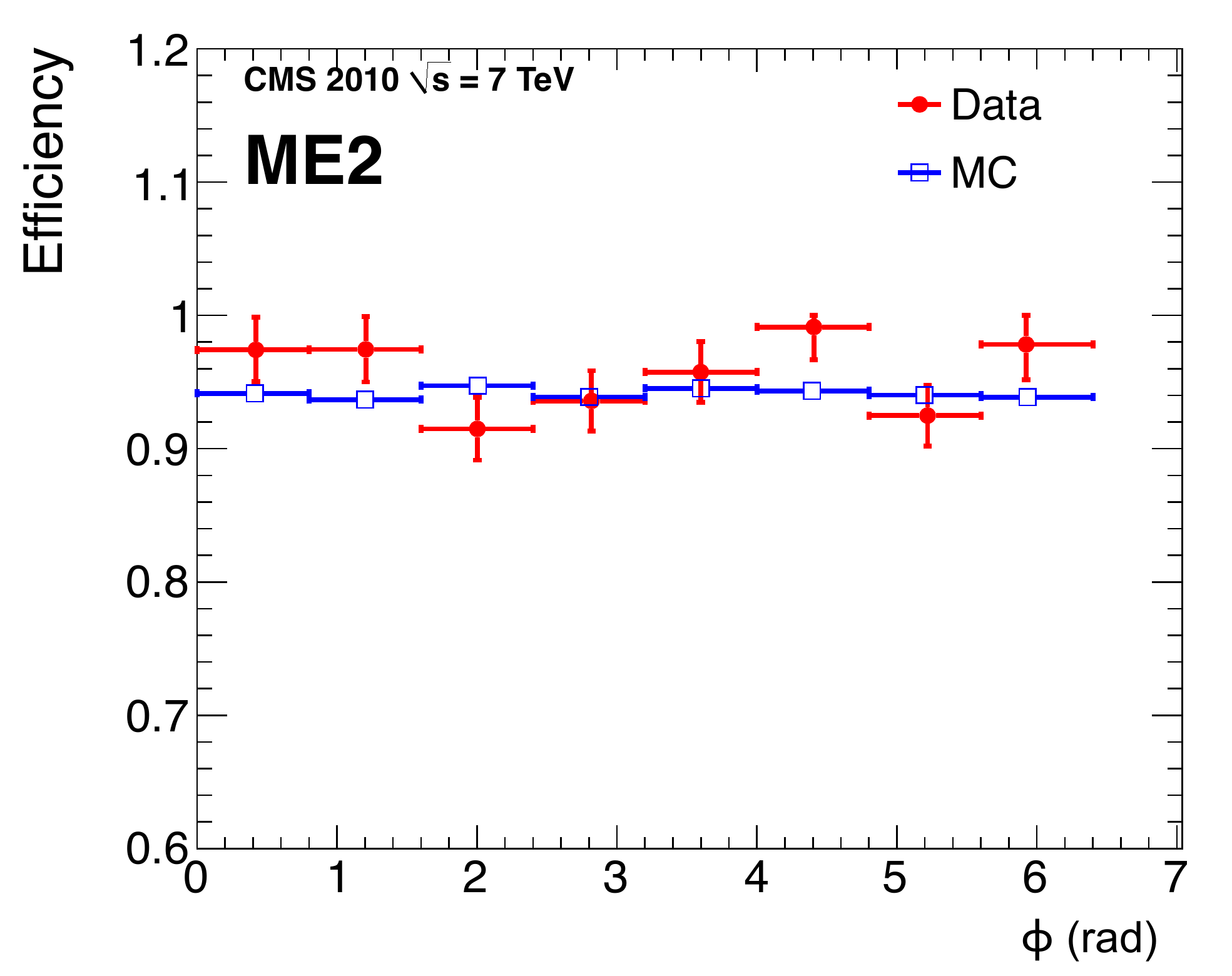}
\includegraphics[height=6.2cm]{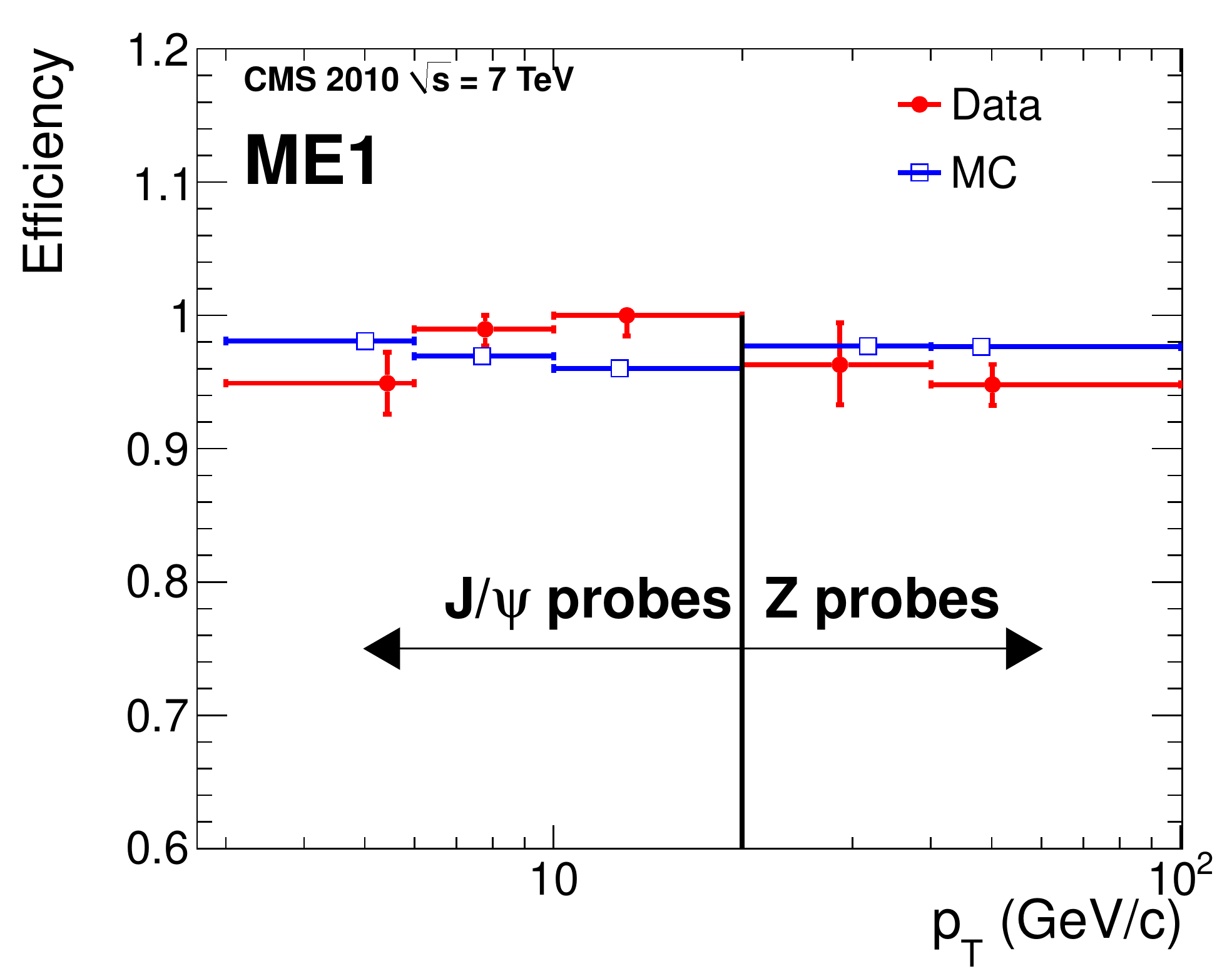}
\includegraphics[height=6.2cm]{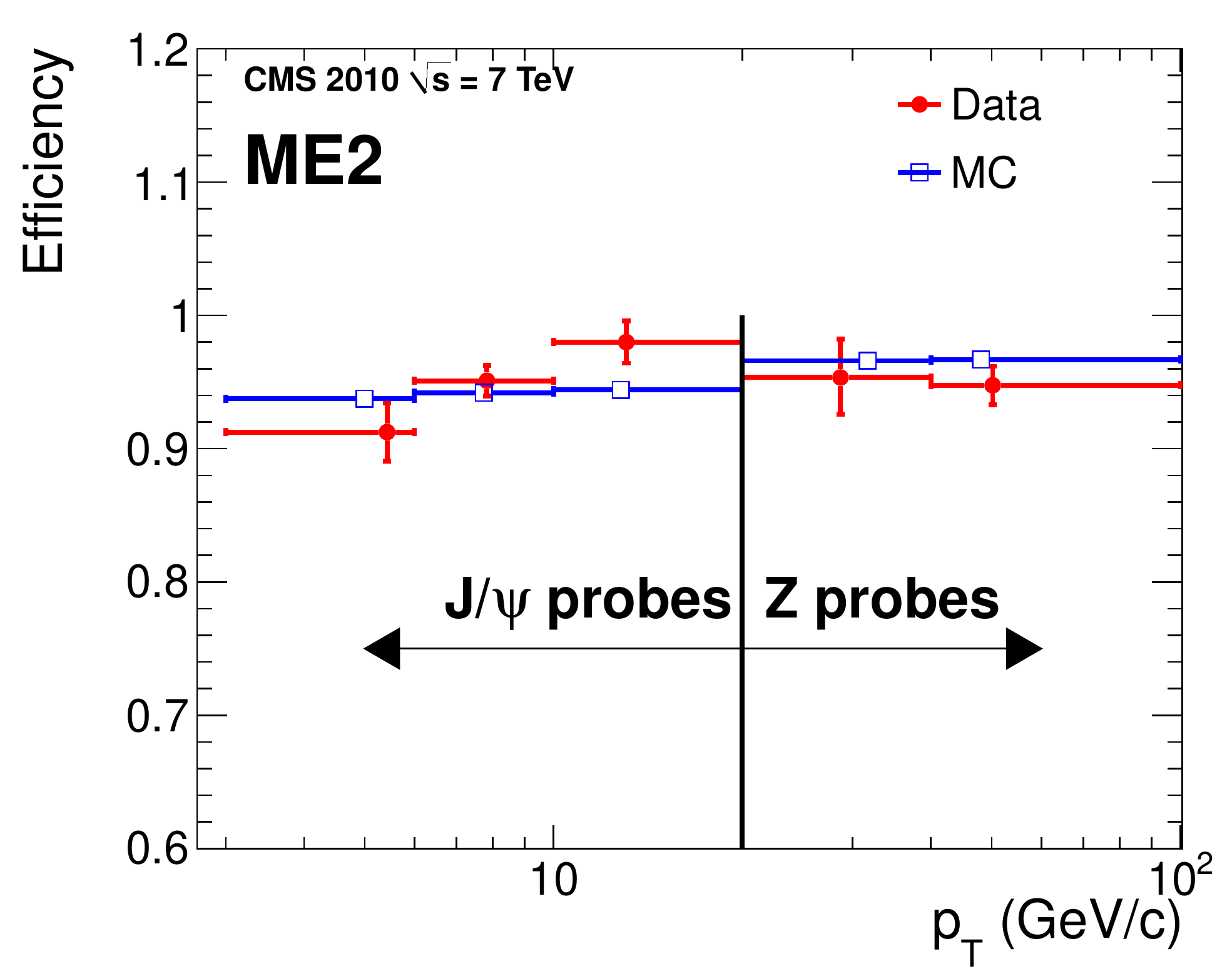}
\caption{ Comparison between CSCLT efficiencies measured using the ``tag-and-probe''  method (red; filled circles) and simulation (blue; open squares)
as functions of muon pseudorapidity $\eta$ (top), azimuthal angle $\phi$ (middle),
and transverse momentum \pt (bottom) for ME1 (left)
and ME2 (right) stations.
The vertical line on the \pt distributions separates the ranges covered by probes
originating from \JPsi (left side) and Z (right side) decays.
The statistical uncertainties are shown as vertical error bars.
The horizontal error bars show the range of each bin, and within each bin the data point is positioned at the weighted average of all values within that bin.}
\label{TP:TPSMLCTEff}
}
\end{figure}

The CLCT efficiency measured using the tag-and-probe method
is compared to simulation in Fig.~\ref{TP:TPSMLCTEff} for the ME1 and ME2 stations
as functions of $\eta$,
$\phi$, and \pt.
Similar results are obtained from the single-track matching method.

\subsection{False local triggers}
\label{FalseTrig}

\begin{figure}[hbtp]
  \begin{center}
\includegraphics[width=7.cm]{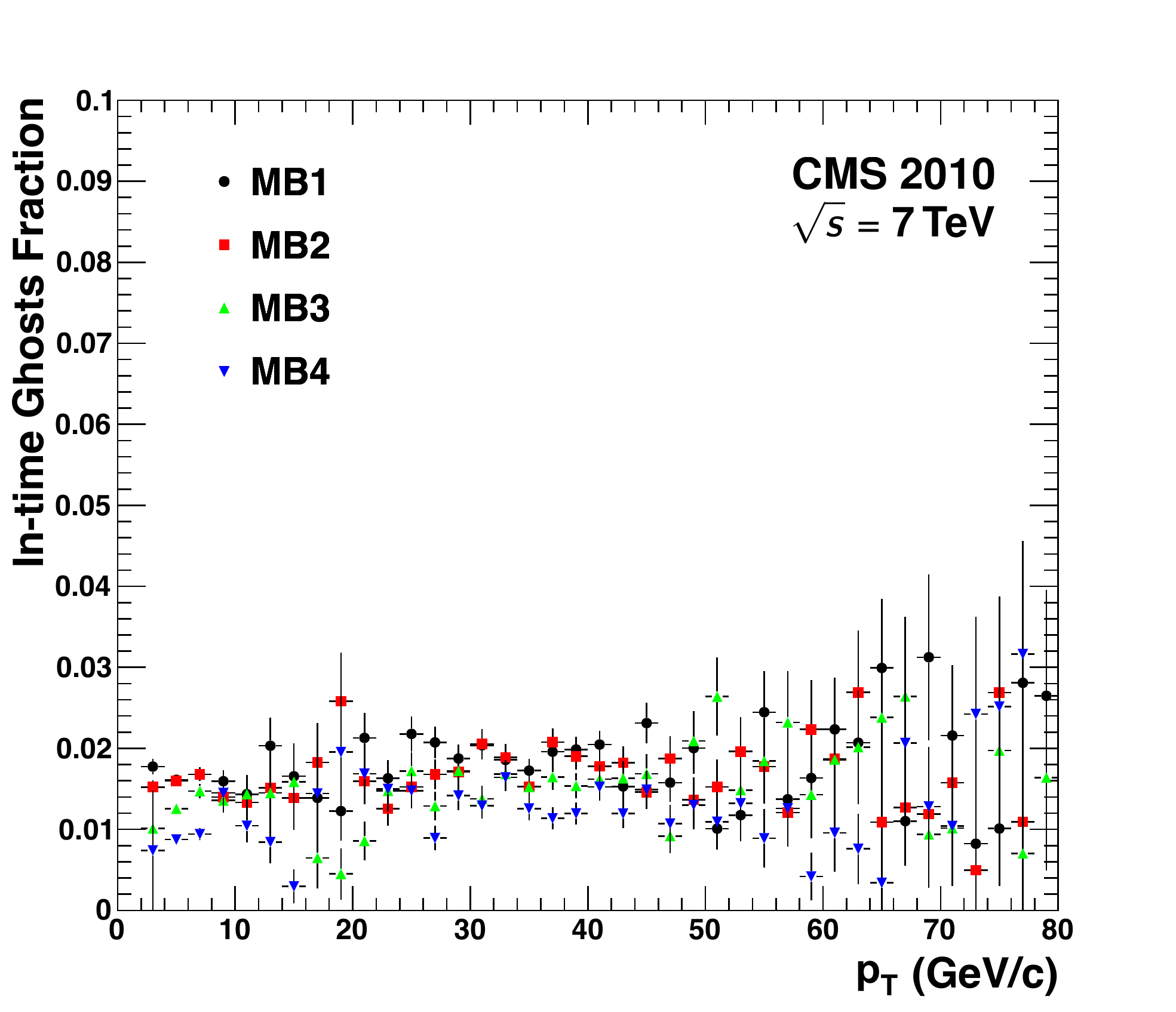}
\qquad
\includegraphics[width=7.cm]{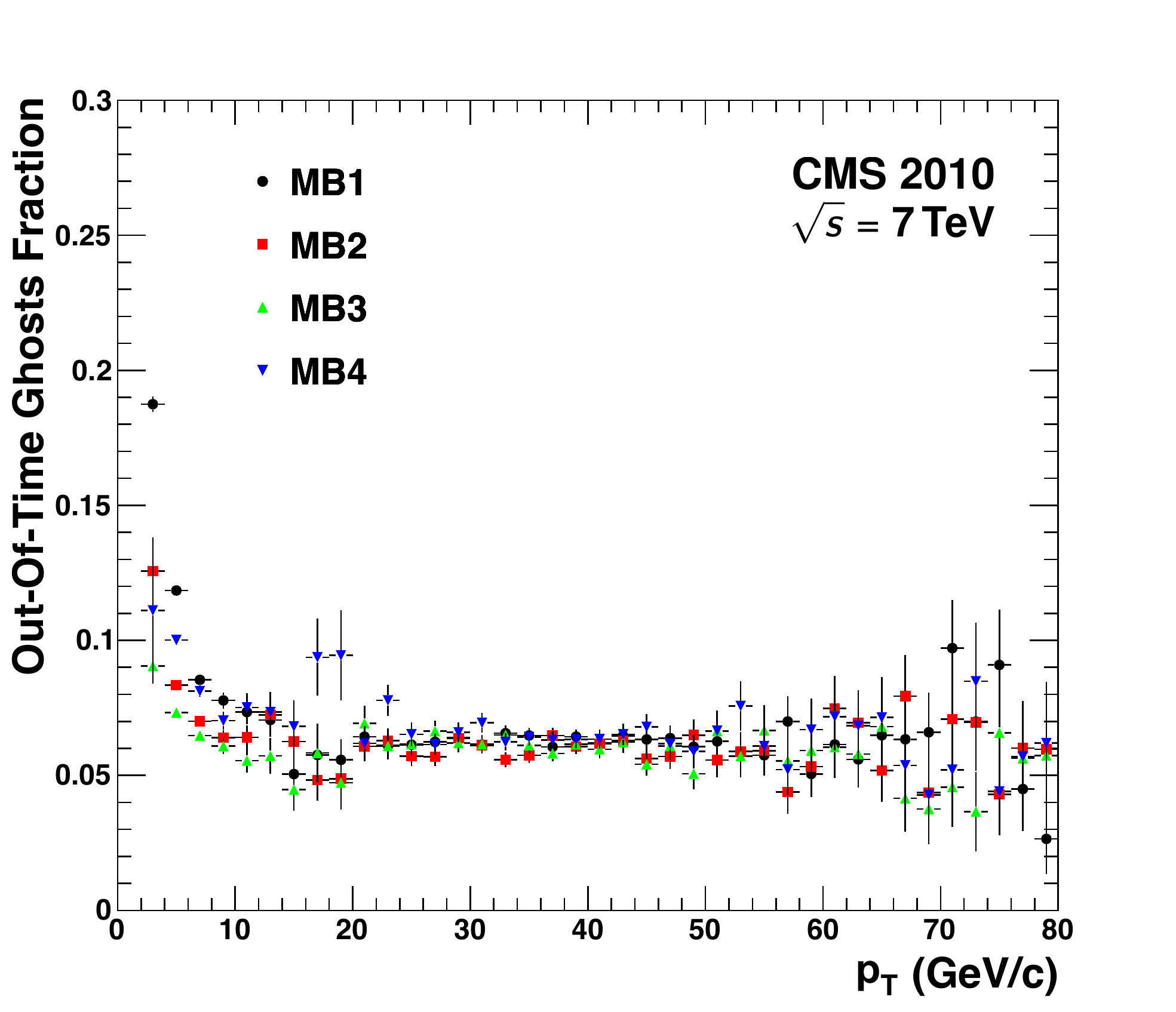}
    \caption{ Left: Fraction of the DTLT in-time ghosts as a function of the muon transverse momentum.
Right: Fraction of the DTLT out-of-time ghosts as a function of the muon transverse momentum. Results for the 4 DT stations are superimposed.}
    \label{fig:DT_ghosts}
  \end{center}
\end{figure}

``Ghost'' trigger candidates can arise from additional misplaced hits around the track, or from the fact that adjacent electronic units share a common DT cell.
The few false copies that pass the ghost-suppression algorithm and occur at the correct BX are
called ``in-time ghosts''. They can produce spurious dimuon trigger signals
if at least 2 of them are matched together by the DTTF.
The probability for the DTLT algorithm to
generate such false
trigger signals in a given station is defined as the number of events
with 2 trigger
primitives in that station,
both associated with the correct BX, divided by the number of
events in which at least 1 trigger primitive is delivered.
The in-time ghost probability determined from the analysis of the data collected in year 2010 is shown in Fig.~\ref{fig:DT_ghosts}
(left) as a
function of the muon transverse momentum, for the 4 DT stations.
The result is in excellent agreement with the Trigger TDR predictions~\cite{trigTDR},
which range from 2\% to 4\% as a function of the muon \pt.

False copies of the trigger primitive assigned to the wrong BX, in addition to
the one that correctly identifies the BX, are called out-of-time ghosts.
If at least 2 such false local triggers are matched by the DTTF, a muon trigger candidate would be associated with the wrong BX.
The probability of out-of-time ghosts in a DT station is defined as the number of events with
2 trigger primitives, one assigned to the correct and the other to the wrong BX in that station,
divided by the number of events in which at least 1 trigger primitive at the correct
BX is present.
The out-of-time ghost probability is shown in Fig.~\ref{fig:DT_ghosts} (right) as a function of the muon transverse momentum, for the 4 DT stations.
In this plot, all out-of-time segments are included, regardless of quality.
This explains why their fraction is almost a factor of 2 larger than in Fig.~\ref{fig:sync}\,(upper left) where only out-of-time segments with a quality higher than that of the segment at the correct BX are considered.
The results are at least a factor 3 better than the Trigger TDR predictions.
However, a direct comparison is not possible in this case because the study in the TDR was performed by using a looser definition of out-of-time ghosts~\cite{trigTDR}.

For the CSCs, the probability of an out-of-time trigger primitive is well below 1\%, as discussed in Section~\ref{TimingAndSync}, and so out-of-time segments are not a concern.
Further aspects related to the timing and synchronization of the DT and CSC systems, which can produce out-of-time triggering, are also discussed in Section~\ref{TimingAndSync}.

\begin{figure}[hbtp]
  \begin{center}
    \includegraphics[width=7.cm]{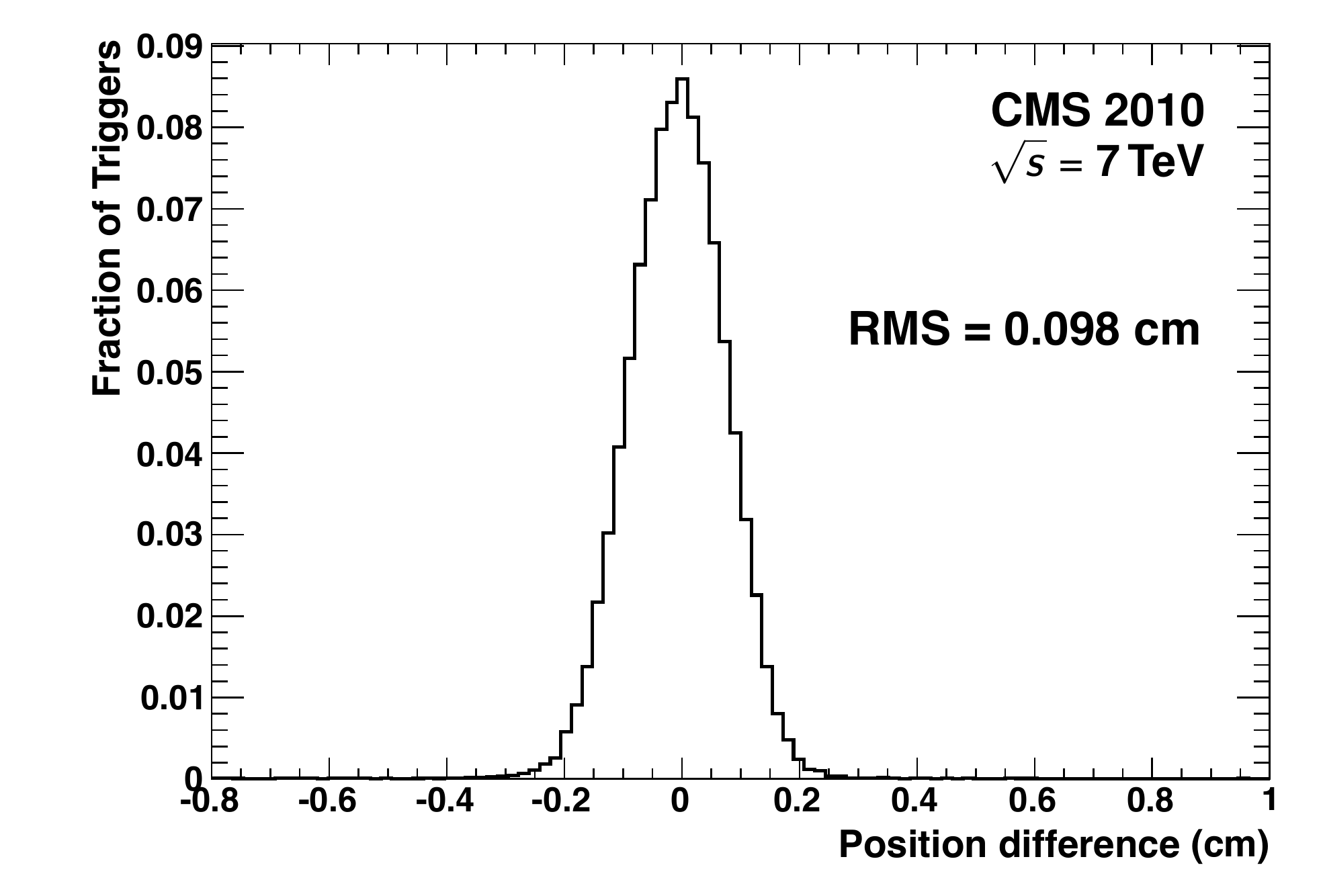} %
\qquad
  \includegraphics[width=7.cm]{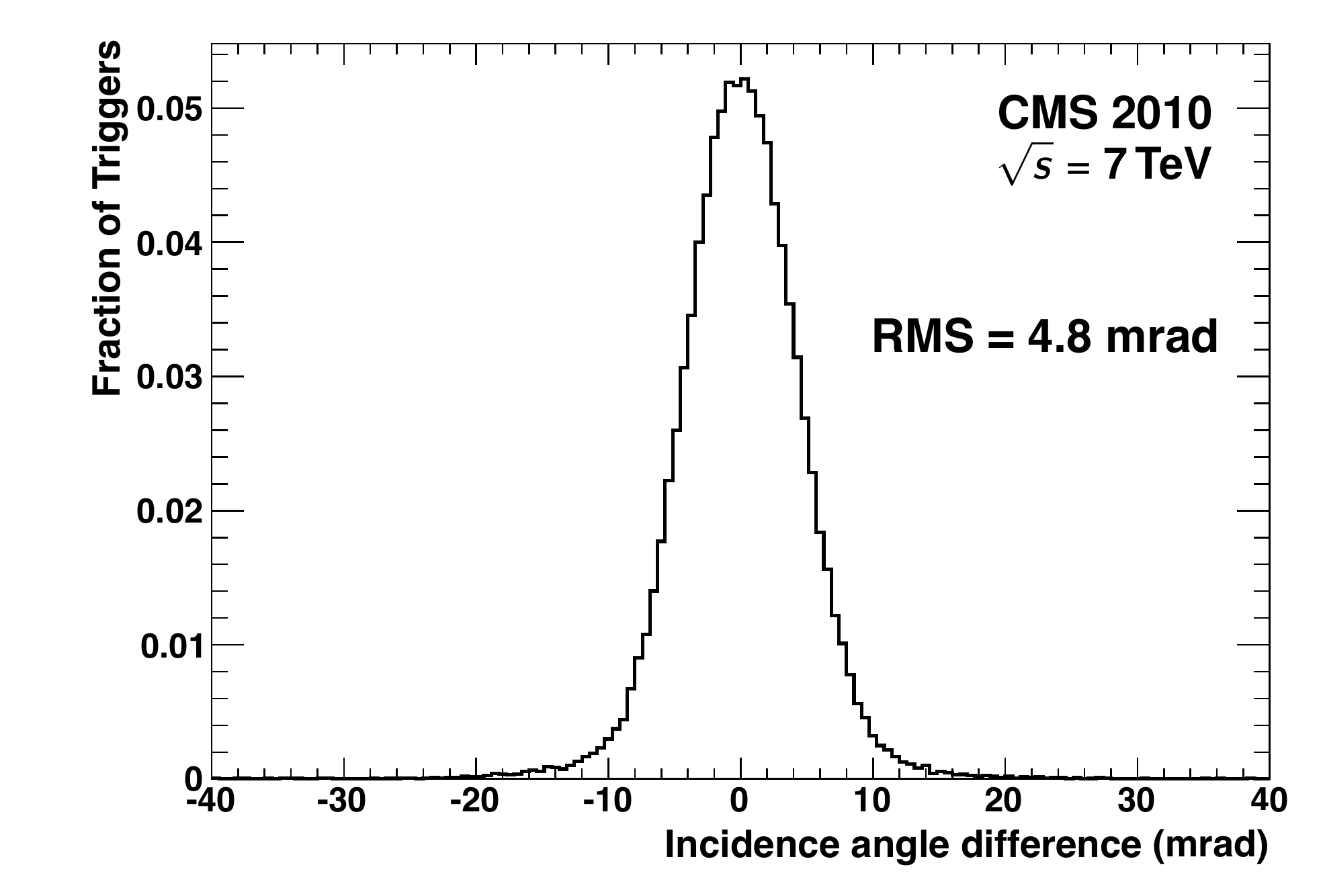}
    \caption{Left: Distribution of the difference between the position of the local track segment and the DTLT
segment. Right: Distribution of the difference between the angle of the local track segment and the DTLT
segment. Results are shown for the MB1 station.}
    \label{fig:resolution}
  \end{center}
\end{figure}

\subsection{Trigger primitive position and angular resolution}
\label{TrigResol}

The track segments obtained by fitting the
TDC information in each DT chamber are
used for offline muon reconstruction, and
provide an accurate determination of the position and incidence angle
of the muon in the DT chamber that is
independent of the DTLT output.
The position and angle of the reconstructed track segments
are compared with the values assigned by the DTLT
to the trigger segments to determine the position and angular resolution of the DTLT primitives.
As an example, Fig.~\ref{fig:resolution} (left) shows the distribution of the difference between the position of the track segment reconstructed offline and the position of the DTLT segment for the MB1 station.
The root-mean-square (RMS) of the distribution is approximately 1\unit{mm}, and is the same for every station type. This result is
in agreement with previous measurements
~\cite{Chatrchyan:2009ig, NIMA534_441}.
Figure~\ref{fig:resolution} (right)
shows the distribution of the difference
between the incidence angle of the reconstructed track and
the DTLT segment in the MB1 DT station.
The RMS of the distribution is 4.8\unit{mrad}.
The result is again in agreement with previous
measurements~\cite{Chatrchyan:2009ig, NIMA534_441}, showing that
the expected performance in terms
of position and transverse momentum resolution at the output of the
Level-1 trigger is achieved~\cite{trigTDR}.

\begin{figure}[hptb]
\centering
\includegraphics[height=6cm]{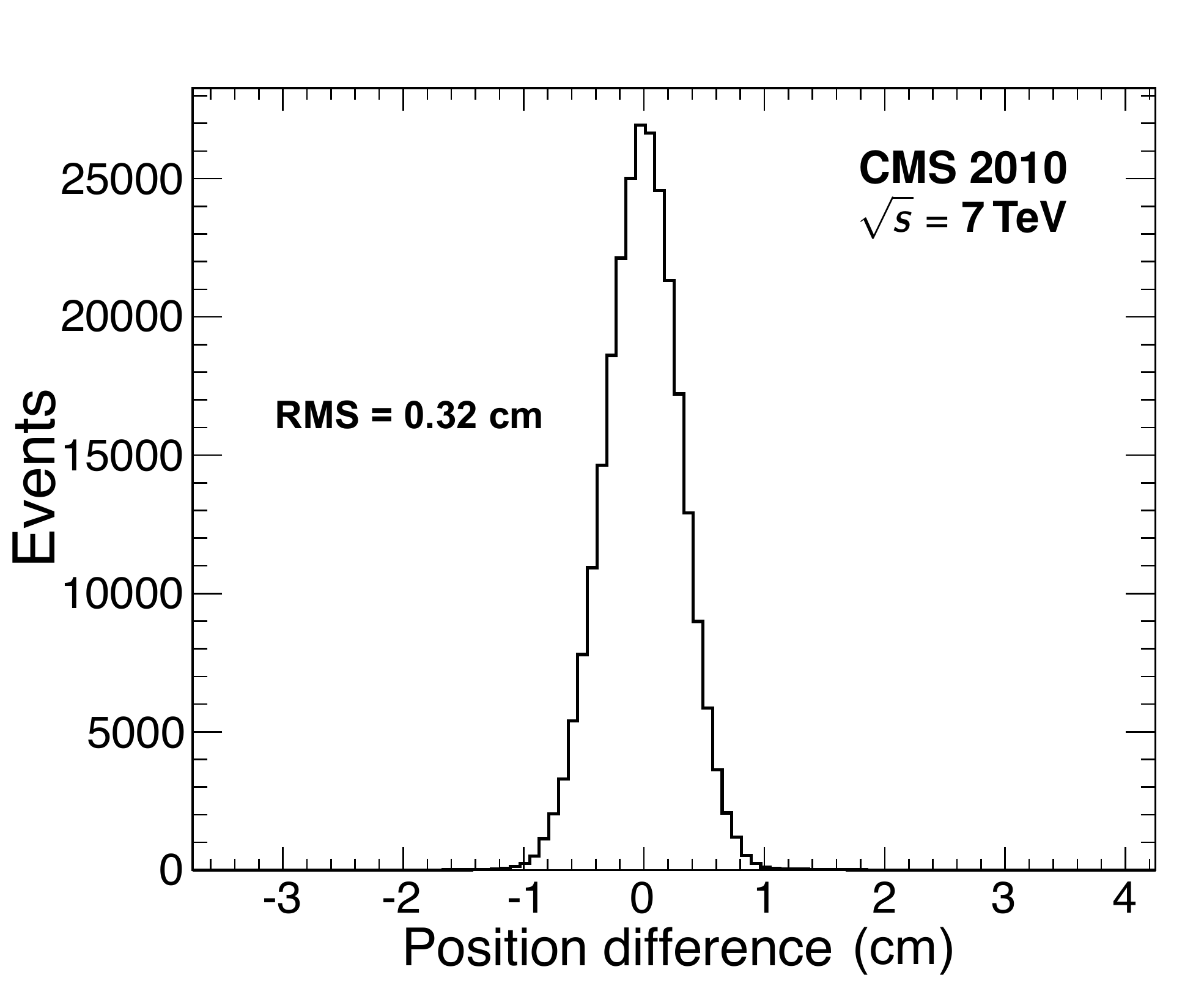}
\includegraphics[height=6cm]{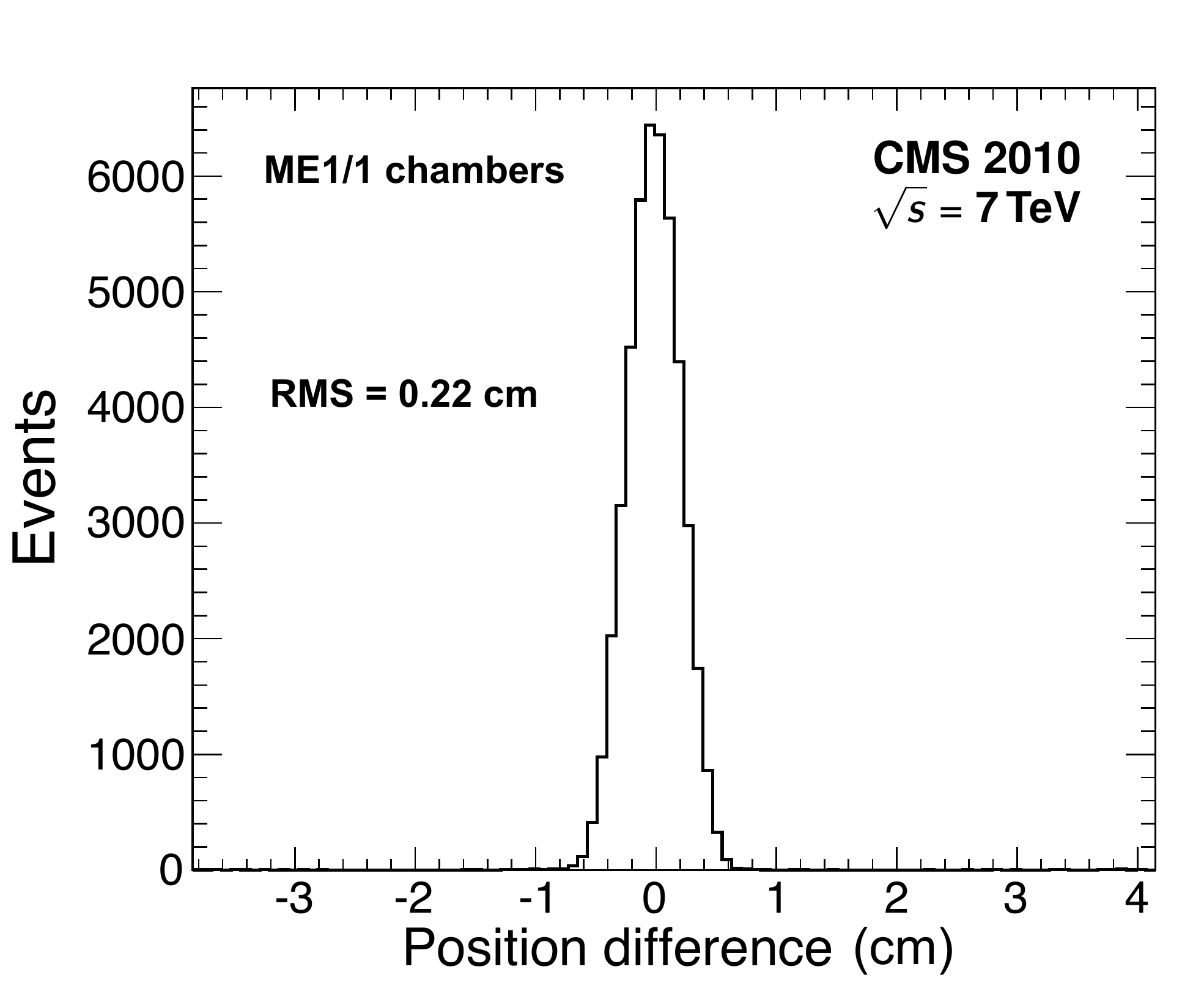}
\caption{Left: Distribution of the position differences measured between trigger primitives and track segments along the strips
for all CSC station types.
Right: The distribution of the same quantity measured for the ME1/1 station type, in which narrower cathode strips are used.}
\label{CSCLT_spaceres}
\end{figure}

The position resolution of the CSCLT primitive is measured by comparing the position of the CSC trigger primitive in the station to the position of the best-matched CSC track segment determined offline.
Figure \ref{CSCLT_spaceres} shows the distribution of the position difference between the CSC trigger primitive and the closest matching track segment in the azimuthal direction corresponding to the strip measurement for all CSC stations and for the ME1/1-type stations separately.
The measured resolution is 3.2\unit{mm}. A smaller value of 2.2\unit{mm} is found for ME1/1, which has narrower cathode strips.
The CSC TF does not depend on CSC trigger primitives for directional information, which is obtained by the position of the primitives relative to the interaction point.

\section{Position resolution}
\label{sec:Resolution}
This section presents measurements of the spatial resolutions of the DT, CSC,
and RPC systems based on data recorded during the first year of LHC collisions in 2010.

Hit resolution is determined from the distribution of hit residuals
with respect to the muon trajectory. This is possible in the DTs and CSCs
with no need of an external reference by using the track stubs
(``segments'') reconstructed with a straight-line fit of the hits in
the different measurement layers.
Therefore, the relative alignment of chambers does not affect the result.

The residual of hits with respect to the reconstructed segment is a biased estimator of the resolution because the hit under study contributes to the segment fit if all available hits are included
in the segment, or because of the uncertainty in segment extrapolation or interpolation if the segment fit is performed after removing the hit under study.
In either case, the bias can be removed by using the
statistical relationship between the width of the residual distribution
for layer $i$ ($\sigma_{R_i}$) and the actual resolution ($\sigma_i$), which
can be obtained from Gaussian error propagation of the explicit
expression of the residual with respect to the straight line obtained
from a least-squares linear fit~\cite{hatMatrix, hatMatrix2}:

\begin{equation}
  \label{eq:layerreso}
\sigma_i=c_i\ \sigma_{R_i},
\end{equation}

where $c_i$ is a factor that depends on the distance of the layer from
the middle of the measurement planes, and is less than 1 in the case where
the hit under study is removed from the segment, and larger than 1 otherwise.
Monte Carlo (MC) studies confirm that such biases can be removed by using these corrections.

Residuals in RPC chambers, which provide a single
measurement of the trajectory, are defined by extrapolating the segment
of the closest DT or CSC chamber.

The following sections describe the details and results for each subdetector.
The measurements are made with a pure sample of high-momentum muons from W and Z decays~\cite{WZcross_section}.
The muons are required to lie within the geometrical acceptance of the corresponding subdetector, and to have $\pt > 20\GeVc$.

\subsection{DT spatial resolution}
\label{sec:dtreso}

In the DT chambers, segments are reconstructed independently in
the 8 layers of the 2 $r$-$\phi$ SLs and, where present, in the
4 layers of the $r$-$z$ SL~\cite{CMSNOTE:2009008}.
All available hits are included in the fit, using
Eq.~(\ref{eq:layerreso}) to obtain the resolution from residuals with
the coefficients shown in Table~\ref{tab:DTresocorrfactors}.

\begin{table}[htb]
  \centering
  \topcaption{Correction factors $c_i$ of Eq.~(\ref{eq:layerreso}) derived
    for the DT geometry and for the case of a segment fit including
    all available hits, separately in the 8 layers of the 2
   $r$-$\phi$ SLs and in the 4 layers of the $r$-$z$ SL.
    The factors depend only on the distance of the layer from
    the middle of the measurement planes, that is, respectively, the
    middle plane of the 2 $r$-$\phi$ SLs and the middle plane
    of the $r$-$z$ SL.
    \label{tab:DTresocorrfactors}
  }
    \begin{tabular}{|c|cccc|}
    \hline
    Layer &  1 & 2 & 3 & 4 \\ \hline
    \hline
    SL1 ($r$-$\phi$) & 1.17 & 1.16 & 1.15 & 1.14 \\
    SL3 ($r$-$\phi$) & 1.14 & 1.15 & 1.16 & 1.17 \\
    \hline
    SL2 ($r$-$z$)    & 1.83 & 1.20 & 1.20 & 1.83 \\
    \hline
  \end{tabular}
\end{table}

For each layer, $\sigma_{R_i}$ is obtained
with a Gaussian fit of the core of the segment residual
distribution. The result is averaged separately for all $r$-$\phi$ and $r$-$z$
layers of a chamber.

Only segments with at least 7 hits in the
$r$-$\phi$ SLs and, in the innermost 3 stations, with 4 hits in
$r$-$z$ SL are used. In addition, segments are required to point towards
the beam line, with a selection on the incidence angle at the chamber in the
transverse plane of $|\psi|<25^\circ$.

\begin{figure}[htb]
  \centering
  \includegraphics[width=0.8\textwidth]{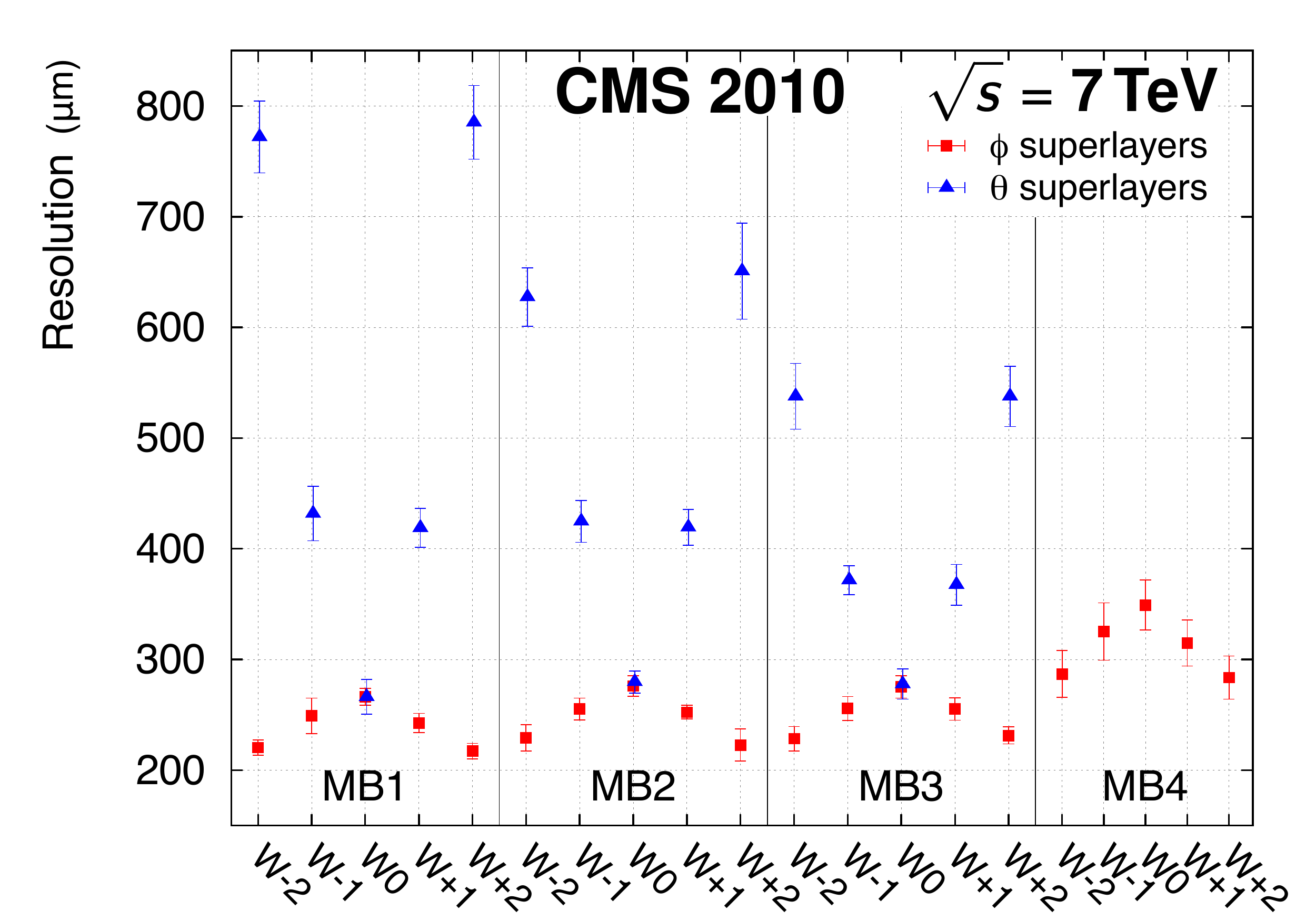}
  \caption{
    Single-hit DT resolution for $r$-$\phi$ and $r$-$z$ layers, averaged
    over all sectors in each barrel wheel and station.
  }
  \label{fig:DTResolution}
\end{figure}

The single-hit resolutions obtained with this method are shown in Fig.~\ref{fig:DTResolution} and are summarized
in Table~\ref{tab:dtreso} separately
for $r$-$\phi$ and $r$-$z$ layers, averaged over all sectors in
each wheel and station. The spatial resolution of the segment fitted
in the whole chamber, obtained as $\sigma/\sqrt{N}$, where $\sigma$ is
the single-hit resolution and $N$ the number of layers included in the
fit, is given in Table~\ref{tab:dtsegmentreso}.

\begin{table}[htb]
  \centering
  \topcaption{Single-hit DT resolution for $r$-$\phi$ and $r$-$z$ layers
    expressed in\,\micron,
    averaged over all sectors in each barrel wheel and station.
    \label{tab:dtreso}
  }
  \begin{tabular}{|c |c c c c c c| }
    \hline
      SL Type  & Station & W-2 & W-1 & W0  & W+1  & W+2  \\ \hline
    \hline
    \multirow{4}{*}{$r$-$\phi$}
   &MB1 & 220$\pm$7 & 249$\pm$16 & 266$\pm$8 & 243$\pm$9 & 217$\pm$7 \\ 
   &MB2 & 229$\pm$12 & 255$\pm$10 & 276$\pm$9 & 252$\pm$6 & 223$\pm$15 \\ 
   &MB3 & 229$\pm$11 & 256$\pm$11 & 275$\pm$10 & 255$\pm$10 & 231$\pm$8 \\ 
   &MB4 & 287$\pm$21 & 325$\pm$26 & 349$\pm$23 & 315$\pm$21 & 284$\pm$19 \\ 
    \hline
    \multirow{3}{*}{$r$-$z$}
   &MB1 & 772$\pm$32 & 432$\pm$25 & 266$\pm$16 & 419$\pm$18 & 785$\pm$33 \\ 
   &MB2 & 627$\pm$26 & 425$\pm$19 & 280$\pm$10 & 419$\pm$16 & 651$\pm$43 \\ 
   &MB3 & 538$\pm$30 & 372$\pm$13 & 278$\pm$14 & 368$\pm$18 & 538$\pm$27 \\
    \hline
  \end{tabular}
\end{table}

\begin{table}[hbt]
  \centering
  \topcaption{The DT chamber resolution in the $r$-$\phi$ and $r$-$z$ projections,
    expressed in \micron,
    averaged over all sectors in each barrel wheel and station.
    \label{tab:dtsegmentreso}
  }
  \begin{tabular}{|c | c c c c c c| }
    \hline
      SL Type  & Station & W-2 & W-1 & W0  & W+1  & W+2  \\ \hline
    \hline
    \multirow{4}{*}{$r$-$\phi$}
  &MB1 & 78$\pm$2 & 88$\pm$6 & 94$\pm$3 & 86$\pm$3 & 77$\pm$2 \\ 
  &MB2 & 81$\pm$4 & 90$\pm$3 & 98$\pm$3 & 89$\pm$2 & 79$\pm$5 \\ 
  &MB3 & 81$\pm$4 & 90$\pm$4 & 97$\pm$4 & 90$\pm$4 & 82$\pm$3 \\ 
  &MB4 & 101$\pm$7 & 115$\pm$9 & 123$\pm$8 & 111$\pm$7 & 100$\pm$7 \\ 
    \hline
    \multirow{3}{*}{$r$-$z$}
  &MB1 & 386$\pm$16 & 216$\pm$12 & 133$\pm$8 & 209$\pm$9 & 393$\pm$17 \\ 
  &MB2 & 314$\pm$13 & 212$\pm$9 & 140$\pm$5 & 210$\pm$8 & 325$\pm$22 \\ 
  &MB3 & 269$\pm$15 & 186$\pm$7 & 139$\pm$7 & 184$\pm$9 & 269$\pm$14 \\
    \hline
  \end{tabular}
\end{table}

Several features can be noted:
\begin{itemize}
\item For both $r$-$\phi$ and $r$-$z$ layers, the resolution of Wheel W+1 (W+2) is
  approximately the same as that of W-1 (W-2), given the
  geometric symmetry of the system.
\item In Wheel 0, the resolution is the same for the $r$-$\phi$ and $r$-$z$
  layers.
\item The resolution changes from inner to outer wheels because of
  the effect of the increased angle of incidence ($\theta$) of
  muons.
  For $r$-$z$ SLs, $\theta$ is the angle in the measurement
  plane; therefore the resolution is significantly degraded in
  external wheels because of the increasing deviation from linearity of
  the space-time relationship (Eq.~(\ref{eq:hit-reconstruction})) with larger angles of
  incidence of the particles.
  For $r$-$\phi$ layers, $\theta$ is the angle in the plane orthogonal
  to the measurement plane; the larger angle in external wheels
  results in longer paths inside the cells
  that increase the number of primary ionizations, causing a
  slight improvement in the $r$-$\phi$ resolution.
\item The poorer resolution of the $r$-$\phi$ layers in MB4 compared to MB1--MB3
  is because in this station, where a measurement of
  the longitudinal coordinate is missing, it is not possible to correct
  for the actual muon time-of-flight and signal propagation time along
  the wire.
  In particular, the signal propagation time along the wires is up to
  about 9.8\unit{ns}, which corresponds to differences in reconstructed
  position of up to about 540\micron.
  This correction can be applied at a later stage,
  during the fit of a muon track using all stations.
\end{itemize}

For comparison, the single-hit resolution obtained in a test beam is about 190\micron for normal incidence on the chamber,
with a deterioration to about 450\micron for an incident angle of
30$^\circ$ in the cell measurement plane and improving to
about 150\micron for an incident angle of 30$^\circ$ in the
orthogonal (non-measurement) plane~\cite{NIMA534_441}.
The observed resolution is that expected from simulation, given the distribution of the incident angle for muons in CMS, and is in agreement with Muon TDR expectations.

\subsection{CSC spatial resolution}
\label{sec:intro}

The spatial resolution of the CSCs is determined by the properties (geometrical and operational) of the chambers, but can vary with the kinematic properties of the detected muons, and can depend on the details of the reconstruction of the hit positions.
 The chamber gas gain (how large a signal results from the passage of an ionizing particle through the gas) is affected by changes in atmospheric pressure, and such changes are also reflected in variation of spatial resolution.
 In the following characterization of the spatial resolution, the measurements are averaged over time periods of months so pressure variations tend to average out.

The precisely measured coordinate in the CSCs is that measured by the strips, since it is in this direction that muons are deflected by the CMS magnetic field, and thus is crucial for input to the L1 trigger for estimation of the \pt of a muon trigger candidate.
In global coordinates this is approximately the azimuthal direction, but the results presented here are expressed in coordinates local to an individual chamber.
To match the endcap CMS geometry, the CSCs have trapezoidal shape so that they can be assembled in concentric rings, and the strips in a CSC are radial, approximately projecting to the beamline ($z$ axis), with each strip subtending a fixed azimuthal angle.
This means that the strip width progressively increases as the radial distance from the beamline increases, as shown in Table~\ref{tab:stripWidths}, and so the resolution within a given CSC depends on the radial position.
It is thus natural to measure the resolution in units of strip width, and the results are quoted in terms of the fractional position $s$ within a strip, where
$-0.5 < s <0.5$.
To convert a resolution from strip width units to standard units the value is multiplied by the mean strip width $\langle w \rangle$ within the CSC; these values are also shown in
Table~\ref{tab:stripWidths}.

\begin{table}[htb]
  \centering
  \topcaption{Selected physical specifications of the CSCs, including the range of strip widths, the average widths, and the strip angular widths. For more information, see Ref.~\cite{CMSdet}.}
  \label{tab:stripWidths}
    \begin{tabular}{| c  c  c  c |}
    \hline
    Ring   & Strip width        & $\langle w \rangle$     & Strip angular \\
                &     $w$ (mm)     &      (mm)   & width (mrad) \\ \hline \hline
    ME1/1 & 4.4--7.6         & 6.0         & 2.96\\ 
    ME1/2  & 6.6--10.4         & 8.5         & 2.33\\ 
    ME1/3  & 11.1--14.9       & 13.0        & 2.16\\ 
    ME2/1  & 6.8--15.6         & 11.2        & 4.65\\ 
    ME2/2  & 8.5--16.0         & 12.2        & 2.33\\ 
    ME3/1  & 7.8--15.6         & 11.7        & 4.65\\ 
    ME3/2  & 8.5--16.0         & 12.2        & 2.33\\ 
    ME4/1  & 8.6--15.6         & 12.1        & 4.65\\ \hline
  \end{tabular}
\end{table}

The spatial resolution in a CSC layer depends on the relative position at which a muon crosses a strip: it is significantly worse for a track crossing the center of a strip than for one crossing near an edge where the induced charge is shared between the strip and its neighbor.
To compensate, alternate layers in a CSC are staggered by half a strip width, except in the ME1/1 chambers where the strips are very narrow and the effect is small~\cite{CMSdet,paper:ME1_a,paper:ME1_b}.
As a measure of the resolution, the Gaussian $\sigma$ resulting from a Gaussian fit to the core region of the residual distribution is used.
For the CSCs with staggered layers, separate residual distributions are formed for hits in the central half of any strip and for hits lying outside this central region, and the distributions are fit separately giving a ``central'' $\sigma_\mathrm{c}$ and an ``edge'' $\sigma_\mathrm{e}$.
Both of these measurements are combined according to

\begin{equation}
  \sigma_{\mathrm{chamber}} = \left( \frac{3}{\sigma_{\mathrm{e}}^2}+\frac{3}{\sigma_{\mathrm{c}}^2} \right)^{-1/2}.
  \label{eqn:chamberRes}
\end{equation}

For the non-staggered ME1/1, the following expression is used:

\begin{equation}
  \sigma_{\mathrm{chamber}} = \left( \frac{6}{\sigma_{\text{layer}}^2} \right)^{-1/2}=\frac{\sigma_{\text{layer}}}{\sqrt{6}}.
  \label{eqn:chamberResME11}
\end{equation}

In Eqs.~(\ref{eqn:chamberRes}) and~(\ref{eqn:chamberResME11}), $\sigma_\mathrm{c}$, $\sigma_\mathrm{e}$, and $\sigma_{\text{layer}}$ are the layer resolutions calculated
by Eq.~(\ref{eq:layerreso}).
The 5-hit fit is used to estimate the width of the residual distributions.
The correction factors $c_i$ for different CSC types are listed in Table~\ref{tab:hatFactors}.

\begin{table}
  \centering
  \topcaption{Correction factors $c_i$ for 5-hit fits in the CSCs.}
  \label{tab:hatFactors}
  \begin{tabular}[htb]{|l| c c c c c c|}
    \hline
    Layer                & 1   & 2   & 3   & 4   & 5   & 6   \\ \hline \hline
    ME1/1 chambers       & 0.69& 0.84&0.905&0.905& 0.84& 0.69\\ 
    ME1/2 edge           & 0.59& 0.73& 0.87& 0.87& 0.73& 0.59\\ 
    ME1/2 center         & 0.80& 0.91& 0.94& 0.94& 0.91& 0.80\\ 
    All others edge      & 0.56& 0.70& 0.86& 0.86& 0.70& 0.56\\ 
    All others center    & 0.83& 0.93& 0.95& 0.95& 0.93& 0.83\\
 \hline
  \end{tabular}
\end{table}

\begin{figure}[htbp]
  \centering
  {\includegraphics[width=0.45\textwidth]{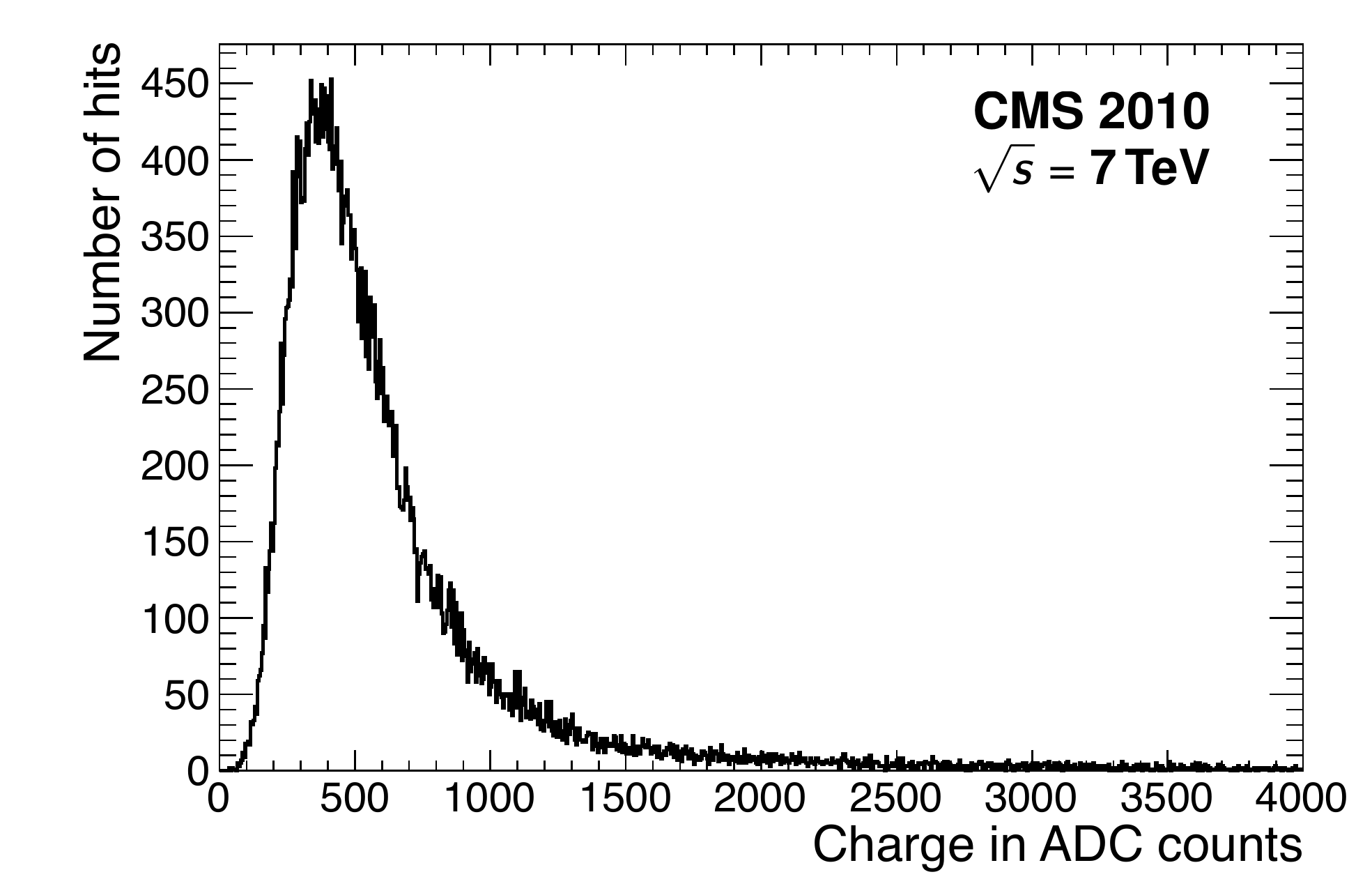}}
  {\includegraphics[width=0.45\textwidth]{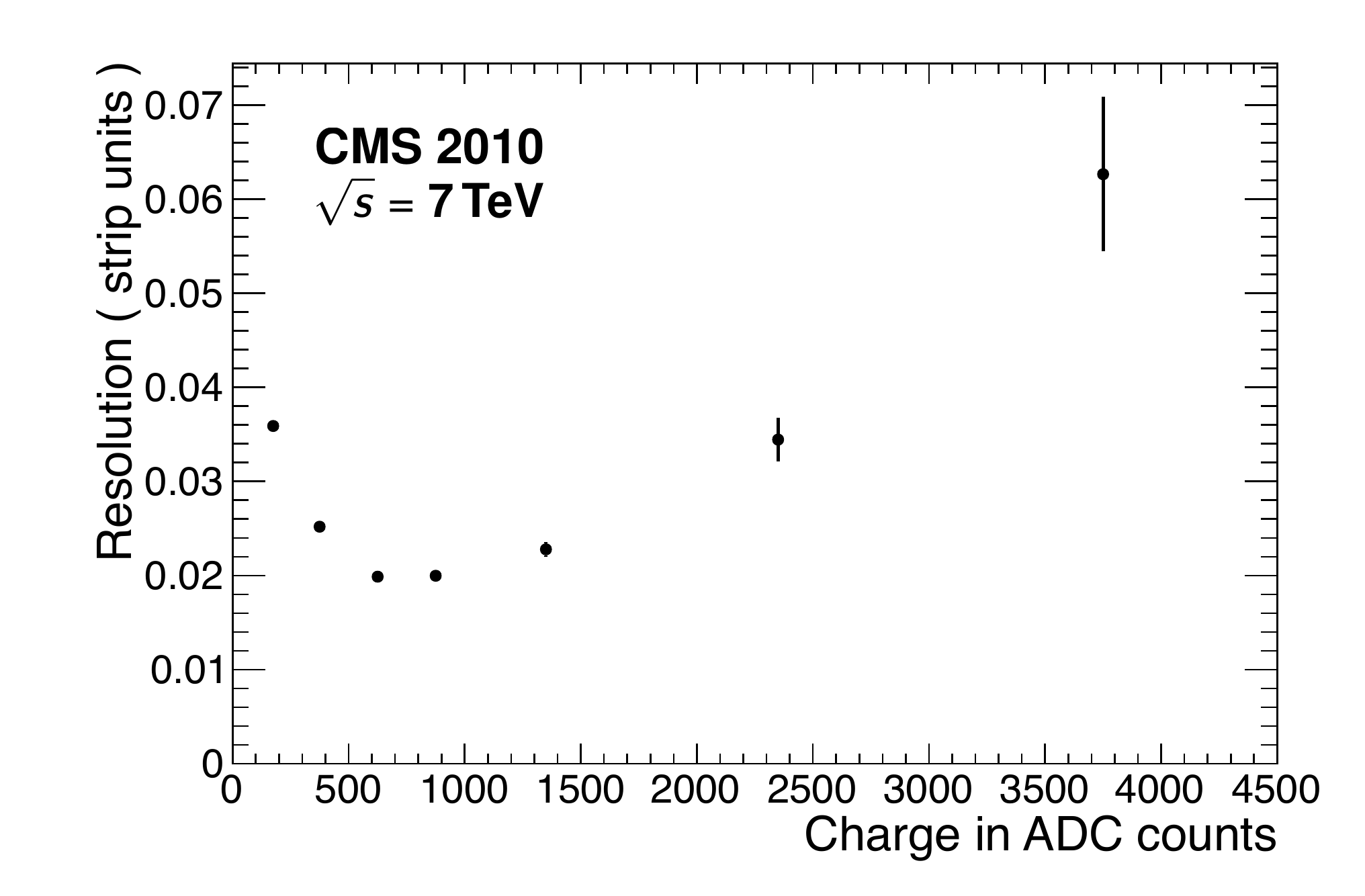}}

  \caption[~Residuals versus cluster charge]{
    Left: Charge distribution $Q_{\mathrm{3x3}}$.
    Right: Variation of the CSC layer resolution as a function of $Q_{\mathrm{3x3}}$.
  }
  \label{fig:adcRange}
\end{figure}

To reduce backgrounds, the following selection criteria were applied to the hits
and segments that were used to measure the resolution:
\begin{itemize}
\item 6 hits on a segment;
\item $\chi^2_{\text{seg}}/\text{dof} < 200/8$ and $\chi^2_{\text{seg strip-fit only}}/\text{dof} < 50/4$, where dof is the number of degrees of freedom;
\item exclude segments with largely displaced hits (leading to
  residuals of $>$0.2 strip widths);
\item segment points roughly towards the interaction point
  ($|dx/dz|<$0.15 and $|dy/dz|<$1.5 in local coordinates, where $y$
  is measured along and $x$ is perpendicular to the strips);
\item reconstructed cluster charge $Q_{\mathrm{3x3}}$, which
  is defined as the sum of the elements of a 3x3 matrix (3 strips x 3 time slices),
   is in a reasonable range ($150 <Q_{\mathrm{3x3}}
  <2000$ ADC counts (Fig.~\ref{fig:adcRange}); hits with high
  charge are often distorted by $\delta$ electrons);
\item for proton--proton collision data, require a well reconstructed muon
  with transverse momentum $\pt > 20$\GeVc
  (reconstructed using muon chamber and central tracker information)
  within $\Delta\eta<0.07$ of the segment.
\end{itemize}
These requirements are similar to those used in previous studies~\cite{CSCPerfCRAFT}.

\label{sec:results}

\begin{figure}[hp]
  \centering
  \includegraphics[width=0.46\textwidth]{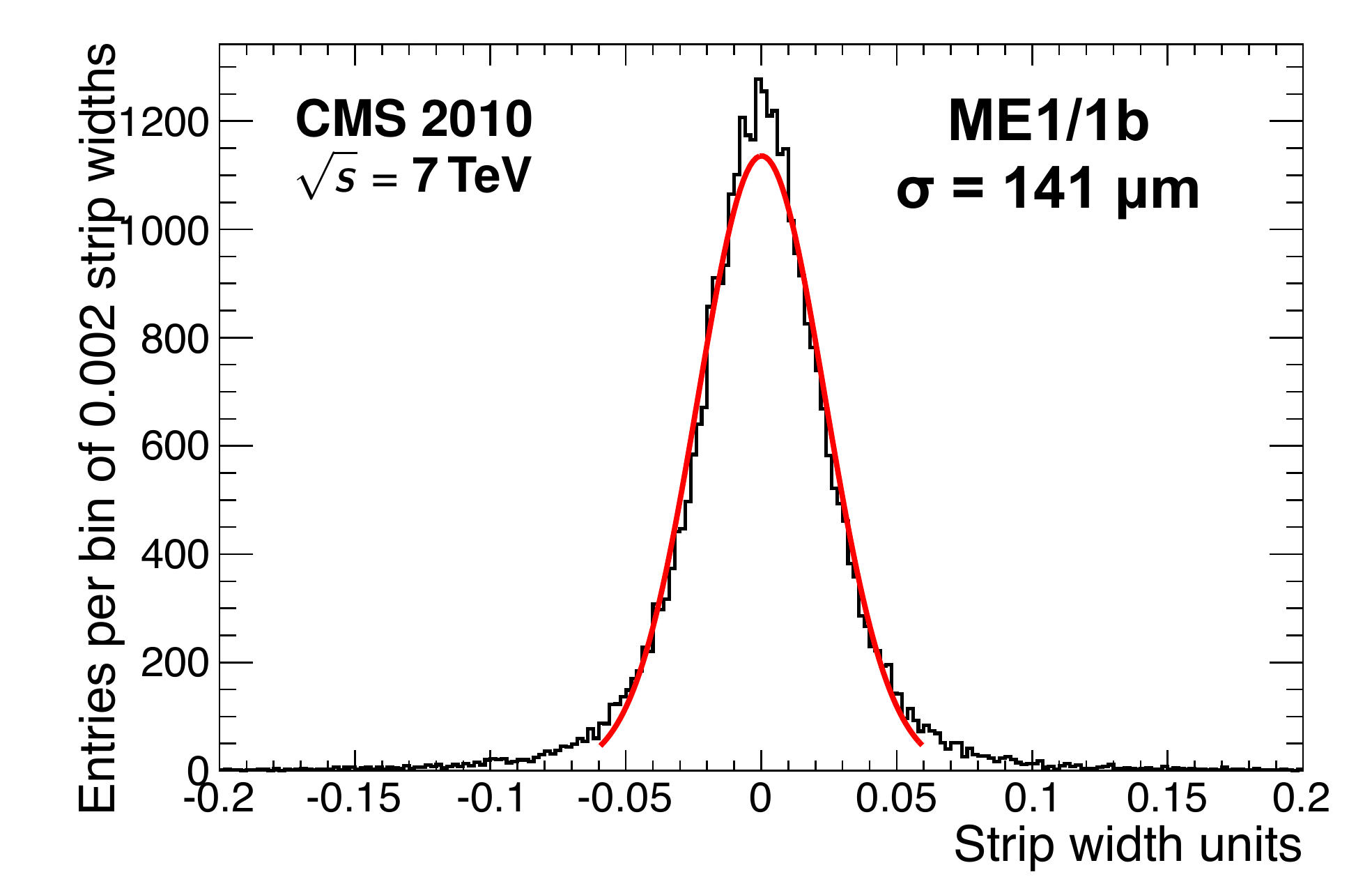}
  \includegraphics[width=0.46\textwidth]{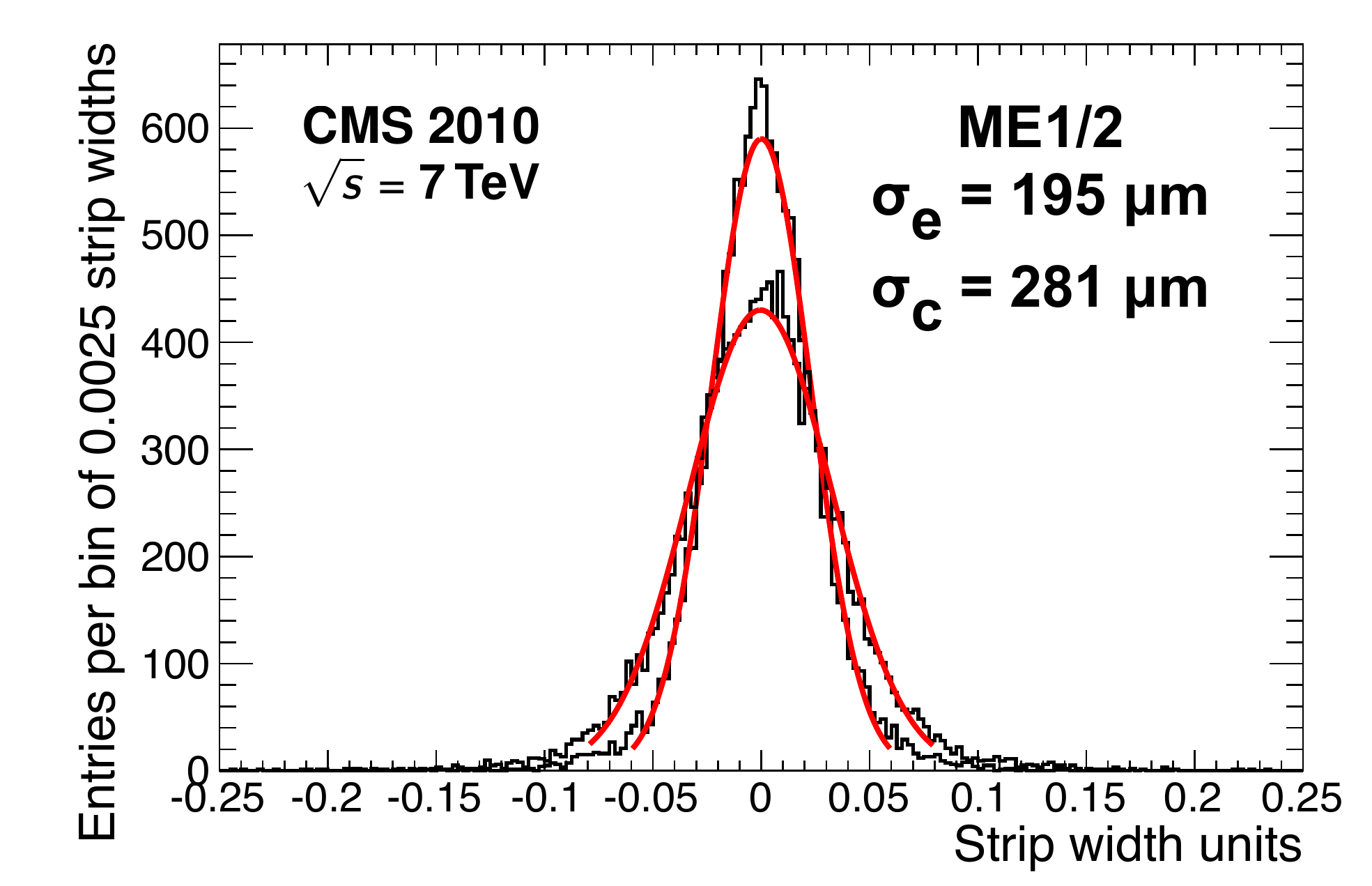}
  \includegraphics[width=0.46\textwidth]{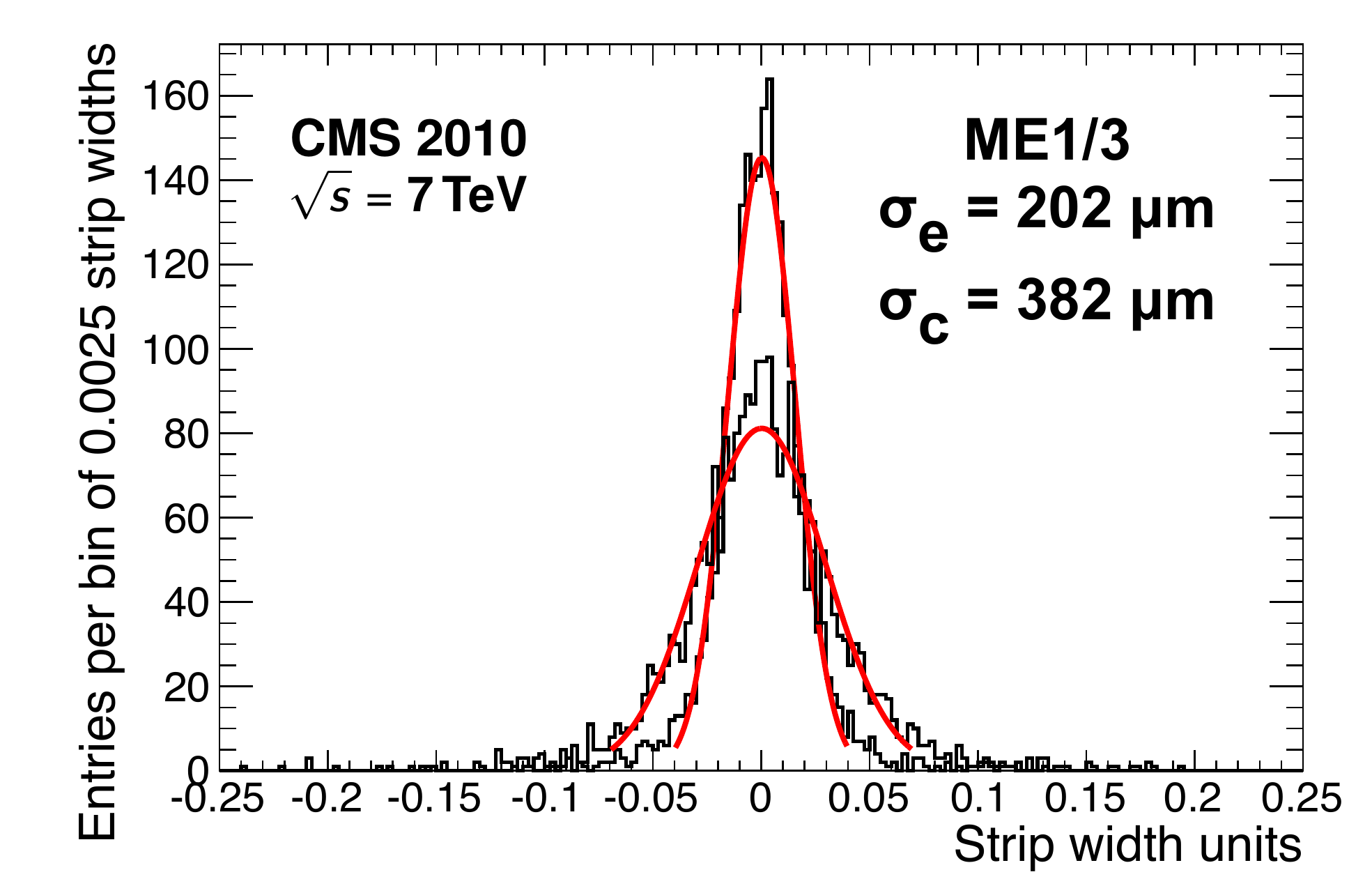}
 \includegraphics[width=0.46\textwidth]{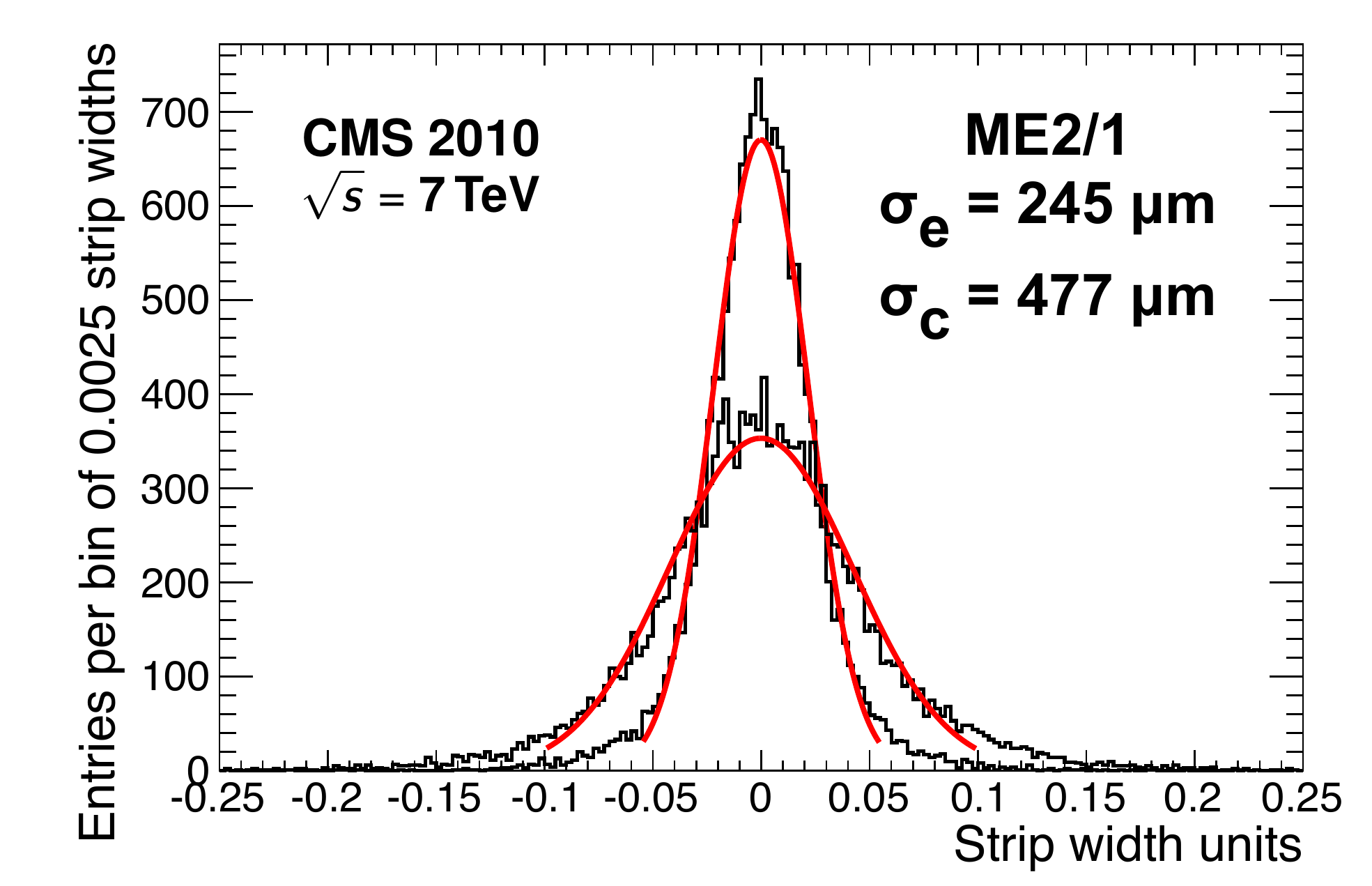}
  \includegraphics[width=0.46\textwidth]{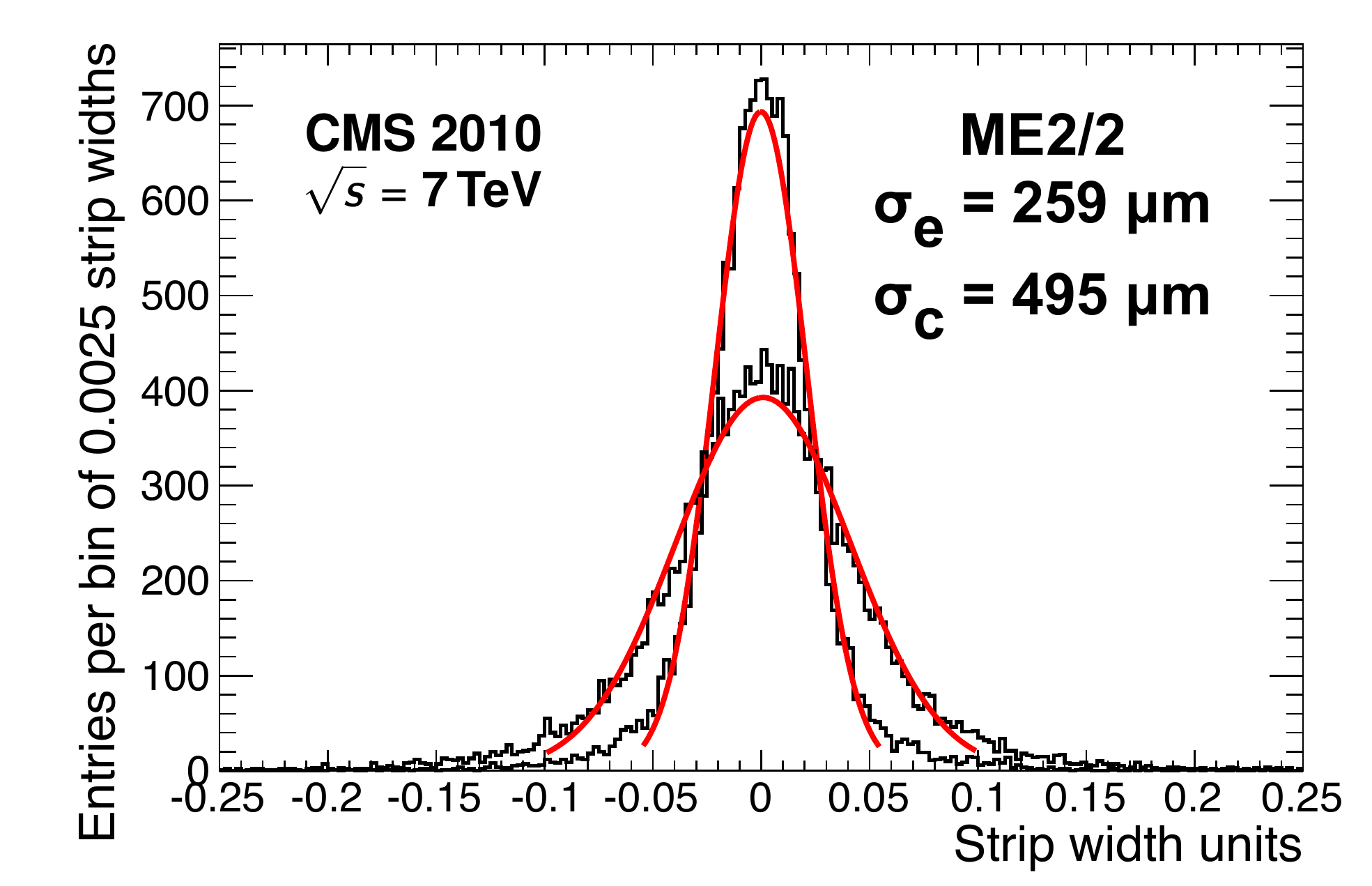}
  \includegraphics[width=0.46\textwidth]{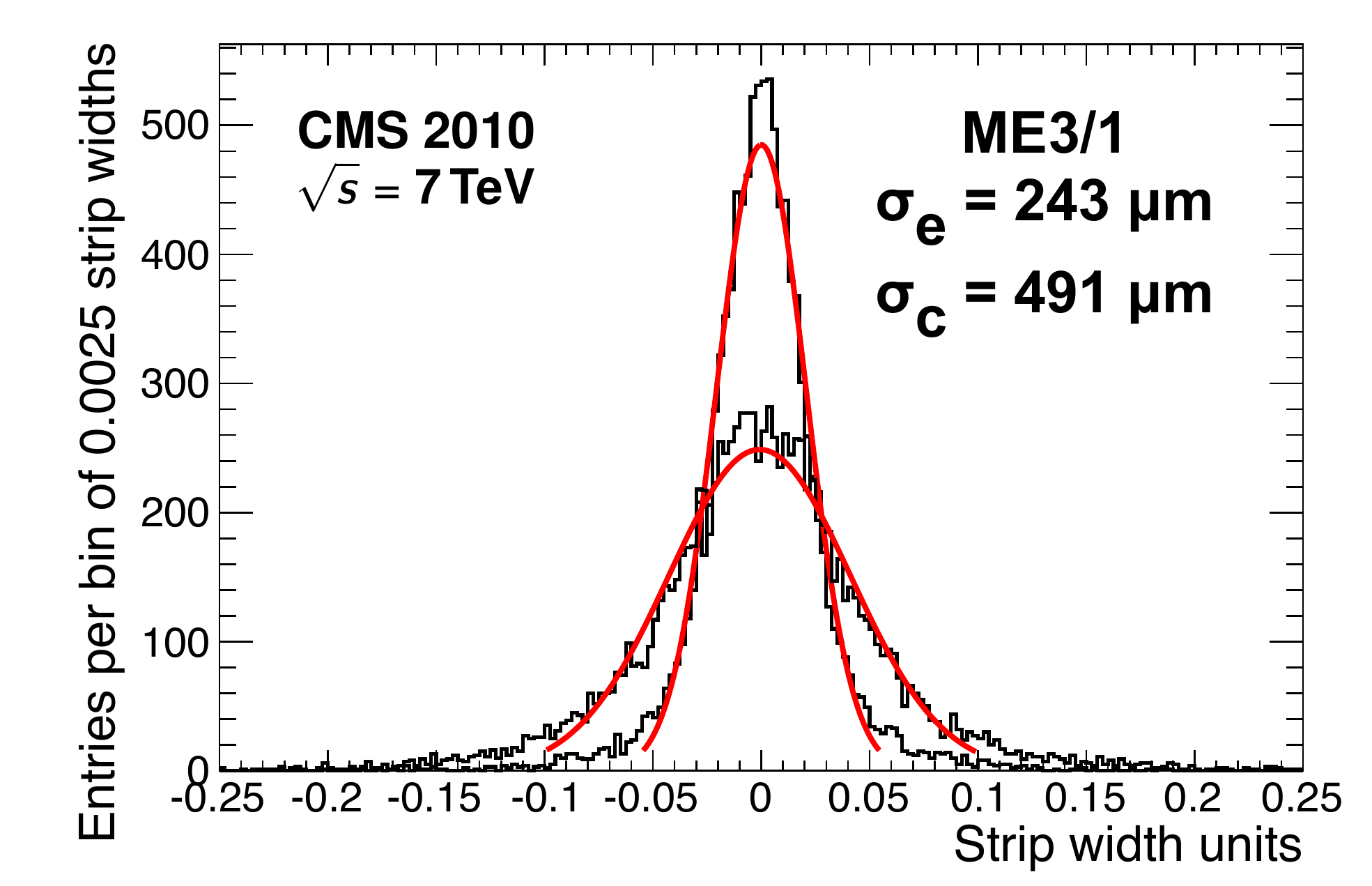}
  \includegraphics[width=0.46\textwidth]{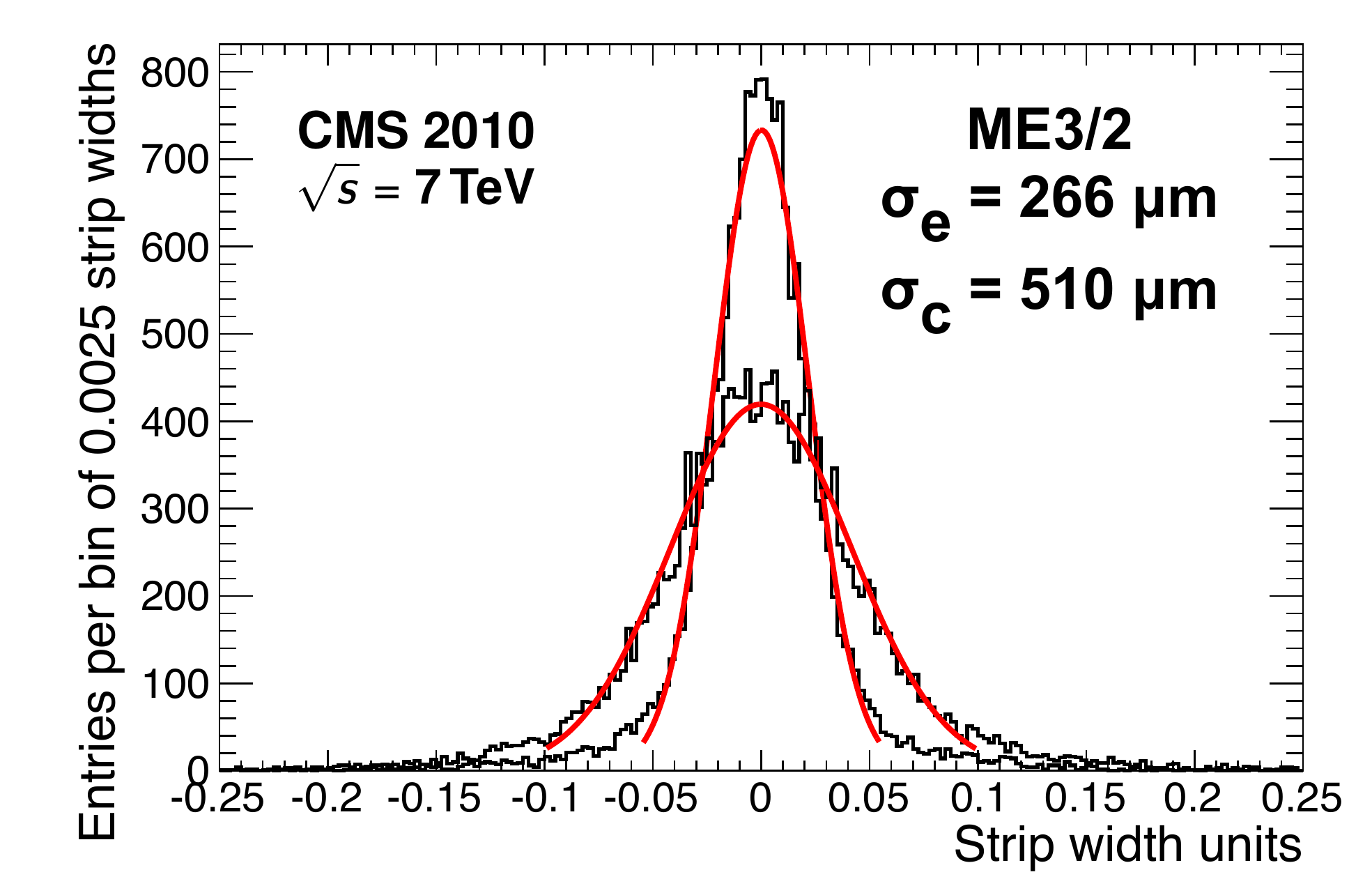}
  \includegraphics[width=0.46\textwidth]{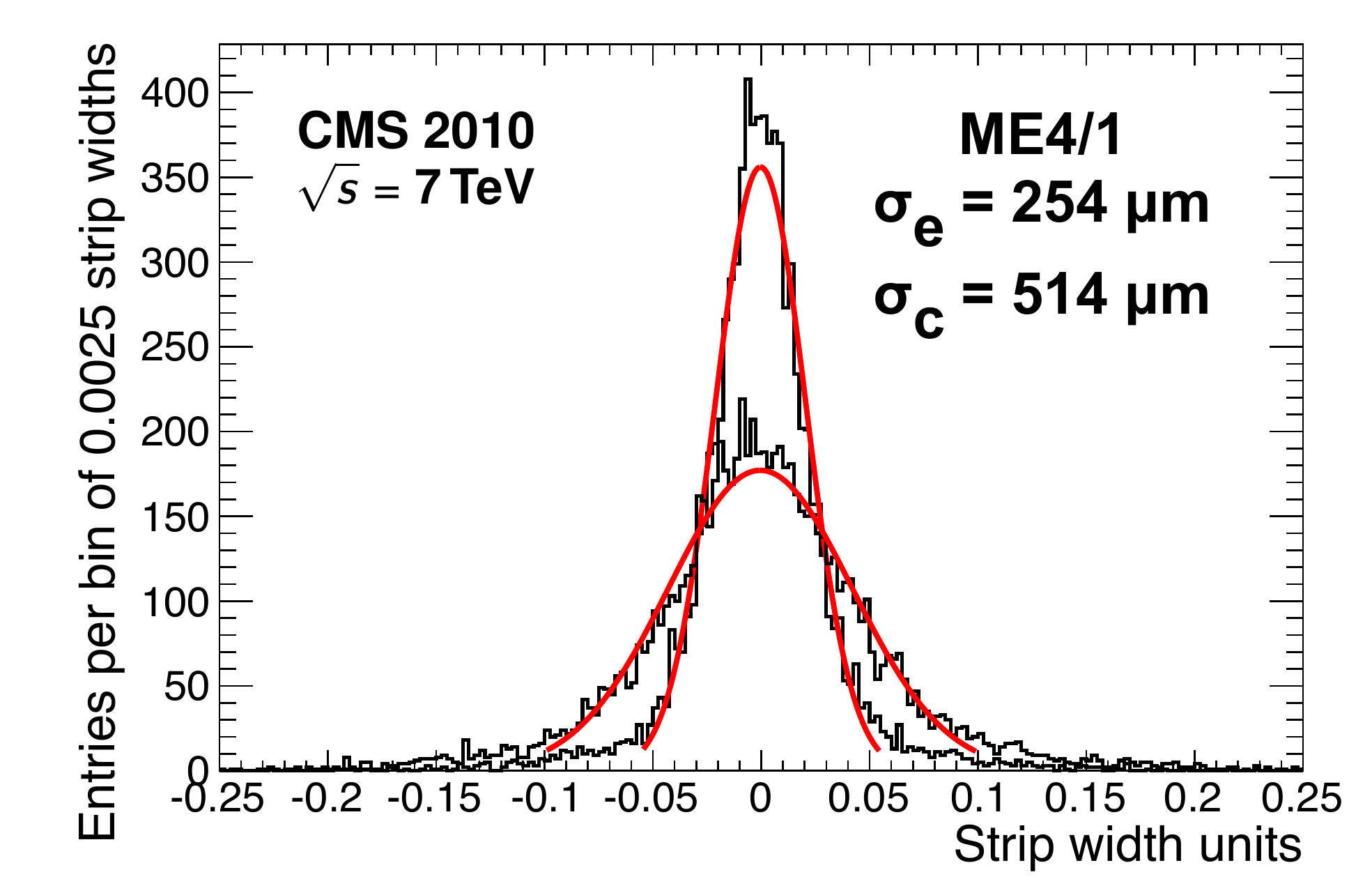}
  \caption[~CSC Chamber residuals distributions]{
    Distributions of spatial residuals  for the different CSC chamber types, corrected per layer according to Eq.~(\ref{eq:layerreso}). The residuals are measured in units of strip width, and for chambers with staggered layers, the distributions for both the center ($\sigma_\mathrm{c}$) and edge ($\sigma_\mathrm{e}$) of the strip are shown. The curves are the results of simple Gaussian fits to the central region (within  ${\pm}2.5\sigma$) of each distribution.
  }
  \label{fig:residualsDistribs}
\end{figure}

The residual distributions for each chamber type are shown in Fig.~\ref{fig:residualsDistribs}.
The distributions agree reasonably well with a single Gaussian distribution in the central region.
The fits shown are used to extract the value of the resolution, which is then scaled by the average strip width per chamber.
The measured chamber resolutions obtained by using Eqs.~(\ref{eqn:chamberRes}) and~(\ref{eqn:chamberResME11}) are summarized in Table~\ref{tab:stationRes}.
The uncertainties on the widths from the Gaussian fits vary from 2\% to 3\%, corresponding to statistical uncertainties of 2--4\micron on the values in the table.

\begin{table}[t!]
  \centering
  \topcaption{Average CSC position resolutions ($\mu$m) for each chamber type comparing cosmic and collision data with Monte Carlo (MC) simulation. The uncertainty on the values is discussed in the text.}
  \label{tab:stationRes}
  \begin{tabular}[t!]{|c|cccccccc|}
    \hline
& \multicolumn{8}{c|}{ Chamber type }      \\
                     \cline{2-9}
Run type/year & ME1/1 & ME1/2 & ME1/3& ME2/1& ME2/2 & ME3/1& ME3/2 & ME4/1 \\ \hline\hline

    Cosmics 2009                        & 73   & 109 & 135 & 147 & 162 & 143 & 200 & 218 \\         
    Cosmics 2010                        & 70   & 110 & 136 & 147 & 164 & 144 & 196 & 205 \\ 
    Collisions 2010                     & 58   & 92  & 103 & 126 & 132 & 126 & 136 & 131 \\ \hline 
    pp MC 2010                           & 37  & 82  & 110 & 121& 152 & 119 & 155 & 119 \\ \hline
  \end{tabular}
\end{table}

Once the chamber gas mixture and operating high voltages are fixed, the specific values for position resolution still depend on the range over which the Gaussian fits are performed, on the momentum spectra of the muons examined, and on the atmospheric pressure.
The datasets used in Table~\ref{tab:stationRes} average over periods of several weeks so that atmospheric pressure effects are also averaged, and the muon momentum spectra are similar.
The similarity of the values from the cosmic datasets in 2009 and 2010 show that this procedure provides a consistent and robust measure of the CSC spatial resolution.
The resolutions obtained from cosmic-ray muons are somewhat worse than those from muons with
$\pt > 20$\GeVc produced in proton--proton collisions.
The cosmic-ray muons are of lower average momentum, arrive uniformly distributed in time, and have larger variation in angles of incidence, all of which tend to lead to poorer spatial resolution.
The agreement between the resolutions determined from simulated collision data and those obtained from real data is reasonable,
considering that the dependence of the chamber gas gain on atmospheric pressure is not simulated.
The MC value for ME1/1 is notably lower than the real value because the MC used the design HV, which was lowered by 4\% during the 2010 running to increase the lifetime of the chambers while still providing the required design resolution.

All measured resolutions are close to and most even exceed the requirements
noted in the CMS Muon TDR~\cite{MUON-TDR}, which called for 75\micron
for the ME1/1 and ME1/2 chambers and 150\micron for the remaining chambers.

\subsection{RPC resolution}

In the CMS muon system, RPCs are used as trigger detectors; in addition, hits are provided for reconstruction and muon identification.
To measure the resolution of the RPC system, DT and CSC track segments were extrapolated with the technique explained in Section~\ref{Efficiency}.
The coordinates of the extrapolated point were then compared with those of the reconstructed RPC hit, \ie, the average coordinates of the strips fired by the muon.

\begin{figure}[htbp]
  \centering
  \includegraphics[width=0.8\textwidth]{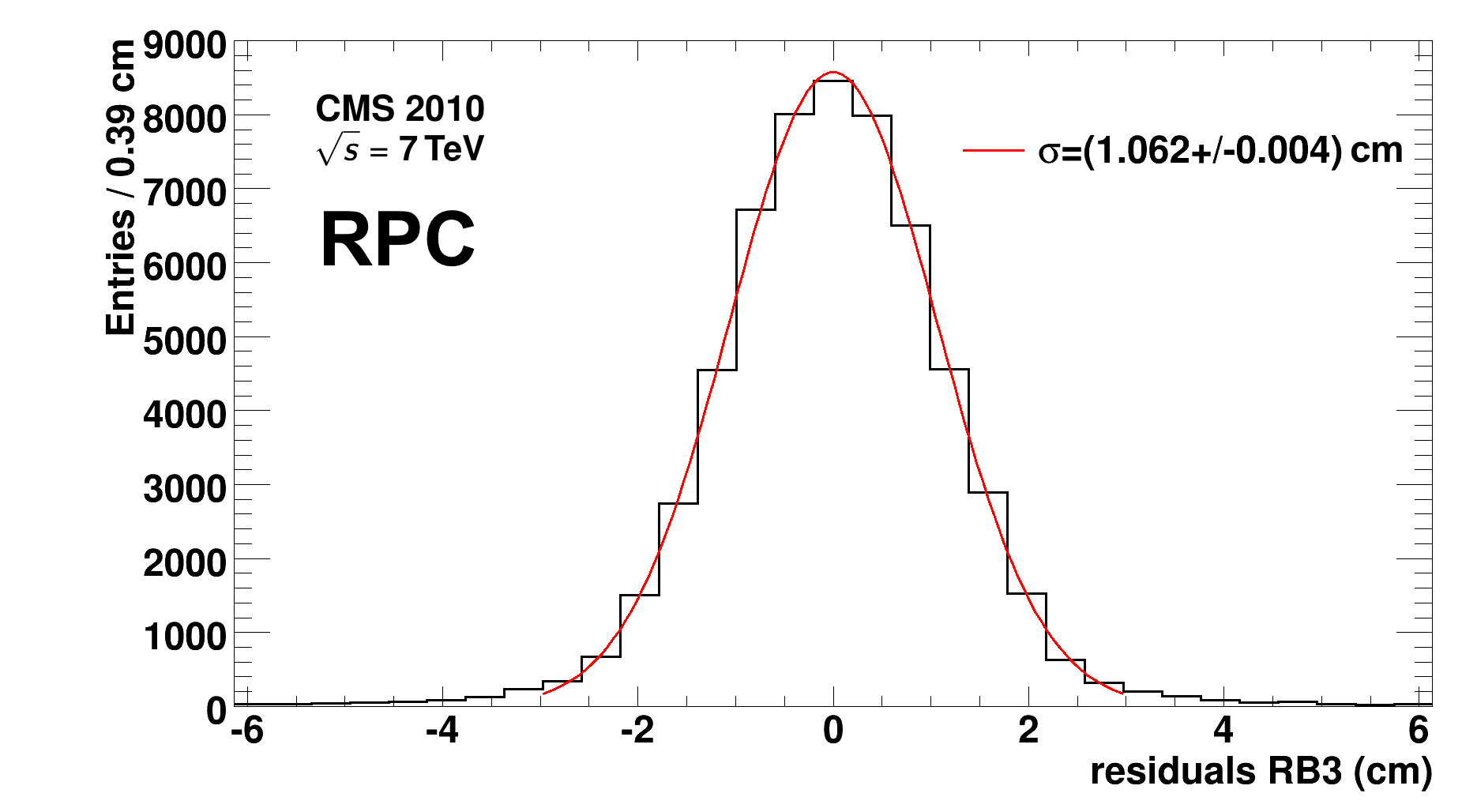}
  \caption{Residuals distribution with a Gaussian fit for RB3 of the barrel RPC system in local $x$ coordinates.}
  \label{resolutionBarrel}
\end{figure}

The RPC hit resolution depends on the strip width (Table~\ref{tab:strippitch}),
the cluster size (Table~\ref{tab:cls}), and the alignment of the RPC chambers.
Since no alignment constants are applied during muon reconstruction
because of the coarse resolution of the RPCs, the RMS
values are shown with and without alignment in Table~\ref{tab:rms}.
The measured spatial resolutions are shown by chamber type in Table~\ref{tab:sigmaali} and are between around 0.8 and 1.3\unit{cm}.
The cluster size measured in strip units (Table~\ref{tab:cls}) decreases for increasing radial distance $r$ from the beam line, following the increasing strip size (Table~\ref{tab:strippitch}).
An example of a residuals distribution fitted with a Gaussian is shown in Fig.~\ref{resolutionBarrel}.
The Gaussian fit results for all the different RPC residuals distributions in CMS (Table~\ref{tab:sigmaali}) constitute the definitive resolution measurements for the RPC system.
As expected, a clear correlation can be seen between RPC resolution and strip width by comparing Tables~\ref{tab:strippitch} and \ref{tab:sigmaali}.

\begin{table}[htbp]
  \centering
  \topcaption{Strip widths for the RPCs. The numbers in parentheses correspond to the disks/stations that have the same strip width, \ie, RE(2,3) means RE$\pm$2 and RE$\pm$3.}
  \label{tab:strippitch}
  \begin{tabular}[t!]{|c|c||c|c|c|c|c|}
    \hline
    \multicolumn{2}{|c||}{Barrel} & \multicolumn{4}{|c|}{Endcaps} \\ \hline
    Layer  & Width (cm)             &  Ring    & Average        &  Ring   & Average \\
           &                        &          & width (cm)     &           & width (cm) \\ \hline \hline
    RB1in  & 2.28                   &  RE1/2/A & 2.38           &    RE(2,3)/2/A & 2.55 \\  
    RB1out & 2.45                   &  RE1/2/B & 2.09           &    RE(2,3)/2/B & 2.23 \\  
    RB2in  & 2.75                   &  RE1/2/C & 1.74           &    RE(2,3)/2/C & 1.95 \\  
    RB2out & 2.95                   &          &                &  RE(1,2,3)/3/A & 3.63 \\ 
    RB3    & 3.52                   &          &                &  RE(1,2,3)/3/B & 3.30 \\  
    RB4    & 4.10                   &          &                &  RE(1,2,3)/3/C & 2.93 \\ \hline
  \end{tabular}
\end{table}

\begin{table}[p]
  \centering
  \topcaption{Measured average cluster sizes in strip units for different RPC strip widths.}
  \label{tab:cls}
  \begin{tabular}[t!]{|c|c||c|c|c|r|c|}
    \hline
    \multicolumn{2}{|c||}{Barrel} & \multicolumn{4}{|c|}{Endcaps} \\ \hline
    Layer   & Cluster  size &  Ring   & Cluster size &  \multicolumn{1}{c|}{Ring}   & Cluster size \\
                  &  (strip units)    &              &  (strip units)   &             &  (strip units)        \\ \hline \hline
    RB1in  & 2.20           &  RE1/2/A & 2.08    &    RE(2,3)/2/A &  1.88\\  
    RB1out & 2.12           &  RE1/2/B & 2.29    &    RE(2,3)/2/B &  2.01\\  
    RB2in  & 1.96           &  RE1/2/C & 2.27    &    RE(2,3)/2/C &  2.46\\  
    RB2out & 1.93           &  &                 &  RE(1,2,3)/3/A & 1.64 \\  
    RB3    & 1.80           &  &                 &  RE(1,2,3)/3/B & 1.57 \\   
    RB4    & 1.63           &  &                 &  RE(1,2,3)/3/C & 1.80 \\ \hline
  \end{tabular}
\end{table}

\begin{table}[p]
  \centering
  \topcaption{Residuals distribution RMS, with and without alignment (Align) for different RPC strip widths.}
  \label{tab:rms}
  \begin{tabular}[t!]{|c|cc||c|cc|r|cc|}
    \hline
    \multicolumn{3}{|c||}{RMS Barrel} & \multicolumn{6}{c|}{RMS Endcaps} \\ \hline
    Layer  & Align  & {no-Align}   &  Ring   & Align & \multicolumn{1}{c|}{no-Align}  &  \multicolumn{1}{c|}{Ring}   & Align & {no-Align} \\
                 & (cm)  &  {(cm)} &            & (cm)  &  \multicolumn{1}{c|}{(cm)} &            & (cm) & {(cm)} \\    \hline \hline
    RB1in  &  1.23    &  1.24             &  RE1/2/A & 1.07 & 1.08 &    RE(2,3)/2/A   & 1.37 & 1.42\\  
    RB1out &  1.32    &  1.36             &  RE1/2/B & 0.99 & 1.00 &    RE(2,3)/2/B   & 1.27 & 1.32\\   
    RB2in  &  1.56    &  1.72             &  RE1/2/C & 1.09 & 1.10 &    RE(2,3)/2/C   & 1.10 & 1.14\\   
    RB2out &  1.54    &  1.55             &  & &                   &  RE(1,2,3)/3/A & 1.70 & 1.77\\  
    RB3    &  1.60    &  1.61             &  & &                   &  RE(1,2,3)/3/B & 1.68 & 1.73\\  
    RB4    &  1.93    &  1.95             &  & &                   &  RE(1,2,3)/3/C & 1.42 & 1.48\\ \hline
  \end{tabular}
\end{table}

\begin{table}[p]
  \centering
  \topcaption{Position resolution ($\sigma$) per chamber type for the RPCs with alignment. The uncertainty on each resolution value is smaller than 0.01\unit{cm}.}
  \label{tab:sigmaali}
  \begin{tabular}[t!]{|c|c||c|c|c|r|c|}
    \hline
    \multicolumn{2}{|c||}{Barrel} & \multicolumn{4}{c|}{Endcaps} \\ \hline
    Layer   & $\sigma$ (cm)  &  Ring   & $\sigma$ (cm) &  \multicolumn{1}{c|}{Ring}   & $\sigma$ (cm) \\ \hline \hline
    RB1in  &  0.81              &  RE1/2/A & 0.94       &  RE(2,3)/2/A &  1.07\\   
    RB1out &  0.90              &  RE1/2/B & 0.88       &  RE(2,3)/2/B &  0.96\\  
    RB2in  &  1.03              &  RE1/2/C & 1.05       &  RE(2,3)/2/C &  0.86\\   
    RB2out &  0.99              &  &                    &  RE(1,2,3)/3/A & 1.11\\  
    RB3    &  1.06              &  &                    &  RE(1,2,3)/3/B & 1.28\\  
    RB4    &  1.32              &  &                    &  RE(1,2,3)/3/C & 1.10\\ \hline
  \end{tabular}
\end{table}

\section{Time resolution}

In addition to a measurement of the track position and direction, the DTs
and CSCs provide a measurement of the arrival time of a muon in a
chamber.
The RPCs also provide a very good time measurement.
The resolution of these measurements is discussed in the
following sections.

\subsection{Time measurement in the DTs}

\begin{figure}[htb]
  \centering
  \includegraphics[height=0.65\textwidth]{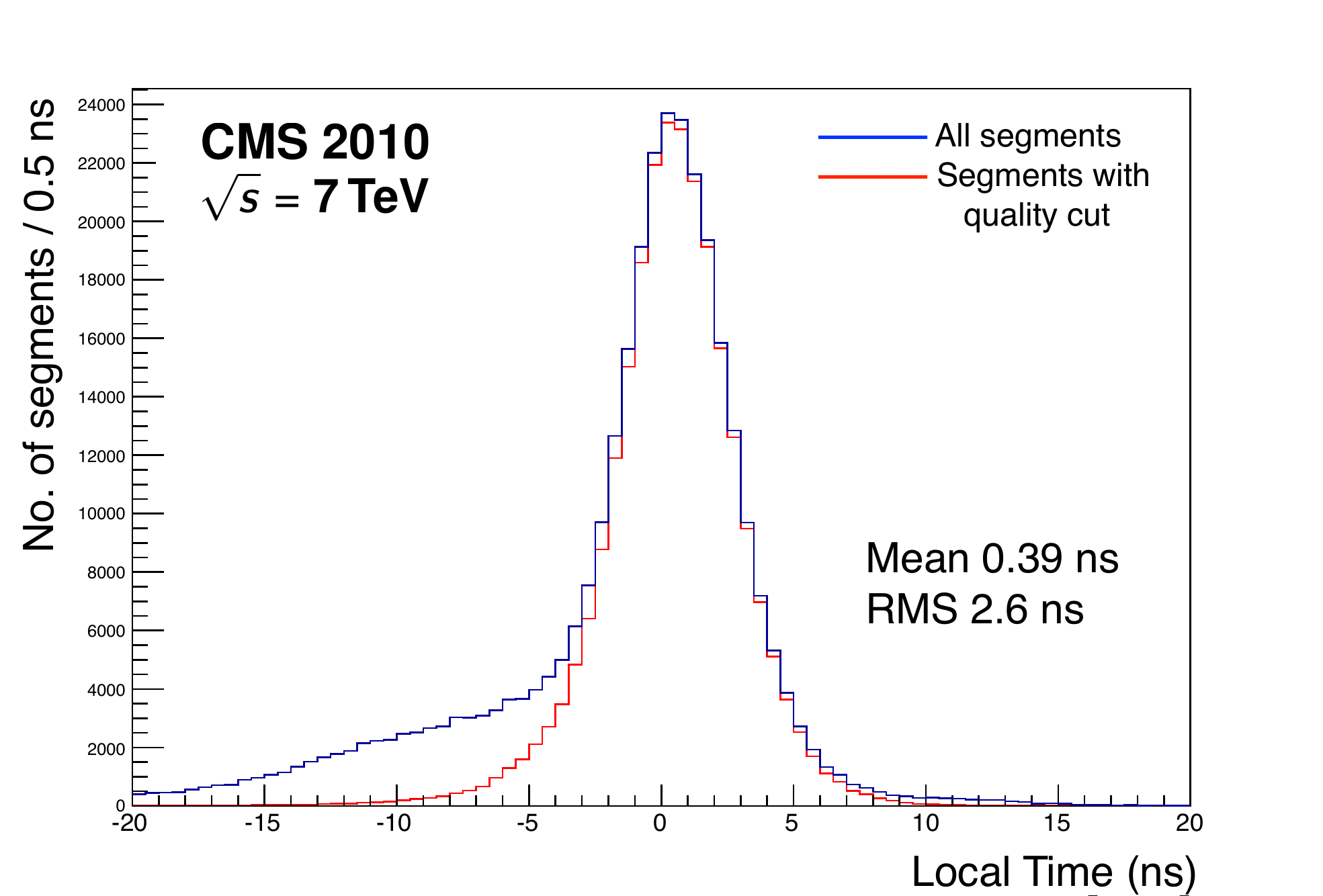}
  \caption{
    Distribution of the local times, as measured in the $r$-$\phi$
    projection of DT chambers in a sample of a high-\pt prompt muon tracks.
    The distribution of all segments (blue line) shows a clear tail from delta-rays; also shown is the distribution of segments after a $\chi^2$ quality cut (red line).
  }
  \label{fig:DTTimeReso}
\end{figure}

The arrival time of a muon track in each DT chamber is reconstructed as follows.
The distance of all hits from the anode wire includes an offset common to all hits, which is taken as a free parameter in the segment fit~\cite{CMSNOTE:2008017}.
Assuming a constant drift velocity, this common displacement corresponds to
a shift in the time of the track, henceforth called local time, with respect to the mean value of the times of
the sample of prompt high-\pt muon tracks used during the
calibration process (cf. Section~\ref{section-dt-calibration}).
This common shift takes into account variations in the arrival time of the muon caused by different bending angles of tracks for different \pt values, and to uncertainties in the calculation of the propagation time of the signal along the wire.
The distribution of these local times, as measured in the $r$-$\phi$ projection in a
sample of high-\pt prompt muon tracks (\pt $>10$\GeVc), is shown
in Fig.~\ref{fig:DTTimeReso}; the overall RMS resolution is better than 2.6\unit{ns}.
For a given track, the spread in the time measurement in different chambers is typically less than 0.2\unit{ns}.
The tail at low values is due to the inclusion of delta-ray hits in the fit,
which may mask the genuine track hit in the same cell.
Hits originating from delta-rays can be removed with a quality cut.

\subsection{Offline CSC timing alignment}
\label{sec:OfflineTimeAlignment}

\begin{figure}[htp]
  \begin{center}
   \includegraphics[width=0.4\textwidth]{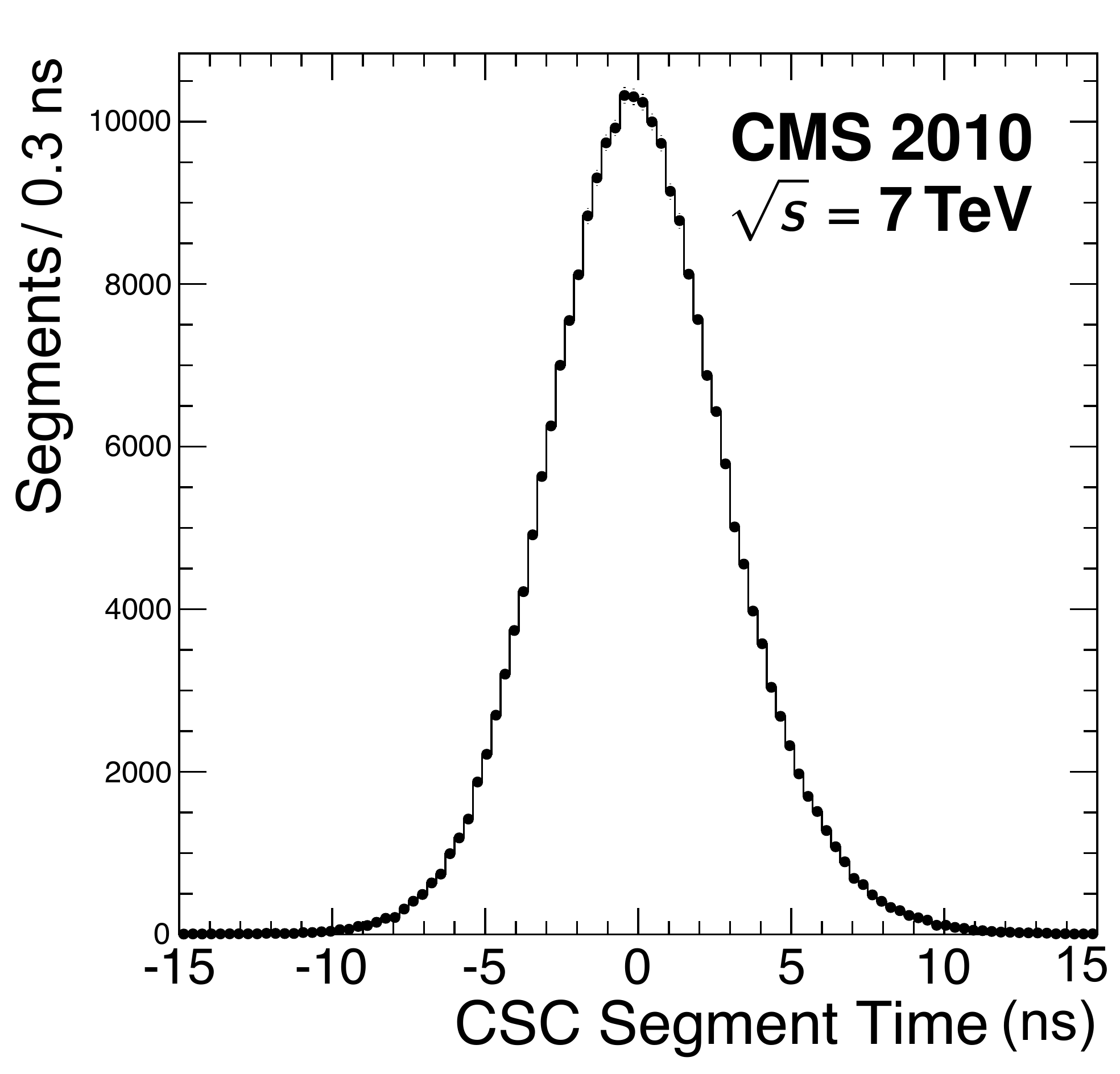}
  \end{center}
  \caption{The CSC segment time distribution for muons with \pt $\ge 20$\GeVc. }
  \label{fig:csc_segtime}
\end{figure}

The CSC hit time is based on the cathode signal, which is amplified, shaped, and then sampled every 50\unit{ns}.
Eight 50-ns samples are saved with the first 2 bins serving as dynamic pedestals~\cite{OSU_NIM}.
The peak time of the pulse is found from a simple comparison of the shape with the known analytical form of the pulse shape delivered by the cathode electronics.
The measured single hit resolution is 5\unit{ns}~\cite{CSCPerfCRAFT}.
Using calibrations and muons from collisions, offsets were derived to shift the average hit time
for each chamber to 0.  These offsets are applied during reconstruction.
To define a CSC segment time, the cathode hit times are combined with the anode hit times as defined in Section~\ref{sec:CSCsync}.
A Gaussian fit to the resulting segment time distribution (Fig.~\ref{fig:csc_segtime}) yields a resolution measurement of about 3\unit{ns}.

\subsection{RPC time measurement capability}

Double-gap RPCs operated in avalanche mode have demonstrated the ability to reach an intrinsic time resolution of around 2\unit{ns}~\cite{RPC_NIM414}.
This has to be folded in with the additional time uncertainty coming from the time propagation along the strip, which contributes about 2\unit{ns}, plus the additional jitter that comes from small channel-by-channel differences in the electronics and cable lengths, again on the order of 1--2\unit{ns}.
These contributions, when added quadratically, give an overall time resolution of better than 3\unit{ns}.
This is much lower than the 25\unit{ns} timing window of the RPC data acquisition system (DAQ) in CMS, which
therefore represents the achievable time precision.

\section{Local reconstruction efficiency}
\label{Efficiency}
The global reconstruction of muons relies on the local reconstruction of
objects inside the individual muon chambers. The 3 muon detector systems (DT, CSC, and RPC) use
different techniques to register and reconstruct signals originating from charged particles traversing them.
Still, in all cases the basic objects are ``reconstructed hits''  or ``rechits'' (\ie, 2D or
3D spatial points with assigned uncertainties)
and segments, obtained by fitting straight lines to the reconstructed hits.

In this paper we do not present reconstructed muon efficiencies~\cite{POG-paper} or muon trigger efficiencies since they convolute detector performance with other aspects of the CMS detector, trigger, and software.
Instead we restrict our measurements to the level of local reconstruction and local trigger.
This means restriction to the reconstruction of hits and track segments for muons in the muon chambers, and to the formation of local trigger objects based on chamber information that are passed on to the central CMS muon trigger system.
Inefficient regions and non-functional electronic channels in the muon system were in general not excluded from the efficiency
calculations, but if an entire CSC failed to provide any rechits at all, it was excluded.
At any time in 2010 about  8 of the total 473 CSCs (1.7\%) in the system were non-operational due to electronic failures like this.
The fraction of non-operating electronic channels  in the CSC system was otherwise small. In the DT system, the fraction of non-operational channels was very small (less than 1\%; see Appendix A).

The track segment reconstruction in the barrel DT chambers proceeds as follows:
first, a hit reconstruction consisting of deriving spatial points from the TDC time
measurements; second, a linear fit of these points in the 2 projections of a chamber (8~$\phi$-layers
and 4~$\theta$-layers), to perform a local pattern recognition
and obtain reconstructed segments.
The first step starts with the calibration of the TDC output to get the real drift times (see Section~\ref{section-dt-calibration-introduction}) of the
ionized charge within the tube. Since a DT cell is 42\unit{mm} wide and has a central wire, the maximum
drift distance is 21\unit{mm}. Then, multiplying the drift times by a known drift velocity, 2 space points (rechits)
are obtained, left and right, at equal distance from the wire. These are the inputs to the linear fit
that attempts to associate the majority of rechits to a segment. Hits that are inconsistent with the fit, yielding high
segment $\chi^2$ values, are discarded. At least 3 rechits from different layers are required to build a segment.

In the endcap CSCs, the rechit reconstruction is based on information from the strips (local $x$ or $\phi$ coordinates) and wires (local $y$ coordinate).
A rechit is built only if signals from both strips and wires are present in a given layer.
The strip width varies between 0.35 and 1.6\unit{cm} for different chamber sizes and locations, and a typical muon signal is contained within 3 to 6 strips.  
The charge distribution of the strip signals is well described by a Gatti function \cite{GattiP,
after_Gatti1, after_Gatti2}, which is the basis of the local $x$ coordinate reconstruction.
The center of gravity of the shower shape induced on a group of contiguous strips is obtained by using a parameterization of the expected Gatti shower shape distribution depending on the measured signals, and on the local strip width and other characteristics of the specific CSC type.
This is considerably faster than an explicit  fit to the expected Gatti shower shape to extract the position, and just as precise.
The uncertainty in the estimated position is also extracted from a parameterization of the associated measurement uncertainties obtained from studies of the CSC response in test beam and collision data.
Crosstalk between neighboring strips is unfolded by using the known crosstalk matrix from the CSC calibration measurements.
The CSC wire signals are read out in groups with widths between 2 and 5\unit{cm}; typically only 1 or 2 wire groups have signals due to a traversing muon.
A rechit is built at each overlap of a strip cluster with a hit wire group (or pair of hit wire groups).
Each rechit has a position in terms of local $\phi$ (equivalent to local $x$) and local $y$.
The known geometrical positions of the chambers allow the transformation of these non-orthogonal local coordinates
(strips and wires are not, in general, aligned
with orthogonal global coordinates ($x$, $y$, $z$) where $z$ is the global $z$ coordinate of a given CSC).
The appropriate covariance matrix is also transformed from local to global coordinates, which in general introduces correlations between the global $x$ and $y$ positions.
This matrix is used both when building track segments from rechits and in full muon track reconstruction.
Segments are built from the available rechits in the 6 layers of each CSC.
The straightest pattern of hits through the chamber is found by a spanning tree algorithm, using only 1 rechit per layer, and at least 3 layers.
A segment is then formed by performing a least-squares fit of the selected rechits to a straight line.
Rechits that are rejected are typically those that have positions distorted by the presence of $\delta$ electrons, or are very close to the edge of a chamber's sensitive region.

The inputs to the RPC local reconstruction are the strips that have signals in a given event.
The strips that are
 next to each other are grouped into a ``strip cluster'', and the average position of the
strips that form the cluster constitutes the reconstructed hit of a given RPC detector.
The uncertainty on the measurement is set to the standard deviation of a uniform distribution
(\ie, the size of the cluster along each direction divided by $\sqrt{12}$).

A crucial aspect of the efficiency measurements is the definition of the probe
used to measure the efficiency of an object.
There are a few basic alternatives that can be explored.

The tag-and-probe technique (see Section~\ref{Trig_Effic}), using muons from the decay of \JPsi, \PgU, and Z resonances, can be used to measure muon efficiencies since such muons can be reliably identified without use of information from the muon system. Trigger and selection requirements limit the size of the samples of such events. An alternative method based on inclusive muons can provide larger samples, especially in the forward region, and at low \pt.

Both methods require the projecting of tracks from the inner tracker region to the muon system through the detector material and magnet steel, and multiple scattering introduces dispersion between the tracks and the associated hits and segments in the muon chambers.
A third method avoids this problem by using the reconstructed segments as probes to measure the efficiency for hit reconstruction in each layer of a muon chamber.

Similarly, for the specific case of the RPC system, where the chambers are firmly attached to the DTs and CSCs, segments reconstructed in the DTs or CSCs are directly
used as RPC probes. They are extrapolated to the RPC layers to measure the hit reconstruction efficiency.

\subsection{Segment reconstruction efficiency in the DTs and CSCs based on the tag-and-probe method}

 The segment reconstruction efficiency was measured by using the tag-and-probe method applied to well-identified muons from \JPsi and Z decays selected from the 2010 collision data.
The same technique was applied to appropriate samples of simulated events.

As discussed in Section~\ref{Trig_Effic}, a single well-identified muon (the tag) of the pair from a resonance decay is required to
satisfy the trigger requirements, and the other muon (the probe) is identified as such just by virtue of forming the resonance
invariant mass with the tag. Since the probe does not make use of any muon system information,
it can then be used to probe the efficiency of the muon detection and reconstruction.
The selected probe tracks were propagated to the muon stations, starting from their point of closest
approach to the interaction point. The propagation procedure allows
the position of the track to be determined at any surface that it crosses.
Uncertainties on the extrapolated position due to multiple scattering agree with the
MC expectations~\cite{POG-paper}.
To reduce the apparent loss of efficiency that might arise from propagation uncertainties,
the point of intersection of a probe track and a chamber was required to be within the
sensitive volume of a chamber, away from the chamber edges by a distance of
at least the uncertainty on the position of this intersection.

The presence of reconstructed segments was checked for each individual chamber
crossed by the probe tracks.
A ``passing probe'' is defined as a probe that matched a reconstructed segment using an appropriate distance criterion.
The segment reconstruction efficiency for each chamber is defined as

\begin{equation}
\label{effTP}
\epsilon = \frac{{{N}}_\mathrm{pp}}{{{N}}_\mathrm{p}},
\end{equation}

where ${{N}}_\mathrm{p}$ and ${{N}}_\mathrm{pp}$ are the number of probes and passing probes,
respectively, obtained after fitting the tag and probe pair to the
\JPsi and Z invariant mass spectra.

The uncertainties on the efficiency are given by a Clopper--Pearson interval~\cite{Clopper-Pearson},
but in cases
where the efficiencies are not close to 0 or 1 we apply the binomial error formula

\begin{equation}
\label{effTP_err}
  \Delta \epsilon = \sqrt{\frac{\epsilon \left(1-\epsilon \right)}{N_\mathrm{p}}}.
\end{equation}

The efficiency was evaluated as a function of the intersection position in the chamber
and of the \pt, \Pgh, and $\phi$ of the traversing probe track. Compared to the \JPsi, the
Z dimuon sample allows higher \pt ranges to be explored. Overall, the efficiencies  determined from the \JPsi and Z samples are consistent.

Because of energy losses in the traversed material, there is a minimum momentum (or \pt at a given \Pgh )
threshold for muons to reach the muon detector.
The \pt threshold is $\approx$1\GeVc for the forward
region, and increases to $\geq$3\GeVc in the central region. To reduce the effect
of multiple scattering on the efficiency measurement and to ensure that the muon has the energy to further
penetrate all the muon stations, a requirement on the minimal \pt  or $p$ of the track probes is
imposed.
In the following, selections of $\pt > 10\GeVc$ in the barrel and  $p>15\GeVc$ in the endcap are applied
(except for the \pt dependent measurement).

\begin{figure}
{\centering
\includegraphics[width=12.7cm,height=12.7cm]{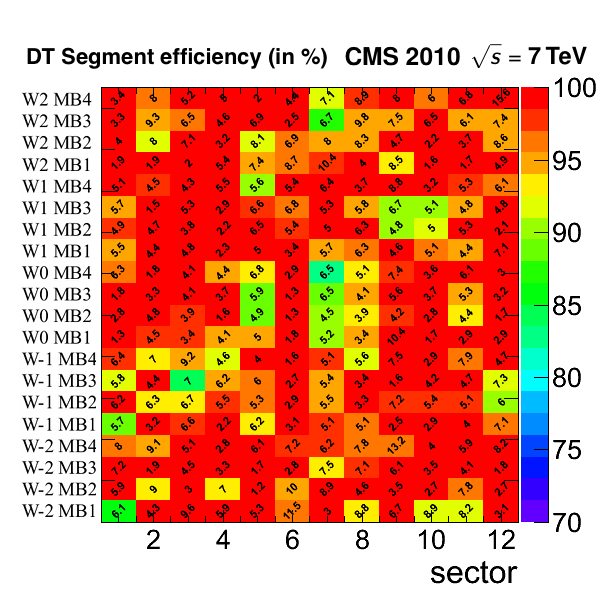}
\caption{\label{EffAllDTdet}
The DT segment reconstruction chamber-by-chamber percentage efficiency (sectors on the horizontal axis, stations and wheels on the vertical axis). Statistical uncertainties are shown as text within the boxes.}
}
\end{figure}

The overall performance of the barrel DT system is summarized in Fig.~\ref{EffAllDTdet}, which shows the typical segment efficiencies for each sector to be better than 95\%.

\begin{figure}
{\centering
\includegraphics[width=7.5cm]{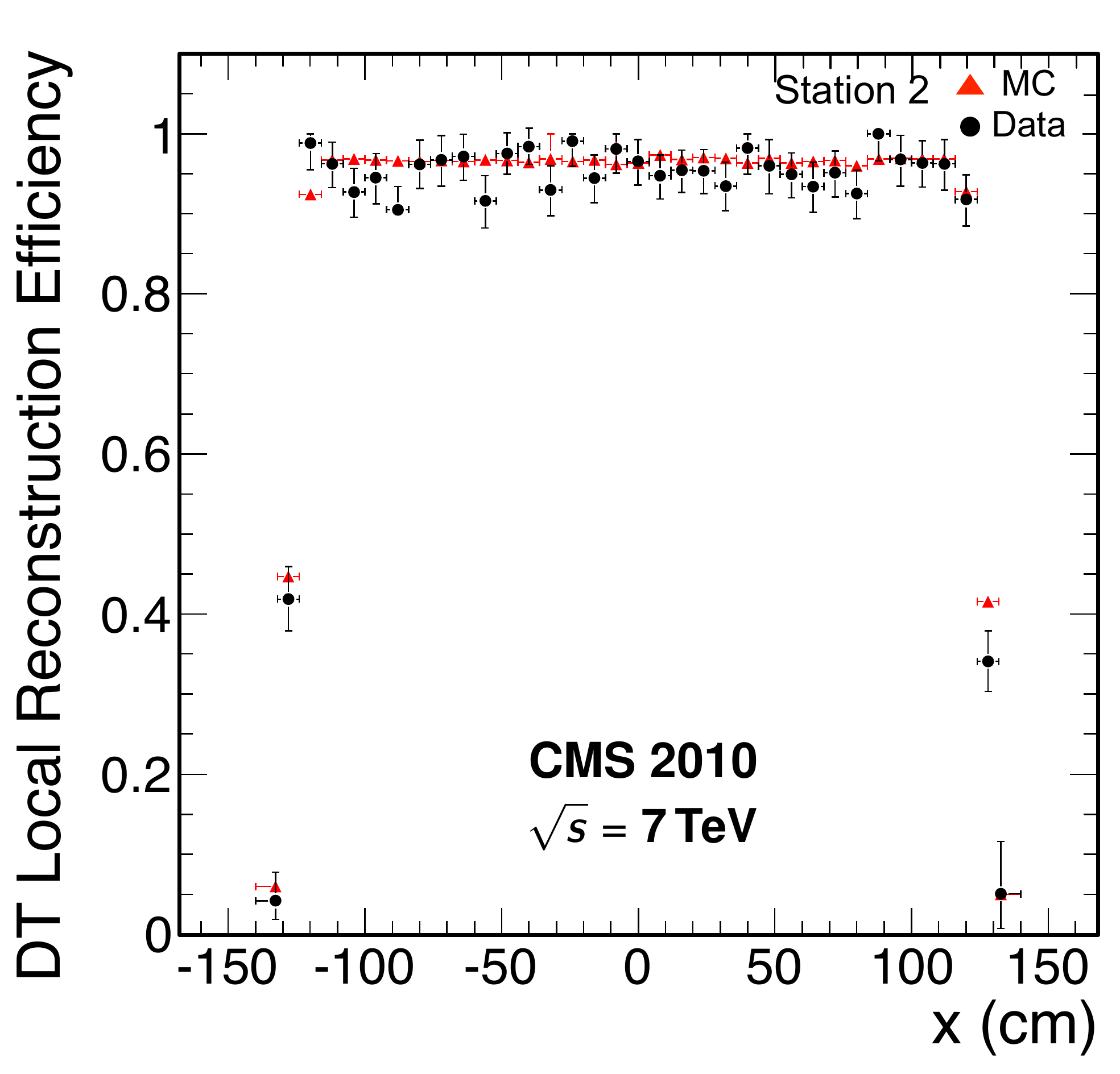}
\includegraphics[width=7.5cm]{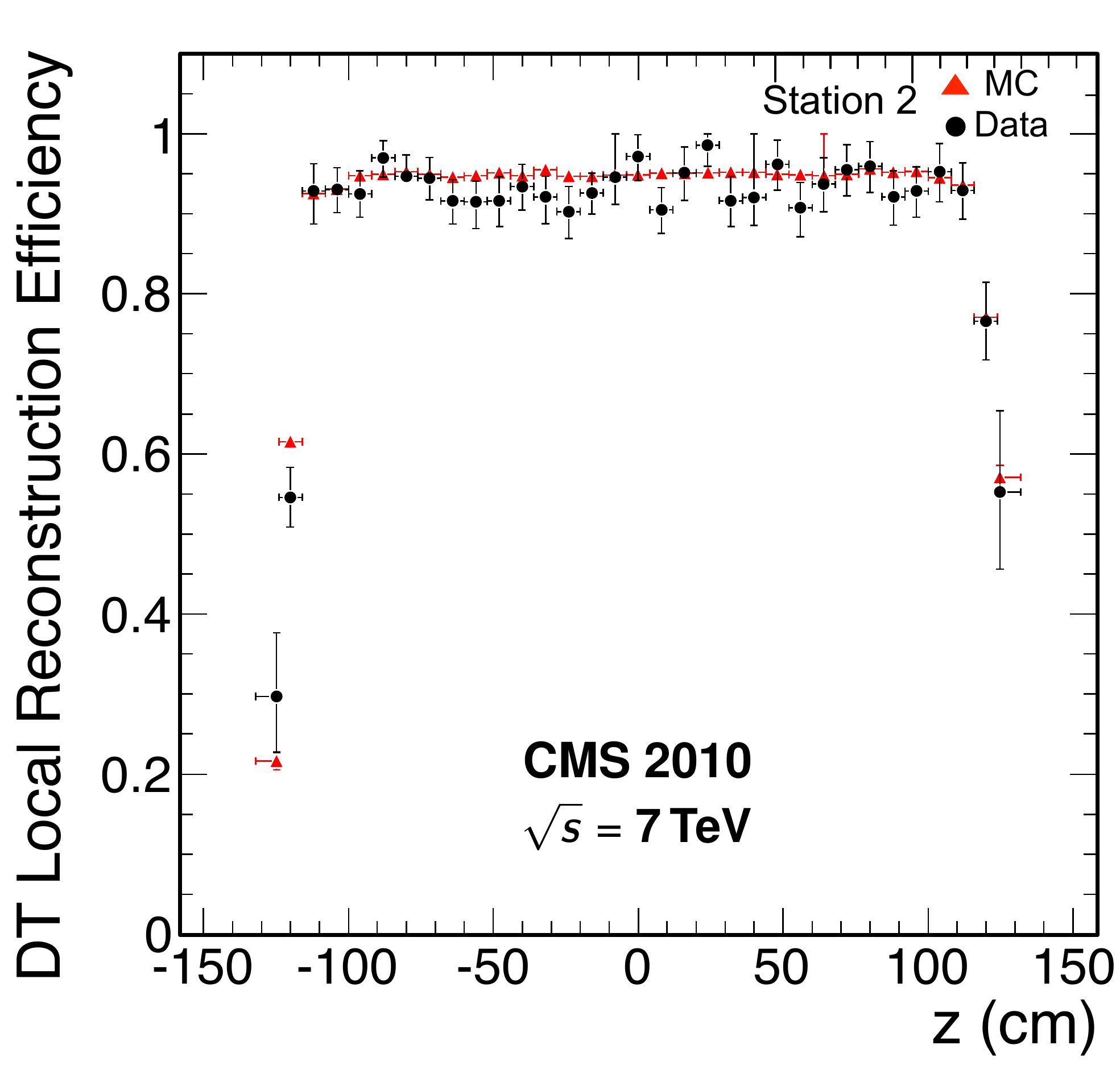}
\caption{\label{Effvslocalpos}
Segment reconstruction efficiency in the MB2 DT station and
comparison with simulated data as a function of local $x$ (left) and local $z$ (right).}
}
\end{figure}

\begin{figure}
{\centering
\includegraphics[width=7.5cm]{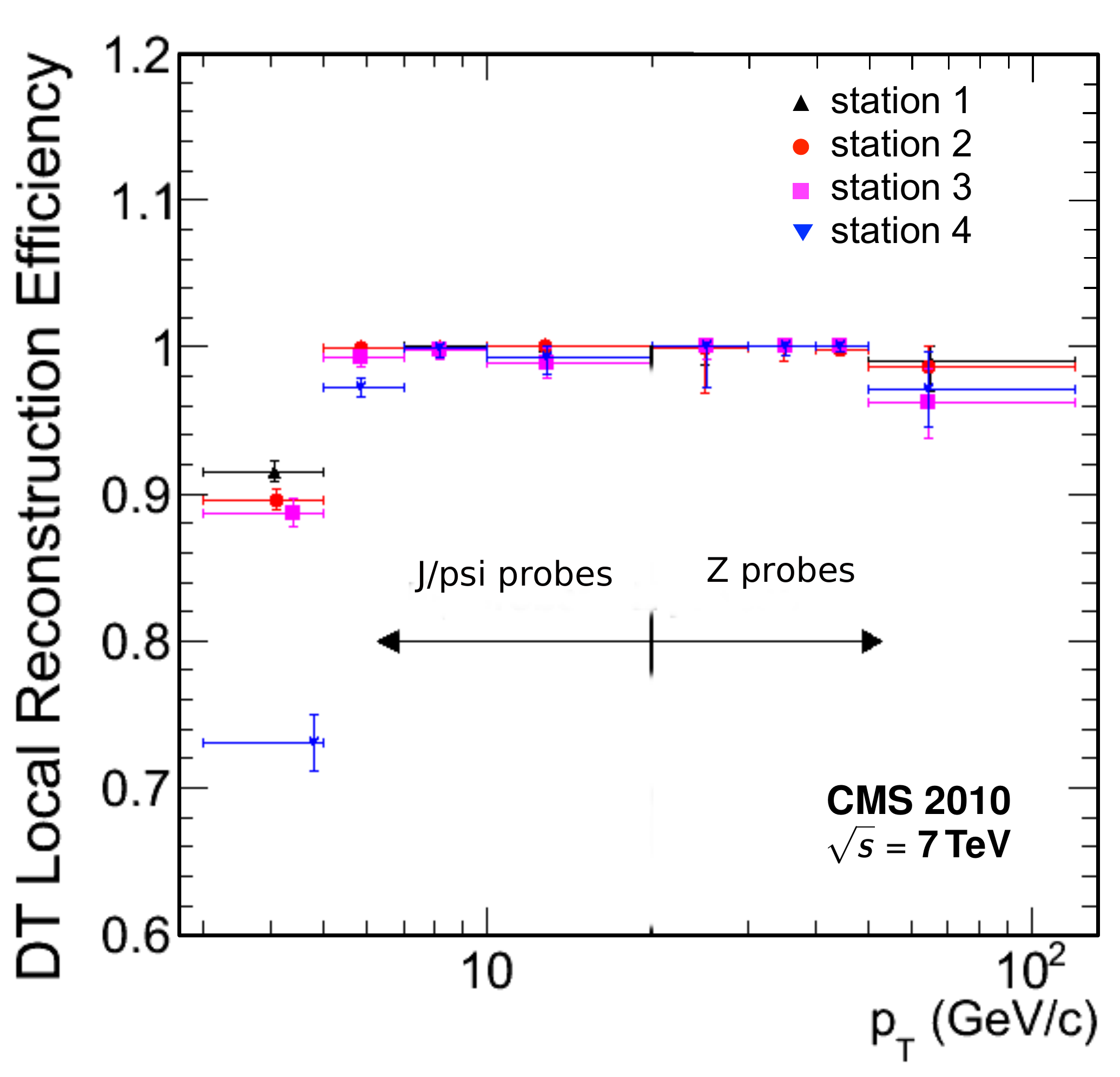}
\includegraphics[width=7.5cm]{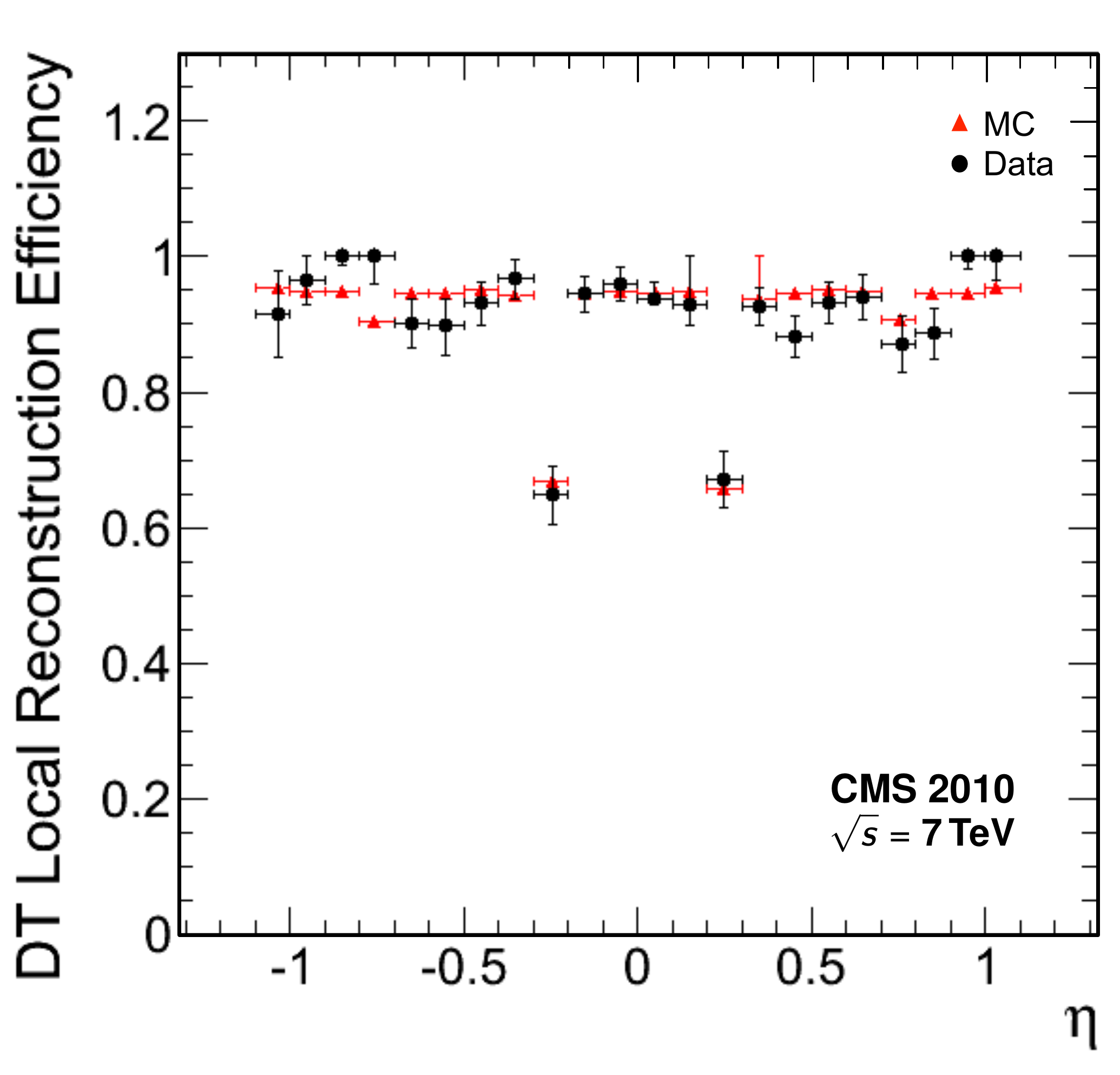}
\caption{\label{EffDTvspt}
Segment reconstruction efficiency as a function of transverse momentum in the 4 barrel DT stations (left). Arrows indicate the ranges covered by probes originating from \JPsi and Z decays, and as a function of $\eta$ in station 2 compared to simulated data (right).}
}
\end{figure}

\begin{figure}
{\centering
\includegraphics[height=6.2cm]{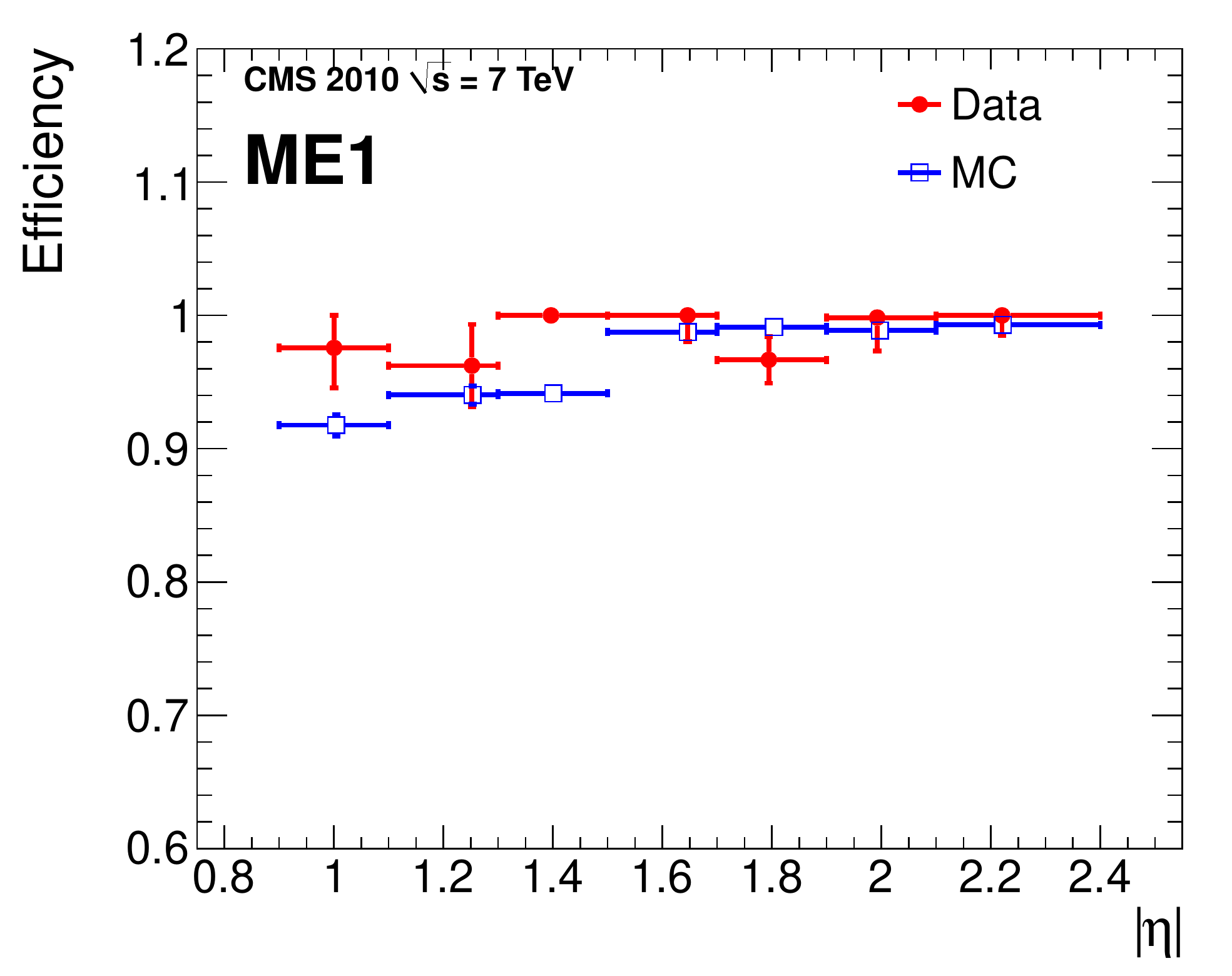}
\includegraphics[height=6.2cm]{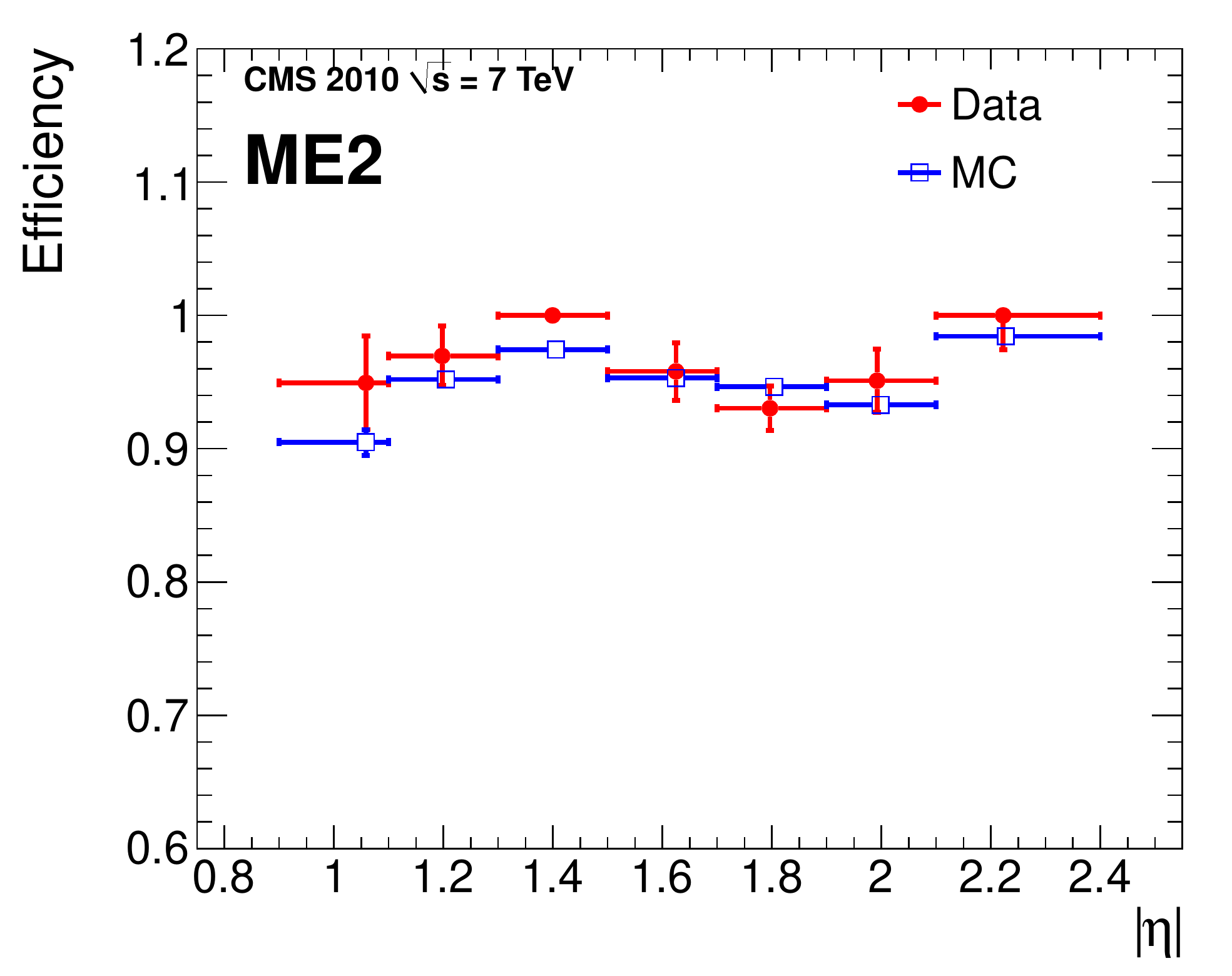}
\includegraphics[height=6.2cm]{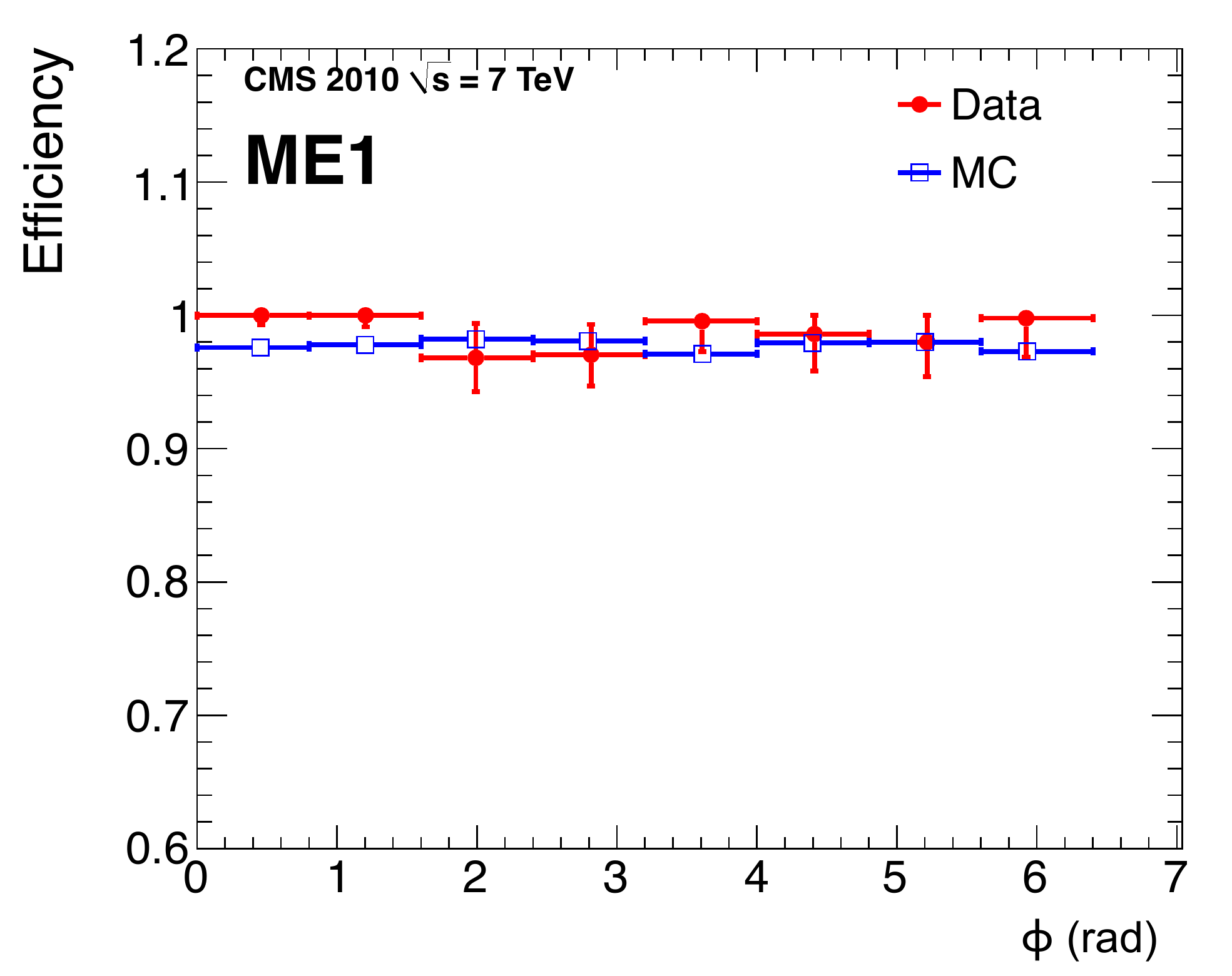}
\includegraphics[height=6.2cm]{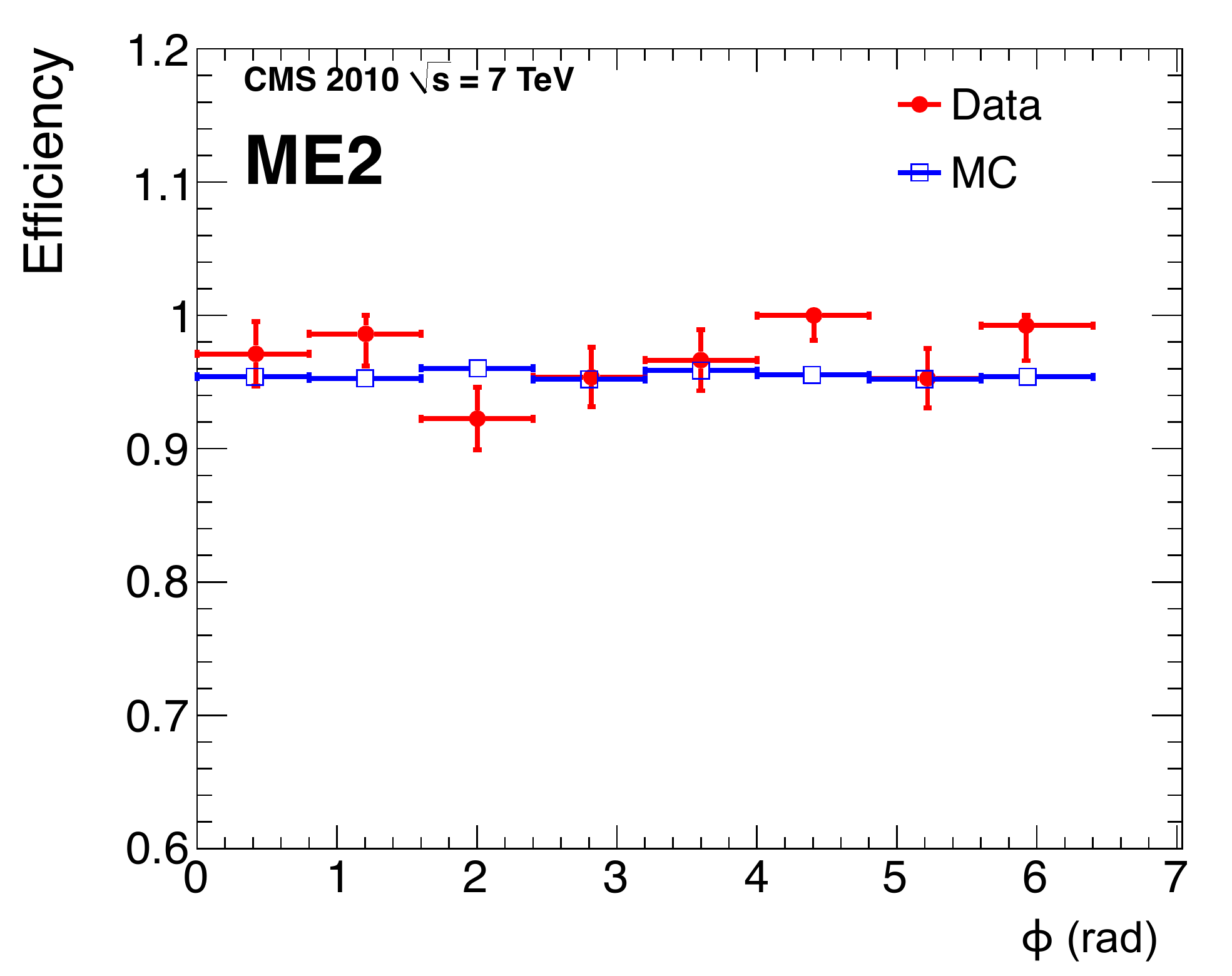}
\includegraphics[height=6.2cm]{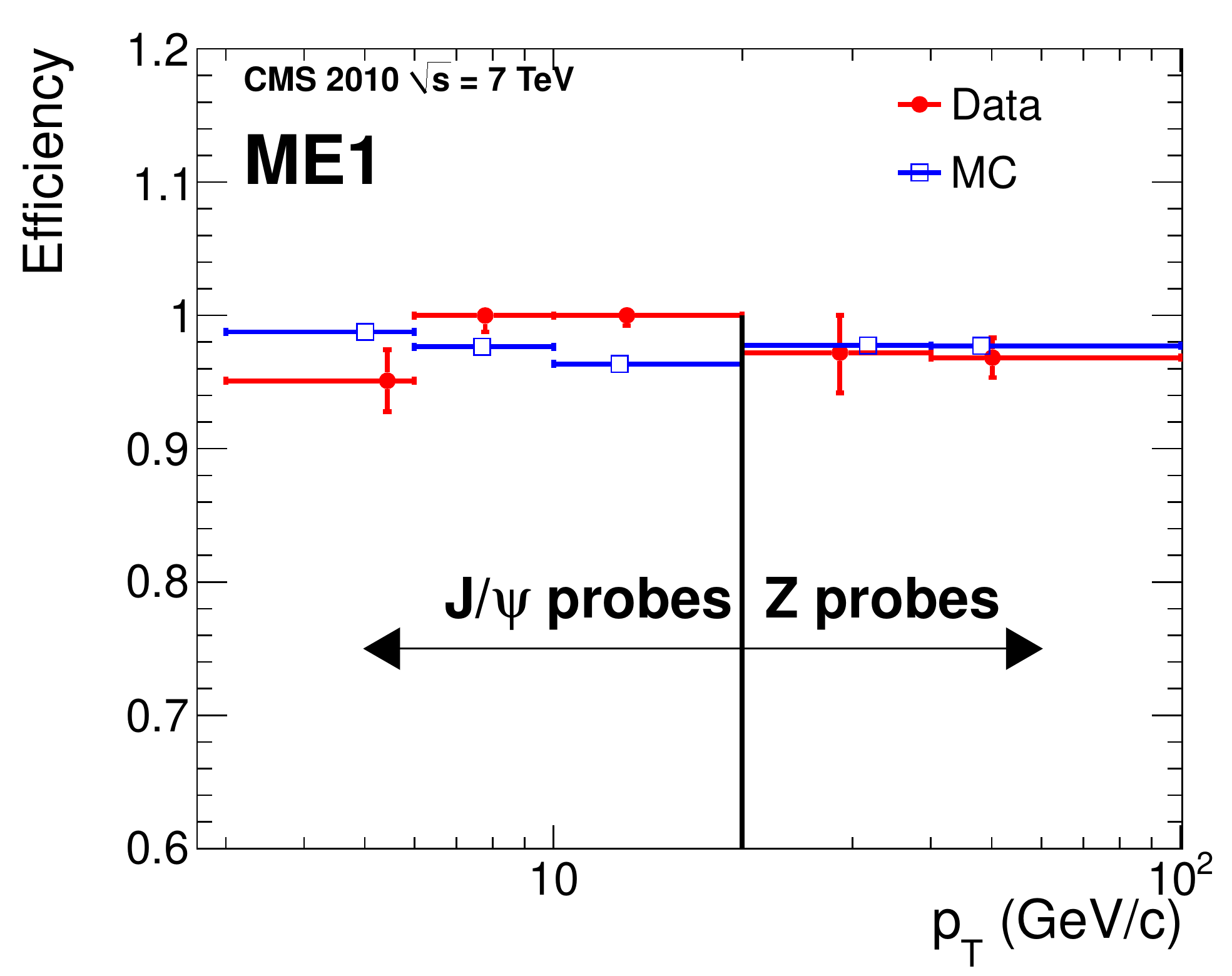}
\includegraphics[height=6.2cm]{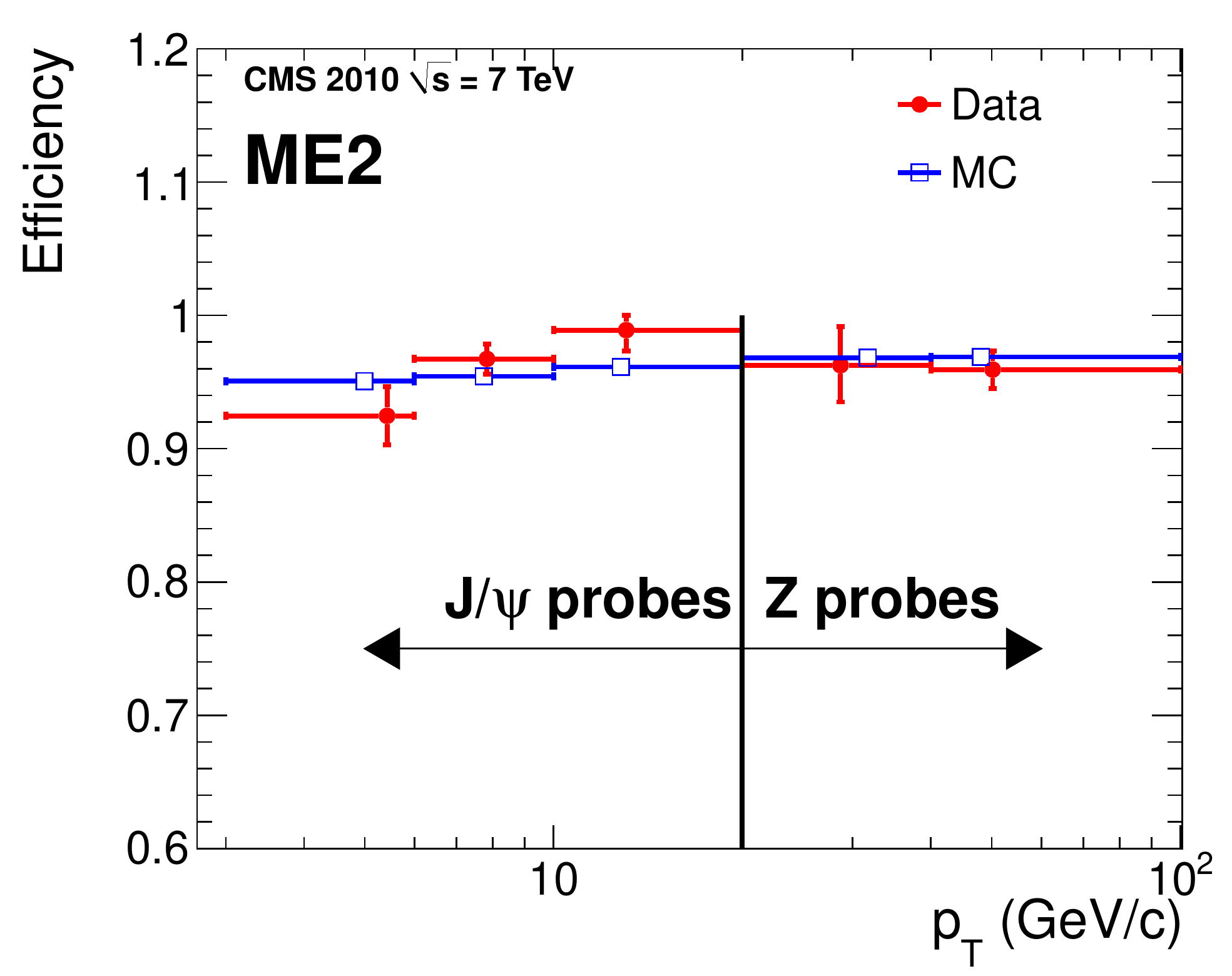}
\caption{\label{EffCSCvspt}
Comparison between measured (red; closed circles) and simulated (blue; open squares) segment reconstruction
efficiencies as functions of \Pgh\ (top), $\phi$ (middle), and \pt (bottom) for endcap CSC stations ME1 (left) and ME2 (right).
The vertical lines in the bottom plots separate the ranges covered by probes
originating from \JPsi and Z decays.
The statistical uncertainties are shown as vertical error bars.
The horizontal error bars show the range of each bin, and within each bin the data point is positioned at the weighted average of all values within that bin.}
}
\end{figure}

Figure~\ref{Effvslocalpos} shows the segment efficiency computed for all barrel sectors and wheels of
the MB2 DT stations as a function of the local $x$ and $z$ coordinates,
respectively (where $x$ is along the layer, normal to the beams, and $z$ is along the beams).
The observed efficiency matches the Monte Carlo expectations all the way to the
edges of the chamber. Similar distributions have been obtained for all other barrel DT stations.
Figure~\ref{EffDTvspt} shows the segment efficiency as functions of the \pt and $\eta$ of the probe track for the 4 barrel DT stations.
Figure~\ref{EffCSCvspt} shows the segment efficiency as functions of \Pgh, $\phi$, and \pt of the probe track for both endcap CSC stations and comparisons with simulation.

 The overall segment reconstruction efficiencies measured in the barrel (DT) and endcap (CSC) muon systems are summarized in Table~\ref{CSCLR:CSCRecoResults}.
 The segment reconstruction efficiencies are susceptible to systematic uncertainties arising from the choice of selection criteria, particularly since multiple scattering can cause deviations between a projected track and the actual position of a segment in a muon chamber.
 In the CSC case, the position matching criteria have been varied over a range that encompasses the average expected deviation of muons attributable to multiple scattering effects (several centimeters), and the resulting efficiencies are stable to within 1\%--2\%.
 We thus assign a systematic uncertainty of 1\%--2\% to the values in the table. Except in station~4, these systematic uncertainties dominate the statistical ones.

There is overall good agreement between data and Monte Carlo simulation (see Appendix~\ref{simulation}) within the uncertainties. For the CSCs, the segment efficiencies closely match the trigger primitive efficiencies of Section~\ref{Trig_Effic}. This is as expected, and is a crosscheck of both independent pathways for CSC data, namely, those sent to the L1 trigger and those sent to offline storage (and the HLT).

In summary, the reconstructed segment efficiency determined using the tag-and-probe method with data is at the level of 95\%--99\% in all muon system stations with a systematic uncertainty of less than 2\%.
The main reasons for segment inefficiencies in the muon chambers are inefficient regions and non-operational electronic channels, although edge effects also affect the measurements, particularly transitions between the rings of CSCs.

\begin{table}[hptb]
\topcaption{Local segment reconstruction efficiency for stations 1--4 of the barrel (DT) and endcap (CSC) muon systems.}
\label{CSCLR:CSCRecoResults}
\begin{center}
\begin{tabular}{|c||cc||cc|}
\hline
\multirow{5}{*}{~} & \multicolumn{2}{c||}{DT Efficiency (\%)} &\multicolumn{2}{|c|}{CSC Efficiency (\%)} \\   \cline{2-5}
                          & Data     &   \multicolumn{1}{|c||}{MC}     & Data   &   \multicolumn{1}{|c|}{MC}    \\ \hline\hline
Station 1                 & 99.2 $\pm$ 0.4 & 98.05 $\pm$ 0.03& 98.9 $\pm$ 0.9  & 97.8 $\pm$ 0.1 \\ 
Station 2                 & 99.0 $\pm$ 0.4 & 98.98 $\pm$ 0.03& 96.8 $\pm$ 0.9  & 95.5 $\pm$ 0.1\\ 
Station 3                 & 99.1 $\pm$ 0.4& 99.08 $\pm$ 0.04& 96.8 $\pm$ 0.9  & 94.1 $\pm$ 0.1 \\ 
Station 4                 & 98.9 $\pm$ 0.6& 99.00 $\pm$ 0.04& 94.9 $\pm$ 1.6  & 91.7 $\pm$ 0.2 \\\hline
\end{tabular}
\smallskip
\end{center}
\end{table}

\subsection{Efficiency measurements from inclusive single muons}

For the CSC system, detailed studies of the reconstruction efficiencies for rechits and segments in each chamber were made by using single prompt muons reconstructed from a data sample collected with jet triggers.
Use of jet-triggered events avoids trigger biases in the muon sample, and a large number of muons could be obtained more rapidly than for the dimuon samples required for the tag-and-probe method.
Stringent criteria, summarized in Table~\ref{TagMuon}, are used to select well-reconstructed muons originating from the proton--proton collision
interaction point, and to reduce the contribution of non-prompt (decay) muons and hadronic punch-through.

The inner track of
the selected muon candidate is propagated through the magnetic field and the detector materials
to the muon chambers.  This defines the probe used to measure the reconstruction efficiencies. If a reconstructed object
(rechit or segment) is found near the track propagation point (within a cone of aperture
$\Delta R=0.01$ around the track) in a given station, the probe is considered efficient.
For stations 1, 2, and 3 the chamber being probed should not be the endpoint of the reconstructed muon.
For chambers in station 4, this requirement is dropped.

\begin{table}
\topcaption{To be selected to contribute to efficiency measurements in the muon endcap, a muon probe must satisfy the following criteria:
 \label{TagMuon} }
\begin{center}
\begin{tabular}{ | l | l |}
\hline
  1 &  be reconstructed separately in both the inner tracker and the muon system\\
\hline
  2 &  have $\pt > 6\GeVc$ and $15 <|p|< 100$\GeVc (to minimize multiple scattering and \\
  &mis-reconstruction)\\
\hline
  3 &  be consistent with originating at the interaction point (to reject decays in-flight \\
  & and mis-reconstruction)\\
\hline
  4 &  have a minimum number (11) of hits in the tracker system (to ensure a good \\
  & momentum measurement)\\
\hline
  5 &  have at least one hit in the muon system even after the combined fit to the tracker \\
  & and muon detector measurements in which hits may be dropped according to \\
  & quality-of-fit requirements (to reject decays in-flight and mis-reconstruction)\\
\hline
  6 &  have good quality fits to both tracker hits alone and to the combination of \\
  & tracker and muon hits \\
\hline
  7 &  have hits in at least 2 muon stations (to reject punch-through by pions, kaons, and jets) \\
\hline
  8 &  be the only muon in the same hemisphere in $z$\\
\hline
\end{tabular}
\end{center}
\end{table}

The CSC system has been designed to be efficient for high-\pt tracks originating from the interaction region, and hence for incidence angles corresponding to straight  lines from the interaction region. Low-\pt tracks can have a variety of incidence angles, because of the effects of the magnetic field and multiple scattering. Thus by requiring higher \pt probe tracks, systematic uncertainties resulting from  incidence angle variations can be reduced, but at the same time increasing statistical uncertainties.

\begin{figure}
{\centering
\includegraphics[width=1.0\textwidth]{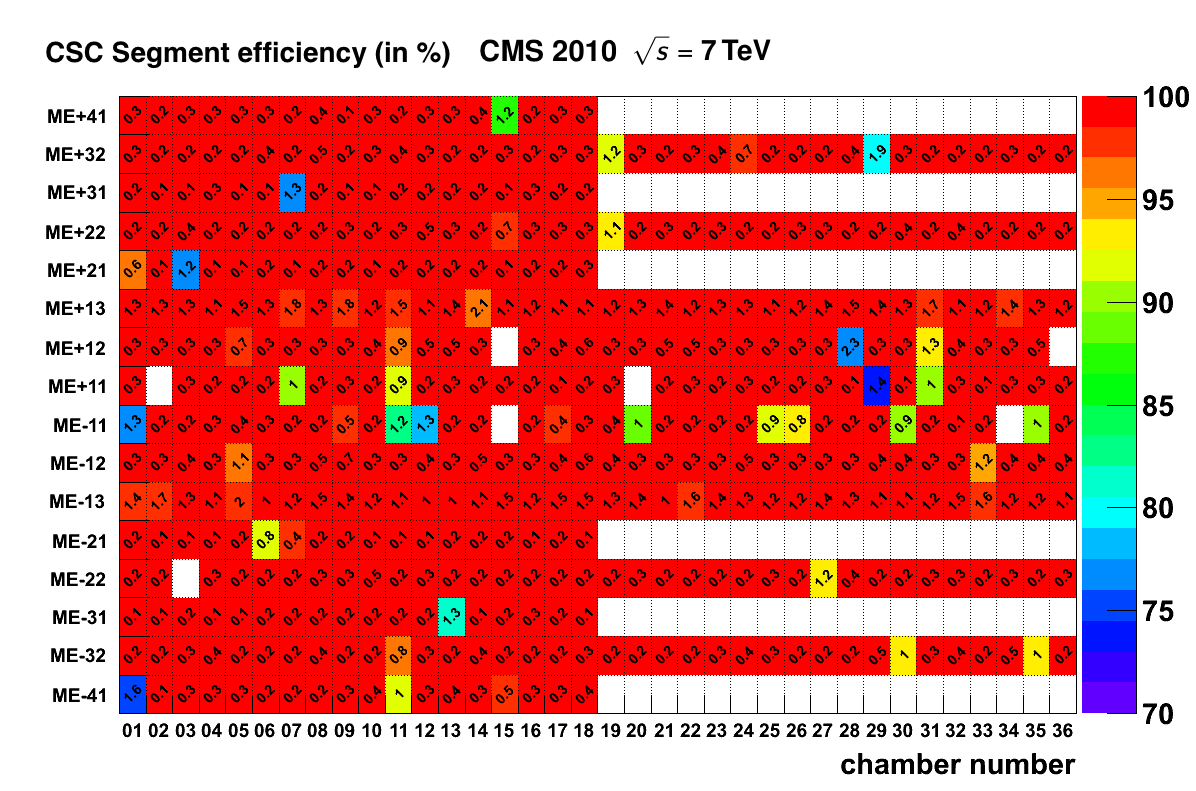}
\includegraphics[width=1.0\textwidth]{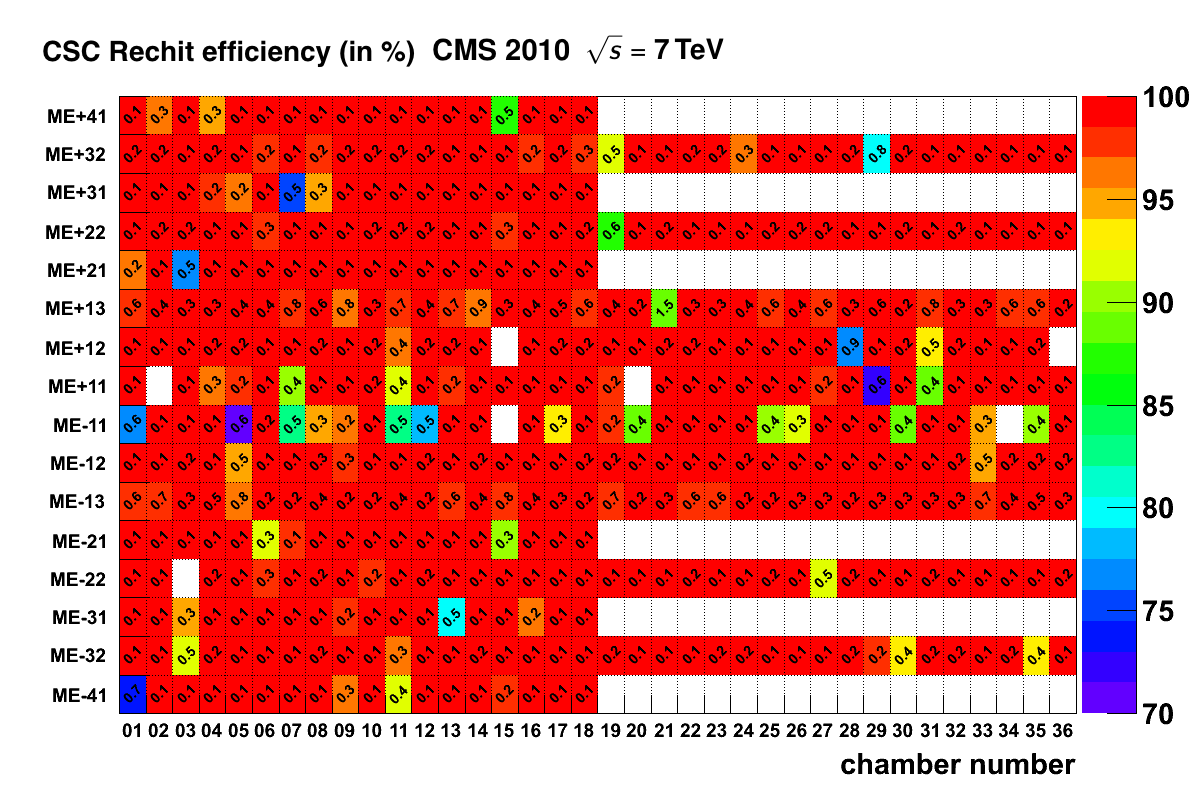}
\caption{\label{Eff_2D_RH_seg}
Reconstruction efficiency in ``active'' CSC regions (per chamber) for
segments (top) and rechits (bottom).
The number in each cell is the estimated statistical uncertainty.
Note that there are only 18 chambers for types ME2/1, ME3/1, and ME4/1 because each chamber subtends 20$^\circ$ in $\phi$, while all the rest cover 10$^\circ$.}
}
\end{figure}

To minimize edge effects, the point at which the probe track intersects a chamber is required to be at least 10\unit{cm} from inefficient regions (high voltage boundaries and geometrical edges).
This requirement
excludes from the measurements strips and wires close to these regions, but should result in the highest intrinsic chamber efficiencies. Figure~\ref{Eff_2D_RH_seg} shows the rechit and
segment efficiencies in these ``active'' regions for the CSC endcap stations and chambers. The segment efficiency is naturally
defined per chamber as segments are built from the information of all 6 chamber layers.
The rechit efficiency is defined per single layer of a chamber. Assuming that no correlation exists between layers
(this assumption depends on the mechanism by which rechits are lost; for example, an inactive HV region could affect a single layer whereas an inactive cathode readout board could affect all 6 CSC layers) the CSC chamber rechit efficiency is defined as

\begin{equation}
\label{eff1}
\bar{\epsilon} = \frac{\sum_i \epsilon_i}{L} =  \frac{\sum_i {n_i}}{N \times L}
\end{equation}
with an estimated uncertainty of
\begin{equation}
\label{err_end}
\Delta \bar {\epsilon} =
        \sqrt{\frac{\bar{\epsilon} \times (1-\bar{\epsilon}) }{L \times N}} ,
\end{equation}
where $L = 6$ is the number of CSC layers,
$\epsilon_i$ is the efficiency in layer $i$ ($i=1,\ldots,6$),
$n_i$ is the number of efficient probes for layer $i$, and
$N$ is the number of all probes traversing the chamber.

With this method, the reconstruction efficiency is
$96.4\pm0.1\%$ (stat.) and $97.0\pm0.1\%$ (stat.) for rechits and segments, respectively, averaged over all endcap CSCs and
restricted to their ``active" regions away from edges and high-voltage supports.
The active region requirement is estimated to result in a systematic uncertainty of up to 0.5\% on each value.
These values also average over chamber regions with non-operational electronic modules,
but exclude chambers that were providing no rechits,
and are in agreement with the tag-and-probe results.

\subsection{Rechit efficiency based on segment propagation}
\subsubsection{DT and CSC}

The hit reconstruction efficiency can also be measured by using reconstructed local segments.
In the absence of major hardware failures (HV faults, gas or readout problems), which may cause
serious malfunctioning or signal losses in single cells or groups of cells, the
signal production in different layers of the same chamber can be considered as a set of
statistically independent processes.
However, to reduce possible biases, loose selection criteria were applied
to the reconstructed segments to discard low-quality segments.

In a barrel DT $\phi$ ($\theta$) chamber, reconstructed segments were required to have at least 5 (3) hits, located in at least 4 out of 8 (3 out of 4) chamber layers,
and a local inclination angle of $\psi < 40^\circ$.
In the endcap CSCs, segments were required to be close ($\Delta R < 1$; defined in Section~\ref{Trig_Effic}) to a probe muon track as described
in Table~\ref{TagMuon}, and have at least 4 out of 6 layers.

Using the set of hits associated with a reconstructed segment, the segment was fitted again, once per layer, ignoring the information for that layer.
Therefore, the
position of the segment in the layer under study is determined in an unbiased way.
Two kinds of efficiencies were considered: the efficiency to find a reconstructed hit within
a cell,
and the efficiency to actually associate a
hit to the segment. The latter efficiency is by definition lower, as it includes the
effects of the calibration and fitting procedures.

\begin{figure}
{\centering
\includegraphics[width=10cm]{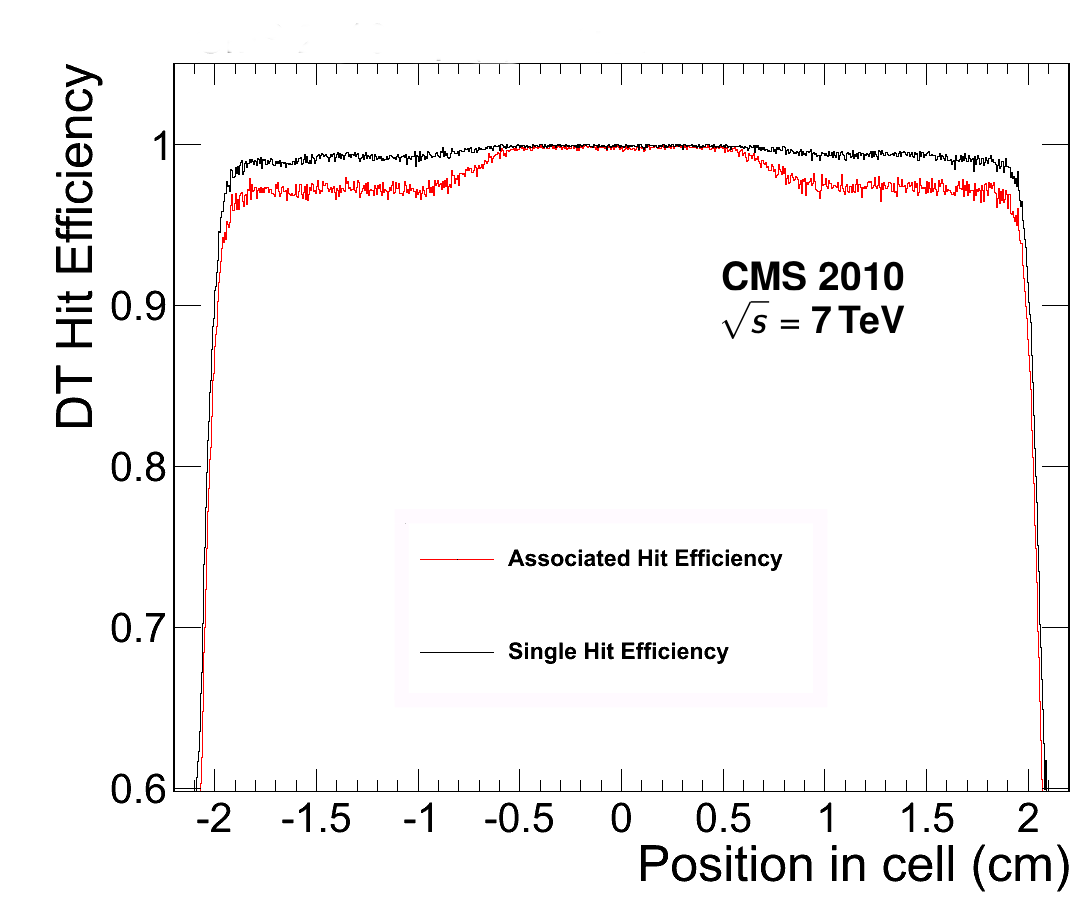}
\includegraphics[width=10cm]{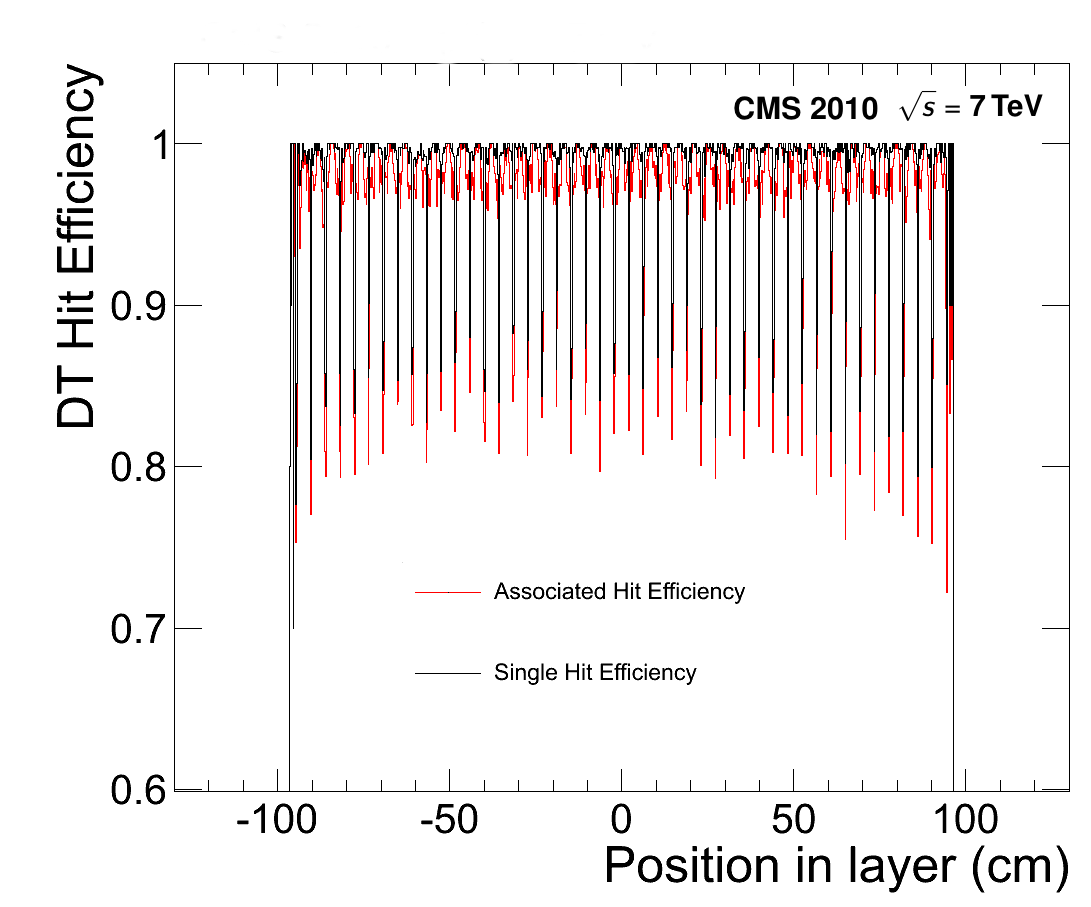}
\caption{\label{hiteffvspos}
Top: hit reconstruction (black) and association (red) efficiencies as a function of the track position in a DT cell.
 Bottom: hit reconstruction (black) and association (red) efficiencies as a function of the track position in a DT MB1 layer.}
}
\end{figure}

Figure~\ref{hiteffvspos} (top) shows both hit reconstruction and hit association efficiency as functions of the
position in a DT cell.
Apart from the known inefficiency induced by the
cathode ``I-beams'' (see Fig.~\ref{fig:dt-chamber-cell}, right) at the edges of the cell \cite{Chatrchyan:2009hg}, the hit reconstruction efficiency is $\ge$99\% everywhere.
The hit association efficiency is, as expected, up to 2\% lower, as it depends on the details of the calibration and contributions from $\delta$ rays.
Indeed, because of the
electronics dead time,  $\delta$ rays may cause an early hit that masks a good one.
In fact, the hit association efficiency matches the reconstruction efficiency in the central region of the cell, where this $\delta$-ray effect is smaller.
Figure~\ref{hiteffvspos} (bottom) shows the hit reconstruction and the hit association efficiencies as functions of the
position in the layer for a subset of DT MB1 chambers.
The efficiency is approximately constant along the layer and the cell structure is clearly visible.
Overall, the hit reconstruction
efficiency in the barrel DT system is on average $\approx$98\%, whereas the association efficiency is $\approx$96\%.

\begin{figure}
{\centering
\includegraphics[width=0.48\textwidth]{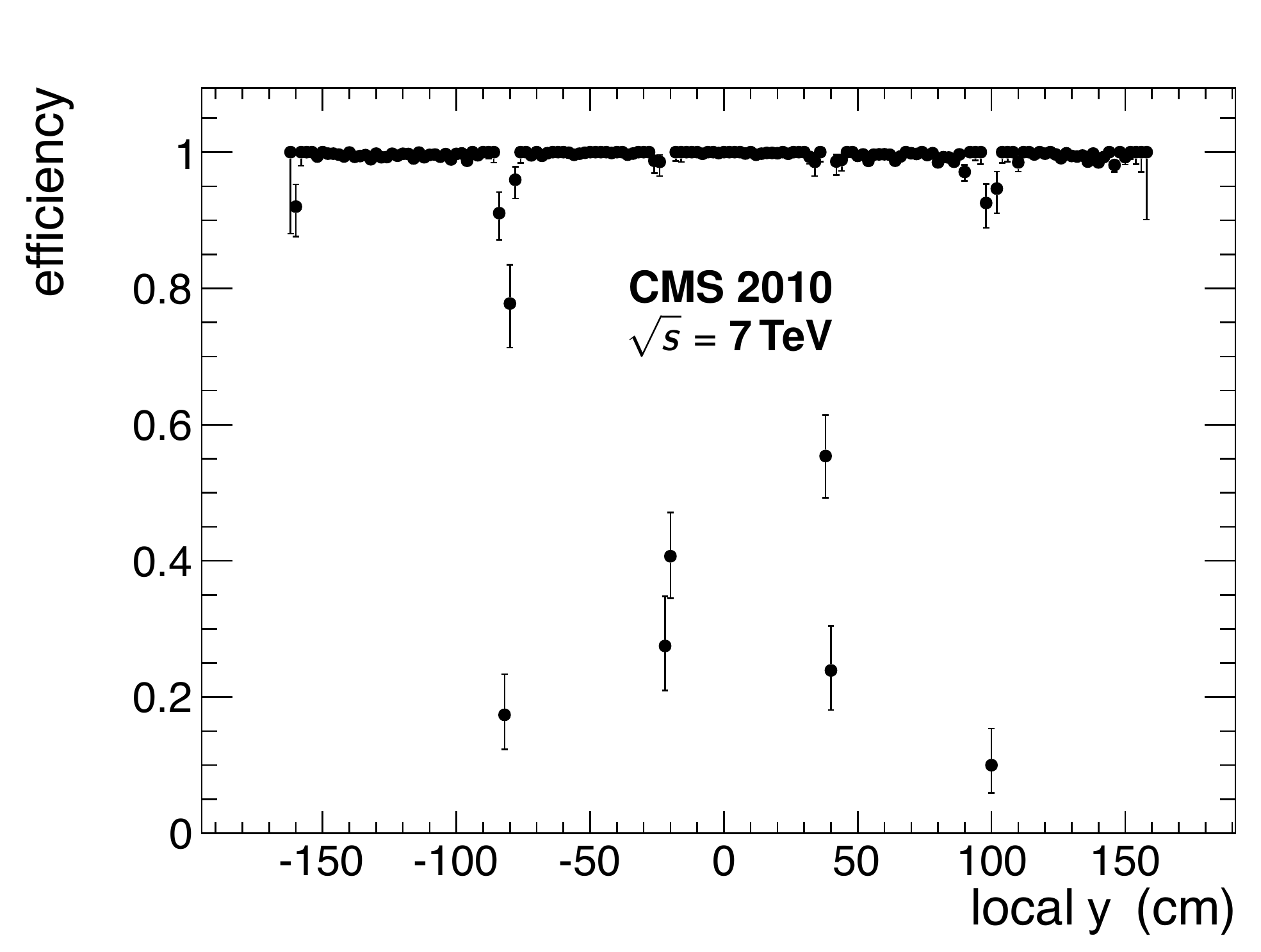}
\includegraphics[width=0.48\textwidth]{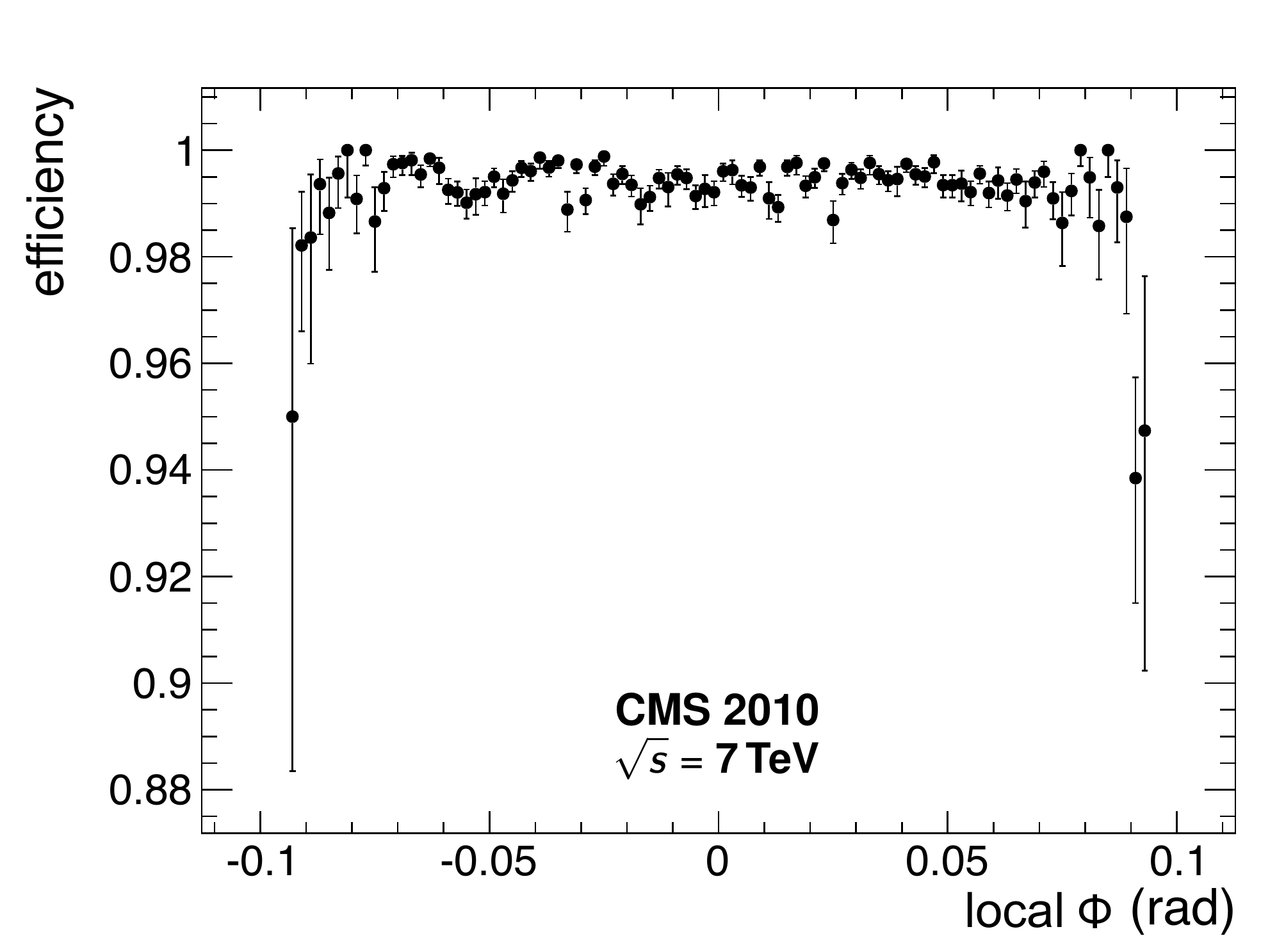}
\caption{\label{Eff_phi_Y}
CSC rechit reconstruction efficiency dependence on the local $y$ coordinate (left) and on the
local $\phi$ coordinate (right). Local $y$ is measured by the anode wires, and local $\phi$ by the cathode strips.}
}
\end{figure}

 Figure~\ref{Eff_phi_Y}
shows the rechit efficiency in the endcap CSCs of station 2, ring 2, for all layers
as a function of the local $y$ coordinate (left) and the strip $\phi$ angle (right). Inefficient chamber regions located between the high voltage supports are clearly visible on the left plot. A slight inefficiency is observed at the boundaries between consecutive cathode readout boards (CFEB) in the $\phi$ efficiency plot (right). The rechit efficiency
in the ``active'' CSC regions is well above 99.5\%.

The association efficiency is not of critical concern in the CSCs due to the redundancy of 6 detection layers per chamber, but was measured to be between $(98.2\pm0.2)\%$ and
 $(98.7\pm0.2)\%$ for muons originating from cosmic rays \cite{CSCPerfCRAFT} and with no layer dependence.
The association inefficiency reflects the specific way $\delta$ rays influence the chamber readout and the subsequent
reconstruction algorithm, and do not need to be the same for DT and CSC chambers.

\subsubsection{RPC}
The barrel and endcap RPC systems are mainly used as trigger detectors; however they also contribute to
the muon reconstruction by providing additional position and time information in the barrel and endcap regions.
Every RPC is located close to a DT or CSC and therefore the extrapolation
of a segment reconstructed by the latter should point to a specific RPC strip and to a particular
location within the strip. In a sense, an RPC can be considered as an additional DT or CSC
layer. This allows the use of reconstructed DT and CSC segments as probes for determining the RPC efficiency and, more generally, for studying the
 hit cluster size, surveying the chamber geometry, performing electronics connectivity tests, and
addressing system alignment issues.

\begin{figure}[ht]
\centering
\includegraphics[width=15cm]{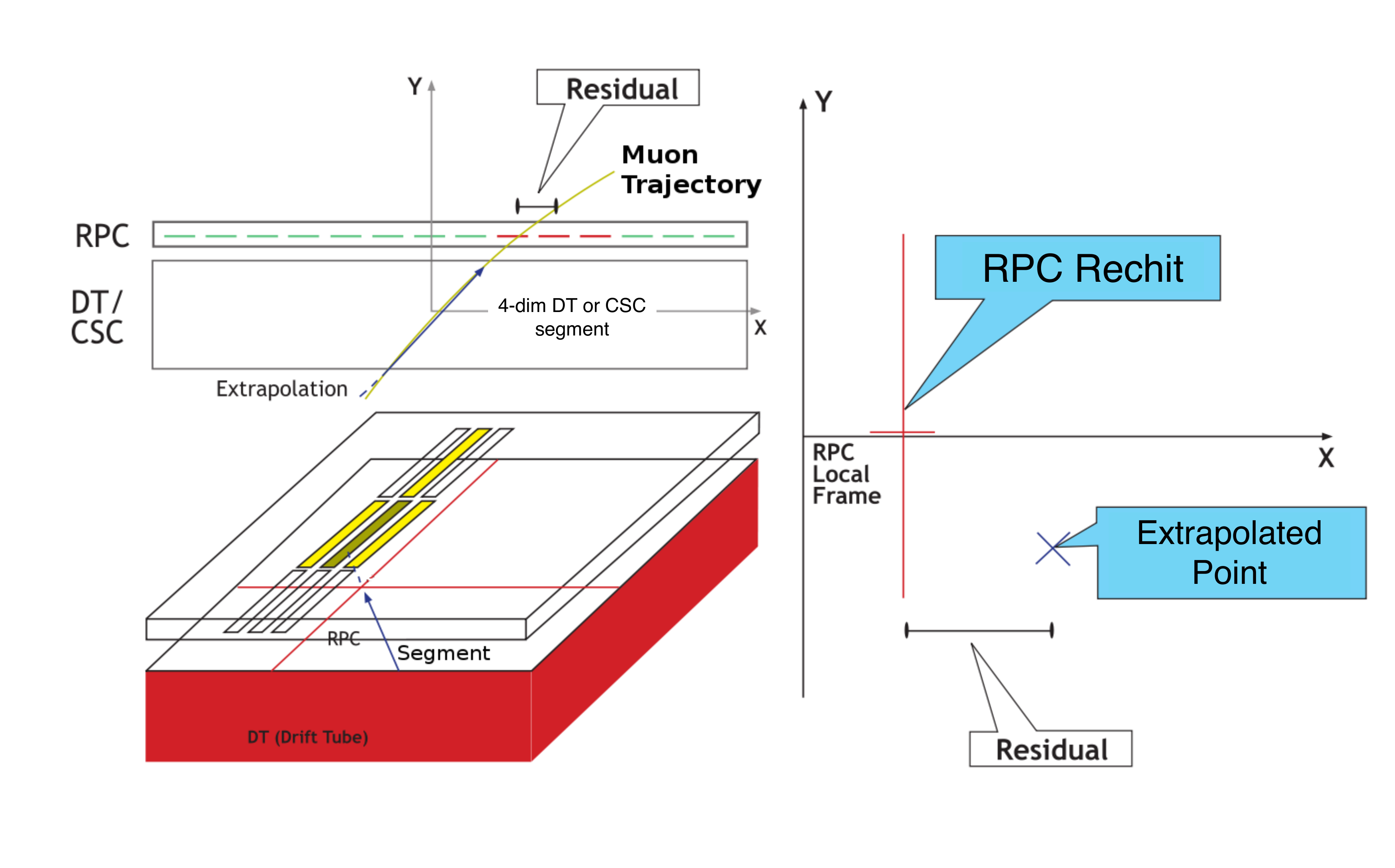}
\caption{Sketch of the segment extrapolation technique.}
\label{SegmentExtrapolationSketch}
\end{figure}

Figure~\ref{SegmentExtrapolationSketch} provides a visualization of the technique explored.
To validate this method, several Monte Carlo
simulations were performed, setting the RPC efficiency to different values and then measuring it
with simulated data.

The RPC hit reconstruction efficiency is defined as the probability of finding an
RPC reconstructed hit when a muon passes through the RPC under study. The
efficiency is computed as the ratio between the number of observed rechits and the expected number of hits, as estimated from segment extrapolation.
A match between the extrapolated DT or CSC segment and the RPC rechit is identified when the distance between the border of the RPC cluster that contains the rechit and the extrapolated point is less than 4 strips (see Table~\ref{tab:strippitch}).
This parameter was tuned to minimize small effects from strip masking (needed for some noisy channels) inside clusters
without introducing significant bias from noise.
To avoid edge effects, only probes that are more than 8\unit{cm} away from the chamber edges are considered.

The efficiency and its uncertainty are defined by Eqs.~(\ref{effTP}) and (\ref{effTP_err}), but the probes here are the (DT, CSC) segments and the passing probes are segments matched to RPC hits.

\begin{figure}[h]
  \centering
  \includegraphics[width=14cm]{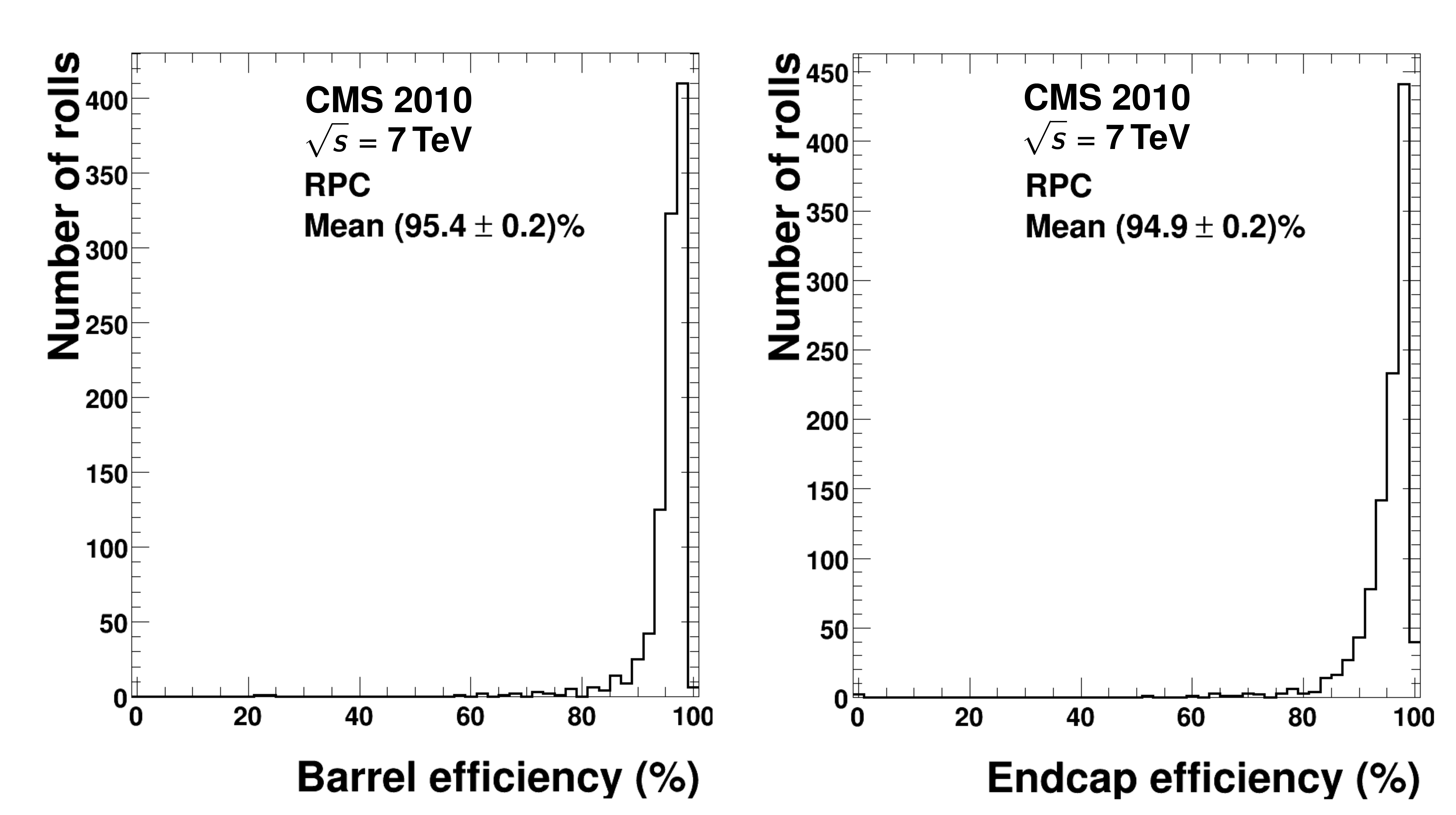}
  \caption{RPC efficiency distributions per chamber roll (see Section~\ref{Intro:RPC}) for the barrel (left) and endcap (right). The percentages of RPCs with efficiencies below 80\% are 1.2\% and 1.3\% in the barrel and endcap, respectively. Chambers operated in single gap mode are excluded from the plot.}
  \label{effrpc}
\end{figure}

The measured efficiency is shown in Fig.~\ref{effrpc}, separately for the barrel (left) and endcap (right) RPCs.
As can be seen, the efficiencies in the barrel and endcap chambers are comparable and around 95\%, which satisfy the TDR requirements.
The tail of lower efficiency chambers is a result of RPCs affected by electronics problems (\eg, a few dead channels) or not operating at the optimal voltage.
Chambers that operated in single gap mode (6 in the barrel and 13 in the endcap) are excluded from the plot.
The percentages of RPCs with efficiencies below 80\% in the barrel and endcap are 1.2\% and 1.3\%, respectively.
For the 2011 data-taking period, a set of calibration runs has been taken to tune the operating voltage chamber-by-chamber.
In addition, a dedicated RPC monitoring stream was set up to detect
any sign of aging effects, and to react quickly
in case of problems by tuning the operating voltage accordingly.

\begin{figure}[ht]
  \centering
  \includegraphics[width=16.5cm]{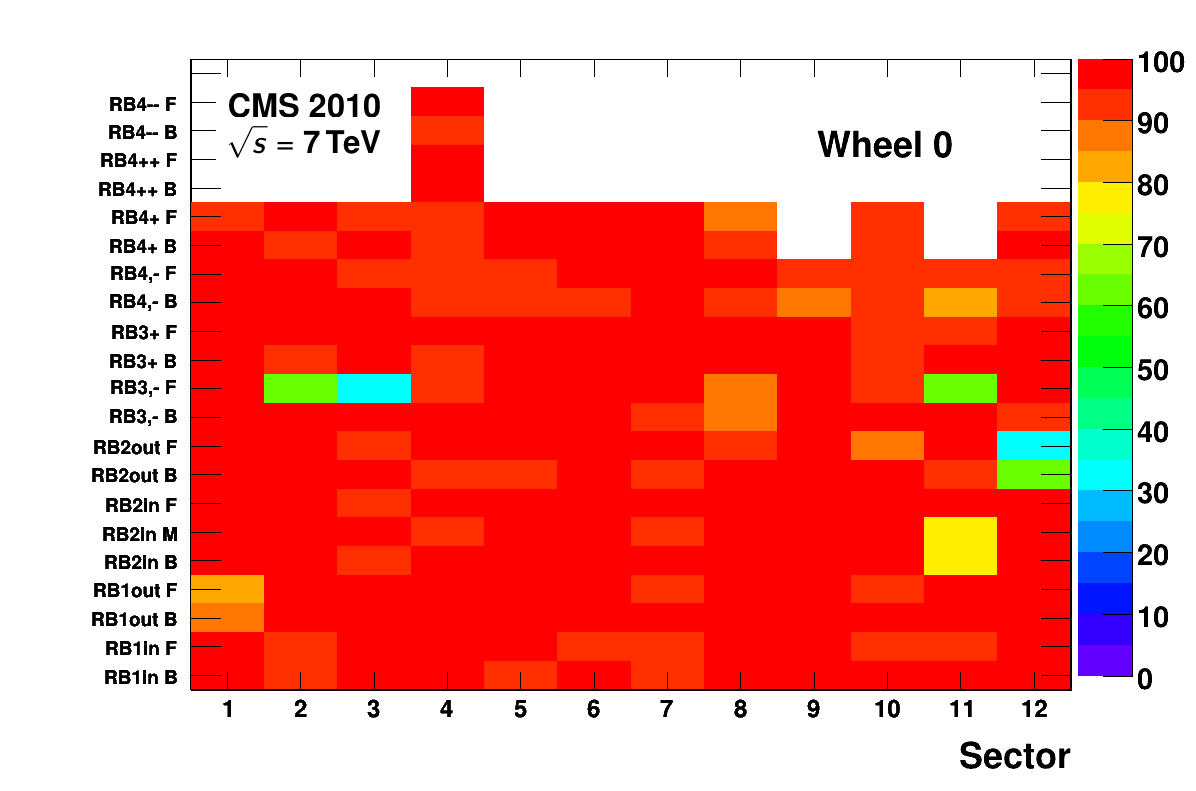}
  \caption{
Percentage efficiency of Wheel 0 of the barrel RPC system.
Chambers are grouped in sectors and stations;
the $x$ axis represents the sectors of the wheel and the $y$ axis shows the chamber trigger sectors~\cite{CMSdet}
 for all layers. The plot has an odd shape because of the complexity of the RPC geometry: station 4 of sector 4 is composed of 4 different chambers, while station 4 of sectors 9 and 11 are composed of a single chamber instead of 2.
  }
  \label{effBarrel_2D}
\end{figure}

\begin{figure}[ht]
  \centering
  \includegraphics[width=16cm]{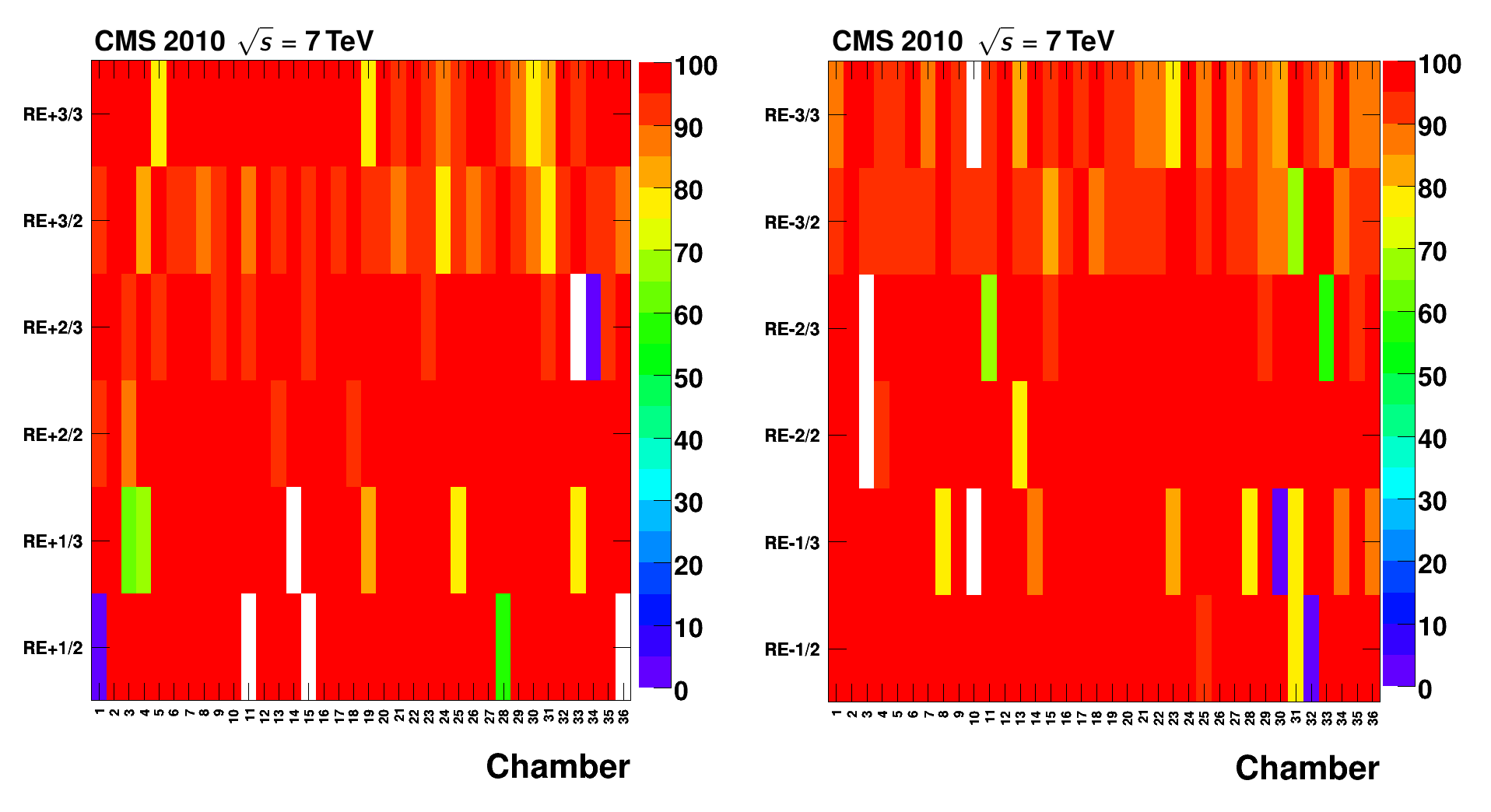}
  \caption{
Percentage efficiency of the endcap RPCs. Chambers are grouped in stations (disks) and rings; chamber numbers are represented on the $x$ axis and rings
on the $y$ axis, for the minus ($z<0$) (left) and  plus ($z>0$) (right) endcaps.
}
  \label{effEndcap_2D}
\end{figure}

\begin{figure}[ht]
  \centering
  \includegraphics[width=14cm]{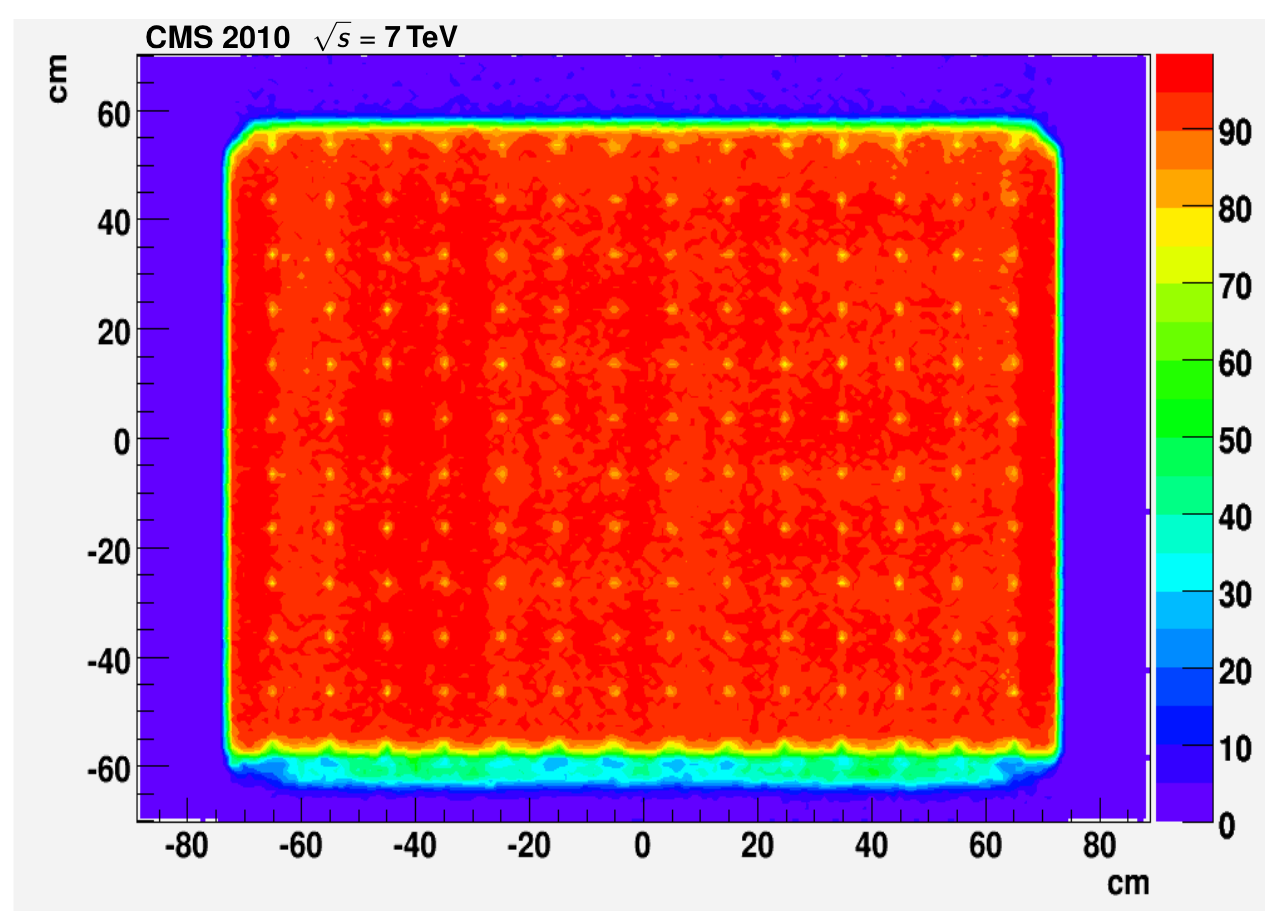}
  \caption{Local percentage efficiency map during the 2010 LHC data-taking period for all RB3 backward barrel rolls. The low-efficiency points correspond to the location of the spacers in the gas gaps.}
  \label{muography}
\end{figure}

Figures~\ref{effBarrel_2D} and \ref{effEndcap_2D} show the hit reconstruction efficiency for individual RPCs in Wheel 0 of the barrel and in the 2 forward/backward endcaps, respectively.
The plot color code shows the percentage efficiencies in 5\% intervals.
The errors, around 1\%, are not shown in the plots.
With the exception of a few non-operating and unstable (low efficiency) chambers, the system has been performing according to expectations.
Finally, Fig.~\ref{muography} shows a high-resolution efficiency map for all the RB3 backward chambers in the barrel.

\section{Radiation background in the muon system}
\label{background}

Background radiation levels in the CMS muon system are an important consideration in its overall performance.
Low-momentum primary and secondary muons, punch-through hadrons, and low-energy \Pgg-rays and neutrons, together with
LHC beam-induced backgrounds (primary and secondary particles produced in the interaction of the beams with collimators,
residual gas, and beam pipe components) could affect the trigger performance and pattern recognition of muon tracks.
In addition, excessive radiation levels can cause aging of the detectors.
The 2010 proton--proton LHC running provides a good opportunity to measure the radiation backgrounds and compare them with
simulation results, which strongly influenced the original system design~\cite{MUON-TDR, trigTDR, Shield}.

\begin{figure}[htb]
\begin{center}
\includegraphics[width=\linewidth,angle=0]{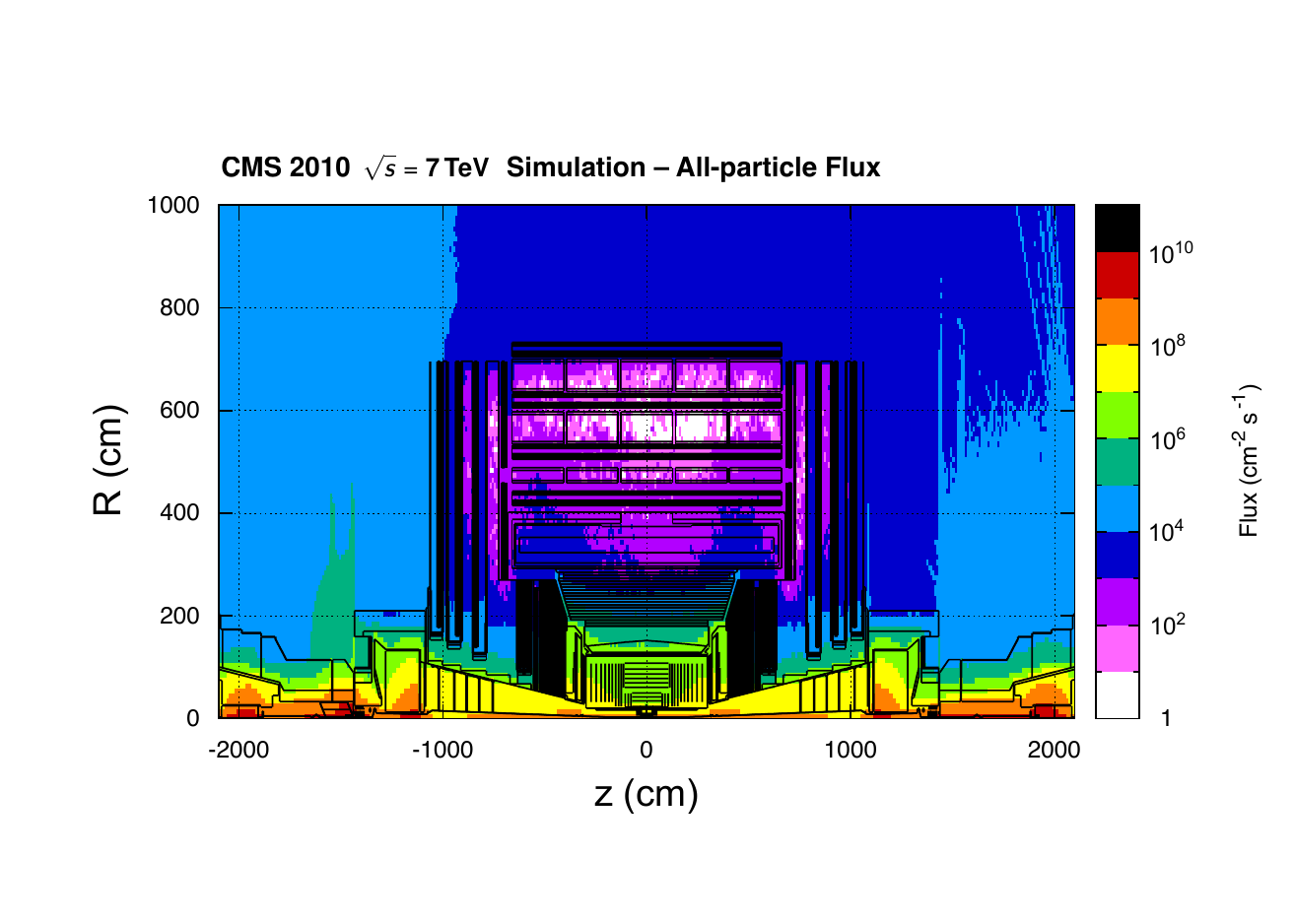}
\caption{ Simulated radiation background in the CMS cavern.
CASTOR is located close to the beam pipe on the minus side ($z < 0$) in the very forward region at around $z = -15\unit{m}$.}
\label{fig1-background}
\end{center}
\end{figure}
The simulated
 radiation background levels inside and around the CMS detector during $\sqrt{s}= 7$\TeV proton--proton collision running in 2010 (Fig.~\ref{fig1-background})
 confirm the main conclusions of previous simulation work and also demonstrate some asymmetry in the background rates around
CMS owing to the presence of the CASTOR very forward calorimeter on the minus side of the detector ($z <0$) at around $z = -15\unit{m}$.

\subsection{Background measurement techniques}
\label{section:techniques}

Each of the muon subsystems employs a different technology and different materials, thus each responds differently to the various backgrounds.
For example, on the one hand, the CSC system has 6 planes per chamber and typically requires 4 out of 6 planes to generate a track segment.
Hence, CSCs are relatively immune to neutrons, which generally affect only a single plane. On the other hand, the CSCs cannot distinguish
punch-through background particles from genuine muon tracks.
The RPC system provides a single hit per chamber
from 2 gaps, and thus triggers on neutron hits.
However, the RPC system has a very tight timing window of about 25\unit{ns} and is relatively unaffected by out-of-time signals.
The DT system has 12 planes per chamber and is mostly immune to neutron hits.
The timing window for the DTs, however, is large, and a high background rate could make track reconstruction difficult, and the event size could become too large for the readout.

As a result of these different technologies, each subsystem measures background rates in a different manner.
The CSC system measures the trigger rate per chamber.
To increase the sensitivity to backgrounds such as neutrons,
the CSC trigger was to run in a special configuration that requires only a single layer coincidence of wire and strip hits  within a time window of 75\unit{ns}
(instead of the nominal coincidence of 4 hits out of 6 layers).
A coincidence of 2 or more hits from different layers that satisfy the single track criteria (a stub of hits pointing to the interaction point) is considered in this configuration as a single trigger event. Since this configuration is different from normal operation, these measurements of radiation load were taken during LHC fills after collisions had been established,
but  before CMS started taking physics data.

  The DTs selected a non-track background sample by requiring patterns with
  only 1 or 2 hits within a superlayer.  By integrating over the full 1.25\mus readout window,
  the DTs measured rates of out-of-time backgrounds originating mainly from slow neutrons
  and punch-through activation, as well as contamination from hits that occurred during other
  bunch crossings (pileup). The time distribution of these
  hits within the full 1.25\mus integration window was checked to be
  flat, in contrast to that for signals generated by prompt muons from proton--proton collisions, which arrive  roughly in a 300 to 700\unit{ns} time interval, corresponding to the drift time across an entire cell.
  (The lower limit of the prompt muon time window is due to time delays, which are accounted for in calibration; see Section~\ref{section-dt-calibration-pedestal}.)
  As a crosscheck, the DTs also took advantage of their long  pipeline
  cycle (1.25\mus) and measured the rate in a 250\unit{ns} time window
  where no signals are expected from in-time particles originating in the proton--proton collision
  that triggered the event. Consistent results were found.

The RPC average strip rate is calculated by using the incremental counts, performed at the level of the RPC DAQ board (link boards), normalized to the strip area.
The noise level is estimated for each run and each chamber separately through a linear extrapolation to a value for an instantaneous
luminosity of 0, which is then subtracted from the chamber rate.
After this noise subtraction, the resulting RPC rate is divided by 2 to account for the 2 RPC gaps. Similarly, the corresponding
CSC rate is divided by 6 to take into account the 6 CSC planes. The hit rate per DT channel is divided by the drift tube area. Results are therefore presented in units of
$\unit{Hz}\unit{layer}^{-1}\unit{cm}^{-2}$ (or for RPCs, $\unit{Hz}\unit{gap}^{-1}\unit{cm}^{-2}$).
This allows comparison among the results of the different technologies and with the Monte Carlo simulation predictions.

The no-collision background rates are not expected to be identical, as they depend on several factors such as intrinsic noise,
signal threshold, natural radioactivity of the chamber constituent materials, and chamber location.
The muon chambers were designed to detect minimum ionizing particles with nearly 100$\%$ efficiency, but they do not necessarily have identical
efficiencies for detection of \Pgg-rays and low energy neutrons, which are the main sources of background.
The detector response to the background depends on the type and thickness of the constituent materials and on the gas mixture, and is a function of the energy of the detected
 particles.
 The DT no-collision background is about $0.0045\unit{Hz}\unit{layer}^{-1}\unit{cm}^{-2}$.
The no-collision background levels seen by the CSCs are about
$0.012\unit{Hz}\unit{layer}^{-1}\unit{cm}^{-2}$ for the chambers located inside the CMS steel
and $0.015\unit{Hz}\unit{layer}^{-1}\unit{cm}^{-2}$ for the chambers located on the outer muon
stations.
The average RPC noise rate during the 2010 data-taking period is of the order of $0.05\unit{Hz}\unit{gap}^{-1}\unit{cm}^{-2}$, in good agreement
with earlier measurements~\cite{CRAFT08}.

\subsection{Background measurement results}

Backgrounds for the CMS detector are expected to be high along the beam pipe and at high $|\eta|$.
They are also expected to decrease with distance away from the beam pipe.
In addition, backgrounds should be somewhat higher on the outside of
CMS because the large amount of steel in CMS provides good shielding for chambers in the interior,
a result that is supported by the background data presented in the next 3 subsections.  For each case
(\ie, dependence on $z$, $r$, and $\phi$) we measure the background
rates as a function of the luminosity, and later we use these data to extrapolate to the luminosities
expected during future running of the LHC.

The CSC rates are averaged over a ring area and the DT rates
are averaged over all chambers in a wheel or station.  The RPC rates, however, are the
maximum observed rates over either the barrel wheels or stations, or the endcap disks or rings.
Because of this and other differences in materials and thresholds, the
RPCs typically report larger rates than the DTs or the CSCs.

\subsubsection{Backgrounds as a function of azimuth and plus--minus asymmetry}

\begin{figure}[htb]
\begin{center}
\centerline{
\includegraphics[width=0.49\linewidth,angle=0]{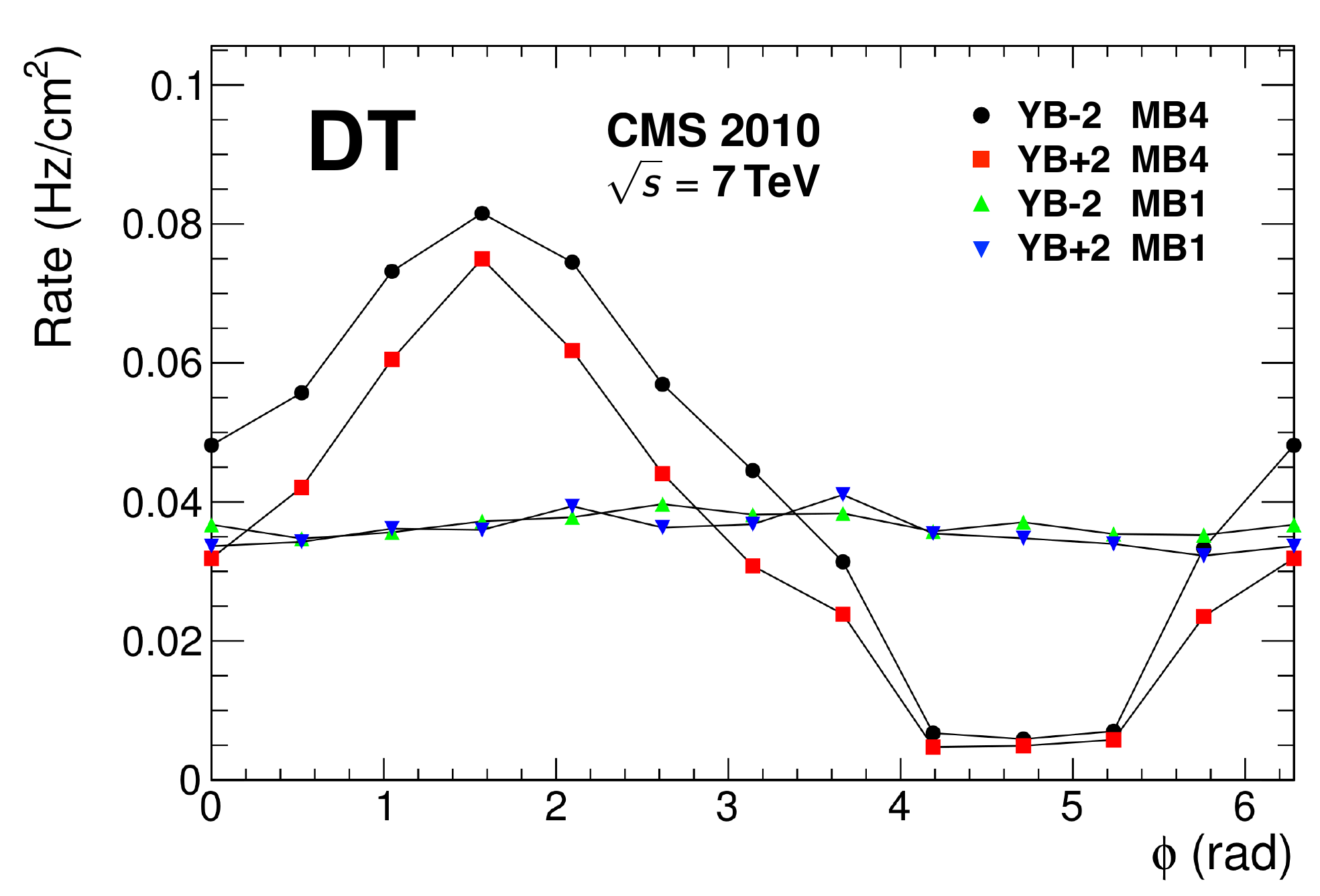}
\includegraphics[width=0.49\linewidth,angle=0]{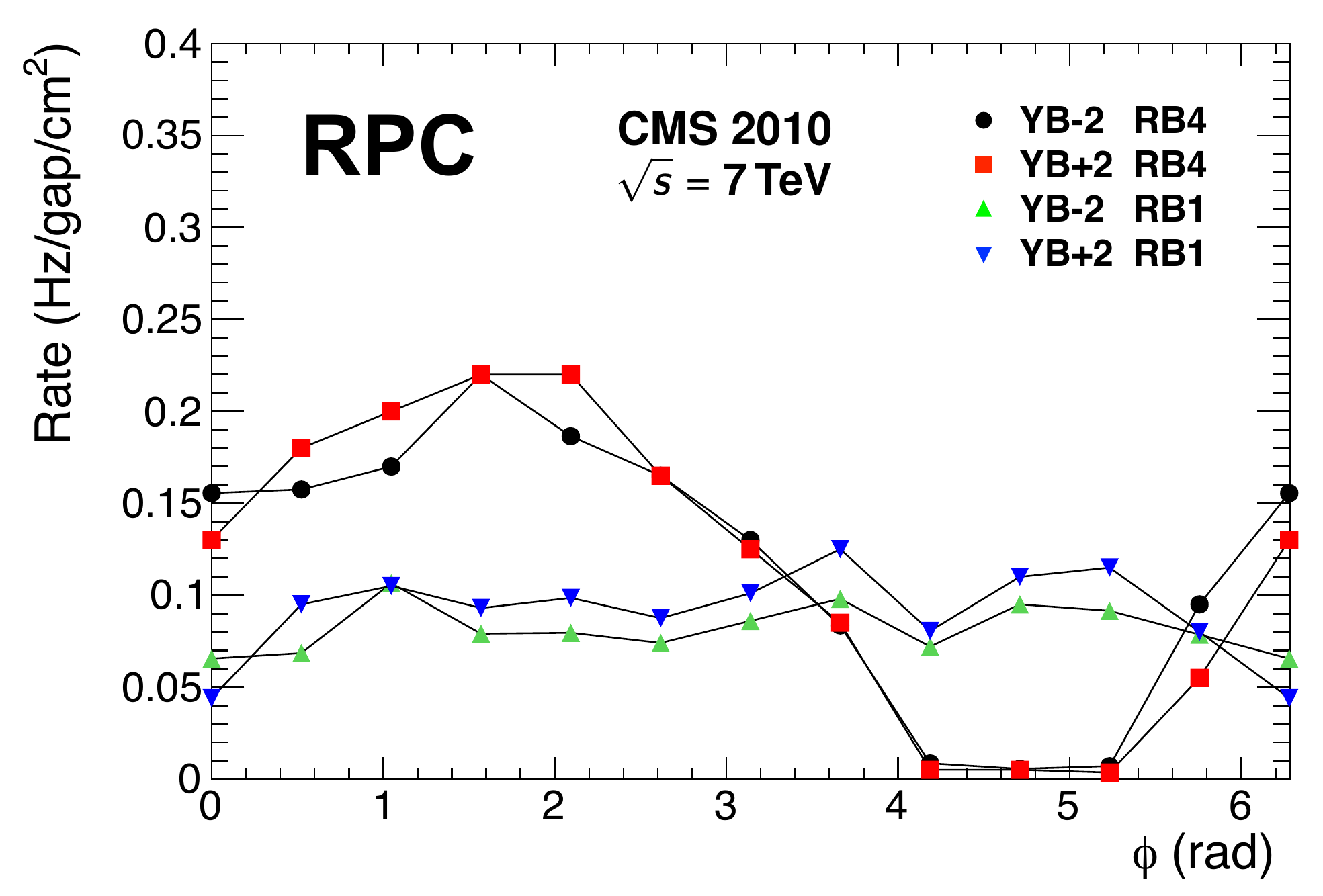}
}
\caption{The background rates of DTs (left) and RPCs (right) located on the inner and outer stations of the external wheels as a function of the chamber azimuthal position at an instantaneous luminosity of 1.5\ten{32}\unit{cm}$^{-2}\unit{s}^{-1}$.
}
\label{fig5-background}
\end{center}
\end{figure}

To first order, CMS is a nearly symmetric detector in both azimuth and $z$, so no azimuthal  dependence nor $+z$/$-z$ asymmetries in background are expected.
Figure~\ref{fig5-background} shows the DT and RPC rates for the inner and outer stations of the barrel Wheel-2 and Wheel+2 (YB-1 and YB+2, respectively)  as functions of azimuthal angle.
 As can be seen in the figure, the rates in the inner rings are
 symmetrical, but those in the outer chambers have a strong azimuthal dependence.
 For both subsystems,
 we observe a difference in rate of approximately a factor 20 between the top ($\phi=1.6$\unit{rad}) and the bottom
 ($\phi=5.2$\unit{rad}). This top--bottom asymmetry is due to non-symmetric features of CMS: the supports for the wheels and disks, and the steel flooring.

\begin{figure}[htb]
\begin{center}
\centerline{
\includegraphics[width=\linewidth,angle=0]{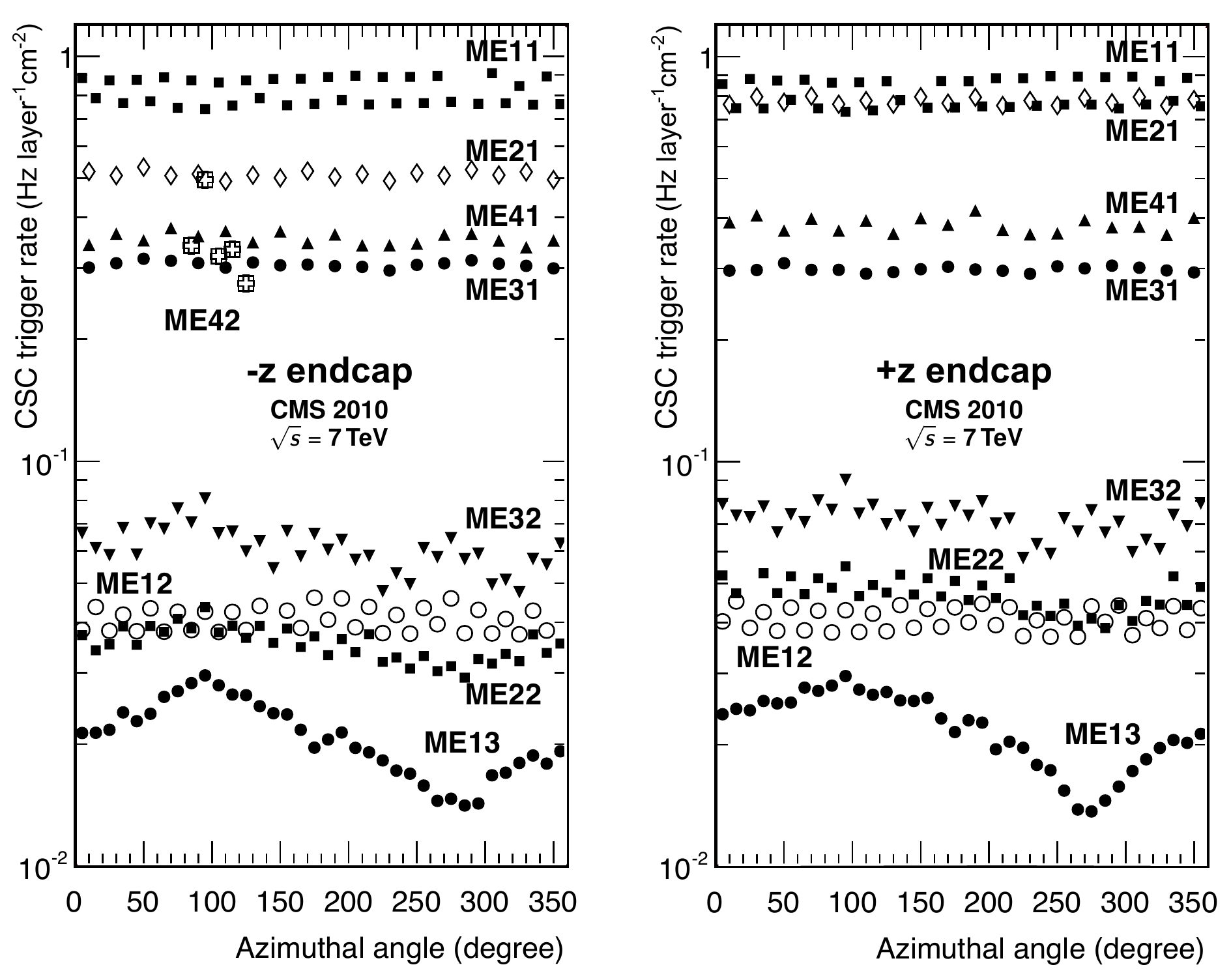}
}
\caption{ The CSC rate at the luminosity 1.9\ten{32}\unit{cm}$^{-2}$\unit{s}$^{-1}$ as a function of the chamber angular position for the minus ($z < 0$) and plus ($z > 0$) endcaps.}
\label{fig6-background}
\end{center}
\end{figure}

The outer DT chambers of the YB-2 wheel observe $\approx$20\% higher background relative to
the chambers located on the YB+2 wheel, while the RPCs do not see any $+z$/$-z$ asymmetry.
The barrel RPC rates are about 2.5 times higher than the DT rates, but we note that, in addition to the different materials and technologies, the DT rates exclude prompt tracks whereas the RPCs do not.
The endcap inner rings (Fig.~\ref{fig6-background}) do not see any $+z$/$-z$ asymmetry, except for the ME2/1 rings, which display an asymmetry at a level of approximately 50\% that is not yet understood.
The outer muon chambers observe a noticeable asymmetry between the plus and minus sides
of  CMS. The CSCs in the minus endcap observe an approximately 20\% larger rate relative to the plus
side for the ME3/2 rings, $\approx$30\% for the ME2/2 rings, and $\approx$10\% for the ME1/3 and
ME4/1 rings. This kind of asymmetry, which is also observed by the outer station barrel DTs
($\approx$20\%), is qualitatively reproduced in \textsc{fluka} simulations
(see Fig.~\ref{fig1-background})
and is related to the presence of the CASTOR detector on the minus side of the CMS detector.

In the endcap region, a complete picture of the rates seen by the CSCs at an LHC luminosity of $1.9\ten{32}\unit{cm}^{-2}\unit{s}^{-1}$
is shown in Fig.~\ref{fig6-background} as a function of the chamber angular position.
The chambers of the inner endcap rings in all stations show no angular dependence, while some is apparent in the outer ring chambers.
The ME2/2 and ME3/2 rings show a small angular dependence.
The most prominent effect appears in the ME1/3 chambers, which are closest to the outer barrel stations in both endcaps.
The plot shows a clear decrease in the rates of the bottom ME1/3 chambers, for $\phi\simeq270^\circ$, near the floor of the cavern.
The azimuthal asymmetry in the endcaps is much smaller relative to that seen in the barrel.
This is because the endcap chambers are located in between steel disks, while the barrel outer chambers are unprotected from the surrounding background.

Adjacent chambers in many rings of both the minus and plus endcaps have different rates (Fig.~\ref{fig6-background}).
Adjacent chambers in a ring overlap to avoid inefficient regions in between neighboring chambers.
Thus there are both ``back" chambers, which are mounted first on the steel disks, and  ``front" chambers, which are mounted on the top of the installed back chambers.
The alternating effect in background rates appears in most rings, but not in the ME3/1 and ME1/3 rings,
the latter which has only a single layer of chambers. The difference in the rates between the back and front
chambers is typically a few percent for most of the rings, but reaches 10\% for the ME1/2 ring and
20\% for the ME1/1 chambers. Background rates oscillate between  front and back chambers, with
higher rates in the chambers located further away from the IP.
This effect is due to the steeply rising background at high $|\eta|$. Chambers farther from the IP also span a slightly higher $|\eta|$ region and thus have higher background rates.

\subsubsection{Backgrounds as a function of $z$}

\begin{figure}[htb]
\begin{center}
\centerline{
\includegraphics[width=0.5\linewidth,angle=0]{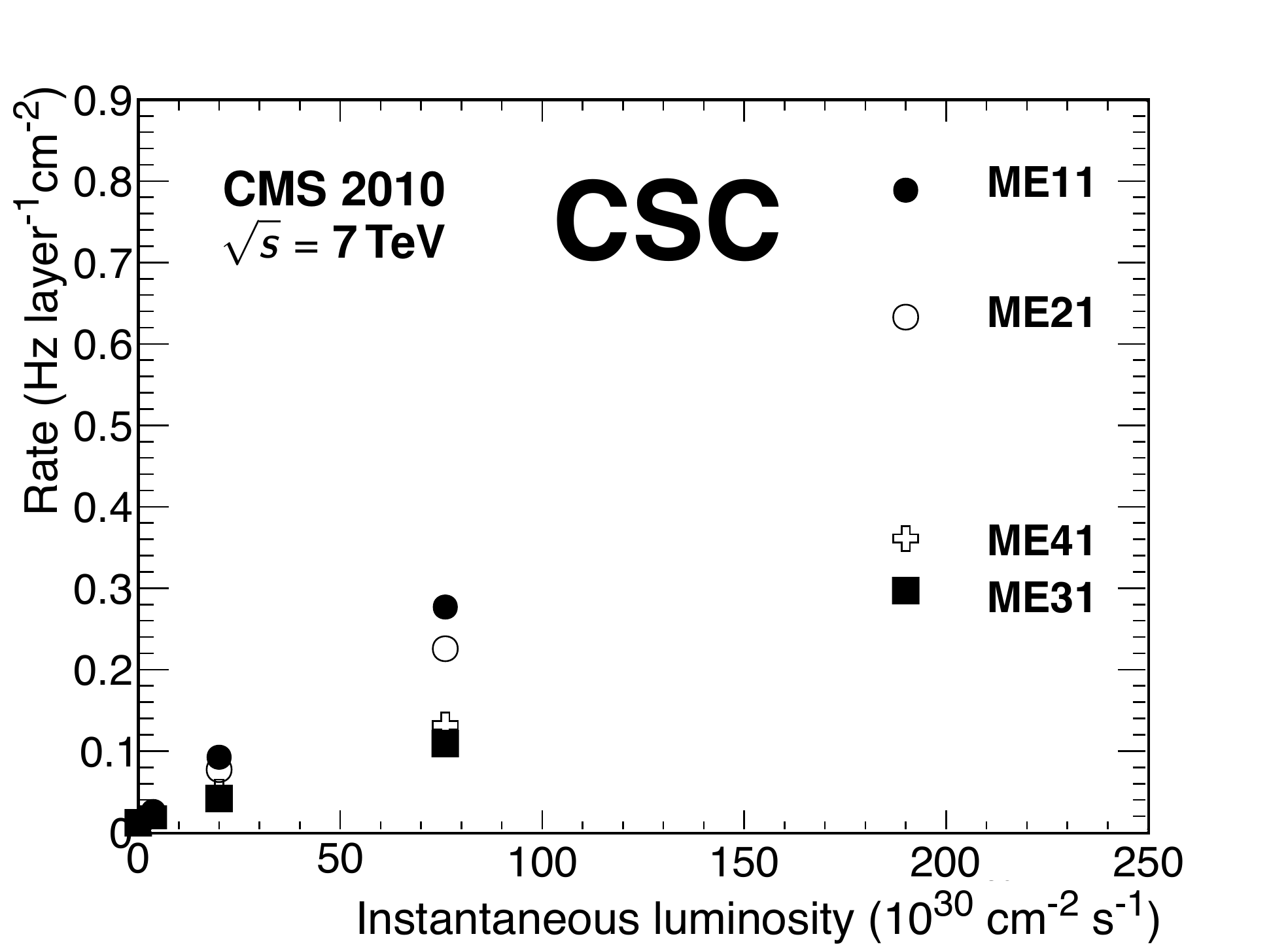}
\includegraphics[width=0.5\linewidth,angle=0]{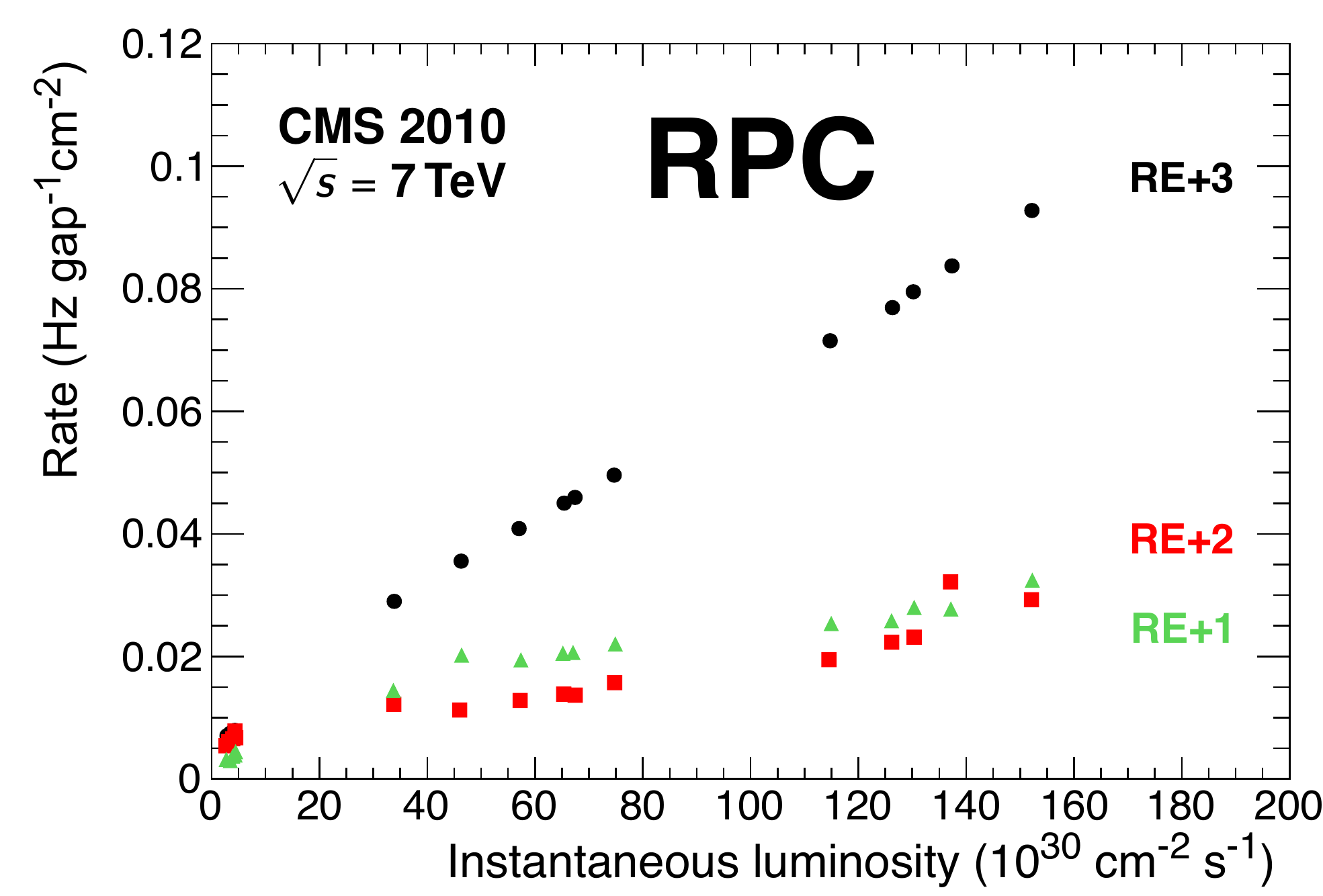}
}
\caption[fig3-background]{Left: The background rates for the CSC inner rings \vs luminosity
 (ME1/1 is closest to the IP; ME4/1 is farthest away). Right: The RPC background rate
  \vs the luminosity for the outer rings of RE (RE1 is closest to the IP; RE3 is farthest away).
  The differing ordering patterns between the CSC and endcap RPC rings reflect the different backgrounds near the beamline and on the outside of CMS.
}
\label{fig3-background}
\end{center}
\end{figure}

The average trigger rates for the inner rings of the endcaps, where  MC simulations
 predict the highest rates, are shown as a function of the LHC luminosity in Fig.~\ref{fig3-background}
(left). As expected, the CSC rates increase roughly linearly with the luminosity of the LHC.  The largest rates are observed for the ME1/1 ring where, at the luminosity 1.9\ten{32}\unit{cm}$^{-2}\unit{s}^{-1}$, the counting rate was up to 60 times the no-collision background rate. The background level drops with increasing distance from the IP and only at the outer station (ME4/1) does it slightly increase.

The maximum rates for the RE stations are also shown versus the LHC luminosity in Fig.~\ref{fig3-background} (right).
The patterns are different because the RPC inner ring 1 has not yet been installed: the inner rings of the CSCs show a mostly decreasing background as $z$ increases, whereas the endcap RPC data show the rates increasing with increasing $z$, with the highest rate
measured in the third (external) disk, in agreement with the outer CSC rings.
This reflects the different types of backgrounds near the beamline and on the outside of CMS. The increase in background for ME4/1 is consistent with this picture.

\begin{figure}[htb]
\begin{center}
\centerline{
\includegraphics[width=0.49\linewidth,angle=0]{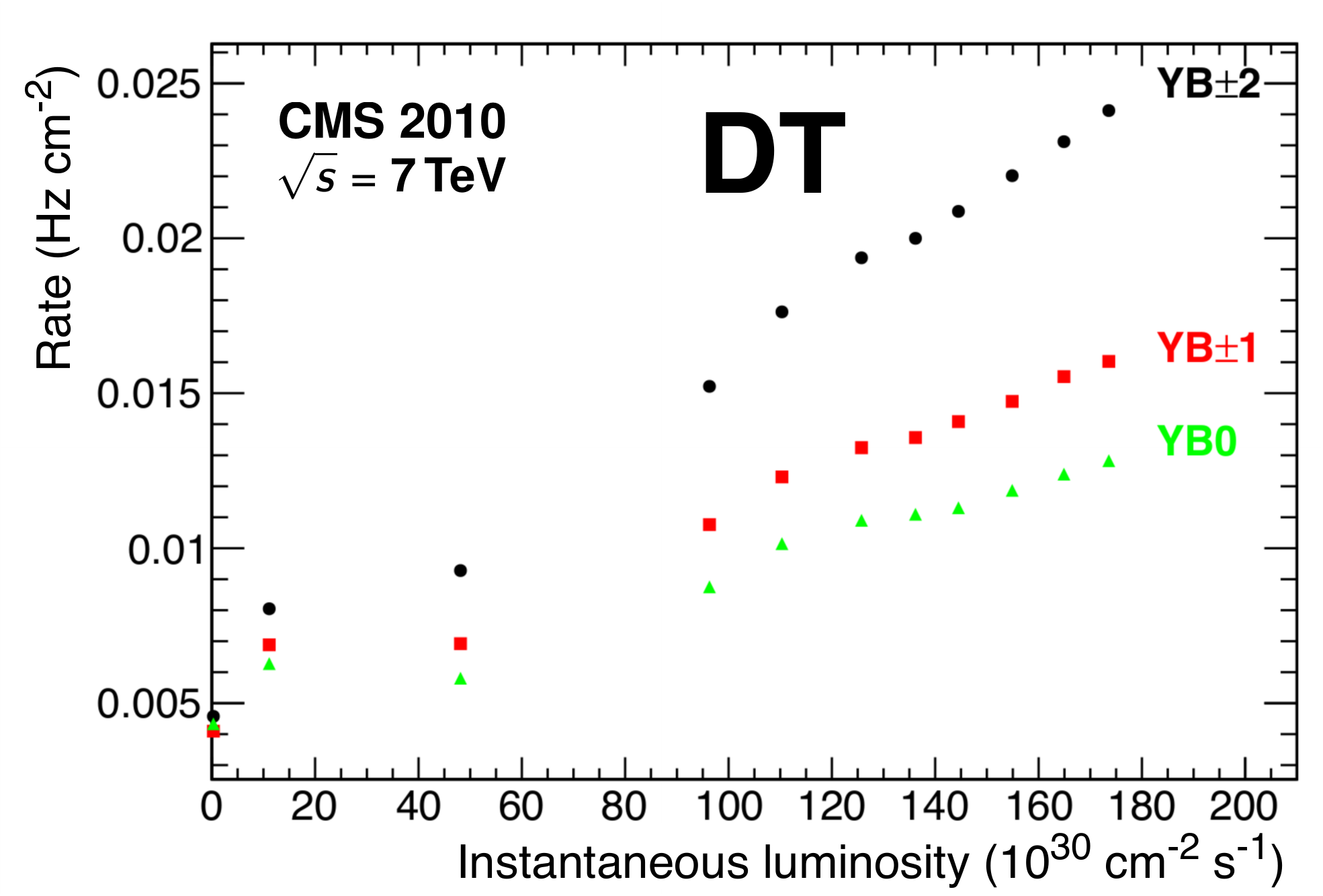}
\includegraphics[width=0.49\linewidth,angle=0]{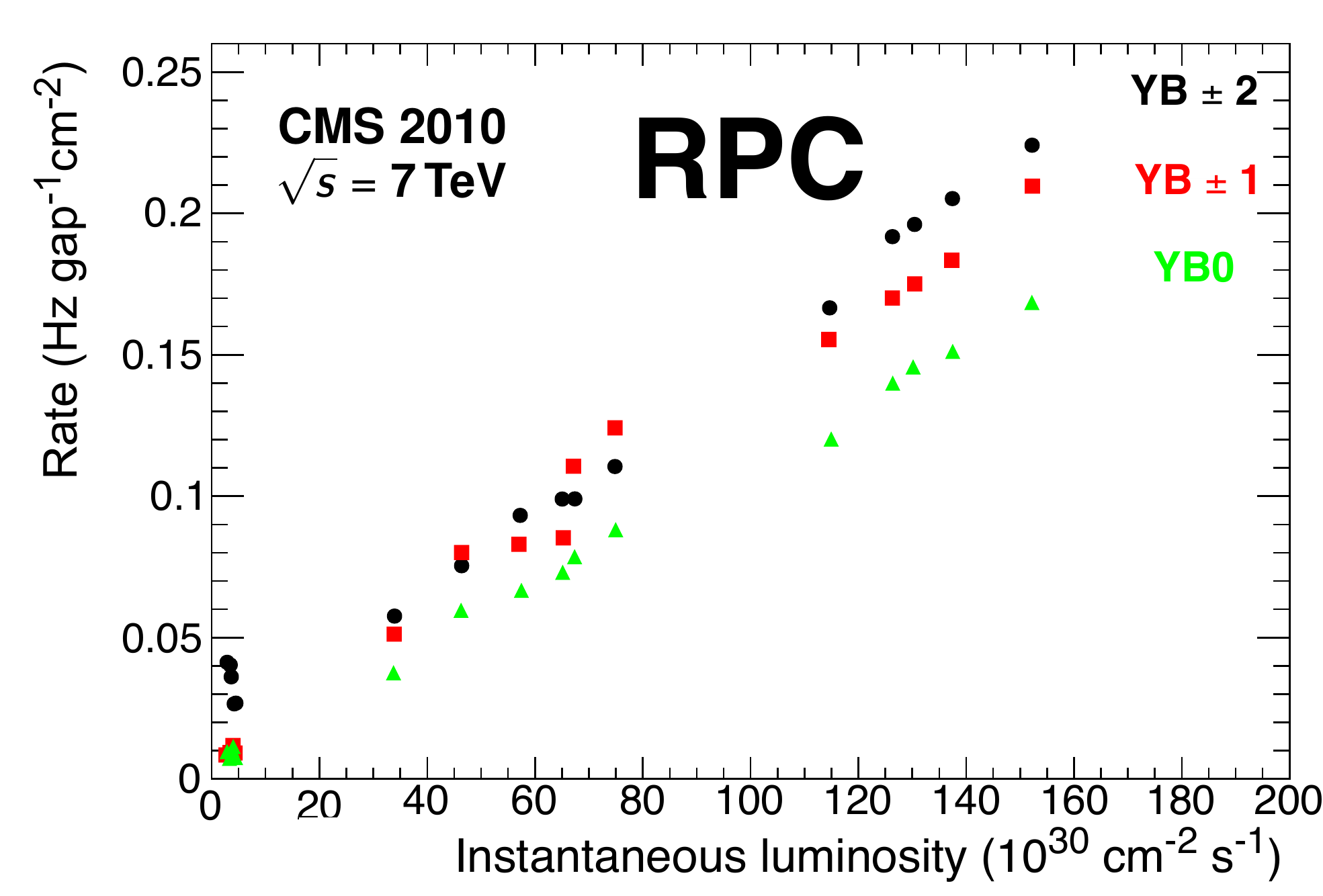}
}
\caption[fig3bis-background]{Left: The DT background rates \vs LHC luminosity for the barrel wheels (YB0 is closest to the IP, YB$\pm$2 are farthest away).  Right: The RB backgrounds \vs luminosity for the barrel wheels.  The patterns are similar, increasing as the distance from the IP increases.  The difference in rates occurs, in part, because the RB rates are maximum rates whereas the DT rates are averages.
}
\label{fig3bis-background}
\end{center}
\end{figure}

In the central $\eta$ region, the behavior of the background is slightly different.  The average DT rates versus luminosity are presented for the barrel wheels (YB0 is closest to the IP and YB$\pm$2 are farthest away) in Fig.~\ref{fig3bis-background} (left).  The right plot shows the corresponding maximum RB rates.  The patterns are roughly the same: rates increase as the chambers are farther from the IP.

\subsubsection{Backgrounds as a function of $r$}

\begin{figure}[htb]
\begin{center}
\centerline{
\includegraphics[height=0.35\linewidth,width=0.49\linewidth,angle=0]{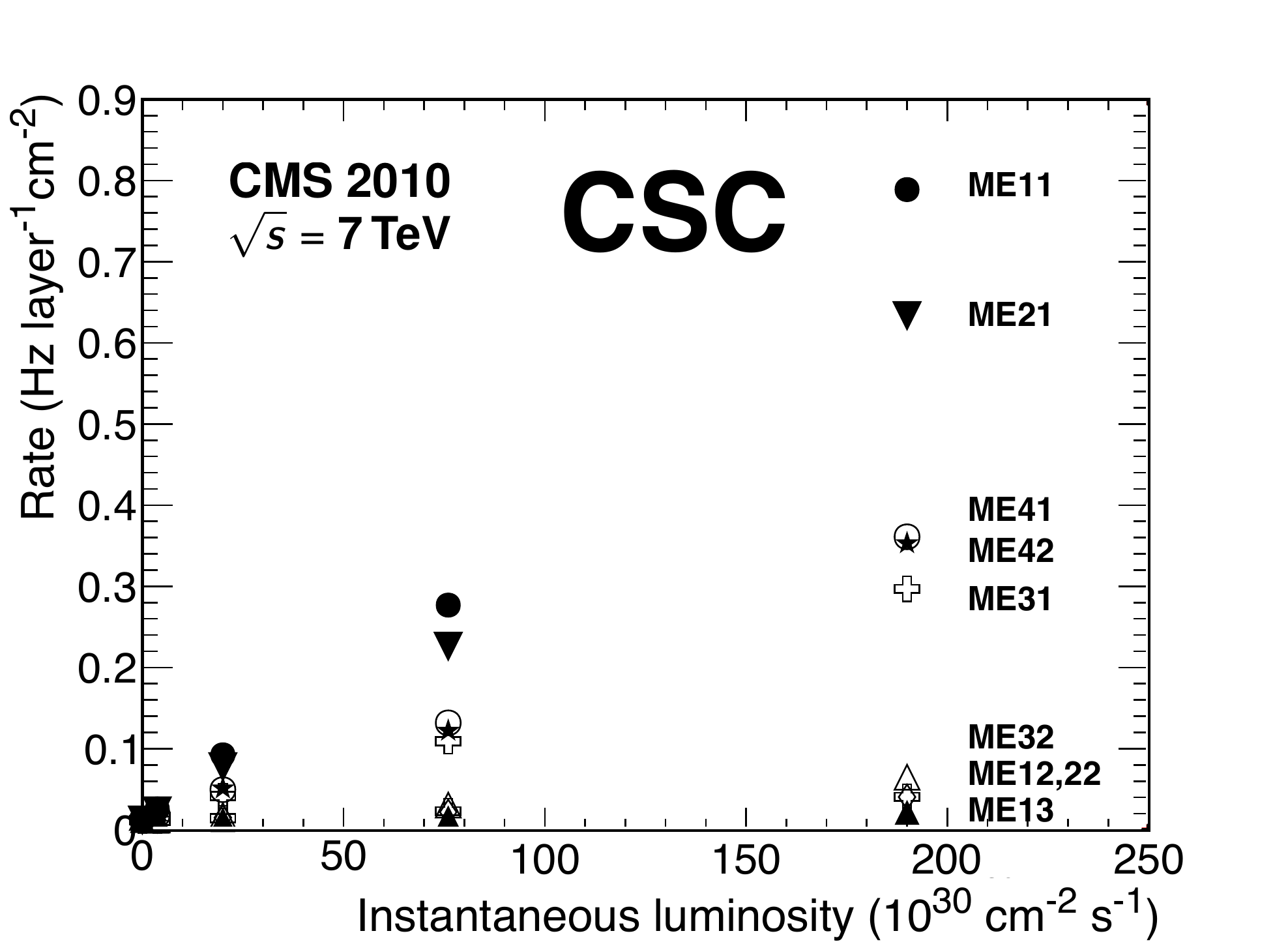}
\includegraphics[width=0.49\linewidth,angle=0]{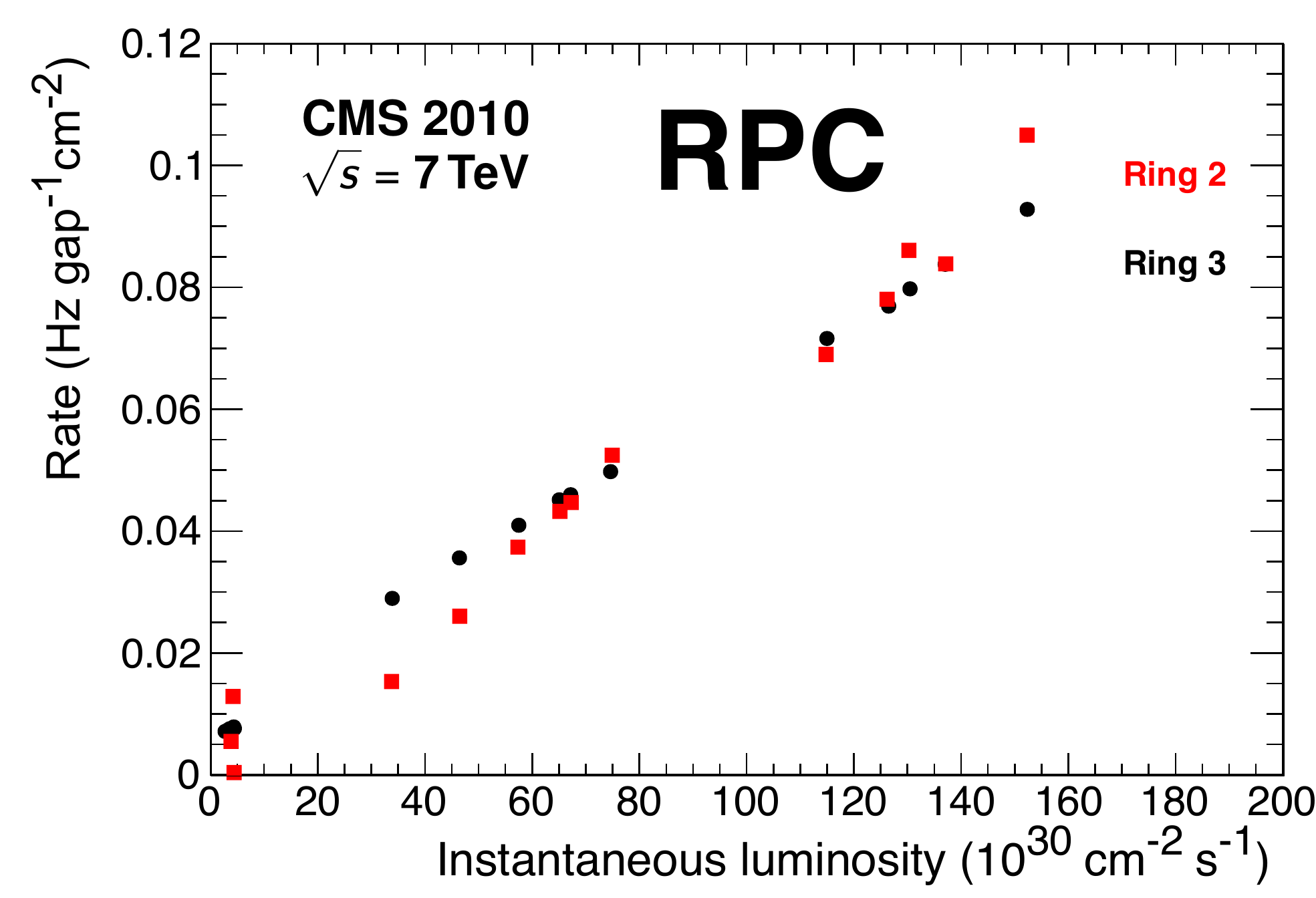}
}
\caption{Left: The average CSC background rates \vs luminosity for all the endcap rings.  Right: The maximum RE rates \vs luminosity for the top trigger sector 10 in the 2 rings of station RE3.
}
\label{fig4-background}
\end{center}
\end{figure}

The background behavior as a function of $r$ is somewhat different.
The CSC inner ring rates are presented together with the rates of the outer CSC rings as a function of the luminosity in Fig.~\ref{fig4-background} (left).
RPC rates were not integrated over $\phi$, so for comparison, the right plot shows the RE3 rates for trigger sector 10 (corresponding to 10\de azimuth at the top of CMS) in the 2 rings of station 3.

The outer CSC and RPC rings show a much smaller increase in rate with luminosity relative to the inner chambers.
Only the ME4/2 chambers show an  increase comparable to the inner ring chambers.
This increase for ME4/2 is predicted by
the simulations and is related to its position on the outside of the detector where the planned outermost shielding disk, called YE4, has
not yet been installed~\cite{MUON-TDR, trigTDR, Shield}.
In spite of their different materials and gas mixtures, the rates seen by the CSCs and RPCs at the same locations are in good agreement. For example, the average ME3/2 rate at a luminosity of $1.9\ten{32}\unit{cm}^{-2}\unit{s}^{-1}$ is about
$0.07\unit{Hz}\unit{layer}^{-1}\unit{cm}^{-2}$ while the RE3/2 and RE3/3 rates are at the level of $0.1\unit{Hz}\unit{gap}^{-1}\unit{cm}^{-2}$.  It should be noted that the CSC rates are averaged over the entire ring, while the RE rates are the maximum rates observed in that ring.

\begin{figure}[htb]
\begin{center}
\centerline{
\includegraphics[width=0.49\linewidth,angle=0]{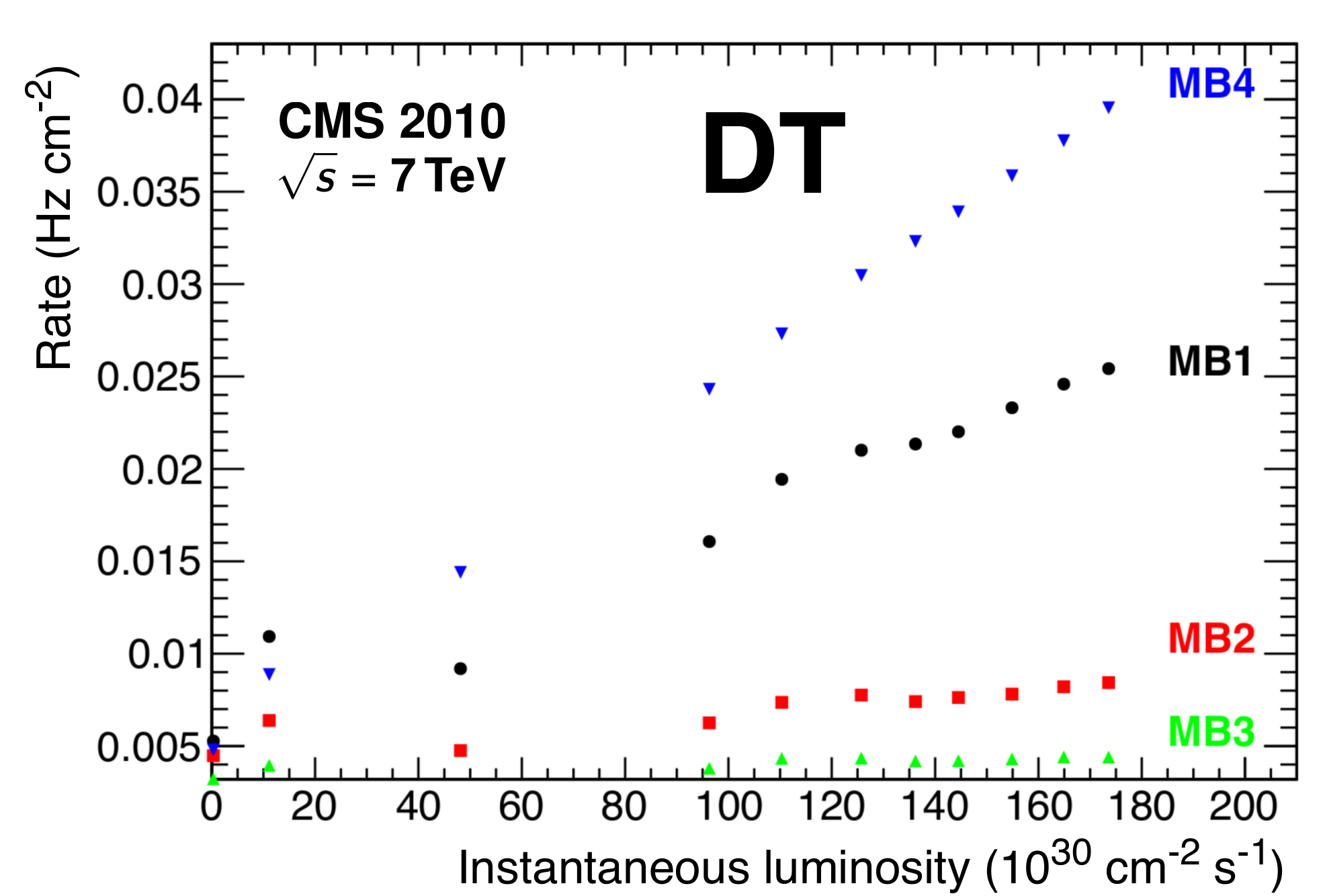}
\includegraphics[width=0.49\linewidth,angle=0]{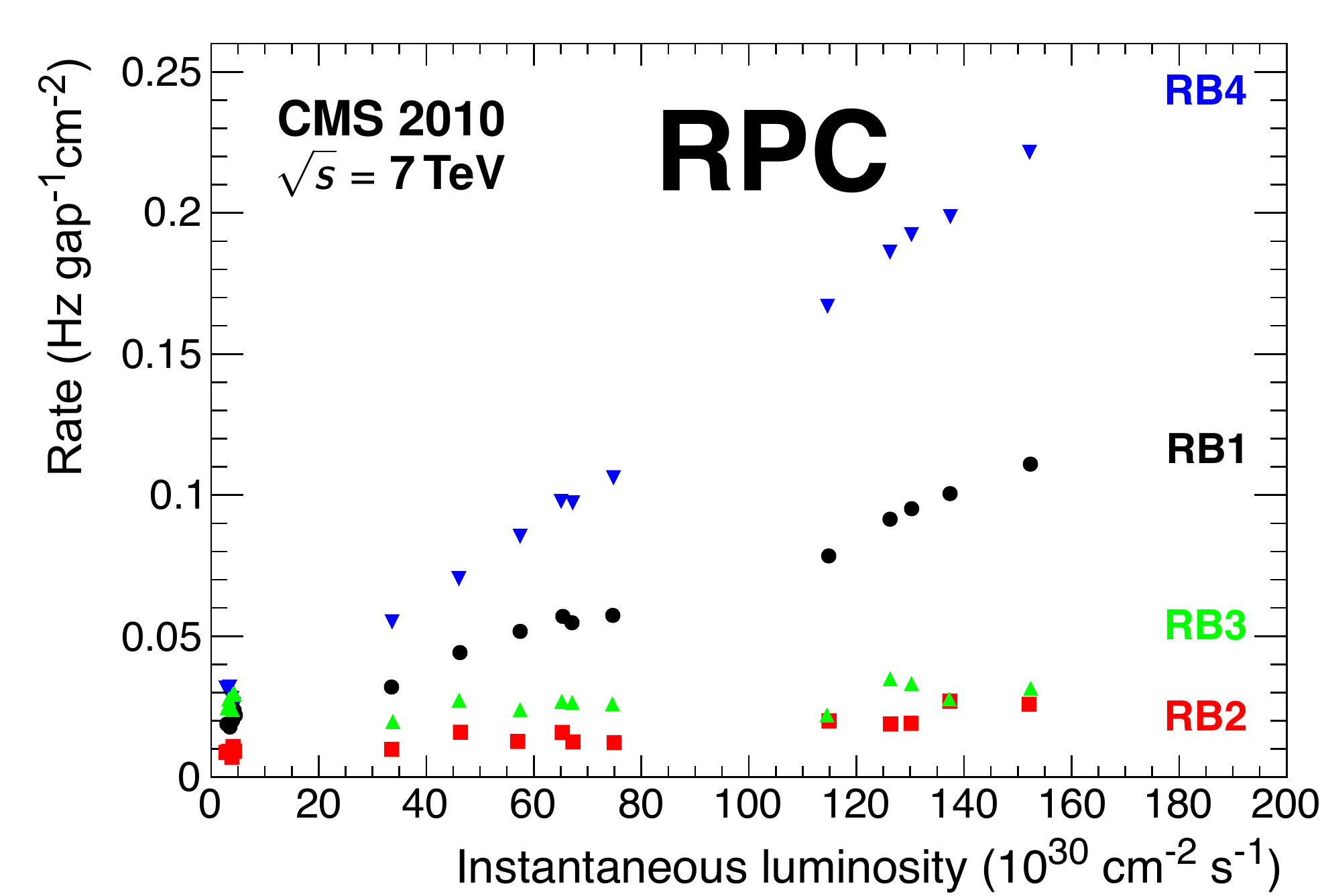}
}
\caption{Left: The DT rates \vs luminosity for each of the 4 stations (MB1 is closest to the beam line and MB4 farthest away).  Right: The RPC background rate \vs the LHC luminosity for the same 4 stations.  Here the background rates are relatively low for RB2 and RB3.  RB1 (closest to the beam line) has a higher rate, but RB4 (farthest from the beam line) has the highest rate because it is on the outside of the detector and measures the backgrounds outside CMS.
}
\label{fig4bis-background}
\end{center}
\end{figure}

For the central region ($|\eta | < 0.9$), the $r$ dependence of the background is shown in Fig.~\ref{fig4bis-background}, which presents the DT and the RPC measured rates in the barrel stations as functions of the LHC luminosity.
The largest rates are observed by the DTs and RPCs for the last and first muon stations.
The chambers of these stations are exposed to different sources of background. The inner chambers are sensitive to activation
caused by leakage of particles from the hadron calorimeter, while the outer station chambers are mostly exposed to the slow neutron gas permeating the cavern.
The DTs and RPCs located in stations 2 and 3 detect much smaller rates.
They are well protected against the above mentioned backgrounds by the steel of the barrel wheels.

\subsection{Extrapolation to higher luminosity}

\begin{figure}[htb]
\begin{center}
\centerline{
\includegraphics[width=\linewidth,angle=0]{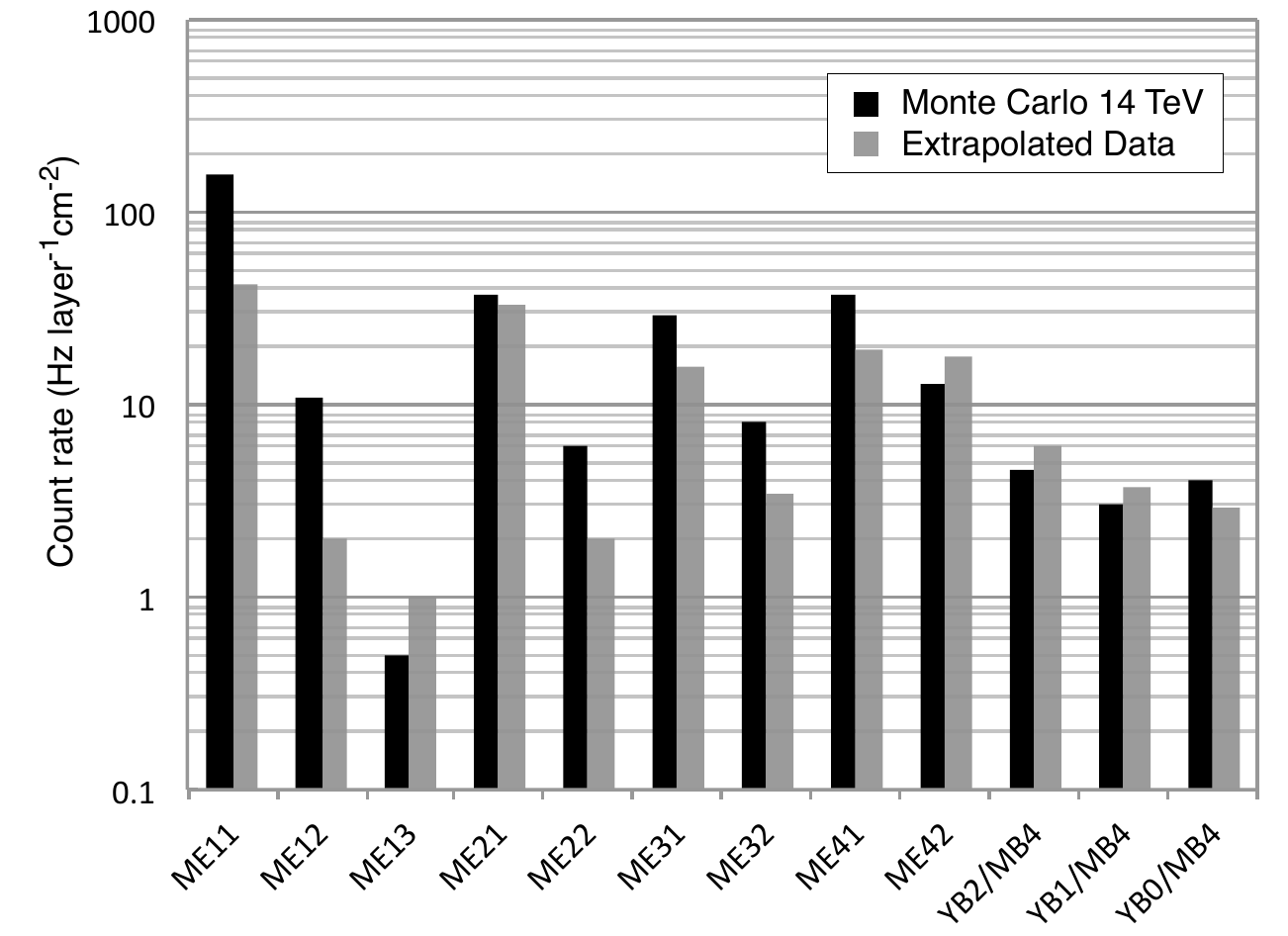}
}
\caption{
Radiation background rates in the CMS muon system projected for an LHC luminosity
of 10$^{34}\unit{cm}^{-2}\unit{s}^{-1}$. The data results are linearly extrapolated from 2010 measurements at LHC luminosities up to 1.8\ten{32}\unit{cm}$^{-2}\unit{s}^{-1}$.}

\label{fig-mctable}
\end{center}
\end{figure}

Based on current measurements of the background levels, the rates seen in the muon system can
be extrapolated to higher luminosities. The results of linear extrapolations to LHC peak instantaneous luminosity values
of 10$^{34}\unit{cm}^{-2}\unit{s}^{-1}$
are presented in Fig.~\ref{fig-mctable} along with Monte Carlo predictions~\cite{trigTDR}.

For the barrel wheels, only the expected neutron background level is taken into consideration.  For the endcap rings, all particle fluences are included.
Note that the Monte Carlo simulations were for an LHC beam energy of 7\TeV, whereas the data were accumulated at 3.5\TeV.

Figure~\ref{fig-mctable} shows that the simulated levels are larger than the
extrapolated measured rates in locations where the background levels are highest.  Although the ME1/1 rate is higher than the others, it is notably less than the prediction.
The extrapolated rates are slightly higher than the predicted levels of background only for the chambers of the outer stations because they are not shielded.
It is expected that the addition of the YE4 shielding disk currently under construction will alleviate the high rates indicated for the ME4/1 and ME4/2 chambers.

\section{Alignment}
\label{sec:Alignment}
Precise measurement of muons up to the\,\TeVc momentum range requires the DTs and
CSCs to be aligned with respect to each other, and to the central
tracking system, with an accuracy of a few hundred microns, comparable to their
intrinsic spatial resolution. The RPCs are already aligned to the
limit of their spatial resolution, which is about 1\unit{cm}.
The muon transverse momentum is measured from the curvature of tracks in
the $r$-$\phi$ plane.
The precision on displacements in the $r$-$\phi$ direction and rotations of
chambers around their local axis parallel to $z$ is therefore directly
related to the momentum resolution. In both of these degrees of freedom, the
alignment system is designed to achieve resolutions of 100--150\micron and 40\unit{$\mu$rad}
for both the MB1 and ME1/1 chambers, which have the largest weight in
the muon momentum measurement. Alignment in other degrees of freedom
affects the momentum measurement as higher-order corrections.
This section describes the current muon alignment procedure and its performance.

To determine the positions and orientations of the muon chambers, the
CMS alignment strategy combines precise survey and photogrammetry information,
measurements from an optical based hardware alignment system, and the results
of alignment procedures based on muon tracks.

The muon hardware alignment system consists of rigid structures in the barrel yoke, in the gaps between the different wheels, and between the barrel and endcap yokes, and of a set of straight line monitors running nearly radially along the surface of the endcap chambers of the forward disks.
The 2 systems are connected to each other and to the central tracker by a link system.
The link system consists of 2 floating rigid carbon fiber rings on the first 2 endcap iron disks close to the gaps that separate the barrel and endcap iron structures at $\eta$ = 0.9--1.2.
These rings are connected via a redundant network of laser lines and proximity sensors to rigid structures on the faces of the tracker and on the outer wheels of the barrel, and to plates that reference the positions of the chambers on the first endcap station.
The link thus provides a relatively robust connection between the 3 detector systems.
Further details about the track and hardware based alignment methods are given in Refs.~\cite{CRAFT08_track_alignment_paper} and \cite{CRAFT08_hw_alignment_paper}, respectively.

\subsection{Muon barrel alignment}
The DTs in the barrel are aligned independently by the hardware alignment
system and by the use of muon tracks. The current CMS reconstruction uses the
results of the hardware alignment, which is based on rigid, radial
carbon fiber structures called modules for alignment of barrel (MAB) that are supported on the faces of the 5 wheels of the CMS steel yoke.
A dedicated reconstruction program called the CMS
object-oriented code for optical alignment (COCOA)~\cite{COCOA} is used to
transform the various optical measurements into DT chamber
positions.

The current implementation of the hardware barrel alignment performs a complete
alignment of all DTs in stations 1, 2, and 3 in a single
computation. The fourth DT station is added in a second step to the resulting
aligned structure from the previous calculation. The motivation for this
factorization is two-fold: the computational problem becomes significantly
simpler, allowing the reconstruction program to run much faster, and the
knowledge of camera positions mounted on the MABs is less precisely known near the zone
of the outer station, and therefore the internal barrel structure is
essentially not affected by excluding station 4.

Once all DTs and MABs are aligned relative to each other, the resulting barrel
structure is treated as a floating rigid body, which must be positioned and
oriented in space with respect to the inner tracker. The positions of the 12 external barrel
MABs (6 on each end of the barrel) are reconstructed independently by the barrel
alignment system (in an arbitrary reference frame) and by the link system.
An initial tracker--barrel ``cross-alignment'' is therefore achieved by fitting
the external MABs of the aligned, rigid barrel to the MAB positions determined
by the link system. This cross-alignment is further refined by a track based
alignment method, which uses internally aligned tracker and muon barrel systems
and obtains their relative position and orientation using only a few tens of
thousands of global muon tracks from proton--proton collisions.

\subsubsection{Validation}
Alignment results must be validated before they can be used for track
reconstruction and data reprocessing. The barrel alignment is validated by
3 independent cross-checks: comparison of photogrammetry measurements with
alignment results obtained from measurements in the absence of magnetic field
(0\unit{T}), residuals of standalone muon segments extrapolated to a
neighboring station, and comparison with an independent track based alignment
of chambers.

\begin{figure}[htbp]
\begin{center}
\includegraphics[width=0.49\textwidth]{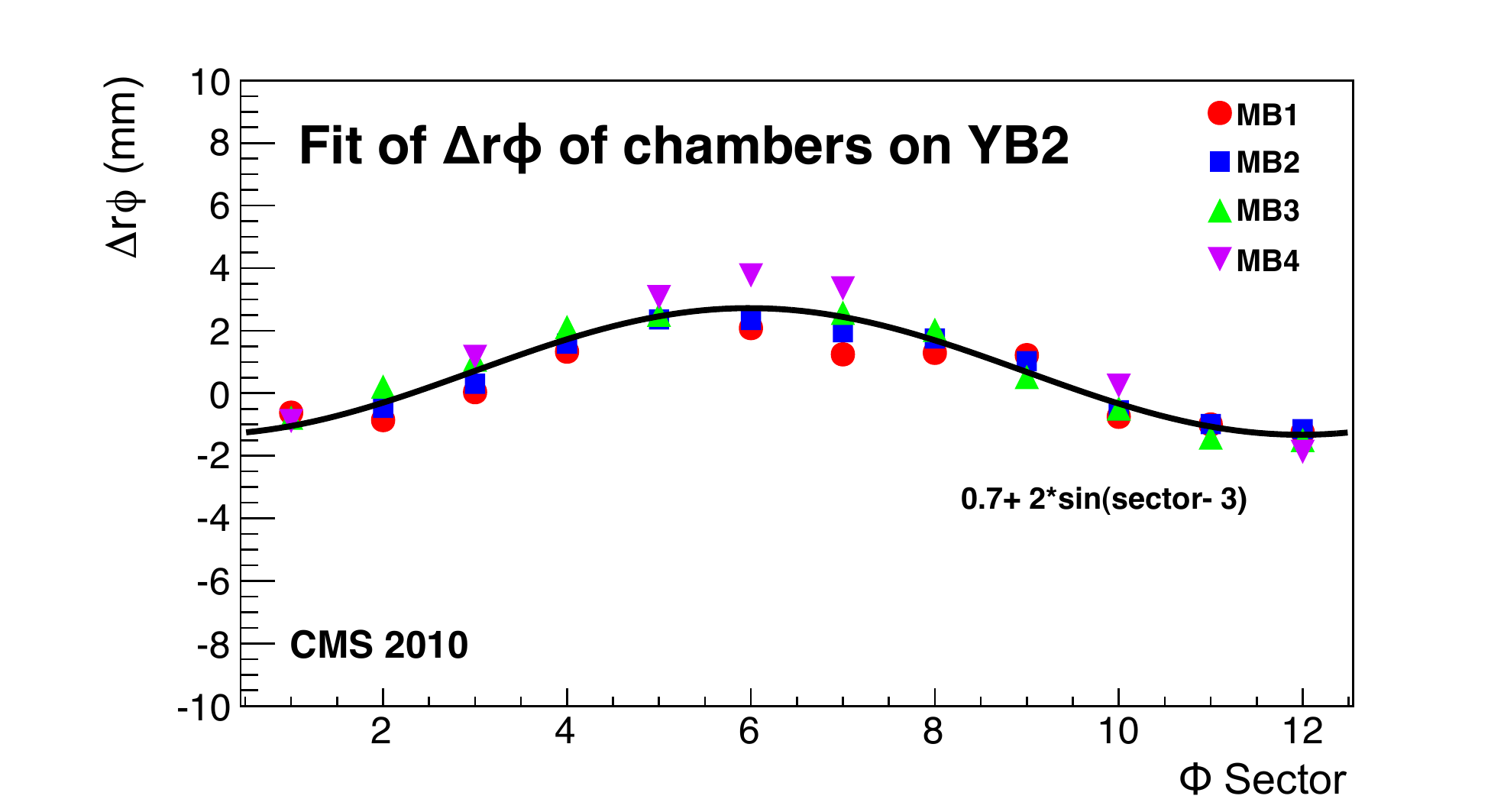}
\hfill
\includegraphics[width=0.49\textwidth]{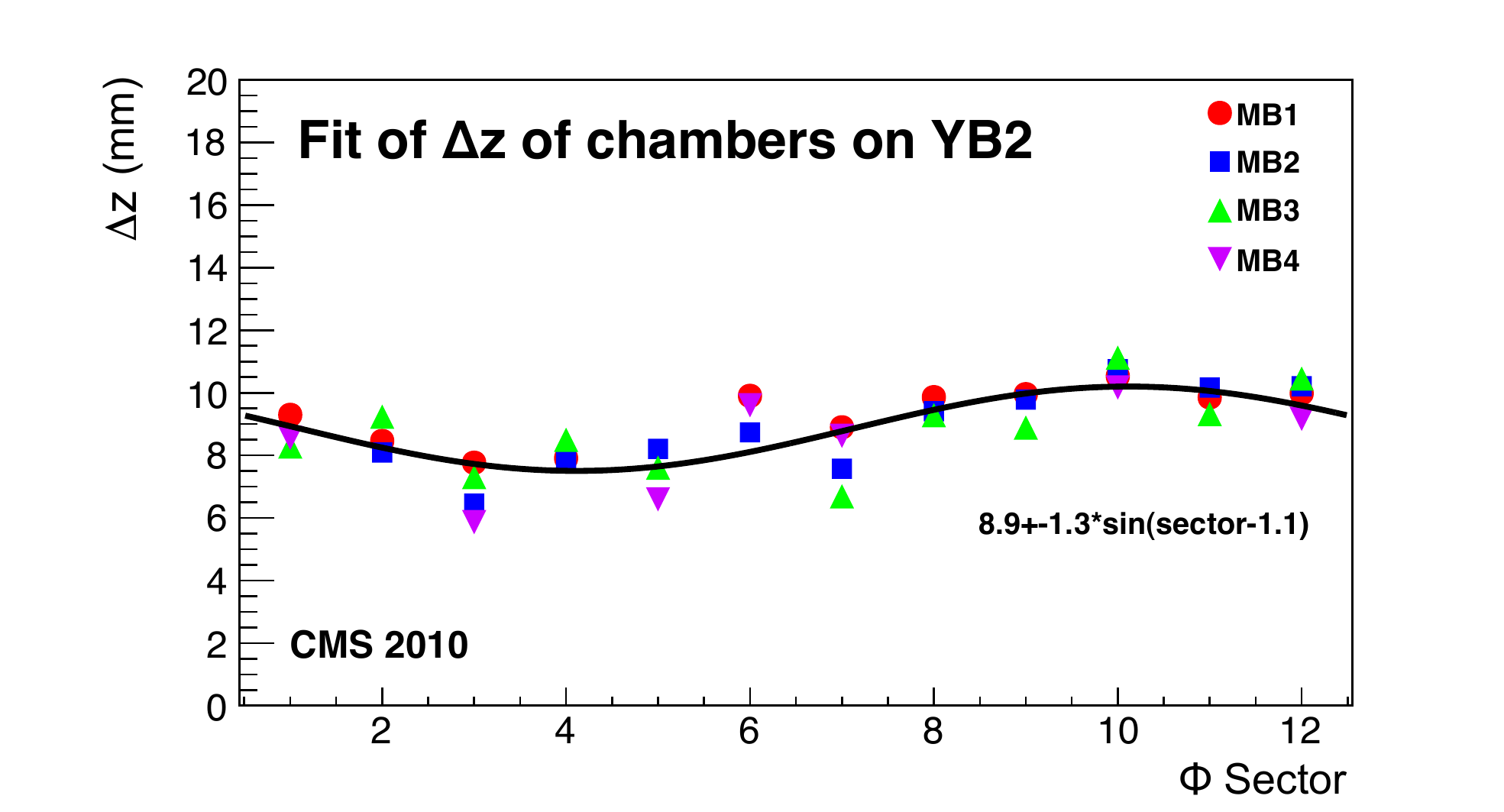}
\includegraphics[width=0.49\textwidth]{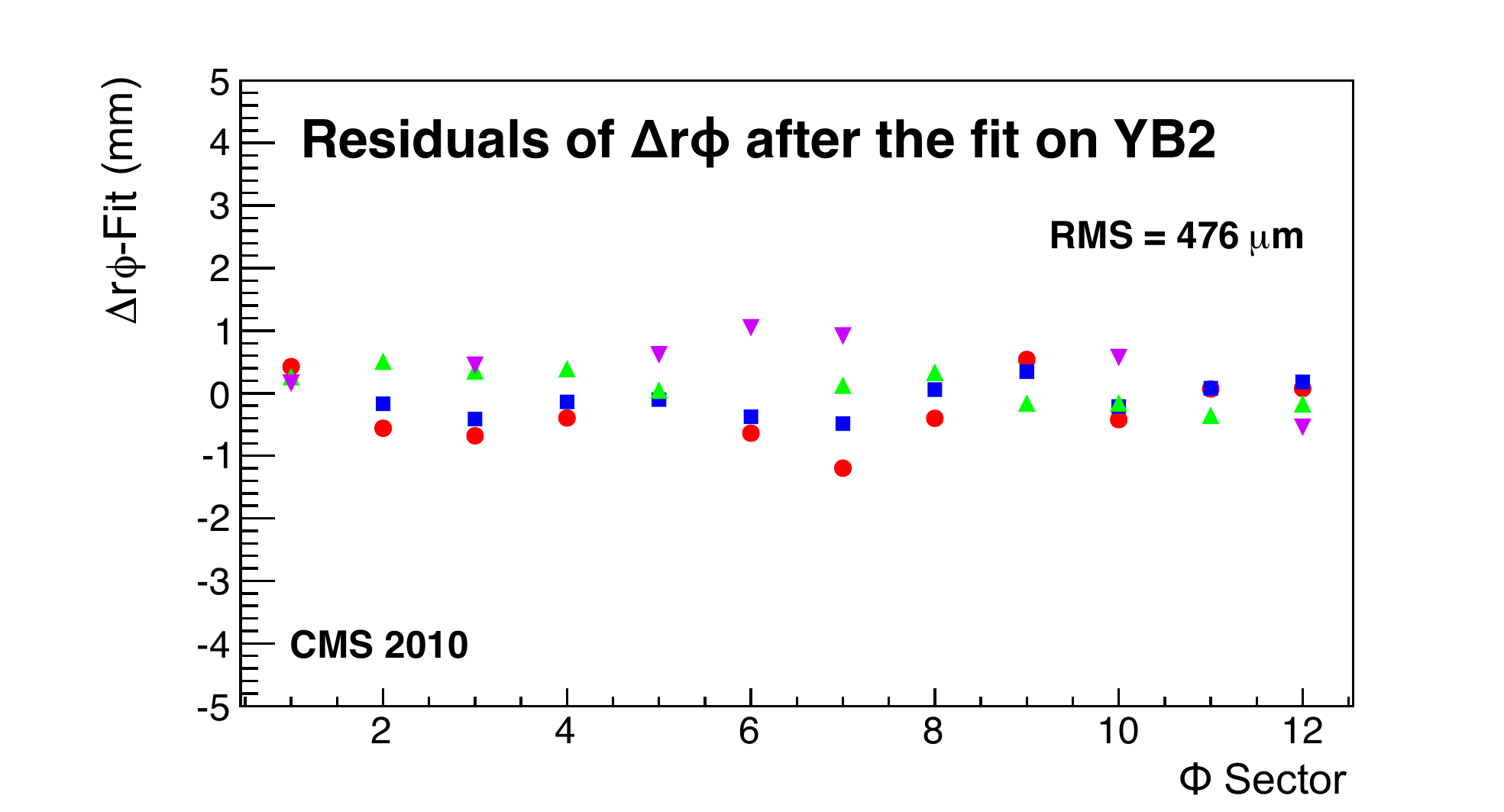}
\hfill
\includegraphics[width=0.49\textwidth]{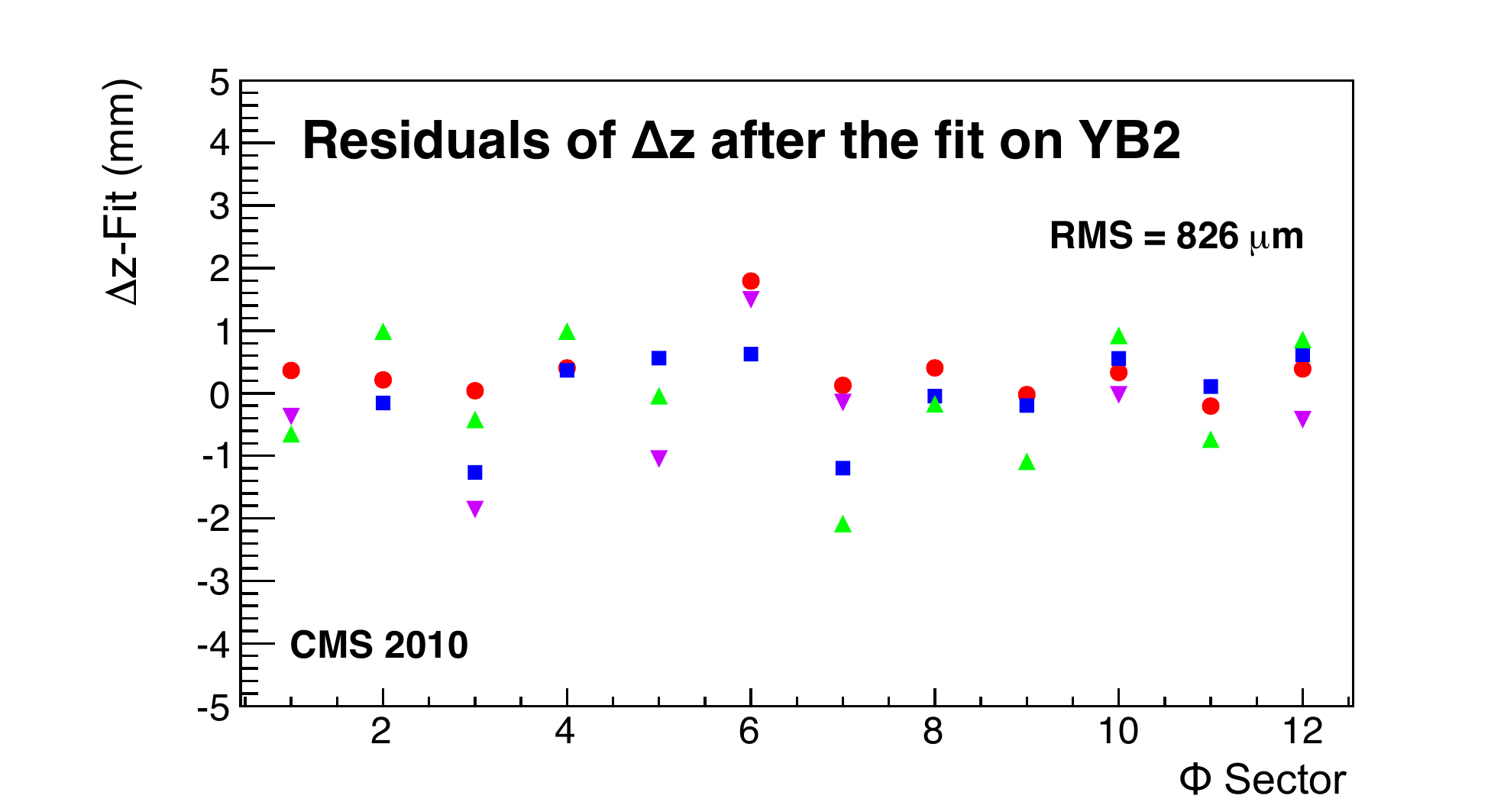}
\caption{Differences in $r$-$\phi$ (left) and $z$ (right) between photogrammetry
measurements and barrel alignment at 0\unit{T} for DTs in YB+2 before (top) and
after (bottom) correcting for the relative overall wheel movement.}
\label{Figures:barrel0TvsPG}
\end{center}
\end{figure}

An estimate of the accuracy of the barrel alignment can be obtained by
comparing the results from geometry reconstruction at 0\unit{T}  with photogrammetry
measurements, in which all MABs and DTs in the same wheel are measured
simultaneously. Care must be taken when comparing photogrammetry measurements,
which are taken with an open detector, to alignment measurements after detector
closing. Since wheels can move and tilt upon closing of neighboring structures,
and since the same MABs are used to measure DTs sitting on different wheels,
only the relative positions of DTs within each wheel can be expected to agree.
For this reason, all comparisons are made independently for each wheel.
Large disagreements between photogrammetry and alignment reveal an overall wheel movement between the
2 sets of measurements, as illustrated in Fig.~\ref{Figures:barrel0TvsPG}
(top) for YB+2. After correcting for this movement, the phi sectors within each
wheel can be compared, as shown in Fig.~\ref{Figures:barrel0TvsPG}
(bottom).
Table~\ref{tab:Barrel0TvsPG} shows the RMS differences between the results of photogrammetry and alignment at 0\unit{T}
in $r$-$\phi$ and $z$ for all DTs in each wheel, after correcting for collective wheel movements.
\begin{table}[htbp]
\topcaption{The RMS of the differences in $r$-$\phi$ and $z$ for all DTs in each wheel
between photogrammetry and barrel alignment at B = 0\unit{T}  after correcting for
overall wheel movements.}
\begin{center}
\begin{tabular}{|l|c|c|} \hline
Barrel Wheel & $\Delta r\phi$ RMS (\micron) & $\Delta z$ RMS (\micron) \\
\hline \hline
YB+2         & 476                     &  826         \\ \hline
YB+1         & 433                     &  1260        \\ \hline
YB0          & 707                     &  989         \\ \hline
YB-1         & 445                     &  822         \\ \hline
YB-2         & 625                     &  847         \\ \hline
\end{tabular}
\end{center}
\label{tab:Barrel0TvsPG}
\end{table}

\begin{figure}[htbp]
\begin{center}
\includegraphics[width=0.45\textwidth]{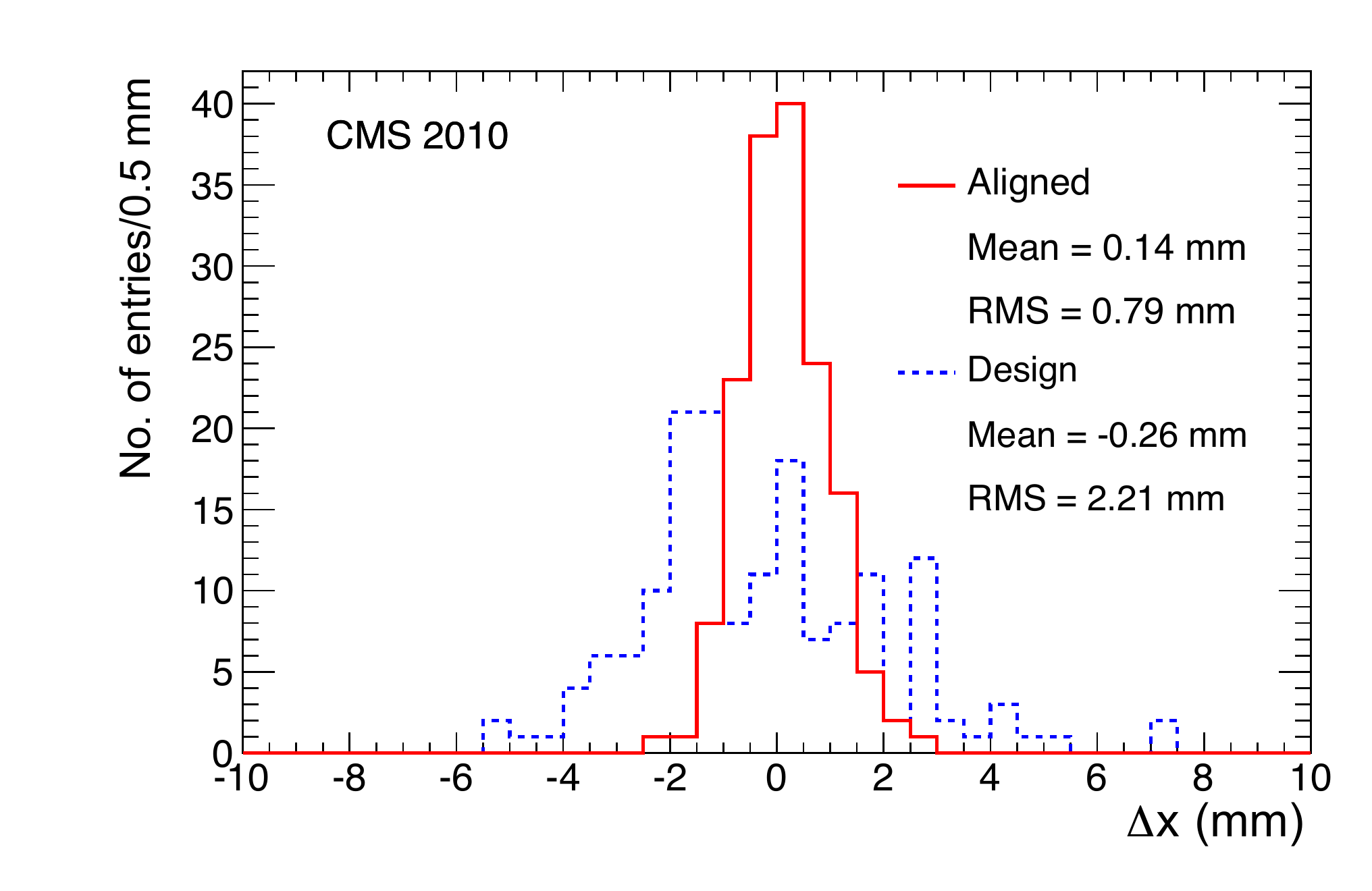}
\includegraphics[width=0.45\textwidth]{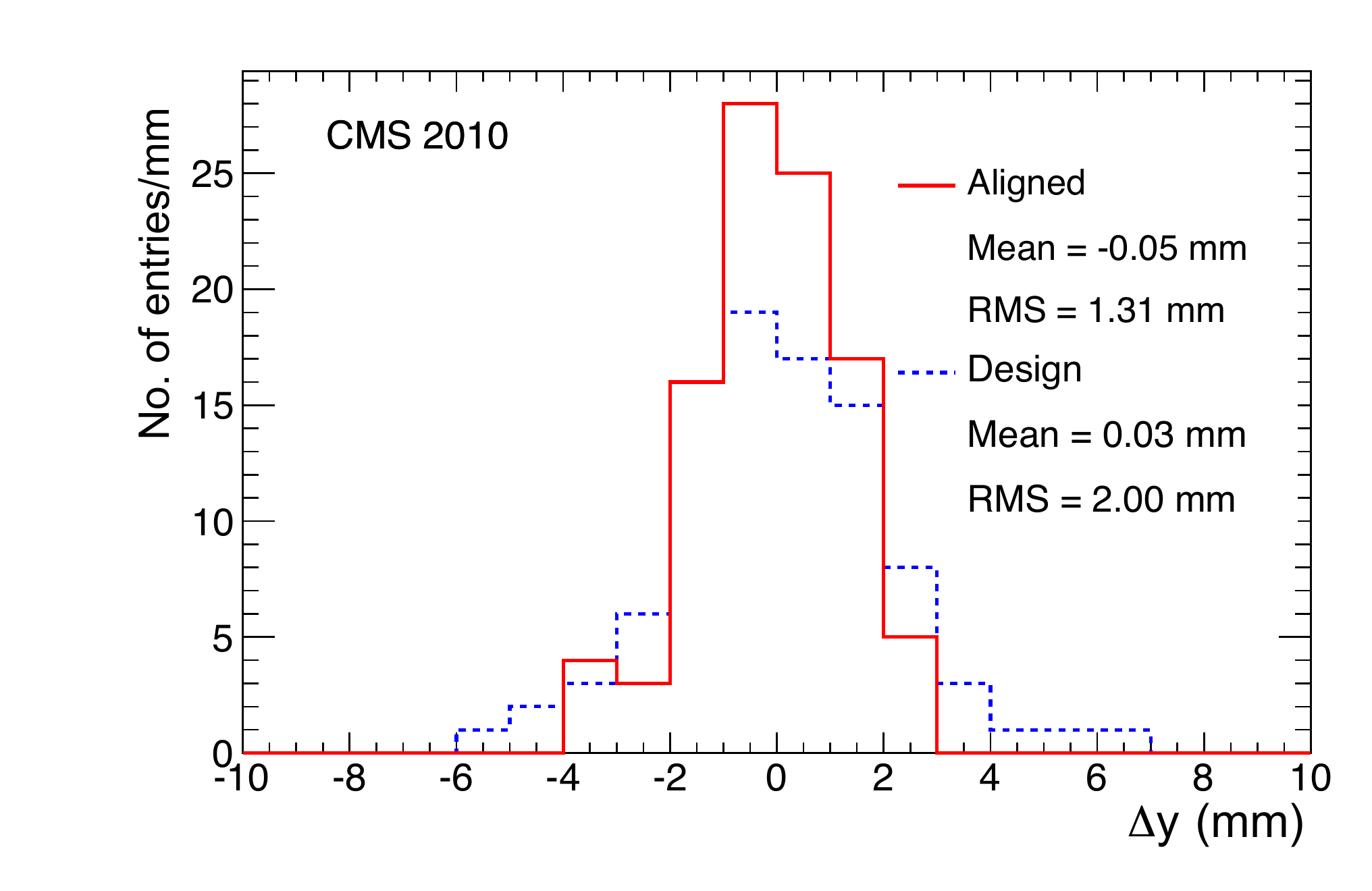}
\includegraphics[width=0.45\textwidth]{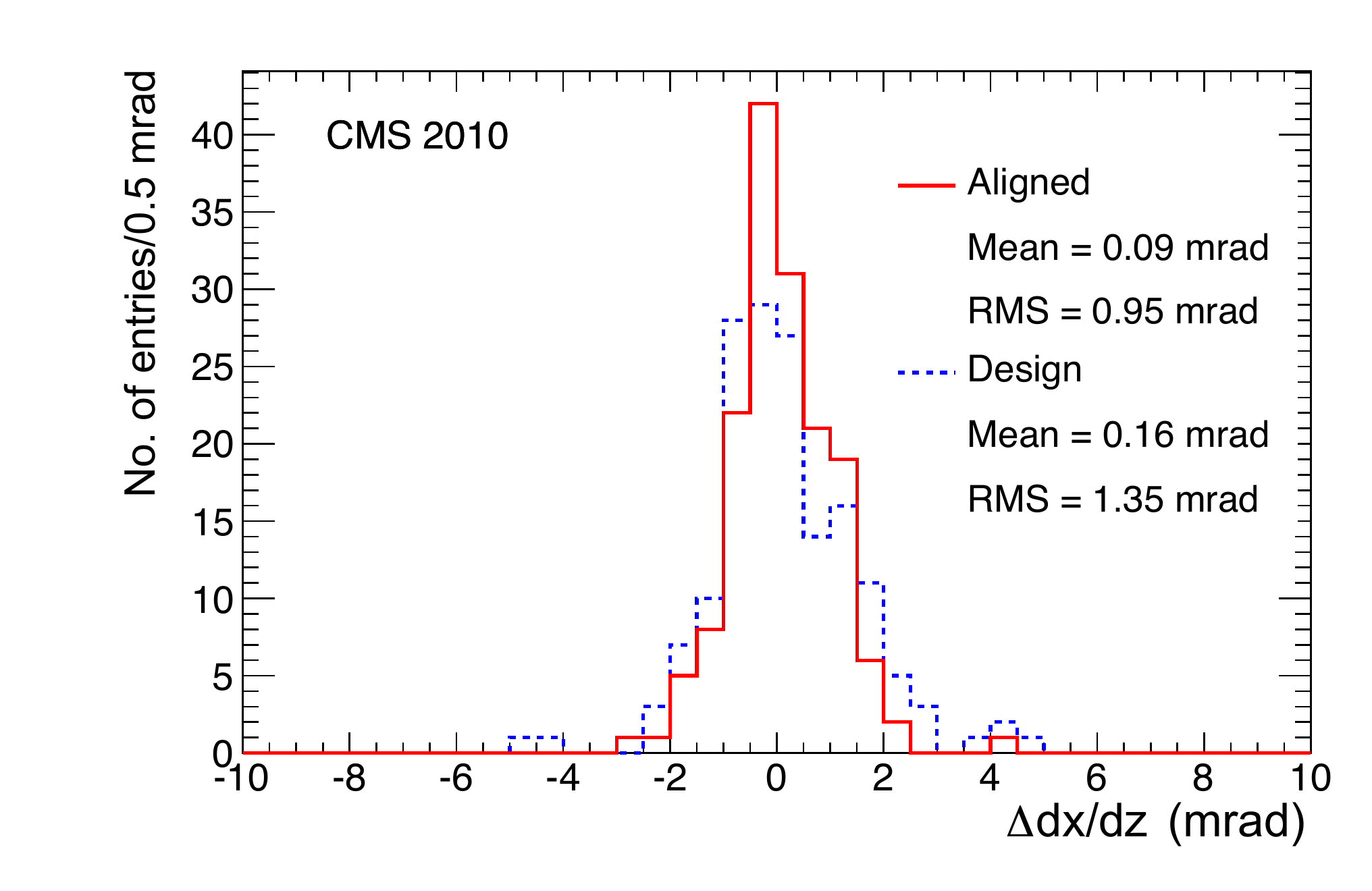}
\includegraphics[width=0.45\textwidth]{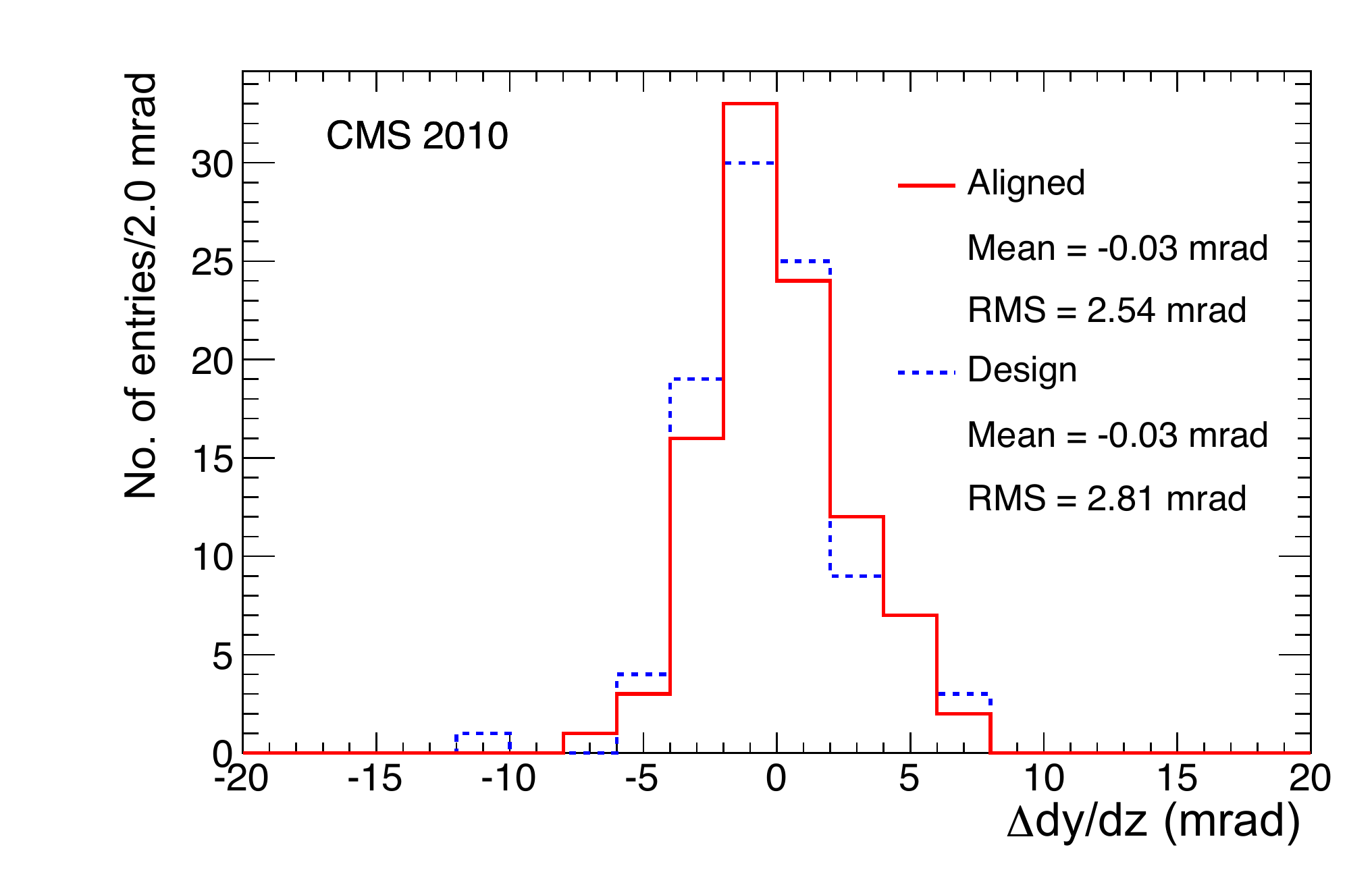}
\caption{The difference before (``design''; blue, dashed lines) and after (red, solid lines) alignment between the measured and
extrapolated stand-alone muon track segments going from one DT station to the
next in the same wheel and sector for all such DT pairs in the 4 coordinates
measured by DTs: local x, y, dx/dz, and dy/dz.}
\label{Figures:SA-residuals}
\end{center}
\end{figure}

Another test of the barrel alignment makes use of stand-alone muon tracks that are reconstructed with chamber positions updated after alignment.
Track segments inside a given DT are extrapolated into the
DT of the next station in the same wheel and sector, and the residuals between
the extrapolated segments and the actual track segments are studied.
Figure~\ref{Figures:SA-residuals} shows the residuals distributions
for all such DT pairs for the 4 coordinates measured by DTs before and
after hardware alignment. In the absence of systematic effects, the mean values of these
residuals are expected to be close to 0, while the RMS of the distributions
get a contribution from the alignment precision (both for the overall chamber
position and for the internal DT alignment) and from other tracking effects.
The improvement of the residual distributions after alignment is clearly visible.

Finally, the hardware and track based alignment results are compared.
Both alignments successfully reproduce the overall, large corrections needed
with respect to the design geometry, in which all chambers are placed at
their design position. These corrections are mostly due to vertical and axial
compression of the wheels due to the huge gravitational and magnetic forces
acting on them. In MB1, both alignments agree in $r$-$\phi$ within
$\approx$750\micron, consistent with the precision expected for the
statistically limited track based alignment. The agreement
for stations 2, 3, and 4 worsens from 1 to 2\unit{mm}, as expected
from multiple scattering effects and longer track propagation in the
track based alignment.

\subsection{Muon endcap alignment}
The muon endcap was aligned by using information from 4 different
sources: photogrammetry, the muon endcap alignment system, tracks from
beam halo muons, and tracks from muons produced in proton--proton collisions.
Some of these sources supply measurements of the same alignment parameters, allowing cross-checks between the sources. The different alignment procedures sometime apply to different subsets of the 3 spatial and 3 angular coordinates that specify each chamber in the system.
For example, alignment of entire disks (rings) of CSCs leads to an overall correction to the coordinates of each CSC in the disk.
To combine the sources and procedures, the alignment corrections must be derived in a sequence of steps carefully chosen to ensure that as the alignment of a given coordinate is improved, the quality of alignment in other coordinates achieved in previous steps is maintained.
For example, local $z$ coordinates of chambers are taken initially from the hardware alignment system and are relatively insensitive to the track-based alignment procedure, but after improving the alignment in the orthogonal coordinates, some improvement in the $z$ alignment can be obtained from the track-based procedure.
Since the corrections from successive steps may be interdependent, the entire process is iterated until the results converge.

\begin{figure}
\begin{center}
\includegraphics[width=0.3\linewidth]{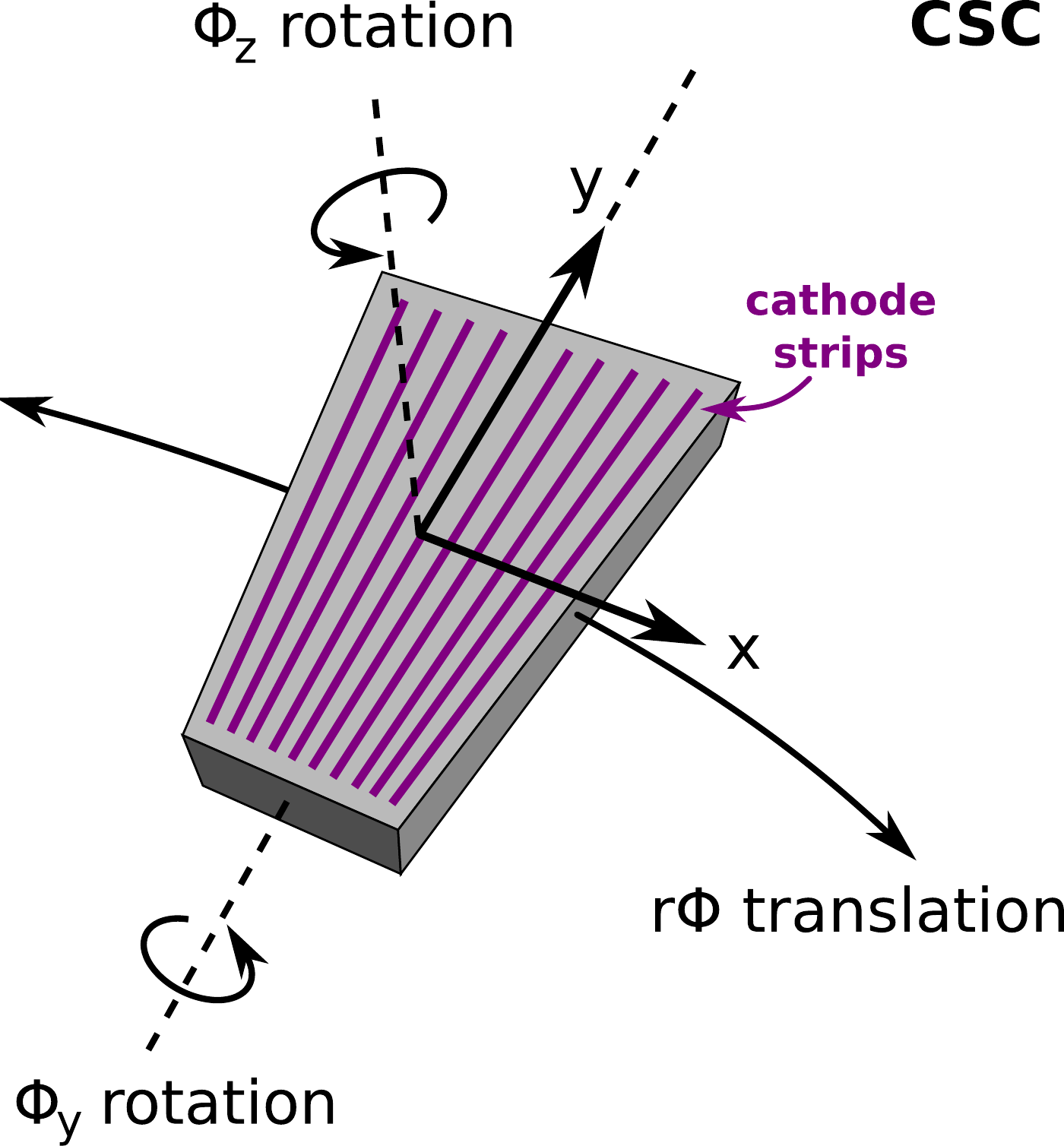}

\caption{Schematic CSC chamber, indicating the local coordinate
  system. \label{Figures:csc_coordinates}}
\end{center}

\end{figure}

\subsubsection{Measurement of disk bending with the muon endcap alignment system}

The muon endcaps suffer a significant deformation when the solenoid is
energized to its operating field strength of 3.8\unit{T}. Some CSC chambers can
move towards the center of the detector by as much as 14\unit{mm}, and they can
rotate around their local $x$ axis by as much as 3.5\unit{mrad}. The hardware endcap
alignment system measures these movements by means of laser beams running
nearly radially across each disk as described in detail in Ref.~\cite{CRAFT08_hw_alignment_paper}.

\subsubsection{Alignment of overlapping chambers using beam halo muons}

The CSC chambers overlap slightly along their edges, and muons passing through these narrow regions provide information about the relative displacement of the neighboring chambers.
To obtain individual chamber positions from the pairwise chamber information, the following objective function is minimized:
\begin{equation}
\chi^2 = \sum_{m_{ij}}^\text{constraints} \frac{(m_{ij} - A_i +
  A_j)^2}{{\sigma_{ij}}^2} + \lambda \left(\frac{1}{N_\text{chambers}} \sum_i^\text{chambers}A_i\right)^2,
\label{eqn:minimizeme}
\end{equation}
where $A_i$ are the chamber coordinates to optimize, $m_{ij} \pm
\sigma_{ij}$ are the pairwise chamber measurements, and $\lambda$ is a
Lagrange multiplier to constrain the floating coordinate system.  Two
types of constraints are used: beam halo tracks and photogrammetry
measurements, with the latter applied only to pairs of chambers that
were missing track data on account of failed readout electronics (14
out of 396 pairs of neighboring chambers).
The alignment proceeds in steps, first aligning $r$-$\phi$ positions (in which case the $A_i$ are interpreted as positions and $m_{ij}$ are residuals), then $\phi_z$ angles (in which case the $A_i$ are chamber angles and $m_{ij}$ are angle residuals), and repeating until the procedure converges.
Definitions of the CSC local coordinates are illustrated in Fig.~\ref{Figures:csc_coordinates}.
The alignment fully converged after a single $r\phi$ pass and $\phi_z$ pass.

\begin{figure}
\includegraphics[width=\linewidth]{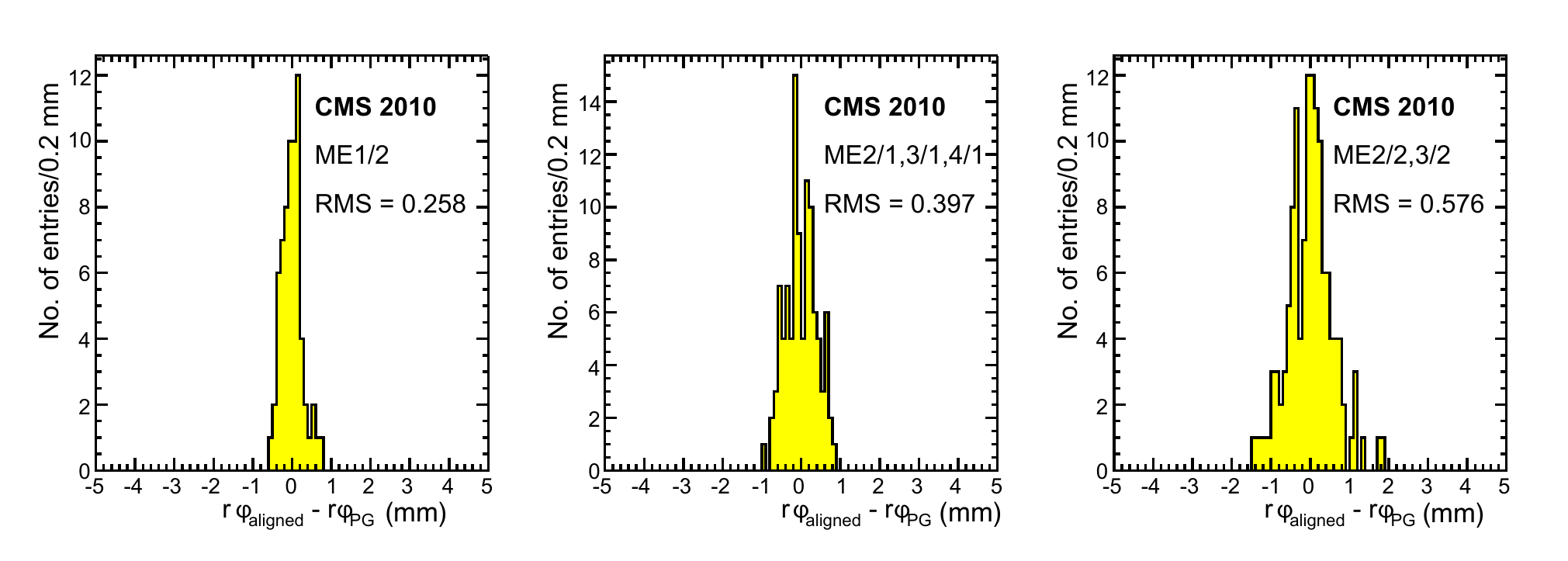}

\caption{Chamber positions after internal-ring alignment compared
  with photogrammetry, split by ring.  (ME1/1 chambers were not
  measured by the photogrammetry method.) \label{Figures:aligned_minus_pg}}
\end{figure}

Although photogrammetry information was used to constrain some
of the chambers, much larger weights were given to the beam halo data,
in inverse proportion to the square of the measurement uncertainties
in the 2 methods.
The level of agreement between the track based technique and photogrammetry is 0.3--0.6\unit{mm} (Fig.~\ref{Figures:aligned_minus_pg}).
This is much smaller than the typical scale of chamber corrections from the design geometry (2--3\unit{mm}).

\subsubsection{Whole-ring placement using muons from proton--proton collisions}

To complete the endcap alignment, the internally aligned rings must be
aligned relative to one another and the tracker.  Tracks from the
tracker were propagated to the muon chambers and whole-ring
corrections were derived from the pattern of $r$-$\phi$ residuals as a
function of global $\phi$.  A constant offset in the residuals is
interpreted as a rotation of the ring in $\phi_z$, while terms
proportional to $\cos\phi$ and $\sin\phi$ are interpreted as
displacements in global $x$ and $y$, respectively.

\begin{figure}
\begin{center}
\includegraphics[width=0.75\linewidth]{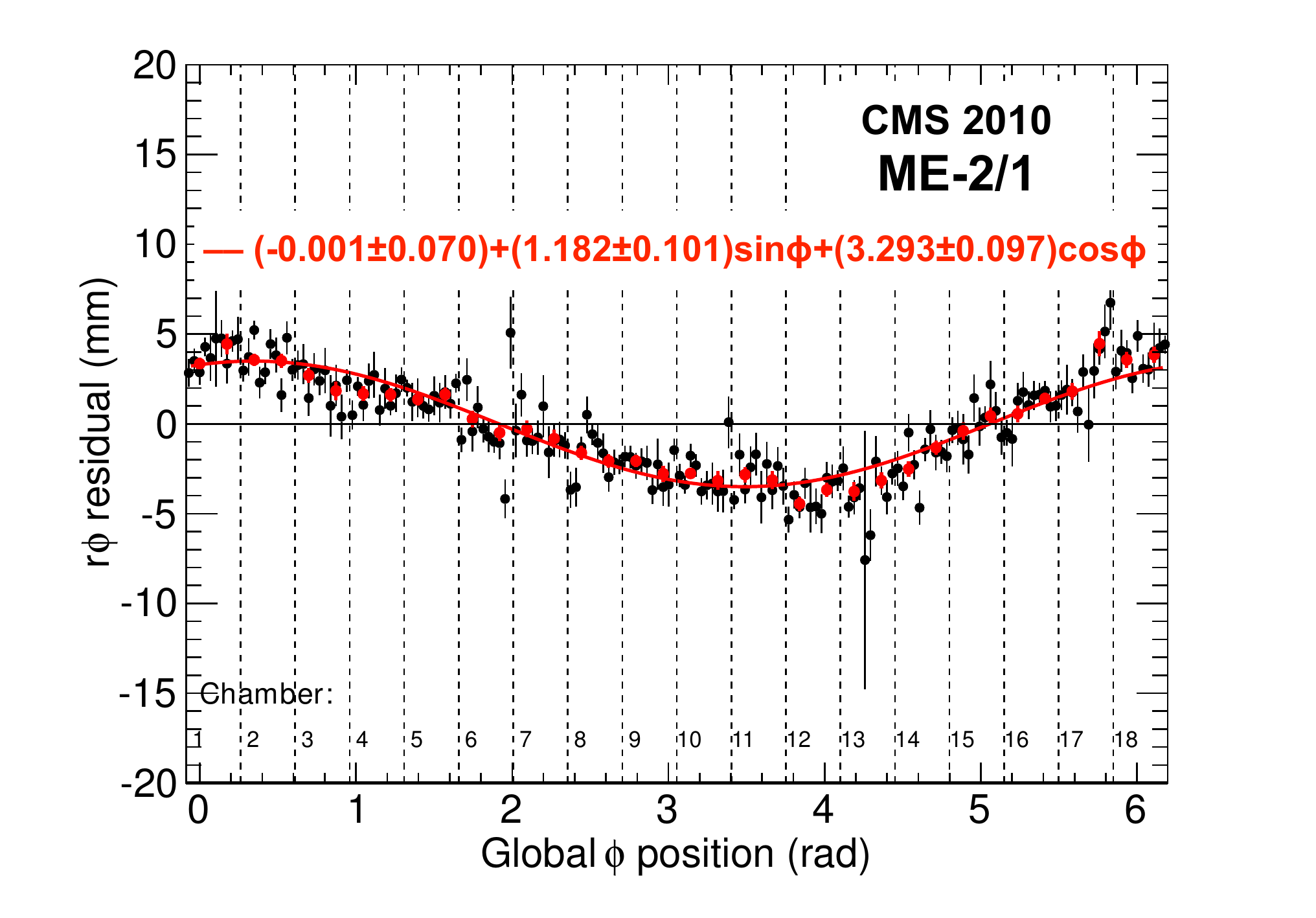}
\end{center}
\caption{Residuals plot used to align a ring before alignment corrections are applied.
  Black points are a profile
  derived from truncated-Gaussian peak fits in each $\phi$ bin, and
  red points are the average of peak fits for \Pgmm and \Pgmp
  separately.  Three parameters of the fitted curve are interpreted as 3 alignment
  degrees of freedom.  Vertical dashed lines indicate the boundaries
  between chambers. \label{Figures:one_and_only_mapplot}}
\end{figure}

Figure~\ref{Figures:one_and_only_mapplot} provides an example of an
alignment fit result for ring ME$-$2/1.  The alignment was performed in
a single pass, with a second iteration to verify self-consistency.

\begin{figure}
\includegraphics[width=0.45\linewidth]{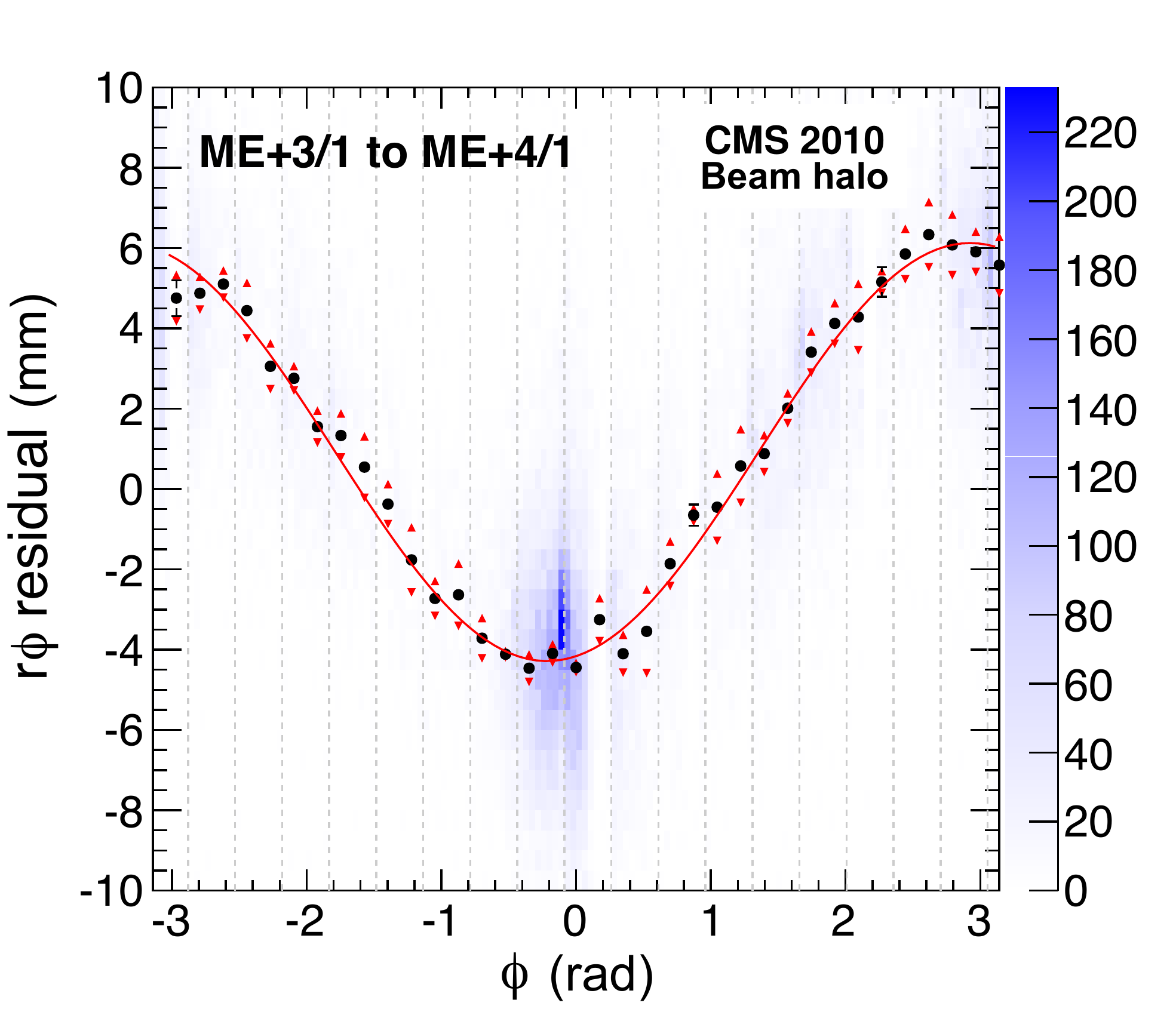} \hfill
\includegraphics[width=0.45\linewidth]{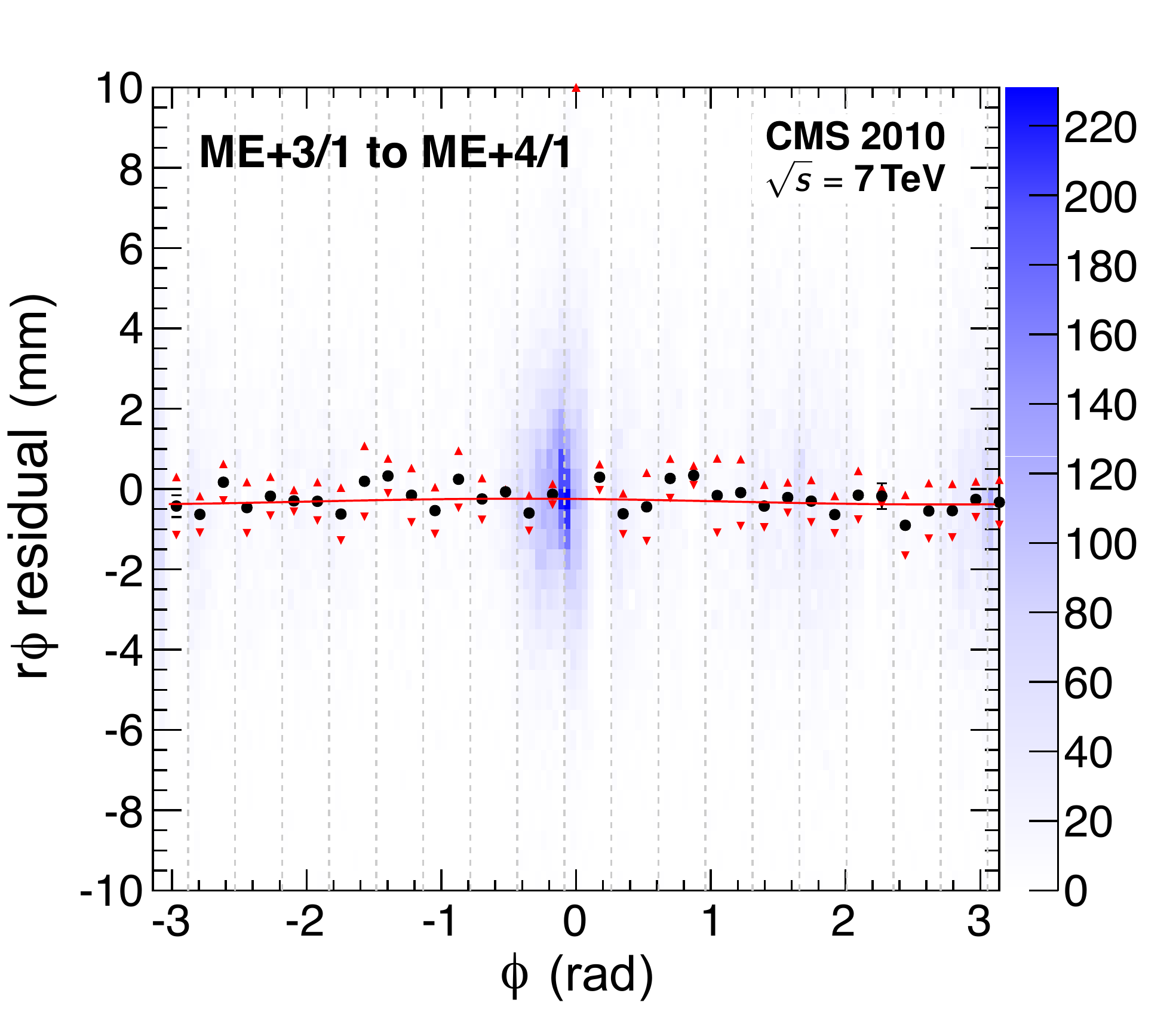}

\caption{Residuals from beam halo tracks used to cross-check the
  alignment performed with collisions.
  The individual residuals are shown as the underlying blue scatterplot, in bins of 2.5$^\circ$ in $\phi$ and 0.5\unit{mm} in distance, with the density indicated by the right-hand scale.
  The symbols in these plots
  have the same meaning as in Fig.~\ref{Figures:one_and_only_mapplot}, though
  residuals were calculated differently (see text).  Left: before
  alignment.  Right: after alignment using collisions (not
  beam halo).  \label{Figures:BHCrossCheck_mep41}}
\end{figure}

To cross-check the alignment with a qualitatively different method,
beam halo tracks crossing an entire endcap (3 or 4 stations,
depending on the distance from the beamline) were used to calculate
residuals by extrapolating segments from one station to another.
Figure~\ref{Figures:BHCrossCheck_mep41} shows an example in which
ME$+$3/1 segments were propagated linearly (no corrections for
material or magnetic field) to ME$+$4/1.  These plots were not used to
perform the alignment, so the fact that the strong $\phi$ trend
observed before alignment is eliminated in the aligned geometry adds
confidence to the result.

\subsection{Alignment impact on physics performance}

The most important test for any calibration or alignment is to study the effect
it has on reconstructed quantities. In the case of muon alignment, higher-level
objects related to muon tracks must be studied. Note that by design the momentum resolution is dominated by the central
tracker for muons with transverse momentum below 200\GeVc.
A well-aligned muon system is therefore expected to induce minor beneficial
changes at reconstruction level for low momentum global muons, and to improve
global muon measurements for very energetic muons.

\begin{figure}[htbp]
\begin{center}
\includegraphics[width=0.48\textwidth]{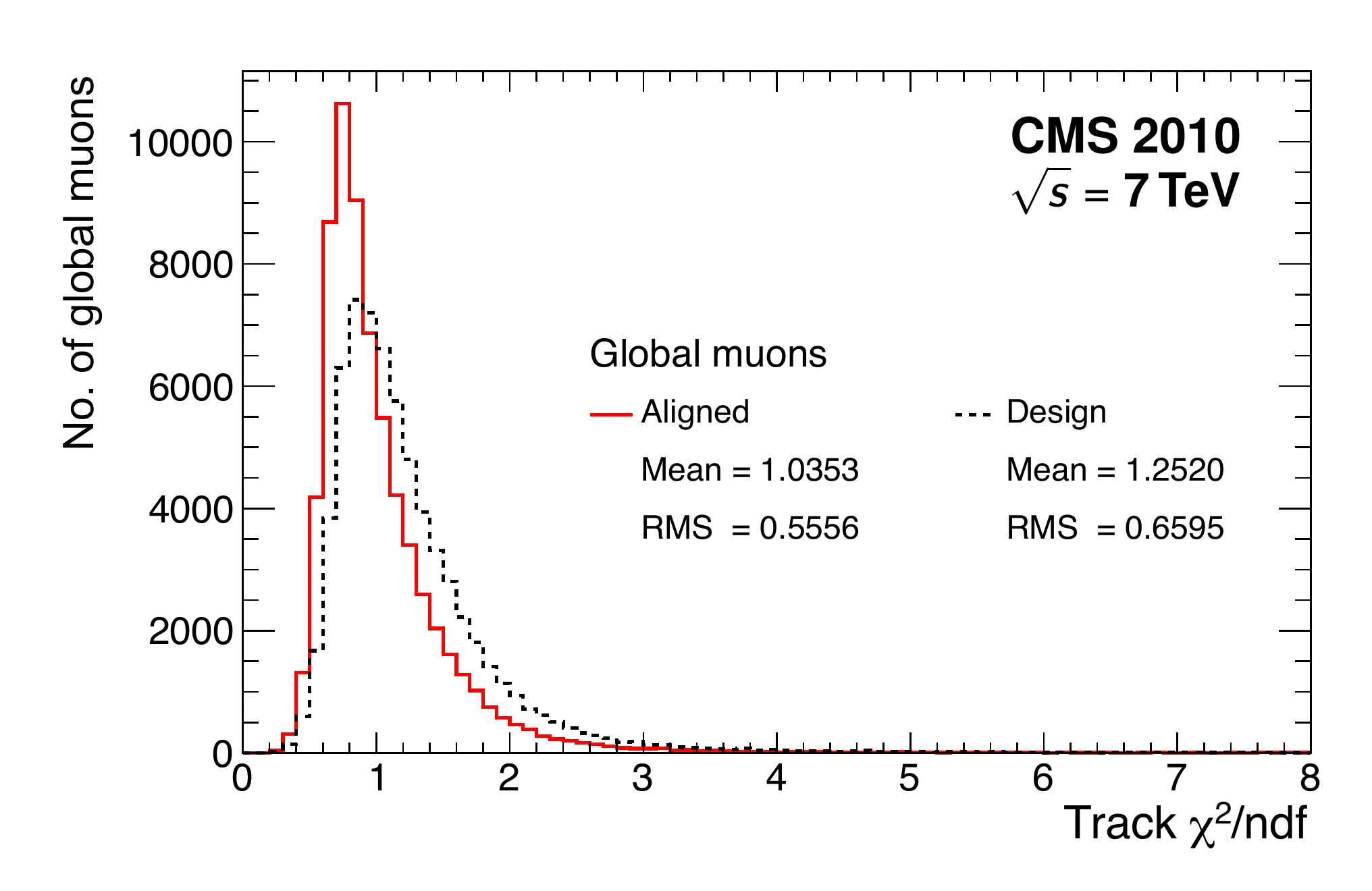}
\includegraphics[width=0.48\textwidth]{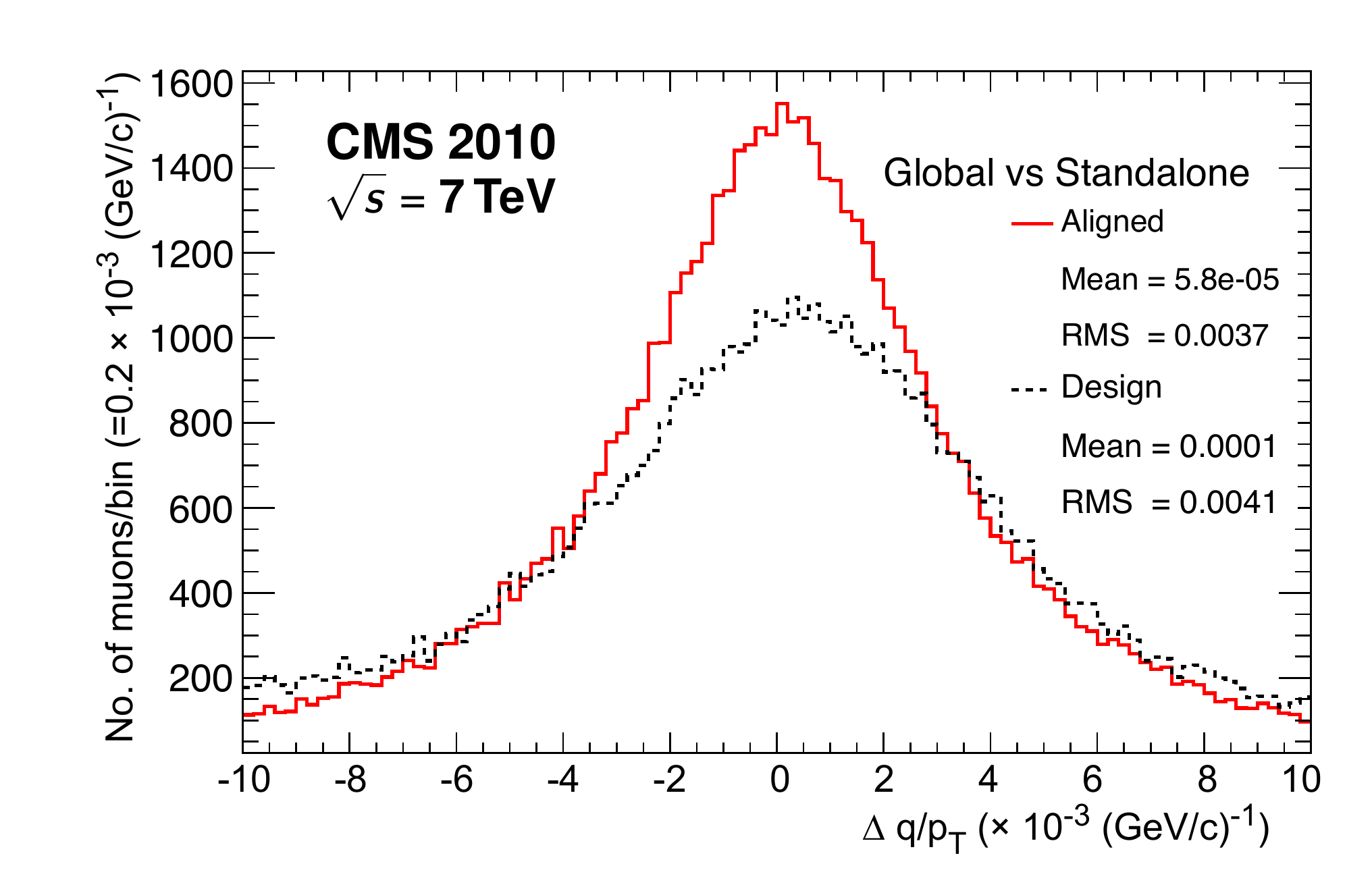}
\includegraphics[width=0.48\textwidth]{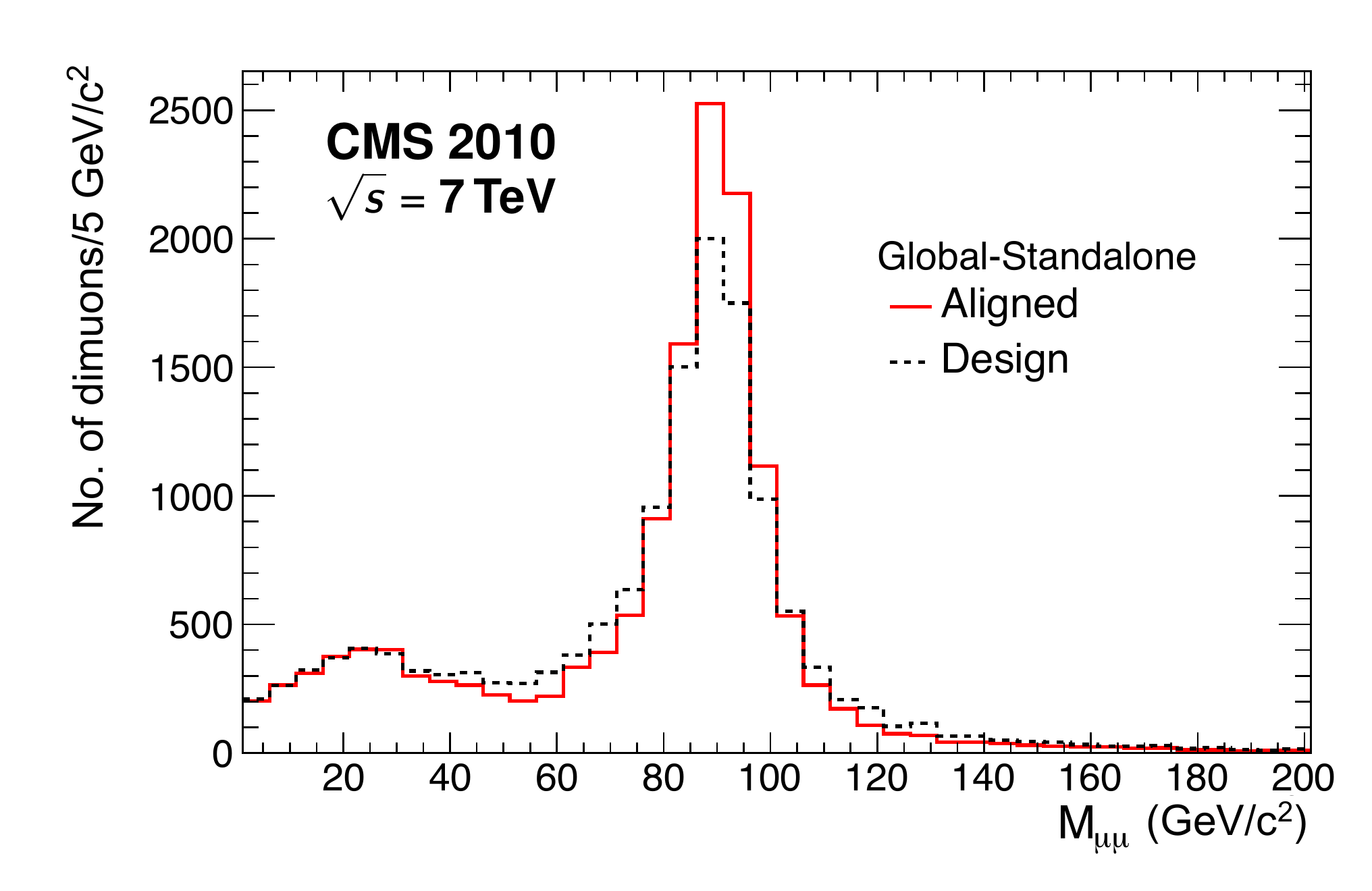}
\includegraphics[width=0.48\textwidth]{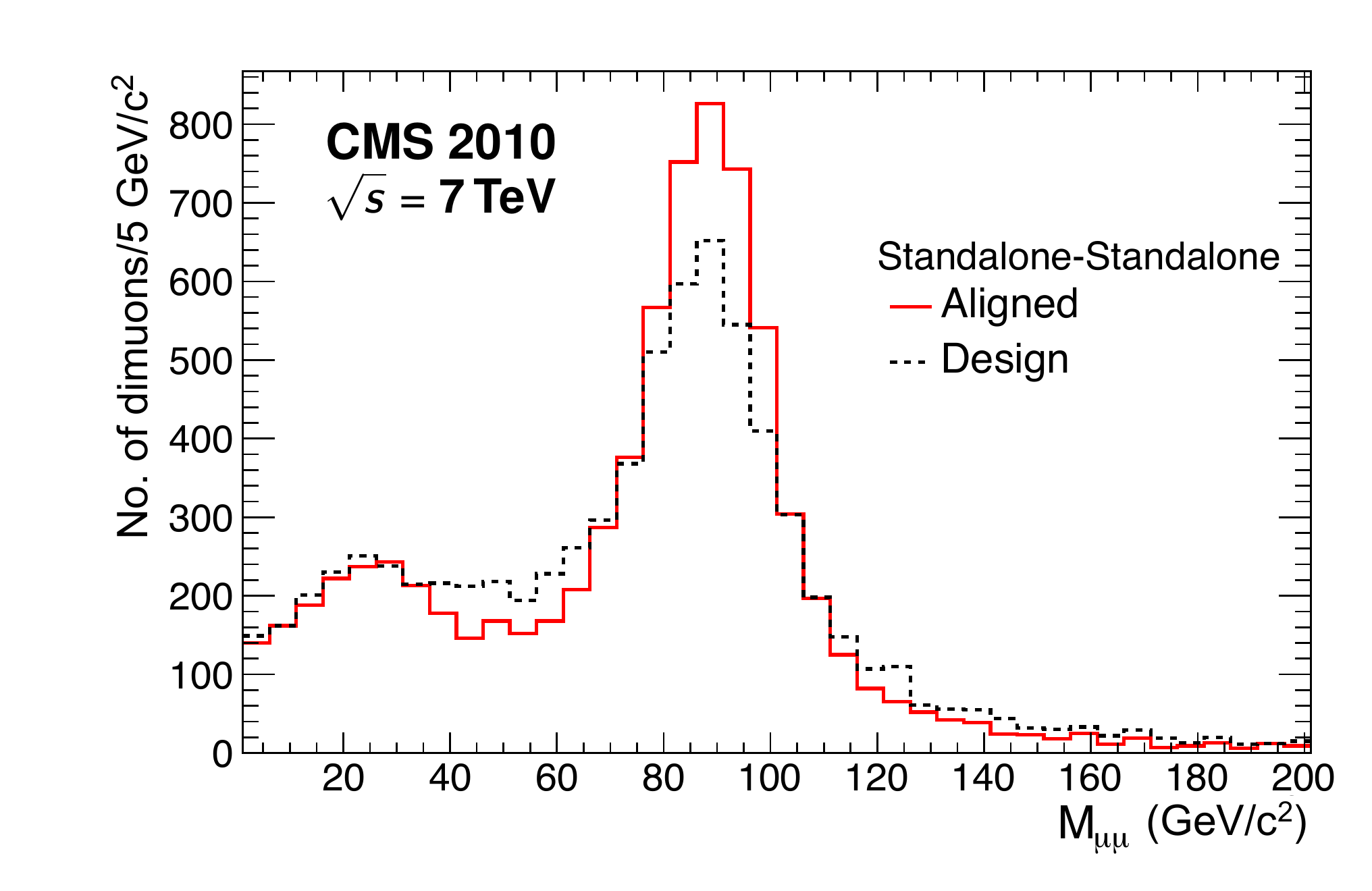}
\caption{Muon track quantities reconstructed with design (black, dashed line) and aligned
(red, solid line) muon chamber positions.
Top-left: normalized $\chi^2$ for global muon tracks.
Top-right: difference in $q/\pt$ between muons reconstructed as global and as
stand-alone.
Bottom: dimuon invariant mass for global-standalone muon pairs (left) and
standalone-standalone muon pairs (right).}
\label{Figures:reco}
\end{center}
\end{figure}

Figure~\ref{Figures:reco} shows distributions of muon-related quantities
for low momentum muon tracks from proton--proton collisions collected during 2010.
The solid red and dotted black distributions correspond to the aligned and
design (no alignment corrections) muon chamber geometries, respectively.
From the top-left figure one can see that the alignment corrections induce an
improvement in normalized $\chi^2$ for global tracks.
The top-right plot shows the difference in track curvature ($q/\pt$) measured
for the same muon when it is reconstructed as a global muon (including
tracker information and therefore dominated by it) and as a standalone muon,
which includes only muon chamber hits. This difference is indicative of the
curvature resolution of the muon spectrometer (assuming the tracker resolution
to be much better for low momentum muons).
The bottom plots show an improvement in the dimuon mass resolution in the
Z region.
At least 1 standalone muon is required to see an improvement, since the invariant mass resolution for pairs of global muons at these energies is controlled by the tracker information.

\begin{figure}[htbp]
\begin{center}
\includegraphics[width=0.75\textwidth]{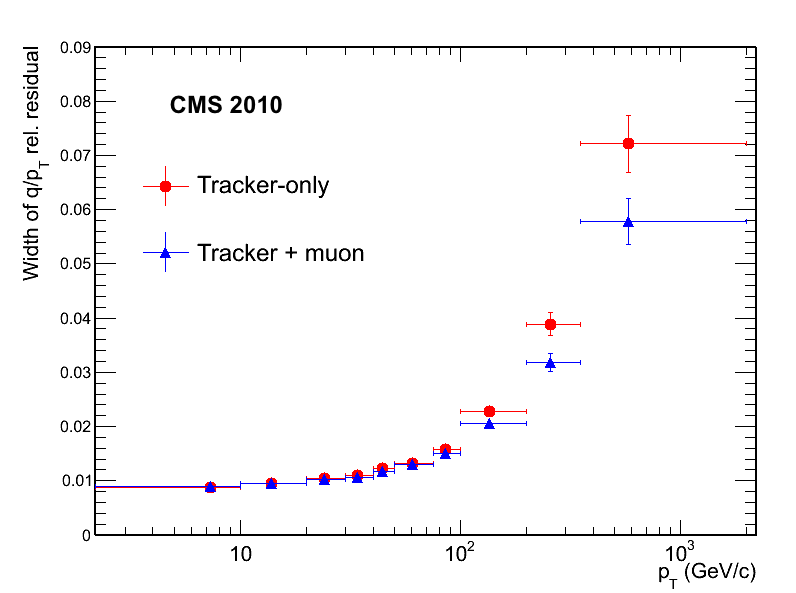}
\caption{Widths of distributions of the difference between the $q/\pt$ values for cosmic tracks reconstructed in the top and bottom parts of the apparatus.
Both the central tracker-only and
the central tracker plus aligned muon chamber cases are shown. The widths were obtained from Gaussian fits.}
\label{Figures:splitCosm}
\end{center}
\end{figure}

To see the effect of the alignment on highly energetic muons, cosmic-ray muons must be used, because there were very few muon tracks
above 200\GeVc from proton--proton collisions in 2010. Cosmic muons collected in 2010
traversing CMS from top to bottom are split into a ``top'' and a ``bottom''
leg, and the momentum resolution is inferred from the difference in momentum
measured for each leg separately. Figure~\ref{Figures:splitCosm} shows the
$q/\pt$ resolution as a function of muon \pt for barrel muons reconstructed
with the tracker plus the aligned muon geometry.
For this figure,
muons are reconstructed by a special algorithm that discards muon hits originating from electromagnetic
showers induced by muon bremsstrahlung, as described in detail in Ref.~\cite{POG-paper}.
Muons reconstructed by using only the tracker are also shown for comparison.
It can be seen that the aligned muon geometry improves
the momentum measurement for values above 200\GeVc, where muons reconstructed by
using muon chamber hits have a better resolution than tracker-only muons.

\section{Data quality monitoring}
\label{sec:DQM}

The primary goal of CMS data quality monitoring (DQM)~\cite{B_DQMFramework1} is to maximize the amount of high-quality data by detecting problems as early as possible.
The DQM for the muon detectors focuses on monitoring the condition and efficiency of the detectors.
Good efficiency is indicated by a uniform distribution of hits across the detectors (allowing for acceptance) and a low rate of readout errors.
Additional monitoring plots are available to investigate issues more deeply.

The DQM checks are made online and offline.
The online checks provide fast response to obvious problems---unpowered regions, problems with data acquisition, or improper calibrations---and a first evaluation of the suitability of the data.
The offline checks use the information available from full event reconstruction with calibration and different trigger paths to provide a more detailed evaluation of the data.
Offline DQM is used to certify the quality of reconstructed data and validate calibration results, software releases, and simulated data.

\subsection{Online monitoring}\label{ssec:onlineMonitoring}

In 2010, operation of the CMS detector moved from constant reliance on sub-detector experts to become the responsibility of a central shift crew.
Sub-detector experts are on call for problems not readily solvable by the shift crew.
With respect to the muon systems, the shift crew is responsible for ramping up the chamber high voltage (HV) after beam injection and bringing down the HV when beams are dumped.
Online monitoring of data quality is also part of the central shift crew duties.
A special stream of events is used to perform DQM operations online~\cite{B_HLT}.
The stream contains detector and trigger raw data, L1 and HLT summary results, in addition to HLT by-products essential for monitoring trigger algorithms.
Events are delivered to data quality monitoring applications at about 10\unit{HZ}.
Delivery speed strongly depends on the rate of event data processing.
There is no event sorting or handling, and no guarantee that parallel applications receive the same events.

The raw-data stream is checked for readout errors, the occupancy of detector channels, and the rates of muon trigger primitives.
Catching readout errors is a vital and unique part of the online monitoring.
For example, format errors in the data stream can indicate that the detector readout is out of synchronization with the L1 trigger, or that a rapid sequence of trigger accepts has overflowed the readout buffers.
These and a multitude of other possible errors are monitored during running.
The readout system can tolerate and recover from periodic errors, but persistent errors indicate a problem that needs expert attention.

There are differences in specifics among the 3 subdetector types of the muon system, but the basics are the same.
At this stage, data are declared good if a large fraction (at least 95\%) of the detector channels register hits.
A small number of readout errors is tolerated.
If there are problems, the shifter can consult more detailed monitoring information to diagnose the problem.
Detector experts are called to resolve potential problems as soon as possible.

Histograms aggregating information from the detectors and procedures for using them to evaluate the muon system operational integrity have been developed for the shift crew.

\subsection{Offline monitoring}\label{SS_OfflineDQM}

The offline DQM runs as part of the reconstruction process at the tier-0 computing center; of the re-reconstruction at the tier-1 centers; and of the validation of software releases, simulated data, and alignment and calibration results.
Here we focus specifically on the DQM performed on newly acquired data.
Offline DQM of reconstructed data is carried out first by offline DQM shifters, then by members of the subdetector groups, normally with a latency period of a few days.

Standard data certification checks for readout errors and measures the detector occupancies again, as done in the online checks, but in greater detail.
We can now look at reconstructed data from the accepted minimum bias stream, single-muon trigger stream, or dimuon trigger stream.
Except at the very low luminosities of early LHC operation, the minimum bias event rate is small, often yielding poor statistics for judging detector performance.

Additional sources of information are available to aid the experts in the final pronouncement on data quality.
These include logbooks for the subsystems, the insights of field managers and hardware experts, and information from full event reconstruction.

In addition, the DT and CSC systems monitor segment related information.
Monitored quantities include the following:
\begin{itemize}
\item the distribution across the detector of the rate of reconstructed segments;
\item the distribution of residuals between hits and reconstructed segments (mean and RMS);
\item the efficiency of matching hits to segments and/or matching segments to stand-alone tracks;
\item timing of hits and segments; 
\item gas gain and noise in the chambers. 
\end{itemize}

\begin{figure}[htbp]
  \begin{center}
  \includegraphics[width=15.5cm]{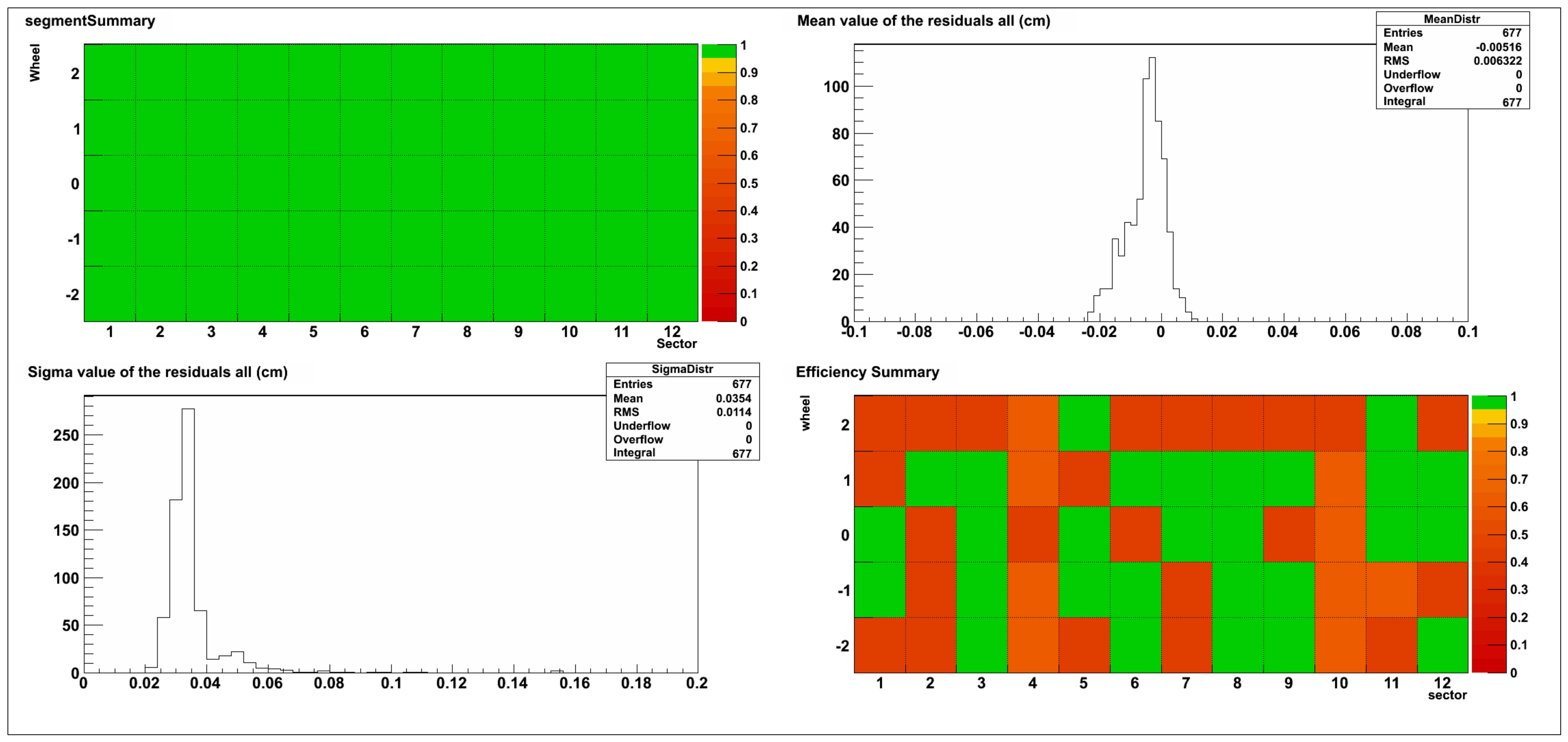}
    \caption{DQM GUI screen shot of a DT frame with DQM plots chosen to determine the quality of the data taken by the DTs for a single run taken during the last period of data taking in 2010.
The plots are as follows: a summary of segment occupancy by wheel and sector (upper left), the means (upper right) and sigmas (lower left) of the segment-hit residuals distributions for each chamber, and the segment efficiency summary (lower right).}
    \label{fig:DToffline}
  \end{center}
\end{figure}

As an example, the overview in Fig.~\ref{fig:DToffline} shows the DT plots to be checked by the off--line shifter for data taken at the end of October 2010.

\subsection{Quality of muon system operation in 2010}

The muon system performed very well during the 2010 data-taking period.
Generally, more than 98\% of the channels were operational, and all the systems produced high-quality data for nearly all collision runs.
These results are summarized in Table~\ref{tab:performance}.
While the problems that affected data quality varied,
the muon systems were responsible for the loss of less than 0.5\% of CMS running time in 2010. The overall loss of CMS running time from all sources was 9\%.

\begin{table}
\begin{center}
\topcaption{Summary of the operational performance of the muon systems during 2010.}
\label{tab:performance}
\begin{tabular}{l | c | c | c |}
\cline{2-4}
& DT & CSC & RPC  \\ \cline{2-4} \hline \hline
\multicolumn{1}{| l |} {Fraction of operating subsystem channels at end of} & 99.8\% & 98.5\% & 98.8\% \\
\multicolumn{1}{| l |} {2010 pp collision running}              &              &               &               \\ \hline
\multicolumn{1}{| l |} {Fraction of recorded pp collision data for which} & 99.3\% & 99.1\% & 99.7\% \\
\multicolumn{1}{| l |} {subsystem was operating}                                       &              &            &        \\ \hline
\multicolumn{1}{| l |} {Fraction of recorded pp collision data certified good} & 99.3\% & 99.0\% & 99.0\% \\
\hline
\end{tabular}
\end{center}
\end{table}

\section{Conclusions}

We studied the performance of the CMS muon system by using a data sample corresponding to an integrated luminosity of
40\pbinv of proton--proton collisions at $\sqrt{s} = 7$\TeV in 2010 at the LHC.
The sustained operation of all 3 subsystems (cathode strip chambers, drift tubes, and resistive plate chambers) at high efficiency demonstrated their robustness and reliability.
The studies reported here show that the performance of the muon system meets the design parameters and
is well reproduced by Monte Carlo simulation:

-- The timing and synchronization of the system exceed the design
specifications, achieving better than 99\% in-time triggering.\\
-- Precise calibration procedures have been established;
they are particularly crucial for the DT system, where they are required for optimal local reconstruction.\\
-- The CSC and DT systems provide fast trigger segments for the muon level-1
trigger, which closely match fully reconstructed segments offline
with an efficiency above 96\%.\\
-- The efficiency for reconstructing local segments and hits is on average
about 95\% per muon station in the CSC and DT systems.
The detection efficiency of the RPC layers is typically between 95 and 98\%. \\
-- Spatial resolutions for hits in all muon subdetector systems reach or
exceed the design requirements in the Muon TDR~\cite{MUON-TDR} .\\
-- Offline reconstructed times for hits and segments in the CSC and DT
systems have resolutions of 3\unit{ns} or better.\\
-- Backgrounds were measured and found to be controllable and
largely in accordance with simulation.\\
-- Alignment results using both hardware and reconstructed track based
techniques are compatible and provide the appropriate precision in
detector positions for the reconstruction of muon tracks~\cite{POG-paper}.

Data from proton--proton collisions at the LHC have enabled us to perform studies of muon system performance under operational conditions, and confirm that this performance matches or exceeds the design specifications.

\section*{Acknowledgements}
\hyphenation{Bundes-ministerium Forschungs-gemeinschaft Forschungs-zentren} We congratulate our colleagues in the CERN accelerator departments for the excellent performance of the LHC and thank the technical and administrative staffs at CERN and at other CMS institutes for their contributions to the success of the CMS effort. In addition, we gratefully acknowledge the computing centres and personnel of the Worldwide LHC Computing Grid for delivering so effectively the computing infrastructure essential to our analyses. Finally, we acknowledge the enduring support for the construction and operation of the LHC and the CMS detector provided by the following funding agencies: the Austrian Federal Ministry of Science and Research and the Austrian Science Fund; the Belgian Fonds de la Recherche Scientifique, and Fonds voor Wetenschappelijk Onderzoek; the Brazilian Funding Agencies (CNPq, CAPES, FAPERJ, and FAPESP); the Bulgarian Ministry of Education and Science; CERN; the Chinese Academy of Sciences, Ministry of Science and Technology, and National Natural Science Foundation of China; the Colombian Funding Agency (COLCIENCIAS); the Croatian Ministry of Science, Education and Sport; the Research Promotion Foundation, Cyprus; the Ministry of Education and Research, Recurrent financing contract SF0690030s09 and European Regional Development Fund, Estonia; the Academy of Finland, Finnish Ministry of Education and Culture, and Helsinki Institute of Physics; the Institut National de Physique Nucl\'eaire et de Physique des Particules~/~CNRS, and Commissariat \`a l'\'Energie Atomique et aux \'Energies Alternatives~/~CEA, France; the Bundesministerium f\"ur Bildung und Forschung, Deutsche Forschungsgemeinschaft, and Helmholtz-Gemeinschaft Deutscher Forschungszentren, Germany; the General Secretariat for Research and Technology, Greece; the National Scientific Research Foundation, and National Office for Research and Technology, Hungary; the Department of Atomic Energy and the Department of Science and Technology, India; the Institute for Studies in Theoretical Physics and Mathematics, Iran; the Science Foundation, Ireland; the Istituto Nazionale di Fisica Nucleare, Italy; the Korean Ministry of Education, Science and Technology and the World Class University program of NRF, Republic of Korea; the Lithuanian Academy of Sciences; the Mexican Funding Agencies (CINVESTAV, CONACYT, SEP, and UASLP-FAI); the Ministry of Science and Innovation, New Zealand; the Pakistan Atomic Energy Commission; the Ministry of Science and Higher Education and the National Science Centre, Poland; the Funda\c{c}\~ao para a Ci\^encia e a Tecnologia, Portugal; JINR, Dubna; the Ministry of Education and Science of the Russian Federation, the Federal Agency of Atomic Energy of the Russian Federation, Russian Academy of Sciences, and the Russian Foundation for Basic Research; the Ministry of Education, Science and Technological Development of Serbia; the Secretar\'{\i}a de Estado de Investigaci\'on, Desarrollo e Innovaci\'on and Programa Consolider-Ingenio 2010, Spain; the Swiss Funding Agencies (ETH Board, ETH Zurich, PSI, SNF, UniZH, Canton Zurich, and SER); the National Science Council, Taipei; the Thailand Center of Excellence in Physics, the Institute for the Promotion of Teaching Science and Technology of Thailand, Special Task Force for Activating Research and the National Science and Technology Development Agency of Thailand; the Scientific and Technical Research Council of Turkey, and Turkish Atomic Energy Authority; the Science and Technology Facilities Council, UK; the US Department of Energy, and the US National Science Foundation.

Individuals have received support from the Marie-Curie programme and the European Research Council and EPLANET (European Union); the Leventis Foundation; the A. P. Sloan Foundation; the Alexander von Humboldt Foundation; the Belgian Federal Science Policy Office; the Fonds pour la Formation \`a la Recherche dans l'Industrie et dans l'Agriculture (FRIA-Belgium); the Agentschap voor Innovatie door Wetenschap en Technologie (IWT-Belgium); the Ministry of Education, Youth and Sports (MEYS) of Czech Republic; the Council of Science and Industrial Research, India; the Compagnia di San Paolo (Torino); the HOMING PLUS programme of Foundation for Polish Science, cofinanced by EU, Regional Development Fund; and the Thalis and Aristeia programmes cofinanced by EU-ESF and the Greek NSRF.

\newpage
\appendix
\section{Electronics performance}
\label{electronics}

In a large experiment, like CMS, that is constructed in a large underground cavern, most of the detector is inaccessible except rarely during long shutdown periods.
Furthermore,  CMS is built in multiple layers and opening it for repairs or modifications takes considerable time and effort.
Moreover, collision running quickly turns the experimental cavern into a highly radioactive environment, which increases the difficulty of servicing detector components.
(In fact, the CMS detector was nearly inaccessible between March 2010 and March 2013 when it was opened again.)
Therefore, each component of CMS must be robust  and have a low failure rate.
The collision running of 2010 presents an opportunity to assess the ability of the electronics to function reliably over many months.

Overall, the reliability of the electronic components was outstanding.
Table~\ref{tab:failure} shows the number of failures that occurred in the on-chamber electronics (nearly all were inaccessible).
For both the DTs and the CSCs, the board failure rate was less than 1\% for all on-chamber electronics over the 2010 running period.

   \begin{table}[hptb]
   \begin{center}
         \topcaption{Failure rates for the various types of on-chamber muon system electronics. The number of failed boards during the 2010 running period is indicated followed by the percentage of the particular type of electronics this number represents. The acronyms identifying the boards are defined in the corresponding sections of the text.}
         \label{tab:failure}
         \begin{tabular} {|l|c|c|c|}
         \hline
 Type           &   DT &  CSC    &RPC    \\ \hline \hline
  Front-end boards     &  Superlayer 1 (0.15\%)        & CFEB 8 (0.3\%) &      0         \\
                                             &  Front-end boards (FEB) 3  (0.03\%)        & AFEB 0 (0.0\%) &                \\
                                             &  Trigger boards (TRB) 1  (0.07\%)      & ALCT 1 (0.2\%) &                \\  \hline
  Low voltage (LV)          &   0                     &  3 LV cables (0.6\%)        & 5 chambers (0.5\%)   \\  \hline
 High voltage (HV)         &   26 channels  (0.01\%) &  5 channels (0.5\%)  & 39 channels (5.6\%)   \\  \hline
          \end{tabular}
   \end{center}
\end{table}

Although the on-chamber electronics failure rate was quite small, this does not mean that maintenance was unnecessary.
Most of the off-chamber electronics was accessible during machine access periods and boards could be replaced or repaired.

The DTs had a variety of failures in accessible modules and all the faulty components were replaced.
There were 11 failures in the low-voltage (LV) connectors, which were fixed by reseating the connector.
There were 15 high-voltage (HV) power supply failures (5.0\%).
Among the readout electronics there were 3 DAQ board failures.
There were the following failures in the trigger sector: 1 trigger selector collector (TSC) optical transmission (optoTX) mezzanine board (1.7\%), 3 TSC input mezzanine boards (1.2\%),  and 2 TSC optical receiver (optoRX) boards (2.4\%).
In addition, 1 crate controller failed (10.0\%).

The major maintenance issue for the CSCs was the high rate of reloading firmware.
The EPROMs needed to be reloaded often; hardly a week went by without a firmware upload.
In the CSC system, DAQ motherboards (DMB) and trigger motherboards (TMB) had to be replaced at a rate of roughly 1 every 2 months.
Another challenging problem was a drop in the 3.3\unit{V} input to the TMB boards.
This was finally traced to corrosion on the TMB fuses, which were replaced with ones with gold contacts.

The RPC maintenance problems during 2010 were relatively minor and easily resolved.
An HV supply trip occurred about once a month.
In the LV system, the failure rates were  3 CAEN A3009 power supply boards (3.2\%), 1  A3016 power supply board (1.6\%), and 1  A1676 branch controller board (5.9\%).
There were the following failures in the readout boards:  11 master link boards (2.2\%), 3 slave link boards (1.0\%), and 4 control boards (3.3\%).
These boards were all replaced during local stops and did not cause major interruptions during data taking.

The DT, CSC, and RPC systems were responsible for 1.1\%, 5\%, and 1.2\% of the total CMS downtime in 2010, respectively.

Most off-chamber maintenance was accomplished during the frequent short accesses associated with LHC problems.
Based on the 2010 running experience, we expect that the increase of luminosity for future running should not cause any notable rise in the failure rates for any muon system.

\section{Detector simulation}
\label{simulation}

Simulation of the CMS detector is crucial in understanding the features, behavior, and appearance of real and hypothesized physics events in the component subdetectors. It was of course also critical in designing and optimizing the detectors, both in geometry and operating characteristics, before construction.

The CMS muon system consists of 3 different technologies: cathode strip chambers (CSC), drift chambers (DT), and resistive plate chambers (RPC).
In each case a muon passing through a sensitive gas volume of the detector causes ionization of the gas. The free electrons drift towards an anode where a signal can be read out, and the positive ions drift more slowly towards a cathode. In the neighborhood of an anode wire, where the electric field reaches high values, gas amplification occurs and an image pulse is induced on the cathode plane. If the cathode is divided into strips, the location of the pulse, and hence the position of the ionizing particle, can be determined from the relative pulse heights on the strips.
The CSCs and RPCs have relatively short drift distances (and hence short drift times) whereas the DTs have longer drift distances (and longer drift times).  The DTs generate a precise hit location using precise measurement of the drift times, whereas the CSCs and RPCs use fine-grained cathode strips to measure the hit location.

The CMS experiment employs the \GEANTfour package v6.2~\cite{GEANT4} for the most precise and detailed simulation of detector operation and performance.
\GEANTfour uses a detailed model of the detector geometry and material composition to simulate the physics processes that occur as particles interact with the detector.
These include energy loss (and energy deposition in the detectors) and multiple scattering, the bending of charged tracks due to the magnetic field, and the production---and subsequent tracking---of secondary particles.
The output from this process is a collection of  ``\GEANTfour simhits'', for each subdetector, which provide information about the passage of each track through a sensitive detector volume.
For example, in a  subdetector sensitive to the passage of charged tracks, like all 3 types of detector in the muon system, a simhit contains the particle type, its energy, the position of entry to and exit from the detector volume, and the energy loss that occurred between the entry and exit points. In the muon system, the sensitive detector volume is a gas volume equivalent to a CSC layer,  a DT cell, or a RPC single-gap module (double-gap RPC modules are approximated with a single active volume, called a ``roll").

A further level of simulation is required to turn simhits into the quantities (\eg, charge, time, and position) that are recorded by the data acquisition system from the actual detectors.
Owing to the hardware differences between the component subdetectors, this level of simulation must be subdetector specific. CMS generically refers to this process as ``digitization'' since it typically involves conversion of simulated analog quantities in the detectors into digitized quantities representing those read out from the electronic channels of the real detector (although these too may be digitized samples of analog quantities.)
The output from digitization is a collection of ``digi'' objects.
These were originally designed to be types of a common CMS-wide data structure intermediate between raw data and reconstructed data, and optimized for use in the reconstruction.
Reconstruction is the process of interpreting the detector information in terms of particles interacting in those detectors; for example, in the muon detectors, the reconstruction includes the process of converting detected signals on electronic channels first to points in space, ``rechits'', at which a muon has crossed a detector, and then the construction of a momentum vector for a muon compatible with giving rise to those hits.

The overall output of the simulation of the muon system, for various physics event processes giving rise to  muons, will be collections of simhits, digis, rechits,  muon track segments (in chambers), and fully reconstructed muon tracks. These  incorporate  charges, times, positions, momenta, and any other quantities that are important to model, which can be compared with those from real data.

The comparison of real and simulated data allows us to monitor and improve the detector operating
characteristics; to define and examine the muon trigger; and to diagnose malfunctioning chambers, faulty firmware,  or  problems arising in software at any stage of the detector operation from hardware to reconstruction.
We can also model the effects of increasing luminosity (\eg, pileup, bunch structure, and dead time) and various potential backgrounds (including halo associated with the accelerated beams, cosmic rays, and effects induced by slow neutrons.)

\subsection {\GEANTfour-level simulation}
All 3 muon subdetectors provide a detailed and realistic description of the idealized geometry of the detectors, incorporated within the overall CMS geometrical description.
This also provides a detailed specification of the materials composing the detector, including---in addition to the materials of the detectors proper---shielding, cables, and electronics modules.
The ideal geometry model we use matches the hierarchical \GEANTfour model, and this allows the local coordinates of the chambers in the system to be related to their global coordinates in the overall global coordinate system of CMS.
Each chamber is typically considered to be a rigid body, with a geometrical specification given in terms of the physical dimensions, together with the global coordinates of a symmetry center, and a rotation matrix specifying its orientation in the global frame.
The process of ``alignment'' of the detector leads to corrections to these ideal values in order that the actual positions of the chambers are known in the global frame.
In either the ideal or real case, these geometrical specifications allow us to transform between the local and global coordinates.

It is important to note that not every aspect of a detector needs to be fully and precisely specified in this geometry.
For example, in the CSC case, the geometries of the cathode strips and anode wires are not specified at this level (\ie, in terms of the global coordinate system).
They are modeled locally only, within each chamber, and then transformation of their associated values to and from the global frame can be made as required using the overall geometry of the parent chamber.

\subsection {Digitization}
Each muon subdetector requires a specific simulation of the electronics and readout of that system. Each of the 3 digitizer packages starts from the \GEANTfour simhits in the detector and produces one or more collections of digis, representing the different quantities read out from the detector:
\begin{itemize}
\item For DTs, each digi consists of a TDC time associated with an individual drift cell;
\item For CSCs, 6 types of digis are created, corresponding to readout channels for strips, wire groups, and di-strip comparators, as well as for level-1 trigger local charged tracks (LCT) in the anode, cathode, and combined views;
\item For RPCs, a digi is a pair of values: the fired strip code and the corresponding bunch crossing.
\end{itemize}
These digis are then packed into raw data format and stored.  Subsequent reconstruction starts from the raw data format, just as in the case of real data.

\subsubsection{DT digitization}

The determination of the position of particles in the DT detectors is based on the measurement of the drift time of ionization electrons. An accurate modeling of the cell response, including the TDC measurement of the drift time information, is essential for a
reliable simulation of physics events. The simulation of the cell response consists of the computation of the drift time of the electrons produced by ionization.

The simulation must reproduce not only the average behavior of the cell as a function of the track parameters and of the magnetic field, but also the smearing of the drift times, which determines the cell resolution. This is achieved with a parameterization of the
drift cell behavior based on a detailed simulation with the \GARFIELD package~\cite{GARFIELD}.
The drift time is parameterized as a function of the distance of the simulated hit from the wire, the incidence angle of the muon, and the magnetic field components parallel and orthogonal to the wire.
This time is smeared according to a Gaussian-based distribution.

Hits from energetic delta rays and any other secondaries
(\eg, \EE from pair production) crossing the whole cell are handled in the same way as muons.
For the case of several hits in the same cell, the drift times are computed independently, but separate digis are created only if the difference in drift times is larger than the configurable dead time.
The $\delta$ rays produced in the gas are ignored, since they are already generated by \GARFIELD and their effect is included (statistically) in the parameterization.
Soft electrons stopping in the gas cannot be handled by the parameterization and are currently ignored. Most of these hits cover a very short path and are not expected to produce a detectable signal.

Finally, since the DTs are time measuring devices, the delays contributing to the TDC measurement must be accounted for.
The time-of-flight of the muons from the interaction point to the cell and the propagation time of the signal along the anode
wire are added to the simulated time.

The emulation of the DT local trigger electronics uses as input the
simulated signals produced as described above.

\subsubsection {CSC digitization}
The CSC digitization simulates the ionization of the CSC gas by generating a sequence of random steps along the line between a simhit entry and exit point, the end of each step corresponding to a collision between the ionizing particle and a gas molecule.
At each collision energy transfer can occur, and if the transfer exceeds the ionization threshold of the gas, ionization occurs and a $\delta$ electron is produced.
If of sufficient energy, such $\delta$ electrons can produce further ionization.
The energy losses and the number of collisions per cm of gas used in the generation are based on tables extracted from \GEANTfour, created within the framework of the photoabsorption ionization model (which in effect extends the usual Bethe--Bloch equation to situations like thin layers of gas, where it is not directly applicable).
This procedure results in a set of clusters of electrons distributed ``randomly'' across the gas gap, along the line defined by the simhit. Typically, about 100 free electrons are produced per gas gap.

These electrons then drift towards the anode plane, driven by the electric field, based on parameterizations of results from more detailed \GEANTfour simulation.
These provide the average drift times and drift distances along the wire direction (due to Lorentz drift in the magnetic field), and their variances, and hence specify Gaussian distributions from which random samplings can be made electron by electron.
The parameterizations are functions of 3 variables: the magnetic field and the distances to the wire in both the direction of the wire plane and the direction perpendicular to the plane.  Two sets of parameterizations were used, one for the ME1/1 chambers, which have smaller gas gaps and operate in high magnetic fields, and one for the others. The ME1/1 chambers have wires that are tilted at 29$^\circ$ to compensate for the relatively large Lorentz drift contribution, which tends to spread the collected charge; tilting the wires minimizes this dispersion. ME1/1 chambers have typical drift times of 20--40\unit{ns}, while the other chambers have drift times of 20--60\unit{ns}.

When an electron reaches a wire, the characteristic MWPC gas amplification is simulated: an avalanche of charge is created, according to a parameterized distribution, with factors to account for charge collection efficiency.
We also account for electron attachment loss during drift through the gas.
The charge on the wire is amplified and shaped into a signal on the readout channel of the corresponding wire group, according to the known specifications (delays, pulse-shaping, and amplification) of the actual hardware.
The charge on the wire also induces image charge on the cathode plane: charge is induced on the 5 nearest cathode strips, according to the Gatti distribution~\cite{Gatti}, which models the expected charge distribution shape.
Here too the charge on each strip is  amplified  and shaped into a signal on the strip readout channel according to the hardware specifications of the real electronics.
The overall signals, which are represented as histograms of charge, are superimposed for all drift electrons and for all simhits contributing to that channel.

 A ``wire digi'' contains 16~bits, each of which represents a 25\unit{ns} time bin.
 A bit is set for the corresponding time bin when the wire pulse exceeds a fixed threshold.
 To model random noise on the signal the simpler approach of applying, for each channel, a random Gaussian fluctuation to this threshold value is used.

The simulation of the cathode readout is more complex since it must take into account pedestal noise, which is correlated between switched capacitor array (SCA) time bins, and crosstalk between signal channels.
The CSC calibration procedure provides a covariance matrix for the time-correlated pedestal noise; this is of dimension $8 \times 8$, since there are 8 SCA time bins per channel.
The diagonal elements are taken from the measured widths of the pedestals in each time bin.
All these measured values are stored in the standard CMS conditions data database, and can be accessed according to a date, or run number, of validity.
A Cholesky decomposition of this covariance matrix then provides the appropriate matrix to apply to a vector of uncorrelated random pedestal noise values to give a vector of pedestal noise values with the appropriate covariance properties of the real CSC system.
Crosstalk between channels is simulated by likewise making use of channel-specific values that represent the crosstalk measured in the periodic calibration procedure of the real CSC detector, and stored in the conditions data databases.
Crosstalk typically results in a 10\% transfer of charge from one strip to a neighbor.
Since the crosstalk has a component proportional to the slope of the neighboring signal, it typically causes neighboring signals to peak earlier than the main signal. The cathode signal is split into 2 paths, with the trigger path resulting in ``comparator digis'', and the data path resulting in ``strip digis''.

The timings of the wire, strip, and comparator signals are tuned by first subtracting a time corresponding to the distance from the center of the readout element to the interaction point, divided by the speed of light.  The speed of signal propagation from the hit position to the readout is measured in data, and was taken into account.
Finally, we fine tune the centering of the timing using configurable constants for each ring of chambers.

The final step of the simulation involves modeling the level-1 trigger. The inputs are the wire and comparator digis, and the outputs are the 3 types of digi containing LCT information. Finally, we pack all the digis into the CSC raw data format.  Just as in the hardware, some zero suppression of the strip signals is applied, based on the
``pre-trigger'' (\ie, early stages of the trigger algorithm). In the simulation this occurs during the cathode LCT simulation.

\subsubsection{RPC digitization}

Each RPC has up to 96 readout strips.
The simulation of the real physical processes taking place inside the RPC is a complex and computer-intensive task.
Therefore the RPC digitization is mostly parameterized.
The digitization task is to assign 0, 1, or more digis to a hit and to simulate the detector noise.

The default digitization algorithm uses the following parameters
from a dedicated database:
\begin{itemize}
\item efficiency parameter for each readout strip, \ie, 96 values per roll;
\item noise rate parameter for each readout strip, \ie, 96 values per roll;
\item timing parameter for each roll, \ie, 1 value per roll;
\item cluster size distribution for each roll, \ie, 100 values per roll.
\end{itemize}

If the roll is hit by an ionizing particle, the digitizing algorithm will assign a digi to it with a probability defined by the efficiency parameter.
The number of adjacent fired strips is calculated using an empirical cluster size distribution for each chamber.
The impact point position is used to decide the fired strip coordinates.
The signal propagation time and the timing parameter are used to decide the bunch crossing. Digis due to noise are simulated according to a Poisson distribution. Inter-roll crosstalk is negligible.
The actual detectors have some dead or masked strips, which are treated in a very simple way by the digitization algorithm: the corresponding efficiency and noise rate parameters are set to 0 in the database.

The simulation parameters are updated regularly. Presently, 3 sets of parameters values are used:
\begin{itemize}
\item the ``ideal'' conditions (\eg, efficiency 95\%);
\item the parameters estimated using the cosmic-ray data;
\item the parameters estimated by the proton--proton collision data.
\end{itemize}

While we lack the amount of data necessary to estimate the efficiency of each strip,  the overall roll efficiency is well estimated. Thus, the efficiency averaged over a roll is assigned to each strip in the roll.
The efficiency parameters are relatively constant over long periods of time, and are updated a few times per year.
The noisy and non-operational strips, however, may change run by run and are constantly monitored and updated in the database.
Every database update is followed by a validation procedure.
All of the simulation parameters are estimated using dedicated detector performance analysis.
The same analysis is used to validate the simulation and compare it to data, with which it is in good agreement.

\bibliography{auto_generated}   

\cleardoublepage \appendix\section{The CMS Collaboration \label{app:collab}}\begin{sloppypar}\hyphenpenalty=5000\widowpenalty=500\clubpenalty=5000\textbf{Yerevan Physics Institute,  Yerevan,  Armenia}\\*[0pt]
S.~Chatrchyan, V.~Khachatryan, A.M.~Sirunyan, A.~Tumasyan
\vskip\cmsinstskip
\textbf{Institut f\"{u}r Hochenergiephysik der OeAW,  Wien,  Austria}\\*[0pt]
W.~Adam, E.~Aguilo, T.~Bergauer, M.~Dragicevic, J.~Er\"{o}, C.~Fabjan\cmsAuthorMark{1}, M.~Friedl, R.~Fr\"{u}hwirth\cmsAuthorMark{1}, V.M.~Ghete, N.~H\"{o}rmann, J.~Hrubec, M.~Jeitler\cmsAuthorMark{1}, W.~Kiesenhofer, V.~Kn\"{u}nz, M.~Krammer\cmsAuthorMark{1}, I.~Kr\"{a}tschmer, D.~Liko, I.~Mikulec, M.~Pernicka$^{\textrm{\dag}}$, D.~Rabady\cmsAuthorMark{2}, B.~Rahbaran, C.~Rohringer, H.~Rohringer, R.~Sch\"{o}fbeck, J.~Strauss, A.~Taurok, W.~Waltenberger, C.-E.~Wulz\cmsAuthorMark{1}
% ...
\vskip\cmsinstskip
\dag:~Deceased\\
1:~~Also at Vienna University of Technology, Vienna, Austria\\
2:~~Also at CERN, European Organization for Nuclear Research, Geneva, Switzerland\\
% ...

\end{sloppypar}
\end{document}